\documentclass[twocolumn]{aastex6}

\usepackage{xspace}
\usepackage{graphicx}
\usepackage{natbib}
\usepackage{amssymb}

\newcommand{\gbonraoblurb}{The Green Bank Observatory and National Radio Astronomy Observatory are
facilities of the National Science Foundation operated under cooperative agreement by Associated Universities, Inc.}
\newcommand{\hide}[1]{}

\newcommand{\gl}{\ensuremath{\ell}\xspace}
\newcommand{\gb}{\ensuremath{{\it b}}\xspace}
\newcommand{\absb}{\ensuremath{\vert\,\gb\,\vert}\xspace}
\newcommand{\absl}{\ensuremath{\vert\,\gl\,\vert}\xspace}

\newcommand{\ra}{\ensuremath{\alpha}\xspace}
\newcommand{\dec}{\ensuremath{\delta}\xspace}

\newcommand{\microns}{\ensuremath{\,\mu{\rm m}}\xspace}

\newcommand{\kpc}{\ensuremath{\,{\rm kpc}}\xspace}

\newcommand{\ghz}{\ensuremath{\,{\rm GHz}}\xspace}

\newcommand{\degree}{\ensuremath{\,^\circ}\xspace}

\newcommand{\microjy}{\ensuremath{{\rm \,\mu Jy}}\xspace}

\newcommand{\msun}{\ensuremath{\,M_\odot}\xspace}     
 
\newcommand{\hii}{{\rm H\,{\footnotesize II}}\xspace}

\shorttitle{Census of Galactic \hii\ Regions}
\shortauthors{Armentrout et al.}

\begin{document}

\title{A VLA Census of the Galactic \hii\ Region Population}

\author{W.~P.~Armentrout\altaffilmark{1,2,3}}
\author{L.~D.~Anderson\altaffilmark{2,3,1}}
\author{Trey~V.~Wenger\altaffilmark{4,5,6}}
\author{Dana~S.~Balser\altaffilmark{5}}
\author{T.~M.~Bania\altaffilmark{7}}

\altaffiltext{1}{Green Bank Observatory, P.O. Box 2, Green Bank, WV 24944, USA}
\altaffiltext{2}{Department of Physics and Astronomy, West Virginia
  University, Morgantown, West Virginia 26505, USA}
\altaffiltext{3}{Center for Gravitational Waves and Cosmology, West Virginia University, Chestnut Ridge Research Building, Morgantown, WV 26505, USA}
\altaffiltext{4}{Astronomy Department, University of Virginia, P.O. Box 400325, Charlottesville, VA 22904-4325, USA}
\altaffiltext{5}{National Radio Astronomy Observatory, 520 Edgemont Road, Charlottesville, VA 22903-2475, USA}
\altaffiltext{6}{Dominion Radio Astrophysical Observatory, Herzberg Astronomy and Astrophysics Research Centre, National Research Council, P.O. Box 248, Penticton, BC V2A 6J9, Canada}
\altaffiltext{7}{Institute for Astrophysical Research, Department of Astronomy, Boston University, 725 Commonwealth Avenue, Boston, MA 02215, USA}

\begin{abstract}
The Milky Way contains thousands of \hii\ region candidates identified by their characteristic mid-infrared morphology, but lacking detections of ionized gas tracers such as radio continuum or radio recombination line emission. These targets thus remain unconfirmed as \hii\ regions. With only $\sim$2500 confirmed \hii\ regions in the Milky Way, Galactic surveys are deficient by several thousand nebulae when compared to external galaxies with similar star formation rates. Using sensitive 9 GHz radio continuum observations with the Karl G. Jansky Very Large Array (VLA), we explore a sample of \hii\ region candidates in order to set observational limits on the actual total population of Galactic \hii\ regions. We target all infrared-identified ``radio quiet" sources from the \textit{WISE} Catalog of Galactic \hii\ regions between $245^{\circ}\geq\gl\geq90^{\circ}$ with infrared diameters less than 80$^{\prime\prime}$. We detect radio continuum emission from 50\% of the targeted \hii\ region candidates, providing strong evidence that most of the radio quiet candidates are bona fide \hii\ regions. We measure the peak and integrated radio flux densities and compare the inferred Lyman continuum fluxes using models of OB-stars. We conclude that stars of approximately spectral type B2 and earlier are able to create \hii\ regions with similar infrared and radio continuum morphologies as the more luminous \hii\ regions created by O-stars. From our 50\% detection rate of ``radio quiet" sources, we set a lower limit of $\sim$7000 for the \hii\ region population of the Galaxy. Thus the vast majority of the Milky Way's \hii\ regions remain to be discovered.
\end{abstract}

\keywords{Galaxy: structure -- ISM: \hii\ regions -- Radio: surveys}


\section{Introduction\label{sec:intro}}

High-mass stars are the archetypical tracers of spiral structure across the Galactic disk. \hii\ regions surround OB-type stars and are among the most luminous objects in the Galaxy at mid-infrared through radio wavelengths. With relatively short lifetimes of $\sim$tens of millions of years, these OB-type stars are zero-age objects compared to the age of the Milky Way. Their \hii\ regions thus reveal the locations where Galactic high-mass stars have formed most recently. The \hii\ region chemical abundances indicate the present state of the interstellar medium (ISM) and reveal the elemental enrichment caused by the nuclear processing of many stellar generations \citep[e.g.,][]{balser15}. Despite considerable effort, the census of Galactic \hii\ regions remains incomplete \citep[e.g.,][]{anderson15b}.


\subsection{H{\footnotesize II} \normalsize{Region Surveys To Date}}

Using the Palomar optical survey plates, \citet{sharpless53} conducts the first Galactic \hii\ region survey. In this study, they detect 142 ``emission nebulae" between Galactic longitudes $105^{\circ}\geq\gl\geq-45^{\circ}$. As with most Galactic plane surveys, the Sharpless regions are constrained to within a few degrees of the Galactic plane. Relying on optical data, these surveys are also severely sensitivity-limited due to interstellar extinction.

The discovery of the H109$\alpha$ radio recombination line (RRL) by \citet{hoglund65} brings spectral line studies of  \hii\ regions into the radio regime. Because the ISM is optically thin at centimeter wavelengths, this enables spectroscopic studies of \hii\ regions across the entire Galactic disk. \citet{lockman89} and \citet{caswell87} conduct \hii\ region RRL surveys in the northern and southern skies, respectively, but their detections are biased toward the largest and brightest \hii\ region complexes.

The goal of the \hii\ Region Discovery Survey \citep[HRDS,][]{bania10, anderson11} is to detect a new population of fainter Galactic \hii\ regions. The HRDS selects its targets based on spatially coincident 24\,\micron\ \citep[{\it Spitzer} MIPSGAL;][]{carey09} and $\sim$20 cm continuum \citep[VGPS and NVSS;][]{stil06, condon98} emission in the range $67^{\circ}\geq\gl\geq-17^{\circ}$, $\absb\/\leq1^{\circ}$. The 24\,\micron\ emission traces hot, small dust grains, a signature of high-mass stars at \hii\ regions' cores. The $\sim$20 cm continuum emission traces thermal bremsstrahlung radiation from the interaction of free electrons and ions in the \hii\ region plasma. This selection criterion proves to be extremely successful: the HRDS detected RRL emission from 95\% of all targets. We find that nearly all of the infrared-identified candidates are confirmed as \hii\ regions \textit{when we also detect spatially coincident radio continuum emission.} 

\begin{figure*}[htb!]
\centering
\includegraphics[width=\textwidth]{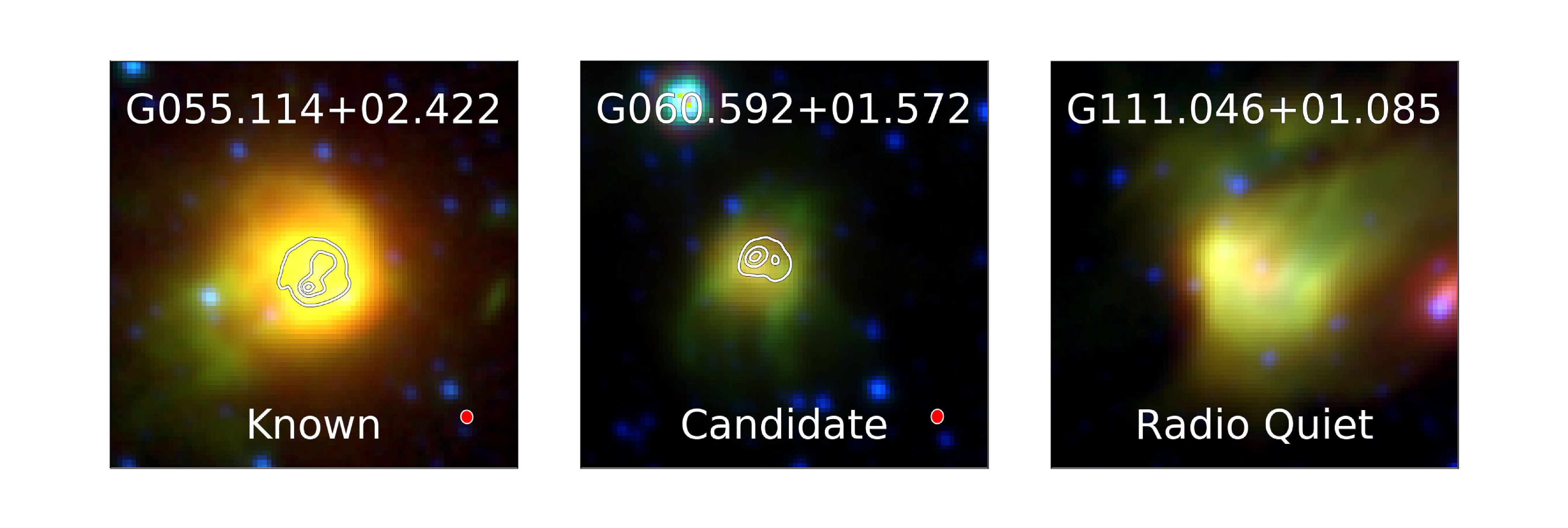}
\caption{Infrared and radio morphologies of known and candidate \hii\ regions. The infrared morphologies of known \hii\ regions and our faint \hii\ region candidates are very similar, though their continuum emission intensities can differ by several orders of magnitude. From left to right, we show a known \hii\ region, a candidate \hii\ region with detected continuum emission, and a ``radio quiet" candidate \hii\ region. Infrared \textit{WISE} bands 
W2 (4.6 \microns\/), W3 (12 \microns\/), and W4 (22 \microns\/) are represented by blue, green, and red respectively. Very Large Array (VLA) continuum contours from 4 minute integrations at X-band in D configuration are overplotted, with contours at 33\%, 66\%, and 90\% of the peak continuum flux. The VLA beam is shown in red in the bottom right corner. Snapshots are 4$^{\prime}$ on a side. One example of an observed source by the VLA through this work is shown in Figure~\ref{fig:contSnapshotsHowManyEx}, while the full figure set is shown in the Appendix in Figure~\ref{fig:contSnapshotsHowMany}.}\label{fig:contSnapshotEx}
\end{figure*}

Even with this high detection rate, there remain many more undiscovered \hii\ regions throughout the Galaxy. This is partly because the \textit{Spitzer} legacy surveys used for HRDS target selection are limited to within $\absb\/\lesssim1^{\circ}$ with $\absl\/\lesssim65^{\circ}$. The all-sky \textit{Wide-Field Infrared Survey Explorer (WISE)} satellite eliminates these restrictions. \citet{anderson14} identifies thousands of additional \hii\ region candidates in the \textit{WISE} Catalog of Galactic \hii\ Regions, extending the \hii\ region candidate sample to cover all Galactic longitudes with a latitude extent of $\pm8\degree$. \textit{WISE} is comparable in wavelength coverage with {\it Spitzer} MIPSGAL (\textit{WISE}: 22\,\micron, \textit{Spitzer}: 24\,\micron), though \textit{WISE} has slightly lower resolution and sensitivity (\textit{WISE}: $12\arcsec$ and 6\,mJy, \textit{Spitzer}: $6\arcsec$ and 1.3\,mJy). \citet{anderson14} show that \textit{WISE} at 22 \micron\ can in principle detect infrared emission from \hii\ regions two orders of magnitude fainter than that expected for an \hii\ region created by a single B0 star at 20 kpc in the Galactic plane. This resolution and sensitivity should be sufficient to detect the mid-infrared emission from {\it all} classical \hii\ regions across the entire Galactic disk (i.e., evolved and not ultra- or hyper-compact). The \textit{WISE} Catalog (V2.2) contains 8412 objects classified as either known \hii\ regions with ionized gas spectroscopic detections (2210 objects) or objects with the characteristic infrared-identified morphology of previously known \hii\ regions (6202 objects). Here, we call all 6202 infrared-identified objects ``candidates," while in previous HRDS papers the term ``candidate" is reserved for objects with both coincident infrared and radio continuum emission detections. \textit{The \textit{WISE} Catalog population shows that all \hii\ regions and candidates appear to have the same characteristic infrared morphology,} with quite different radio continuum emission intensities (Figure~\ref{fig:contSnapshotEx}).

To date, the HRDS has discovered 887 \hii\ regions, bringing the total number of known Galactic \hii\ regions over the survey range to 1603, more than doubling the previous census \citep{bania10, anderson11, anderson14, anderson15b, anderson18}. The Southern \hii\ Region Discovery Survey (SHRDS) extends this work into the Southern hemisphere with the Australia Telescope Compact Array, detecting 295 new \hii\ regions to date with an additional $\sim$100-200 more detections expected in the next data release \citep{wenger19}. Across the entire Galaxy, the full \textit{WISE} Catalog (V2.2) census now includes 2210 known \hii\ regions.


\subsection{The Missing Galactic H{\footnotesize II} \normalsize{Regions}}

Extragalactic observations of \hii\ regions suggest that Galactic \hii\ region surveys are still incomplete. For instance, the galaxy NGC 628 (M74) has 4285 \hii\ regions identified through H$\alpha$ observations \citep{rousseau17}. This archetypical Grand Design spiral galaxy has a star formation rate of 0.68 \msun\ yr$^{-1}$ \citep{kennicutt11}. While NGC 628 is producing stars at a slower rate than the Milky Way \citep[$\sim$1.9 \msun\ yr$^{-1}$ for the Milky Way,][]{chomiuk11}, it has approximately double the number of identified \hii\ regions. Additionally, many faint or clustered \hii\ regions in NGC 628 would be confused and indistinguishable from each other at the survey resolution of 35 pc in \citet{rousseau17}. The tally of \hii\ regions in NGC 628 could thus be even greater, dwarfing the number of known \hii\ regions in our Galaxy. Where are the missing Milky Way \hii\ regions?

The \hii\ region candidates in the \textit{WISE} Catalog represent some of these missing nebulae. Of the $\sim$6000 Galactic \hii\ region candidates, over 2000 have coincident radio continuum emission while 4000 are only apparent in the infrared, with no detected radio continuum emission (hereafter ``radio quiet''). The 2000 candidates with coincident radio continuum emission are \textit{very likely} to be \hii\ regions, since previous HRDS surveys confirmed that 95\% of our infrared-identified candidates with detected coincident radio continuum emission were \hii\ regions. We simply have not yet searched for RRL emission from these candidates to confirm that they are indeed \hii\ regions. The 4000 radio quiet objects could also represent a significant population of yet to be identified \hii\ regions. Radio quiet candidates do not necessarily lack the radio continuum signature of \hii\ regions, but may have radio continuum emission fainter than the sensitivity limit of current Galactic plane surveys.

Higher sensitivity observations can determine whether there is associated radio continuum emission from these radio quiet candidates. In \citet{armentrout17}, we target 65 \hii\ region candidates with the National Science Foundation's Karl G. Jansky Very Large Array (VLA). These candidates are all coincident with the Outer Scutum-Centaurus Spiral Arm \citep[OSC,][]{dame11} in the first Galactic quadrant, and 47 of the 65 targets are marked radio quiet. We detect radio continuum emission in 43\% of the formerly radio quiet sources. This high detection rate suggests that many radio quiet candidates have the expected radio continuum emission signature of \textit{bona fide} \hii\ regions. If these faint sources truly are \hii\ regions, they could either be produced by B-stars, with fewer photons able to ionize the surrounding ISM, or they could be especially distant. Either case would result in radio fluxes below the sensitivity limits of extant radio surveys of the Galactic disk. An \hii\ region created by a single B1.5 star at a heliocentric distance of 22 kpc, the farthest detected distance for a Galactic \hii\ region to date, would have an integrated 9 \ghz\ flux density of $\sim$300 \microjy\ \citep{armentrout17}. This is below the 3$\sigma$ sensitivity limits of existing Galactic plane surveys. Future Galactic plane surveys will well surpass this sensitivity, including the VLA Sky Survey with RMS~$\simeq$~69 \microjy\/ \citep[VLASS,][]{condon15, vlass16}. With a largest resolvable angular scale of 58\arcsec\/, VLASS will be able to detect the most compact \hii\ regions in the \textit{WISE} Catalog, but it will be insensitive to extended, low surface brightness regions.


\section{VLA Observations and Data Reduction}

Here, we study a more complete sample of radio quiet \hii\ region candidates than in \citet{armentrout17}. We target all infrared-identified \textit{WISE} Catalog radio quiet sources in $245^{\circ}\geq\gl\geq90^{\circ}$ with infrared diameters less than 80$^{\prime\prime}$. Determining how many of these radio quiet sources are \textit{bona fide} \hii\ regions will allow us to estimate the size of the \hii\ region population of the Milky Way. By systematically targeting all compact regions in this range, we will characterize the Galaxy-wide radio quiet population and infer a lower limit for the number of \hii\ regions in the Milky Way.  We also estimate the stellar types of stars ionizing these faint \hii\ regions, placing limits on the lowest mass stars able to produce an \hii\ region with our characteristic infrared and radio morphologies.

We choose this longitude range in the second and third Galactic quadrants ($245^{\circ}\geq\gl\geq90^{\circ}$) because the outer Galaxy has significantly less star formation than the inner Galaxy, so there are fewer sources to observe as well as less source confusion. Additionally, heliocentric distances are limited to within $\sim$10 kpc so their fluxes are on average higher compared to inner Galaxy sources, which can be twice as far away. The \hii\ region size constraint ($<$80$^{\prime\prime}$) was chosen because the VLA filters out large-scale emission, dependent on telescope configuration and observing frequency. Over one quarter (26.6\%) of the radio quiet \hii\ regions in our survey range are smaller than 80$^{\prime\prime}$. Over the full Galaxy, 71.7\% of radio quiet sources are smaller than 80$^{\prime\prime}$, since sources in the first and fourth Galactic quadrants are further away on average than those in our survey range in the second and third Galactic quadrants.

Observing the smallest class of radio quiet candidates biases our results slightly towards detecting \hii\ regions surrounding later spectral type stars and more distant \hii\ regions, both of which would appear angularly smaller on the sky. One of our main science goals is to characterize the latest type star able to produce a classical \hii\ region, and the bias does not affect this result.


\subsection{VLA Observations}

We observe the 9 \ghz\ ($\sim$3 cm) continuum emission from 146 compact (infrared diameter $<$80$^{\prime\prime}$) radio quiet \hii\ region candidates with the VLA in D-configuration (Project ID: VLA/16A-371). We follow the observing procedures of \citet{armentrout17} who use the VLA to search for distant \hii\ regions in the OSC Arm. Our configuration has two overlapping 2 \ghz\ bands, covering 8.012 to 10.041 \ghz\/. We use the X-band receiver and the Wideband Interferometric Digital ARchitecture (WIDAR) correlator. These observations were conducted between February and May 2017, split into two-hour scheduling blocks by source longitude. 3C48 is used as both a bandpass and flux density calibrator for all observations. Our gain calibrators vary based on RA, including J0154+4743, J0512+4041, J0559+2353, J0730-1141, J2137+5101, and J2148+6107. The observations are summarized in Table~\ref{tab:observeParameters}.

We observe all 146 radio quiet sources with 4 minute snapshots per source. We reobserve 21 of the 146 sources between $173^{\circ}\geq\gl\geq128^{\circ}$ for an additional 12 minutes each. The main purpose of these longer observations is to see if deeper integrations recover flux from even more radio quiet \hii\ regions. We also reobserve four sources to increase signal to noise and confirm that measured flux values from our initial snapshots do not change significantly as we fill out the \textit{u-v} plane with longer integrations.

\begin{figure*}[!hbt]
\includegraphics[width=\textwidth]{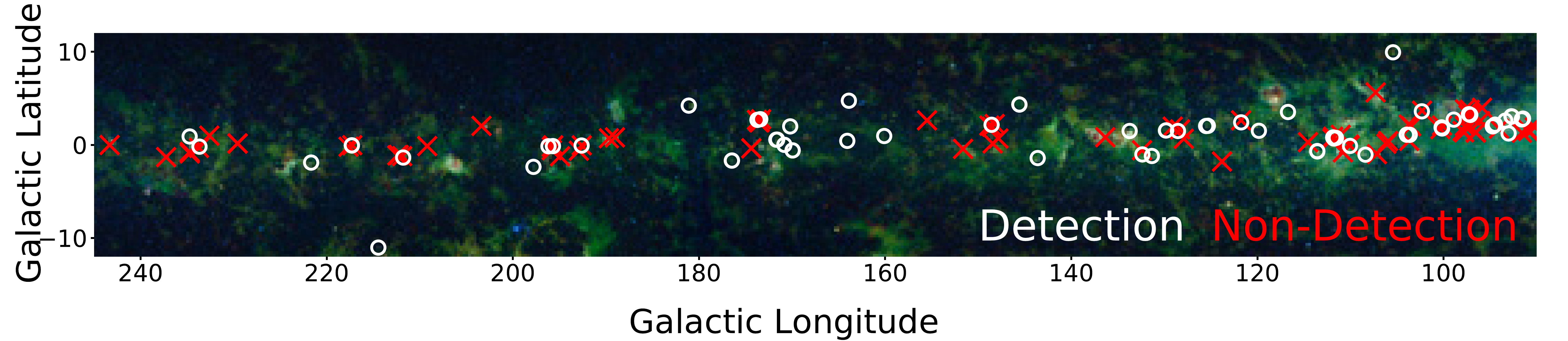}
\caption{Observed radio quiet \hii\ region candidates between Galactic longitudes of 90$\degree$ $-$ 245$\degree$. We observe every radio quiet candidate in this longitude range that has a measured infrared diameter $<$80$^{\prime\prime}$ (expected radio diameter $<$40$^{\prime\prime}$). Detections are shown by white circles while non-detections are marked by red ``x"s. Of our 146 VLA pointings, we detect radio continuum emission from 73 sources, giving us a detection rate of 50\%. The background image is from \textit{WISE}, showing bands W2 (4.6 \microns\/), W3 (12 \microns\/), and W4 (22 \microns\/) in blue, green, and red, respectively. The 12 \microns\/ emission traces polycyclic aromatic hydrocarbons (PAHs) and dominates the sky coverage of \textit{WISE} in the outer Galaxy. \label{fig:radioQuiet_Targets}}
\end{figure*}

\begin{figure*}[htb!]
\centering
\includegraphics[width=0.73\paperwidth]{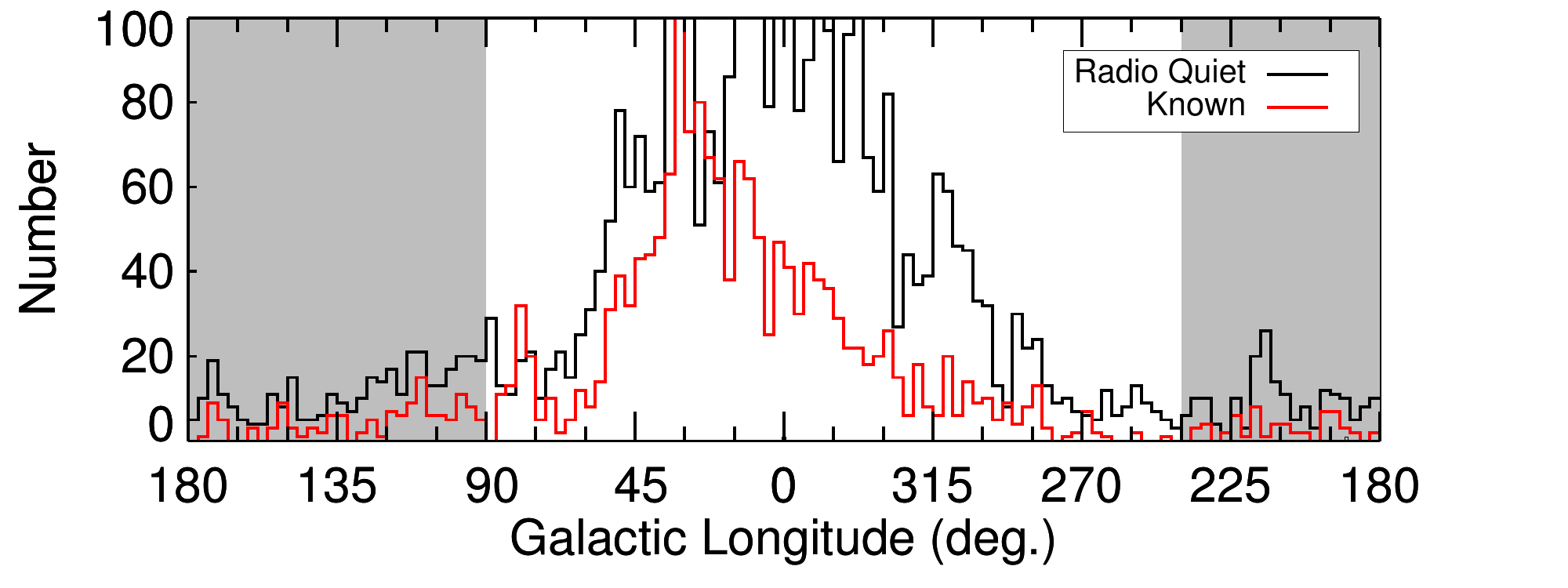}
\caption{Distribution of known \hii\ regions (red) and radio quiet candidates (black) as identified through \textit{WISE} infrared emission. There are over 4000 radio quiet candidates distributed throughout the Galactic plane, with $\sim$15\% of the radio quiet population in our survey range, marked in grey. This plot also shows the relative lack of known \hii\ regions in the Southern sky, which is caused by a dearth Southern hemisphere \hii\ region surveys. This will be rectified in part by \citet{wenger19} and ongoing surveys with the Australian Telescope Compact Array.  \label{fig:wise_catalog_lb_hist}}
\end{figure*}

We match the target sizes to the maximum size of recoverable flux in the VLA D-configuration.  The VLA data have a typical angular resolution of $10^{\prime\prime}$ and are sensitive to angular scales of up to 70$^{\prime\prime}$. The mid-infrared emission traces the diffuse photo-dissociation region surrounding the fully ionized thermal radio continuum emission. \citet{bihr16} show that \textit{WISE} infrared \hii\ region sizes are on average twice as large as the sizes measured through radio observations. Therefore, an \hii\ region with an infrared diameter of 80$^{\prime\prime}$ would have an expected radio diameter of $\sim$40$^{\prime\prime}$. Our targets are shown in Figure~\ref{fig:radioQuiet_Targets}, overlaid on \textit{WISE} infrared emission. The distribution of all \textit{WISE} radio quiet candidates is shown in Figure~\ref{fig:wise_catalog_lb_hist}; about 15\% of all radio quiet candidates in the \textit{WISE} Catalog are in our survey range in the outer Galaxy.

\begin{deluxetable}{lc}
\tabletypesize{\scriptsize} 
\tablecaption{Observation Summary \label{tab:observeParameters}}
\tablecolumns{2}
\tablewidth{0pt}
\startdata
Number of Targets & 146 \\
Longitude Range & $245^{\circ}\geq\gl\geq90^{\circ}$ \\
Frequency Range & 8.012$-$10.041 \ghz\ \\
Resolution & $\sim$10\arcsec\ \\
Largest Angular Scale & $\sim$70\arcsec\ \\
Integration Time & 4$-$16 minutes \\
Peak RMS & $\gtrsim$ 10 $\mu$Jy beam$^{-1}$ \\
\enddata
\end{deluxetable}

\subsection{VLA Data Reduction\label{sec:analysis}}
Our data are automatically calibrated by the Common Astronomy Software Applications \citep[CASA; ][]{mcmullin07} VLA pipeline \citep{kent18}\footnote{VLA Data Processing Pipeline, https://science.nrao.edu/facilities/vla/data-processing/pipeline}. This pipeline applies bandpass, gain, and flux calibrations to our VLA data and flags suspected corrupted data. The pipeline does not produce imaged data products.

We therefore also use CASA to image and analyze our VLA data. We use the CASA \textit{clean} task to image our data. We produce 300$\times$300 pixel images with 1$^{\prime\prime}$ cells. The interactive process completes 10,000 \textit{clean} cycles with Briggs weighting and a robust parameter of 0.5. Typically, we end the \textit{clean} process after one full iteration of 10,000 cycles. Within CASA, we also correct for the primary beam (\textit{impbcor}) and regrid the images into Galactic coordinates (\textit{imregrid}, template=`GALACTIC').

With these cleaned and regridded images, we smooth all of our maps with a 15$^{\prime\prime}$ boxcar filter and then identify, or mask, emission by hand in each continuum snapshot. We averaged data from the two orthogonal polarizations, and calculate the RMS noise, given as \textit{$\sigma$}, outside of any masked regions in the image. We calculate peak and integrated fluxes inside each masked region using the \textit{imstat} routine. In the case of no detection, there is no masked region and \textit{$\sigma$} is therefore the noise across the entire image. VLA continuum parameters for the 15$^{\prime\prime}$ spatial resolution images are given in the Appendix in Table~\ref{tab:contParamsHowMany}, which lists integrated and peak flux densities, the RMS noise for integrated and peak fluxes, and the area of each source mask.

Integrated noise values are calculated according to $\sigma_{\rm int} = \sqrt{\sigma_{\rm peak}^2~N_{\rm beams} + (0.1~S_{\rm int})^2}$, where $N_{\rm beams}$ is the number of synthesized beams contained within the masked region, or the mask area divided by beam area, $\sigma_{\rm peak}$ is the RMS noise of the image, and $S_{\rm int}$ is the integrated flux within the masked region. This form assumes 10\% flux calibration error and includes the contribution for the variance in the sum of brightnesses within a region, as detailed in \citep{wenger19b}. Since our targets are located towards the center of the primary beam in each snapshot image, we assume noise does not vary across the source and are able to used the simplified form of Equation~3 from \citet{wenger19b}.


\section{Results}

We detect 73 of the 146 sources (50\%) in 9 \ghz\ radio continuum from 4 minute integrations with the VLA. We require peak continuum fluxes~ $\gtrsim$ 3$\sigma$ and for the continuum emission to be coincident with and morphologically similar to the infrared emission. Coupled with their common infrared morphologies, detection of radio continuum emission \textit{strongly} suggests an \hii\ region candidate is an \hii\ region. This 50\% detection rate is slightly higher than the detection rate of radio quiet sources in the first quadrant in \citet{armentrout17} where we detect 43\% of radio quiet sources.  In the first quadrant survey, we target especially distant ($\sim$20 kpc) \hii\ regions, which are fainter on average. One example VLA detection is shown in Figure~\ref{fig:contSnapshotsHowManyEx}, while the full figure set (146 images) is shown in the Appendix in Figure~\ref{fig:contSnapshotsHowMany}.

We detect continuum emission from 10 of our 21 ``long integrations" of 16 minutes each. Four of these are also detected in shorter 4 minute observations, while most (17) of the 21 sources are not detected with 4 minute integrations. We detect 6 new sources (of the 17 previously undetected) from 16 minute observations. This represents an additional detection rate of 35\% for 16 minute integrations which were not detected previously in 4 minutes (6/17). While this is a small sample size, it indicates that deeper observations will continue to uncover even fainter continuum emission across our survey range. We mainly focus on our sample of 4 minute observations in this work to derive limits on the total population of Galactic \hii\ regions, because it is a larger, more complete sample. 

\begin{figure}[htb!]
\centering
\includegraphics[width=0.4\paperwidth]{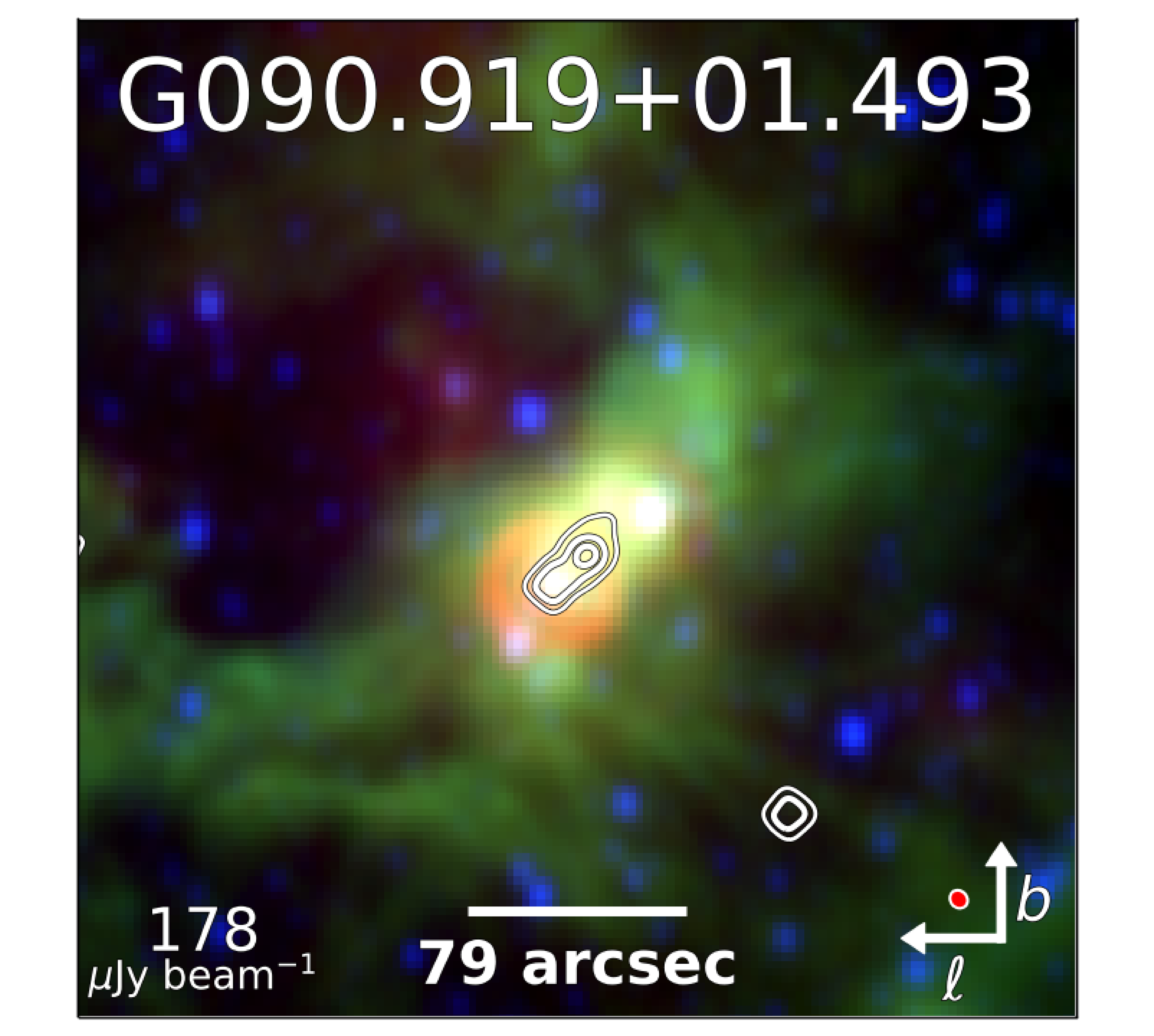}
\caption{\textit{WISE} composite images of a VLA snapshot observation for a previously ``radio quiet" candidate. The complete figure set (146 images) is shown in the Appendix. Bands W2 (4.6 \microns\/), W3 (12 \microns\/), and W4 (22 \microns\/) are represented by blue, green, and red, respectively. VLA radio continuum contours at X-band in D configuration are overplotted for regions with newly detected continuum emission. Contours are placed at 33\%, 66\%, and 90\% peak radio continuum flux for sources detected in radio continuum. Peak radio continuum intensities are shown in the bottom left of each panel (in $\mu$Jy) for each detected source and infrared sizes are shown by a scale bar. The VLA beam is shown in red in the bottom right of each image. Most integrations were 4 minute snapshots, but 16 minute integrations are available for 21 sources. These 16 minute integrations are indicated by ``Deep Integration" labels. Each image is 6$^{\prime}$ on a side, and scale bars represent the regions' infrared angular diameters as cataloged in the \textit{WISE} Catalog of Galactic HII Regions \citep{anderson14}. Measured radio diameters are on average half that of infrared diameters. We smoothed radio continuum emission with a 15$^{\prime\prime}$ boxcar filter for display. Since these objects are all exceedingly faint, there are often much brighter targets in the field that are saturated. When there are multiple candidates within one frame, we only report continuum parameters for the central region.}
\label{fig:contSnapshotsHowManyEx}
\end{figure}

\subsection{Comparison with Known H{\footnotesize II} \normalsize{Regions}}

\begin{figure}[htb!]
\centering
\includegraphics[width=0.4\paperwidth]{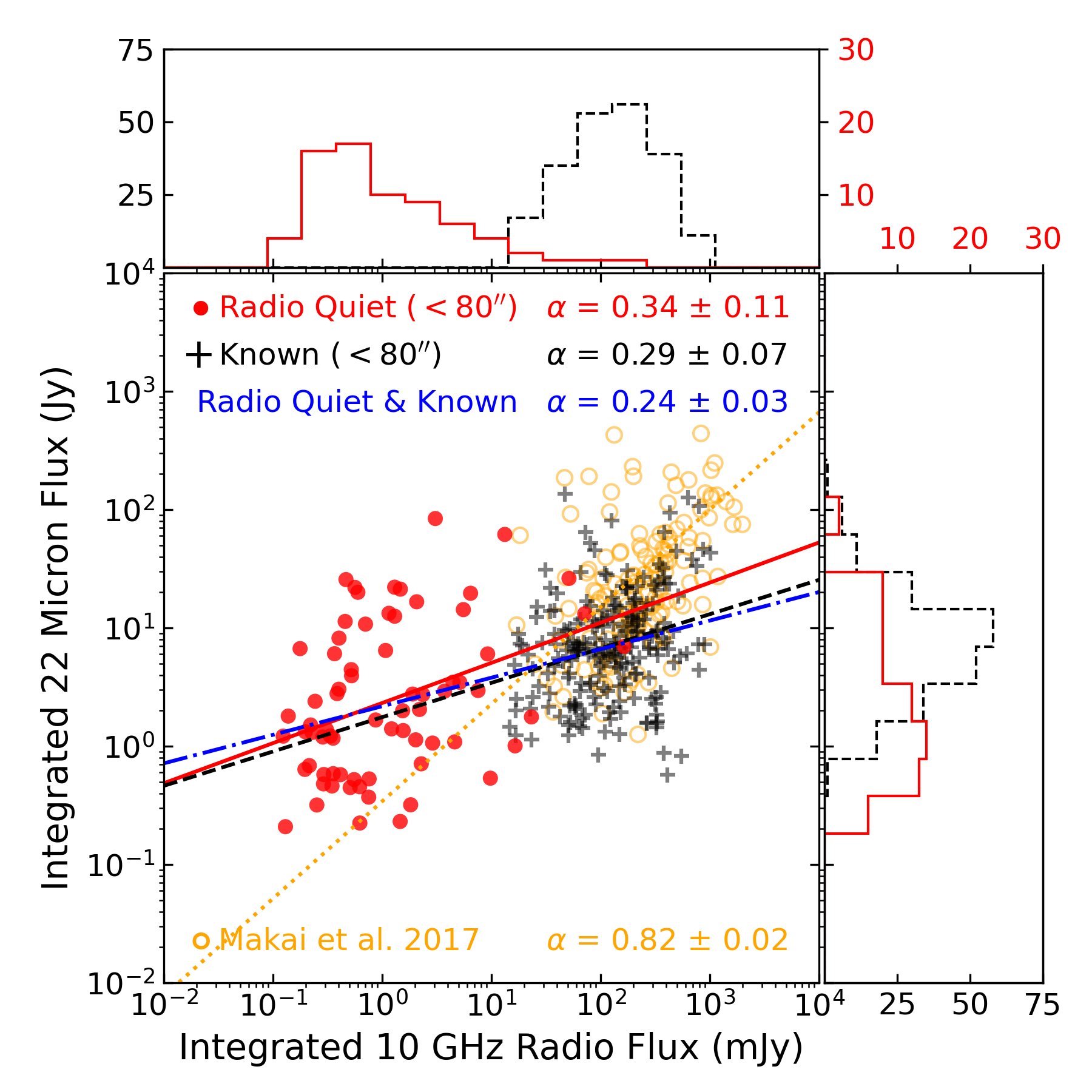}
\caption{Comparison of the integrated 10 \ghz\ continuum flux density and integrated \textit{WISE} 22 \microns\ flux density for radio quiet candidates in this study ($<$80$^{\prime\prime}$, red circles) and the known \hii\ regions from the \textit{WISE} Catalog of Galactic \hii\ Regions smaller than 80$^{\prime\prime}$ (black crosses). The red solid line and black dashed line represent power law fits to the radio quiet and known samples, respectively, while the blue dashed line represents a power law fit to both samples together. Also shown are the known \hii\ regions that are larger than 80$^{\prime\prime}$ (orange circles). The Makai et al. fit shown in orange includes all of the known regions (both the black and orange points). The distribution of flux densities for radio quiet and known \hii\ regions smaller than 80$^{\prime\prime}$ have similar power law slopes, although with significant scatter. When larger, more luminous known \hii\ regions are also included, the power law fit steepens.
 \label{fig:fluxflux}}
\end{figure}

We show a comparison of the radio continuum emission versus the mid-infrared emission in Figure~\ref{fig:fluxflux} for both this work and the \textit{WISE} Catalog of Galactic \hii\ Regions, including every source with a recorded flux density at both 22 \,\micron\ and 10 \ghz\/. The mid-infrared flux densities for the known \hii\ regions are taken from \citet{makai17}. We calculate the mid-infrared flux densities for radio quiet sources by integrating the 22 \,\micron\ \textit{WISE} emission within a circular mask determined by the \textit{WISE} Catalog region size, and we determine the background level within a 5$^{\prime\prime}$-wide annulus surrounding each region.

Figure~\ref{fig:fluxflux} shows that there is significant scatter in the 22 \,\micron\ and 10 \ghz\ flux densities of both the radio quiet and known samples, and that these flux densities span five orders of magnitude at both infrared and radio wavelengths. \citet{makai17} describe the scatter at low flux densities as likely due to photometric uncertainties from background subtraction. While the samples do overlap, the radio quiet sources naturally lie at lower radio and infrared fluxes. We expect ``radio loud" candidates from the \textit{WISE} Catalog to fill in the space between the radio quiet and known samples in this plot, but accurate 10 \ghz\ flux densities are not available for the radio loud candidate sample.


To compare our known and radio quiet samples, we split the angular size of known \hii\ regions shown in Figure~\ref{fig:fluxflux} into those larger and smaller than 80$^{\prime\prime}$. We do not have distance estimates for our radio quiet sources and therefore do not know the physical sizes of these regions. We perform a least squares fit to each sample, according to the form $F_{22\,\mu\text{m}} = A~\times~(F_{10\,\text{GHz}})^{\alpha}$, where $F_{22\,\mu\text{m}}$ is the integrated 22 $\microns$ flux density, $F_{10\,\text{GHz}}$ is the integrated 10 GHz flux density, A is a scaling factor, and $\alpha$ is the power law slope of the distribution. The power law fits for our radio quiet \hii\ region sample ($<$80$^{\prime\prime}$, red in Figure~\ref{fig:fluxflux}) and known \hii\ regions smaller than 80$^{\prime\prime}$ (black in Figure~\ref{fig:fluxflux}) are consistent with each other, although they have low individual coefficients of determination, or R$^2$ values. Radio quiet \hii\ regions have a power law slope of $\alpha$ = 0.339 $\pm$ 0.105 (A = 2.319 Jy $\pm$ 0.379, R$^2$ = 0.131), while compact known \hii\ regions have a power law slope of $\alpha$ = 0.290 $\pm$ 0.073 (A = 1.760 Jy $\pm$ 0.631, R$^2$ = 0.070). When both the compact known and radio quiet \hii\ region samples are fit together (all red and black points), the distribution has a power law slope of $\alpha$ = 0.241 $\pm$ 0.027 (A = 2.177 Jy $\pm$ 0.268, R$^2$ = 0.216). This fit is shown in blue in Figure~\ref{fig:fluxflux}.

If we include \hii\ regions larger than 80$^{\prime\prime}$, the power law fit steepens. \citet{makai17} find a power law fit of $\alpha$ = 0.822 $\pm$ 0.017 (A = 0.082 Jy $\pm$ 0.001, R$^2$ = 0.526) when comparing 24 $\microns$ \textit{Spitzer} data with 1.4 \ghz\ radio continuum data for all Galactic \hii\ regions in the \textit{WISE} Catalog. We scale this \citet{makai17} fit from 1.4 \ghz\ to 10 \ghz\/, assuming an \hii\ region spectral index of $-$0.1 and plot it in Figure~\ref{fig:fluxflux} in orange. When \citet{makai17} restrict their analysis to regions larger than 1 pc, the power law steepens to $\alpha$ = 1.096 $\pm$ 0.024 (A = 0.069 Jy $\pm$ 0.001, R$^2$ = 0.716). It seems that compact \hii\ region samples have a shallower infrared-radio correlation than larger, and typically more luminous, \hii\ region samples. The \hii\ region luminosity function is further explored in \citet{mascoop21}.

The radio quiet and known \hii\ region samples smaller than 80$^{\prime\prime}$ have very similar power law fits, and the overall fit to both samples together is consistent with the slope of each independent sample. The radio quiet and known \hii\ region samples may be drawn from the same parent population, but the significant scatter in their flux density measurements precludes us from making that determination.


\subsection{VLA Continuum Properties}

Following our analysis in \citet{armentrout17}, we determine the spectral type of the ionizing stars in \hii\ regions. We first calculate source luminosities, \textit{L$_{\nu}$}, using:
\begin{equation}
L_{\nu}~\big[\mathrm{W~Hz^{-1}} \big] = 1.2\times10^{14}~\bigg[\frac{D}{\mathrm{kpc}}\bigg]^{2}~\bigg[\frac{S_{\mathrm{int}}}{\mathrm{Jy}} \bigg],
\label{eq:luminosity}
\end{equation}
where D is the distance to the source and $S_{\mathrm{int}}$ is the integrated continuum flux density.

We do not have distance measurements for these sources. While we can examine integrated flux density distributions from the sample, we cannot provide an estimate of individual source luminosities without distance measurements. Instead of providing luminosity estimates for individual sources, we analyze the distribution assuming several uniform distances. The maximum distance of sources in the 2nd and 3rd Galactic quadrants is approximately 10 kpc. The majority of our targets are likely in the Perseus arm, which ranges in distance from 2 to 4 kpc from the Sun across our survey range \citep{reid2019}.

We then determine the number of Lyman continuum photons emitted from the central ionizing source, $N_{\mathrm{Ly}}$, assuming thermal and ionization equilibrium (LTE) \citep{rubin68, condon2016} :
\begin{equation}
N_{\mathrm{Ly}}~\big[\mathrm{s}^{-1} \big] = 6.3\times10^{32} \bigg[ \frac{T_e}{10^4\:\mathrm{K}} \bigg] ^{-0.45} \bigg[ \frac{\nu}{\ghz\ } \bigg] ^{0.1} \bigg[ \frac{L_\nu}{\mathrm{W}\:\mathrm{Hz}^{-1}} \bigg] .
\label{eq:lymanPhots}
\end{equation}
In LTE, the electron temperature, \textit{T$_e$}, is a measure of the Maxwellian kinetic energy distribution of colliding electrons in the \hii\ region. We assume an average \hii\ region \textit{T$_e$} of $\sim$10$^4$ K. This value is consistent with the measured \textit{T$_e$} distribution for R$_{Gal}$ of 14 kpc in \citet{balser15}, and it is slightly above the average Galactic \textit{T$_e$} derived through that work ($\sim$9000~K). Our central observing frequency, \textit{$\nu$}, is 9 \ghz\/.

We estimate the spectral type of the central ionizing sources directly from the number of Lyman continuum photons using the stellar models of \citet{smith02} and \citet{martins05}, detailed in Table~\ref{tab:spectralTypeValues}. Both of these models assume expanding stellar atmospheres and convert from Lyman-continuum photon count to spectral type using non-LTE line-blanketing. Where the models overlap, their Lyman continuum fluxes are consistent to within $\sim$50\% on average for solar metallicity and luminosity class V, differing at most by $\sim$150\%. We use \citet{smith02} for B-type stars and \citet{martins05} for O-type stars, as indicated by bold font in the table. We also include \citet{panagia73} in Table~\ref{tab:spectralTypeValues} as an example of trends for later-type stellar ionization fluxes. However, \citet{panagia73} differs by up to an order of magnitude from the other models for B- and late-type O-stars, so we do not use the model for stellar type determination.

\begin{deluxetable*}{lccccclccc}
\tabletypesize{\footnotesize} 
\tablecaption{Single-Star \hii\ Region Parameters \label{tab:spectralTypeValues}}
\tablecolumns{10}
\tablewidth{0pt}
\tablehead{\colhead{Spectral} & \multicolumn{3}{c}{Log$_{10}$[$N_{\mathrm{Ly}}$]\,\,(s$^{-1}$)}& && \colhead{Spectral} & \multicolumn{3}{c}{Log$_{10}$[$N_{\mathrm{Ly}}$]\,\,(s$^{-1}$)}  \\ \cline{2-4} \cline{8-10}
\colhead{Type} & \colhead{Panagia} & \colhead{Smith} &\colhead{Martins} &&&\colhead{Type}  & \colhead{Panagia} &  \colhead{Smith} &\colhead{Martins} }
\startdata
B3 & 43.91&$-$&$-$&&&O8&48.59&48.50&\textbf{48.29}\\
B2 & 44.89&$-$&$-$&&&O7.5&48.70&48.70&\textbf{48.44}\\
B1.5&$-$&\textbf{46.10}&$-$&&&O7&48.86&49.00&\textbf{48.63}\\
B1&45.52&\textbf{46.50}&$-$&&&O6.5&49.02&$-$&\textbf{48.80}\\
B0.5&46.50&\textbf{47.00}&$-$&&&O6&49.24&$-$&\textbf{48.96}\\
B0&47.63&\textbf{47.40}&$-$&&&O5.5&49.50&$-$&\textbf{49.11}\\
O9.5&48.08&$-$&\textbf{47.56}&&&O5&49.71&49.20&\textbf{49.26}\\
O9&48.32&47.90&\textbf{47.90}&&&O4&49.93&49.40&\textbf{49.47}\\
O8.5&48.45&$-$&\textbf{48.10}&&&O3&$-$&49.50&\textbf{49.63}\\
\enddata
\tablecomments{Results from \citet{panagia73}, \citet{smith02}, and \citet{martins05} for the conversion between number of Lyman continuum photons and spectral type for Luminosity Type V stars. The Smith and Martin models assume expanding stellar atmospheres and non-LTE line-blanketing. We use Smith for B-type stars and Martins for O-type stars, as indicated by bold font in the table. Panagia models later type stars than Smith or Martin, but the Panagia results are up to an order of magnitude discrepant from the more recent models of Smith and Martins. We do not use Panagia for spectral type determination, only as an example of trends in earlier spectral types.}
\end{deluxetable*}

\begin{figure}[htb!]
\centering
\includegraphics[width=0.4\paperwidth]{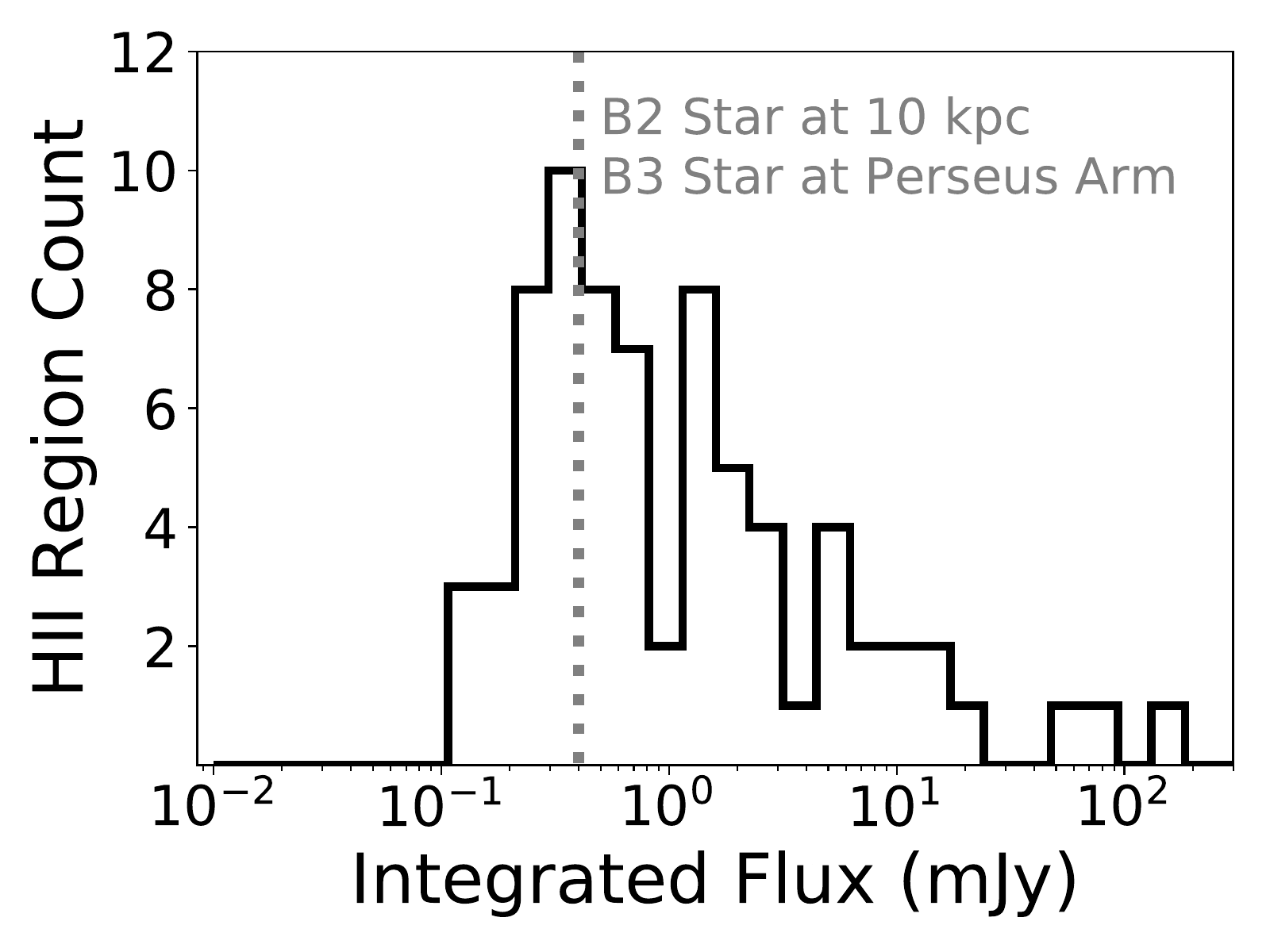}
\includegraphics[width=0.4\paperwidth]{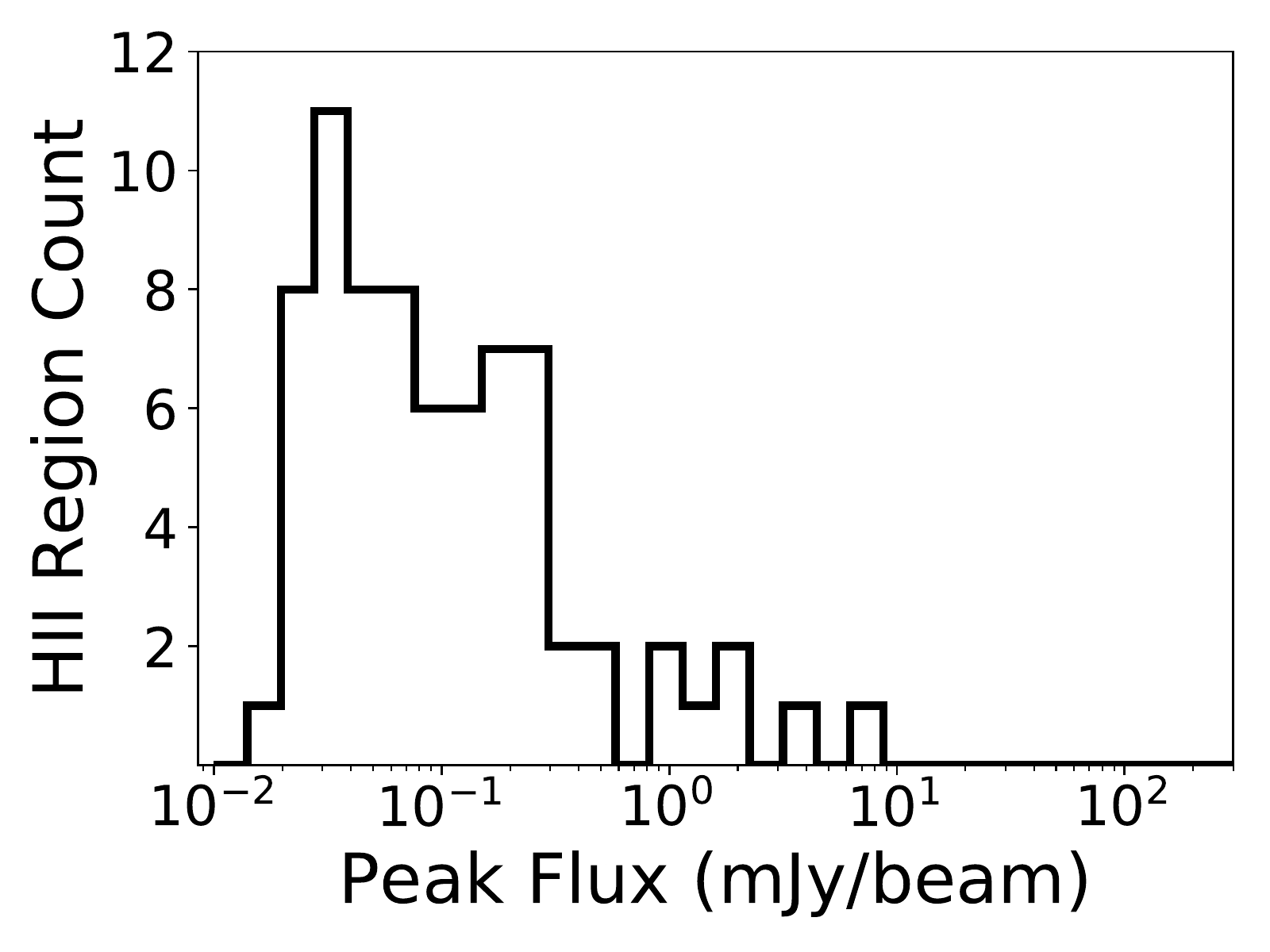}
\caption{Distribution of integrated and peak fluxes for \hii\ regions detected by the VLA in 4 minute snapshots. Each of these regions was identified as a radio quiet candidate based on its characteristic infrared morphology but lack of detected coincident radio continuum emission in Galactic plane surveys. The vertical dashed line shows the expected integrated flux for a B2 star at a distance of 10 \kpc\ or a B3 star at distances of 2-4 \kpc\ coincident with the 400 \microjy\ peak of the integrated flux distribution for our radio quiet sample.
 \label{fig:intFluxHist}}
\end{figure}

The distribution of integrated and peak radio continuum flux densities for our detections are shown in Figure~\ref{fig:intFluxHist}. The peak of the integrated 10 \ghz\ flux density histogram for our sample (Figure~\ref{fig:intFluxHist}) is 400 $\mu$Jy beam$^{-1}$. Using Equations~\ref{eq:luminosity} and \ref{eq:lymanPhots} we can determine average luminosities for varying source populations at the Perseus spiral arm and at the maximum expected distance of outer Galaxy sources, as detailed in Table~\ref{tab:stellartypedist}.

Spectral type determinations for sources with total log$_{10}$[$N_{\mathrm{Ly}}$] of less than 46.10 are especially uncertain because of lack of intermediate-mass star modeling. \citet{smith02} shows a decrease in Lyman continuum photons of log$_{10}$[$N_{\mathrm{Ly}}$] = 0.9 between type B0 and B1 stars. The spacing between adjacent spectral types is not quite uniform and becomes steeper at lower spectral types, as shown in all three models. An additional decrease of 0.9 from B1 to B2 would indicate that the B2 cutoff should be lower than log$_{10}$[$N_{\mathrm{Ly}}$]  = 45.60 (i.e. 0.9 lower than the log$_{10}$[$N_{\mathrm{Ly}}$] = 46.50 estimate for a B1 star in the Smith model). These trends in the stellar models suggest that our predicted ionizing flux of log$_{10}$[$N_{\mathrm{Ly}}$] $\simeq$ 45.57 from the peak of the distribution in Figure~\ref{fig:intFluxHist} and the assumed maximum distance of 10 kpc roughly corresponds to a spectral type B2 star, while the distance of the Perseus arm across our survey range (2-4 kpc) likely corresponds to a B3 star.

\begin{deluxetable}{ccc}
\tabletypesize{\footnotesize} 
\tablecaption{Peak Lyman Continuum Luminosity at Various Distances \hii\ Region Parameters \label{tab:stellartypedist}}
\tablecolumns{10}
\tablewidth{0pt}
\tablehead{\colhead{Distance} & \colhead{Log$_{10}$[$N_{\mathrm{Ly}}$]}& \colhead{Spectral Type} \\
\colhead{(kpc)} & \colhead{(s$^{-1}$)} & \colhead{ }  }
\startdata
2 kpc & 44.18 &B3\\
3 kpc & 44.53 &B3\\
4 kpc & 44.78 &B3\\
10 kpc & 45.58 &B2\\
\enddata
\tablecomments{Using Equations~\ref{eq:luminosity} and \ref{eq:lymanPhots} and assuming an integrated 10 \ghz\ flux density of 400 $\mu$Jy beam$^{-1}$, we determine average luminosities for distances of 2, 3, 4, and 10 kpc. The Perseus arm ranges from 2--4 kpc across our survey area, while the maximum expected distance for a star formation region in this part of the Galactic plane is 10 kpc.}
\end{deluxetable}

\citet{panagia73} lists a B2 star ionizing flux of log$_{10}$[$N_{\mathrm{Ly}}$]  = 44.89. The discrepancies between \citet{panagia73} and the models of \citet{smith02} and \citet{martins05} are possibly due to more thorough modeling of line-blanketing and winds in non-LTE models of expanding stellar atmospheres. A more accurate spectral type assignment will require more complete stellar modeling of later type B-stars using these updated methods and distances to individual sources in the sample.

Approximately one third of our 16 minute ``deep integrations" without previously detected radio continuum emission yield detections (6 of 17 sources), with flux densities at the $\sim$200 \microjy\ level. Assuming a distance of 10 kpc, these regions would produce ionizing radiation of at least log$_{10}$[$N_{\mathrm{Ly}}$]  = 45.27. The spectral types of these lower ionizing flux regions are especially uncertain because of the lack of recent, non-LTE line-blanketing stellar models for stars later than type B1.5. VLA continuum parameters for these deep integrations are given in the Appendix in Table~\ref{tab:contParamsHowMany}. If we expect approximately 6/17 all of our non-detections from 4 minute integrations to be detected with deeper integrations, we might expect 68\% of all of our 146 radio quiet candidates to have detected radio continuum emission at the 200 \microjy\ level or above (50\% + 50\% $\times$ 35\% $\approx$ 68\%).  Since our deep integrations represent a small sample size, we will use the 50\% detection rate from the larger sample for future estimates of the total population of Galactic \hii\ regions in this paper, but it is clear that our population estimate is a lower limit.

\textit{It is very likely that the objects we detect with the VLA are \hii\ regions.} Over 95\% of our \hii\ region targets with this characteristic infrared morphology and coincident continuum emission were confirmed to be \hii\ regions through their RRL emission in \citet{anderson11}. In most cases, the VLA continuum emission we detect is morphologically similar to infrared emission, indicating that the continuum emission is associated with the infrared sources and not just spatially coincident along the line of sight. These radio quiet sources may be consistent with the same distribution of objects as the known \hii\ regions in the WISE catalog of Galactic \hii\ regions, as shown in Figure~\ref{fig:fluxflux}, but the significant scatter in our flux density measurements prevents us from making that determination. Detecting RRL emission for the faintest radio quiet sources would be challenging, but evidence suggest that given enough sensitivity we could do so.


\section{Discussion}

\subsection{How Many \textit{WISE} Catalog Objects are \hii\ Regions?}

We can estimate a lower limit for how many \hii\ regions there are in the Galaxy based on the proportion of radio quiet regions we detect with the VLA. The \textit{WISE} Catalog (V2.2) contains 2210 known \hii\ regions, 2422 candidate regions with coincident radio continuum emission, and 3780 radio quiet candidates. Extrapolating our 50\% detection rate from 4 minute VLA integrations, we expect 1890 of the 3780 \textit{WISE} Catalog radio quiet \hii\ region candidates to have detectable radio continuum emission. If we assume that all \hii\ region candidates with detected thermal radio continuum emission are \hii\ regions, we would expect there to be 6522 Galactic \hii\ regions in the \textit{WISE} Catalog (2210 + 2422 + 1890).

This prediction already dwarfs the number of known Galactic \hii\ regions, and it is likely an underestimate. Most \hii\ regions produced by later type stars (i.e. $\sim$B2) are likely not in the \textit{WISE} Catalog because we only include radio quiet objects that are angularly resolved by \textit{WISE}; point sources are excluded. Because young \hii\ regions and those created by later type stars are preferentially more compact, the true number of Galactic \hii\ regions is likely far greater still. The \textit{WISE} Catalog may in fact be missing some compact radio quiet \hii\ regions created by distant, later type stars as suggested by the ``yellowball" analysis in \citet{kerton15}. We have also shown that deeper integrations recover radio continuum emission from more radio quiet candidates. We explore our non-detections further in Section~\ref{sec:nondetection}.

To estimate the total population of Galactic \hii\ regions, we need population synthesis models of Milky Way star formation (W.~Armentrout et al., \textit{in prep}). Such models rely on the Galactic initial mass function (IMF), the high-mass end of the cluster mass function (CMF), and our catalogs of Galactic star formation. 

A natural link to this work is the study of Galactic OB stars, since the full population of Galactic OB stars clearly sets the upper bound for the total number of Galactic \hii\ regions. There is not, however, a well defined conversion between the number of Galactic OB stars and the number of \hii\ regions, for reasons including OB-clustering and the population of isolated runaway stars. The majority ($\gtrsim$80\%) of OB stars form in close binary systems \citep{chini12}. \citet{oey04} show that this OB clustering appears to follow a power law, but there are large uncertainties. \textit{Gaia} observations of many young, nearby clusters have begun constraining high-mass star cluster assembly and dispersion processes. Future \textit{Gaia} data releases are expected to improve astrometry measurements in many more nearby clusters \citep{kuhn19}. The proportion of ``runaway" OB stars (i.e. stars ejected from their natal systems) is also uncertain. Early interpretation of \textit{Gaia} data suggests that 10-12\% of O-stars are runaways \citep{maiz18}, while historical estimates put the runaway fraction between 20-30\% \citep{blaauw61, stone79}. These high-velocity stars in low-density intercloud regions were thought able to produce compact \hii\ regions, ionizing new gas reservoirs along the stars' trajectories \citep{elmegreen76b}. More recent simulations confirm this and show that runaway stars can create roughly spherical \hii\ regions, with the star near the center of the \hii\ region \citep{raga97}. The \hii\ regions produced by these runaway stars would always be dynamically young, leading to their compact, spherical morphology \citep{mackey13}. It is still unclear how many Galactic \hii\ regions may be caused by isolated runaway O-stars.

The VLASS should be able to image an additional $\sim$2500 radio quiet candidates across the Galaxy (i.e. expected radio size $<$58$^{\prime\prime}$ in the VLASS range). In addition, the GLOSTAR survey \citep{medina19} will cover an additional $\sim$400 radio quiet candidates between $85^{\circ}\geq\gl\geq-2^{\circ}$ and $|\gb|\leq2^{\circ}$. Based on our results here, we expect to detect continuum emission from roughly half of our current radio quiet candidates in these ranges, greatly expanding the number of faint \hii\ region candidates for which we have radio observations. Because infrared point sources are excluded from the \textit{WISE} Catalog if they do not also have detected coincident radio continuum emission, we can also add new \hii\ region candidates to the \textit{WISE} Catalog that are infrared point sources. This potentially adds a large new sample of candidates to the catalog.  A more complete census of Galactic \hii\ regions will give us a global view of high-mass star formation, allow us to constrain the Milky Way luminosity function, and refine Galactic structural details, together with many other Milky Way-wide parameters. 


\subsection{Radio Quiet Contribution to the Total Ionizing Radiation of the Milky Way}

\hii\ regions created by intermediate-mass central stars may dominate the total population of Galactic \hii\ regions by number, but they represent a small percentage of the total Lyman continuum emission of the Milky Way. This will be dominated by mostly large, well-known \hii\ regions and the warm ionized medium (WIM).

Through Galactic \hii\ region population synthesis modeling (W.~Armentrout et al. 2021, \textit{in prep}) and work on the Galactic \hii\ region luminosity function \citep{mascoop21}, we are also able to estimate the total ionizing radiation of Galactic B-stars compared to the rest of the Milky Way. From this work, under 1\% of the Milky Way's ionizing photons come from \hii\ regions with a central stellar type of B2 or later. Under 5\% of the Milky Way's ionizing photons come from \hii\ regions with a central stellar type of B0 or earlier.


\subsection{VLA Continuum Non-Detections\label{sec:nondetection}}

We hypothesize that the infrared sources that we do not detect with the VLA represent either early stages of star formation or less massive star formation regions. Radio continuum emission from earlier stages of star formation has steep, positive spectral indices ($\alpha$$\sim$1) near 9 \ghz\/, which is not characteristic of optically thin \hii\ regions ($\alpha$$\sim-$0.1). \citet{rosero19} explores 70 radio continuum sources associated with dust clumps that are thought to be areas of young high-mass star formation, most of which have positive spectral indices between 5 and 23 \ghz\/. These sources may be optically thick, pressure confined \hii\ regions. The favored creation scenario, however, is that the positive spectral indices are the results of shock ionized jets associated with new star formation.

In the lower-mass regime (e.g. mid-type B-stars), \citet{cyganowski11} studies a sample of Extended Green Objects (EGOs), thought to be massive young stellar objects. These are morphologically similar to our sample, though they are ``greener" than our candidates in general. The authors found the two EGOs with the highest Lyman continuum photon flux in their sample to be \hii\ regions created by early B-type stars. These regions also have fewer emitted Lyman continuum photons (log$_{10}$[$N_{\mathrm{Ly}}$] $\lesssim$ 45) and would be below our current sensitivity limits for 10 \kpc\ distances. \citet{lundquist14} also studies a population of intermediate-mass star forming regions with strong 12\microns\ emission. These are again ``greener" than our candidates here. The most massive central stars in this sample are spectral type B2 or later.

There is also significant source overlap between our radio quiet \hii\ regions and the yellowballs cataloged in \citet{kerton15} from the Milky Way Project \citep{simpson12}, which could be a mix of early stages of star formation and less massive regions. These yellowball sources are identified by citizen science volunteers as \textit{Spitzer} GLIMPSE/MIPSGAL targets that appear compact and yellow from their 4.5-, 8-, and 24-\microns\ emission shown in blue, green, and red respectively. \citet{kerton15} find that 65$\%$ of their inner Galaxy ($\absl\leq65^{\circ}, \absb\leq1^{\circ}$) sample are spatially coincident with sources in the \textit{WISE} Catalog. These are split between known, candidate, group, and radio quiet sources in the catalog, but radio quiet sources have the largest overlap. In total, 208 of their 928 yellowballs (24$\%$) are coincident with a \textit{WISE} radio quiet source. The authors interpret the yellowballs as a mix of less evolved and less massive \hii\ regions, which matches our interpretation of the radio quiet sources in our sample. They also suggest that the \textit{WISE} Catalog may be missing a significant number of radio quiet sources based on the low-luminosities of the overlapping yellowball-radio quiet sources (B-type stars). While the Milky Way Project does not cover the same Galactic longitudes as our current survey, many outer Galaxy radio quiet sources also have compact yellow morphologies and would likely have been identified as yellowballs. This could include G093.015+01.235, G096.546+01.358, G097.131+03.225 and many more compact objects for our analysis.

Based on the above discussion, it appears that all of our infrared-identified candidates are tracing star formation regions, but our radio quiet \hii\ regions provide a rough boundary between high- and intermediate-mass star formation, with early-type B stars serving as the dividing spectral types. In general, stars less massive than approximately type B2 produce faint and compact \hii\ regions that we can not easily see in radio continuum surveys of the Galactic plane. 


\section{Summary \label{sec:summary}}

The \textit{WISE} Catalog of Galactic \hii\ Regions contains nearly 4000 \hii\ region candidates that lack detected radio continuum emission. Based on their infrared morphologies, we hypothesize that the majority of these candidates are bona fide \hii\ regions, but that they are less luminous populations of distant late-type OB stars or that they represent earlier stages of high-mass star formation. We observe all such compact radio quiet candidates (infrared diameters $< 80^{\prime\prime}$) between $245^{\circ}\geq\gl\geq90^{\circ}$ with the VLA at 9 \ghz\/. We detect radio continuum emission from 50\% of these 146 \hii\ region candidates with 4 minute integrations. From a smaller sample of longer integrations, we might expect to detect as many as $\sim$68\% of our radio quiet sources with 16 minute integrations. These radio continuum detections suggest that there is a large population of uncataloged, faint Galactic \hii\ regions and they suggest that central stars of approximately type $\sim$B2 create \hii\ regions with the characteristic infrared and radio continuum morphologies common to those in the \hii\ Region Discovery Survey papers \citep{bania10, anderson11, anderson15b, anderson18, wenger19}. Future sensitive radio Galactic plane surveys will allow us to identify radio continuum emission from many more radio quiet \hii\ regions across the Galactic disk, improving the completeness of this work. We here estimate a lower limit for the population of Galactic \hii\ regions to be $\sim$7000.


\begin{acknowledgments}
\gbonraoblurb\ We thank the staff of the Very Large Array for providing server space for on-site data reduction. The anonymous referee provided valuable feedback and helped us craft a more robust paper. This work was supported by NSF grant AST1516021 to LDA. TMB is supported in part by NSF grant AST-1714688. TVW was supported by the NSF through the Grote Reber Fellowship Program administered by Associated Universities, Inc./National Radio Astronomy Observatory, the D.N. Batten Foundation Fellowship from the Jefferson Scholars Foundation, the Mars Foundation Fellowship from the Achievement Rewards for College Scientists Foundation, and the Virginia Space Grant Consortium through the bulk of this work. This research has made use of NASA's Astrophysics Data System Bibliographic Services, Astropy, a community-developed core Python package for Astronomy \citep{astropy13}, and also APLpy, an open-source plotting package for Python \citep{robitaille12}.
\end{acknowledgments}


\facility{VLA}
\software{Matplotlib \citep{hunter07}, Astropy \citep{astropy13, astropy18}, APLpy \citep{robitaille12}, CASA \citep{mcmullin07}}


\bibliographystyle{aasjournal}
\bibliography{ref}


\appendix

Continuum emission parameters for all 4- and 16-minute integrations are listed in Table~\ref{tab:contParamsHowMany}. The table includes positions, integrated and peak flux densities, RMS noise, and the area of each source mask for our 146 sources. Parameters for both 4- and 16- minute integrations are listed where available.

Snapshot images of each source observed with the VLA are shown in the Figure Set~A1. Peak fluxes are listed for detected sources. We only show the 16-minute snapshot image for sources with ``deep integrations." When there are multiple candidates within one frame, we only report on the central region. There are occasionally multiple ``radio loud" candidates or known regions within one field. Those regions are outside the scope of this work and often add noise or interferometer relics to the images.

The \textit{WISE} Catalog of Galactic \hii\ Regions\footnote{http://astro.phys.wvu.edu/wise/} contains an interactive map of the Galactic plane, showing all detected and candidate \hii\ regions cataloged to date, along with detected source velocities and known parameters.

\setcounter{table}{0}
\renewcommand{\thetable}{A\arabic{table}}


\newpage
\begin{deluxetable*}{rrrcccccc}
\tablecaption{VLA Radio Continuum Parameters \label{tab:contParamsHowMany}}
\tablecomments{Parameters in this table stem from the CASA routine \textit{imstat} with user-defined masks. Images were smoothed to a common resolution of 15$^{\prime\prime}$ with a boxcar filter before these parameters were calculated. Entries filled with ``$\ldots$" have no detections. Areas were determined from hand drawn CASA masks for each detected region. Most integrations were 4 minute snapshots, but 16 minute integrations are available for 21 sources.}
\tabletypesize{\footnotesize} 
\tablecolumns{8}
\tablewidth{0pt}
\tablehead{ 
\colhead{Name} &\colhead{$\ra_{\rm J2000}$} & \colhead{$\dec_{\rm J2000}$} & \colhead{$S_{\rm int}$} & \colhead{$\sigma_{\rm int}$} & \colhead{$S_{\rm peak}$} &  \colhead{$\sigma_{\rm peak}$} & \colhead{Region} &  \colhead{Integration}\\
\colhead{} &\colhead{} & \colhead{} & \colhead{} & \colhead{} &  \colhead{} & \colhead{} & \colhead{Area} &  \colhead{Time}\\
& \colhead{(hh:mm:ss)} & \colhead{(dd:mm:ss)} & \colhead{(mJy)} & \colhead{(mJy)} & \colhead{(mJy beam$^{-1}$)} & \colhead{(mJy beam$^{-1}$)} & \colhead{(arcsec$^2$)} &  \colhead{(min.)}
}
\startdata
G090.192+01.887&21:04:19.84&49:44:54.1&0.524&0.093&0.021&0.008&4350&4\\
G090.912+01.599&21:08:39.66&50:05:14.7&2.195&0.261&0.165&0.019&2880&4\\
G090.919+01.493&21:09:10.58&50:01:16.2&2.068&0.225&0.178&0.017&1260&4\\
G091.467+02.826&21:05:17.29&51:19:26.3&$\ldots$&$\ldots$&$\ldots$&0.009&$\ldots$&4\\
G091.582+01.303&21:12:52.54&50:22:30.2&$\ldots$&$\ldots$&$\ldots$&0.006&$\ldots$&4\\
G091.590+02.742&21:06:13.20&51:21:31.2&$\ldots$&$\ldots$&$\ldots$&0.011&$\ldots$&4\\
G092.658+03.054&21:09:23.19&52:21:21.1&$\ldots$&$\ldots$&$\ldots$&0.251&$\ldots$&4\\
G092.677+03.053&21:09:28.69&52:22:10.0&$\ldots$&$\ldots$&$\ldots$&0.290&$\ldots$&4\\
G092.707+03.072&21:09:31.03&52:24:16.1&$\ldots$&$\ldots$&$\ldots$&0.346&$\ldots$&4\\
G093.015+01.235&21:19:29.95&51:21:25.2&$\ldots$&$\ldots$&$\ldots$&0.007&$\ldots$&4\\
G093.370+02.604&21:14:44.11&52:33:58.8&$\ldots$&$\ldots$&$\ldots$&0.014&$\ldots$&4\\
G094.045+01.746&21:21:52.04&52:26:47.7&1.219&0.144&0.038&0.007&5700&4\\
G094.474+02.178&21:21:51.60&53:03:19.5&$\ldots$&$\ldots$&$\ldots$&0.006&$\ldots$&4\\
G094.712+01.994&21:23:50.72&53:05:31.2&$\ldots$&$\ldots$&$\ldots$&0.008&$\ldots$&4\\
G095.868+03.990&21:19:47.88&55:19:12.7&$\ldots$&$\ldots$&$\ldots$&0.019&$\ldots$&4\\
G096.076+02.562&21:27:47.70&54:26:56.1&0.752&0.108&0.036&0.009&3810&4\\
G096.538+01.314&21:35:54.73&53:50:57.2&3.720&0.811&0.091&0.081&3990&4\\
G096.546+01.358&21:35:44.92&53:53:16.0&6.445&1.277&0.280&0.098&6050&4\\
G097.105+03.165&21:30:05.50&55:35:38.2&1.822&0.214&0.045&0.011&4960&4\\
G097.131+03.225&21:29:55.93&55:39:19.1&$\ldots$&$\ldots$&$\ldots$&0.006&$\ldots$&4\\
G097.186+03.768&21:27:32.90&56:05:10.8&0.286&0.050&0.025&0.009&1000&4\\
G097.240+03.300&21:30:07.74&55:47:03.6&23.253&2.777&0.256&0.091&27270&4\\
G097.276+03.298&21:30:19.86&55:48:27.6&$\ldots$&$\ldots$&$\ldots$&0.136&$\ldots$&4\\
G097.284+03.179&21:30:56.92&55:43:36.8&0.415&0.062&0.065&0.011&1700&4\\
G097.325+03.226&21:30:56.03&55:47:18.7&1.556&0.215&0.453&0.053&790&4\\
G097.693+01.994&21:38:42.39&55:07:33.1&$\ldots$&$\ldots$&$\ldots$&0.020&$\ldots$&4\\
G097.701+03.821&21:29:57.99&56:28:46.7&0.759&0.093&0.055&0.008&4870&4\\
G097.704+02.048&21:38:30.97&55:10:25.7&1.154&0.148&0.137&0.017&3090&4\\
G097.786+01.373&21:42:02.64&54:43:12.9&4.614&0.502&0.145&0.021&9260&4\\
G097.802+01.350&21:42:14.00&54:42:48.9&0.124&0.051&0.048&0.023&490&4\\
G097.974+01.494&21:42:28.58&54:56:05.8&164.067&17.096&7.997&0.901&3080&4\\
G098.908+02.716&21:41:48.36&56:28:12.1&$\ldots$&$\ldots$&$\ldots$&0.009&$\ldots$&4\\
G100.161+01.766&21:53:11.12&56:32:30.7&$\ldots$&$\ldots$&$\ldots$&0.022&$\ldots$&4\\
G100.168+02.082&21:51:46.89&56:47:34.0&$\ldots$&$\ldots$&$\ldots$&0.057&$\ldots$&4\\
G100.213+01.883&21:52:57.40&56:39:57.2&$\ldots$&$\ldots$&$\ldots$&0.026&$\ldots$&4\\
G100.338+01.691&21:54:31.89&56:35:34.9&1.460&0.149&0.475&0.011&760&4\\
G102.311+03.677&21:56:57.46&59:22:21.3&0.359&0.062&0.102&0.021&560&4\\
G102.334+03.608&21:57:25.99&59:19:55.6&$\ldots$&$\ldots$&$\ldots$&0.010&$\ldots$&4\\
G103.483+01.998&22:12:02.04&58:42:52.8&0.243&0.039&0.039&0.008&1350&4\\
G103.640+01.087&22:16:56.28&58:03:01.0&$\ldots$&$\ldots$&$\ldots$&0.008&$\ldots$&4\\
G103.690+00.434&22:19:56.87&57:32:01.0&7.547&0.942&0.212&0.064&7250&4\\
G103.745+02.182&22:12:53.10&59:00:56.6&9.235&1.051&0.299&0.053&8640&4\\
G103.954+01.097&22:18:52.72&58:13:55.8&$\ldots$&$\ldots$&$\ldots$&0.011&$\ldots$&4\\
G105.433+09.949&21:42:46.27&66:10:48.7&$\ldots$&$\ldots$&$\ldots$&0.009&$\ldots$&4\\
G105.962+00.420&22:34:40.10&58:42:40.6&2.355&0.261&0.094&0.014&4890&4\\
G106.142+00.129&22:36:58.70&58:32:53.3&0.871&0.103&0.072&0.009&2750&4\\
G107.156$-$00.988&22:47:49.02&58:02:49.7&51.334&5.327&2.100&0.171&5110&4\\
G107.298+05.638&22:21:27.04&63:51:33.8&0.703&0.074&0.237&0.010&470&4\\
G108.394$-$01.046&22:56:27.66&58:32:34.8&$\ldots$&$\ldots$&$\ldots$&0.351&$\ldots$&4\\
G110.054$-$00.107&23:05:15.81&60:05:01.2&$\ldots$&$\ldots$&$\ldots$&0.155&$\ldots$&4\\
G110.094$-$00.064&23:05:25.36&60:08:18.7&1.301&0.401&0.308&0.100&1060&4\\
G110.160+00.040&23:05:34.69&60:15:36.6&16.48&12.479&1.667&2.312&2110&4\\
G110.812$-$00.799&23:12:58.07&59:44:10.3&1.900&0.242&0.213&0.033&1510&4\\
G111.046+01.085&23:08:55.66&61:34:06.2&0.521&0.060&0.112&0.009&830&4\\
G111.567+00.751&23:14:01.62&61:27:19.3&$\ldots$&$\ldots$&$\ldots$&1.857&$\ldots$&4\\
G111.774+00.689&23:15:49.87&61:28:21.4&$\ldots$&$\ldots$&$\ldots$&0.033&$\ldots$&4\\
G111.860+00.800&23:16:10.13&61:36:26.0&$\ldots$&$\ldots$&$\ldots$&0.048&$\ldots$&4\\
G111.870+00.881&23:16:00.05&61:41:12.4&0.138&0.042&0.046&0.015&540&4\\
G111.941+00.677&23:17:10.37&61:31:17.8&0.177&0.036&0.060&0.012&520&4\\
G113.566$-$00.698&23:33:37.82&60:45:06.1&$\ldots$&$\ldots$&$\ldots$&0.047&$\ldots$&4\\
G113.569$-$00.657&23:33:33.24&60:47:30.6&$\ldots$&$\ldots$&$\ldots$&0.188&$\ldots$&4\\
G114.569+00.290&23:39:17.26&61:59:04.5&1.074&0.152&0.134&0.029&1020&4\\
G116.719+03.536&23:50:38.02&65:40:20.3&$\ldots$&$\ldots$&$\ldots$&0.011&$\ldots$&4\\
G119.844+01.518&00:22:59.34&64:13:10.4&$\ldots$&$\ldots$&$\ldots$&0.008&$\ldots$&4\\
G121.750+02.423&00:40:08.62&65:16:08.5&$\ldots$&$\ldots$&$\ldots$&0.008&$\ldots$&4\\
G121.755+02.640&00:40:06.01&65:29:09.0&$\ldots$&$\ldots$&$\ldots$&0.008&$\ldots$&4\\
G123.809$-$01.781&00:58:41.39&61:04:43.9&0.599&0.066&0.061&0.006&1390&4\\
G125.331+02.050&01:14:00.34&64:49:00.7&$\ldots$&$\ldots$&$\ldots$&0.006&$\ldots$&4\\
G125.513+02.033&01:15:41.59&64:47:01.2&0.228&0.035&0.039&0.007&930&4\\
G127.916+00.657&01:35:43.13&63:05:59.4&0.355&0.051&0.035&0.008&1250&4\\
G128.315+01.667&01:40:53.01&64:01:15.7&0.621&0.086&0.056&0.008&3280&4\\
&&&0.537&0.064&0.055&0.005&3280&16\\
G128.539+01.497&01:42:35.46&63:48:38.4&0.252&0.043&0.024&0.006&2110&4\\
&&&0.208&0.027&0.014&0.003&2110&16\\
G129.101+01.965&01:48:28.91&64:09:07.7&0.347&0.079&0.033&0.016&1240&4\\
G129.806+01.561&01:53:52.16&63:35:48.6&$\ldots$&$\ldots$&$\ldots$&0.006&$\ldots$&4\\
G129.830+01.549&01:54:03.74&63:34:46.2&$\ldots$&$\ldots$&$\ldots$&0.006&$\ldots$&4\\
&&&$\ldots$&$\ldots$&$\ldots$&0.004&$\ldots$&16\\
G131.343$-$01.188&02:00:39.20&60:32:57.6&$\ldots$&$\ldots$&$\ldots$&0.013&$\ldots$&4\\
&&&$\ldots$&$\ldots$&$\ldots$&0.009&$\ldots$&16\\
G132.361$-$01.021&02:08:57.98&60:25:27.2&$\ldots$&$\ldots$&$\ldots$&0.007&$\ldots$&4\\
&&&0.303&0.048&0.020&0.006&2470&16\\
G132.399$-$00.575&02:10:20.76&60:50:19.2&0.356&0.053&0.029&0.008&1460&4\\
G133.733+01.498&02:26:41.98&62:21:19.9&$\ldots$&$\ldots$&$\ldots$&0.229&$\ldots$&4\\
&&&$\ldots$&$\ldots$&$\ldots$&0.242&$\ldots$&16\\
G136.348+00.823&02:44:40.35&60:42:39.1&5.140&0.539&0.095&0.016&6010&4\\
G143.611$-$01.410&03:24:18.32&55:12:09.5&$\ldots$&$\ldots$&$\ldots$&0.008&$\ldots$&4\\
&&&0.594&0.076&0.027&0.006&4170&16\\
G145.573+04.336&04:01:14.66&58:35:52.2&0.553&0.084&0.025&0.007&4890&4\\
&&&0.426&0.056&0.021&0.004&4885&16\\
G145.583+04.347&04:01:21.35&58:35:57.5&$\ldots$&$\ldots$&$\ldots$&0.007&$\ldots$&4\\
&&&$\ldots$&$\ldots$&$\ldots$&0.003&$\ldots$&16\\
G147.832+00.694&03:56:30.51&54:21:47.4&0.624&0.079&0.028&0.006&3590&4\\
G148.225+02.241&04:05:36.55&55:16:28.0&0.196&0.031&0.027&0.007&690&4\\
G148.449+00.160&03:57:22.55&53:33:25.1&$\ldots$&$\ldots$&$\ldots$&0.007&$\ldots$&4\\
G148.533+01.952&04:05:51.56&54:51:12.4&2.278&0.242&0.120&0.012&2690&4\\
G148.566+02.162&04:07:00.42&54:59:13.6&$\ldots$&$\ldots$&$\ldots$&0.006&$\ldots$&4\\
&&&0.152&0.024&0.015&0.004&1364&16\\
G148.809+02.073&04:07:50.63&54:45:30.3&0.292&0.041&0.023&0.005&1870&4\\
G151.633$-$00.431&04:10:24.68&51:00:15.4&9.787&1.104&0.972&0.106&1350&4\\
G155.503+02.640&04:41:21.84&50:22:24.4&$\ldots$&$\ldots$&$\ldots$&0.011&$\ldots$&4\\
&&&$\ldots$&$\ldots$&$\ldots$&0.007&$\ldots$&16\\
G160.104+00.958&04:51:33.02&45:46:56.0&$\ldots$&$\ldots$&$\ldots$&0.019&$\ldots$&4\\
&&&$\ldots$&$\ldots$&$\ldots$&0.016&$\ldots$&16\\
G163.899+04.748&05:21:47.64&45:01:42.9&$\ldots$&$\ldots$&$\ldots$&0.006&$\ldots$&4\\
&&&$\ldots$&$\ldots$&$\ldots$&0.004&$\ldots$&16\\
G164.069+00.436&05:03:02.51&42:21:23.4&$\ldots$&$\ldots$&$\ldots$&0.007&$\ldots$&4\\
&&&0.265&0.036&0.017&0.004&2002&16\\
G169.936$-$00.589&05:16:53.45&37:01:41.3&$\ldots$&$\ldots$&$\ldots$&0.006&$\ldots$&4\\
&&&$\ldots$&$\ldots$&$\ldots$&0.003&$\ldots$&16\\
G169.951$-$00.587&05:16:56.65&37:01:02.3&$\ldots$&$\ldots$&$\ldots$&0.006&$\ldots$&4\\
G170.204+01.990&05:28:23.72&38:16:05.9&$\ldots$&$\ldots$&$\ldots$&0.010&$\ldots$&4\\
&&&0.172&0.044&0.015&0.007&1668&16\\
G170.830+00.009&05:21:54.08&36:38:19.5&$\ldots$&$\ldots$&$\ldots$&0.007&$\ldots$&4\\
&&&$\ldots$&$\ldots$&$\ldots$&0.005&$\ldots$&16\\
G171.602+00.563&05:26:20.47&36:18:46.7&$\ldots$&$\ldots$&$\ldots$&0.008&$\ldots$&4\\
&&&0.686&0.085&0.023&0.005&5292&16\\
G171.692+00.564&05:26:35.75&36:14:20.5&$\ldots$&$\ldots$&$\ldots$&0.008&$\ldots$&4\\
&&&$\ldots$&$\ldots$&$\ldots$&0.007&$\ldots$&16\\
G173.328+02.754&05:40:06.26&36:03:28.1&0.386&0.056&0.049&0.009&1070&4\\
G173.418+02.760&05:40:22.09&35:59:06.4&$\ldots$&$\ldots$&$\ldots$&0.007&$\ldots$&4\\
&&&$\ldots$&$\ldots$&$\ldots$&0.005&$\ldots$&16\\
G173.482+02.445&05:39:12.93&35:45:48.9&1.466&0.344&0.277&0.073&980&4\\
G173.485+02.430&05:39:09.62&35:45:10.9&1.301&0.201&0.195&0.042&720&4\\
G173.583+02.444&05:39:28.63&35:40:40.1&0.366&0.059&0.025&0.007&2360&4\\
&&&0.403&0.052&0.024&0.005&2360&16\\
G173.722+02.693&05:40:52.74&35:41:30.7&$\ldots$&$\ldots$&$\ldots$&0.152&$\ldots$&4\\
G173.779+02.683&05:40:59.02&35:38:16.9&2.894&0.326&0.234&0.033&1240&4\\
G174.376$-$00.363&05:30:05.60&33:29:30.1&0.221&0.033&0.024&0.006&940&4\\
G174.384$-$00.371&05:30:04.98&33:28:50.5&0.215&0.038&0.018&0.006&1520&4\\
G176.462$-$01.674&05:30:20.81&31:01:43.7&$\ldots$&$\ldots$&$\ldots$&0.039&$\ldots$&4\\
G176.475$-$01.685&05:30:20.12&31:00:41.8&$\ldots$&$\ldots$&$\ldots$&0.026&$\ldots$&4\\
G181.101+04.223&06:04:56.65&30:06:45.8&$\ldots$&$\ldots$&$\ldots$&0.007&$\ldots$&4\\
G189.032+00.809&06:08:46.74&21:31:44.8&0.466&0.067&0.121&0.020&370&4\\
G189.684+00.715&06:09:46.88&20:54:47.8&0.308&0.041&0.026&0.006&1280&4\\
G192.601$-$00.048&06:12:54.31&17:59:22.6&13.264&4.228&3.733&0.975&1090&4\\
G192.617$-$00.067&06:12:52.03&17:57:59.1&$\ldots$&$\ldots$&$\ldots$&0.854&$\ldots$&4\\
G192.908$-$00.627&06:11:23.57&17:26:31.8&71.533&7.296&1.269&0.126&7870&4\\
G194.935$-$01.222&06:13:16.34&15:22:42.2&0.563&0.063&0.096&0.008&770&4\\
G195.666$-$00.043&06:19:00.88&15:17:47.0&$\ldots$&$\ldots$&$\ldots$&0.038&$\ldots$&4\\
G195.699$-$00.127&06:18:46.50&15:13:40.5&$\ldots$&$\ldots$&$\ldots$&0.095&$\ldots$&4\\
G195.811$-$00.215&06:18:40.25&15:05:14.2&$\ldots$&$\ldots$&$\ldots$&0.010&$\ldots$&4\\
G195.822$-$00.566&06:17:24.75&14:54:40.5&0.401&0.058&0.038&0.010&1020&4\\
G196.178$-$00.142&06:19:39.65&14:47:52.6&$\ldots$&$\ldots$&$\ldots$&0.017&$\ldots$&4\\
G197.792$-$02.356&06:14:47.14&12:19:40.3&$\ldots$&$\ldots$&$\ldots$&0.103&$\ldots$&4\\
G203.373+02.040&06:41:12.88&09:26:06.8&4.452&0.482&0.185&0.031&2320&4\\
G209.208$-$00.127&06:44:11.16&03:15:20.9&0.130&0.024&0.036&0.007&540&4\\
G211.790$-$01.354&06:44:32.29&00:23:56.4&$\ldots$&$\ldots$&$\ldots$&0.012&$\ldots$&4\\
G211.896$-$01.203&06:45:16.06&00:22:24.3&5.535&0.594&0.196&0.027&4310&4\\
G212.270$-$01.079&06:46:23.58&00:05:50.8&0.336&0.042&0.084&0.009&500&4\\
G212.420$-$01.126&06:46:29.77&$-$00:03:28.3&2.027&0.217&0.183&0.016&1600&4\\
G214.456$-$11.026&06:14:46.74&$-$06:20:26.0&$\ldots$&$\ldots$&$\ldots$&0.006&$\ldots$&4\\
G217.258$-$00.032&06:59:13.67&$-$03:51:49.7&0.402&0.057&0.035&0.009&1410&4\\
G217.302$-$00.059&06:59:12.76&$-$03:54:54.8&$\ldots$&$\ldots$&$\ldots$&0.018&$\ldots$&4\\
G217.307$-$00.042&06:59:16.84&$-$03:54:43.7&$\ldots$&$\ldots$&$\ldots$&0.013&$\ldots$&4\\
G217.657$-$00.193&06:59:23.05&$-$04:17:32.5&$\ldots$&$\ldots$&$\ldots$&0.007&$\ldots$&4\\
G221.678$-$01.897&07:00:40.59&$-$08:38:48.9&$\ldots$&$\ldots$&$\ldots$&0.020&$\ldots$&4\\
G229.569+00.154&07:23:02.23&$-$14:41:22.5&0.458&0.049&0.163&0.008&460&4\\
G229.598+00.159&07:23:06.83&$-$14:42:45.2&0.721&0.083&0.040&0.006&4120&4\\
G232.620+00.995&07:32:09.59&$-$16:58:12.3&3.066&0.344&0.991&0.061&630&4\\
G233.676$-$00.186&07:29:56.64&$-$18:27:50.1&$\ldots$&$\ldots$&$\ldots$&0.050&$\ldots$&4\\
G233.727$-$00.318&07:29:33.67&$-$18:34:17.8&1.539&0.198&0.064&0.015&5630&4\\
G234.717$-$00.913&07:29:21.31&$-$19:43:33.2&0.198&0.028&0.033&0.006&890&4\\
G234.729$-$00.752&07:29:58.90&$-$19:39:32.6&0.509&0.064&0.061&0.010&1190&4\\
G234.732+00.907&07:36:08.45&$-$18:51:34.7&$\ldots$&$\ldots$&$\ldots$&0.183&$\ldots$&4\\
G237.261$-$01.306&07:33:07.18&$-$22:08:41.3&$\ldots$&$\ldots$&$\ldots$&0.035&$\ldots$&4\\
G243.335$-$00.012&07:51:18.61&$-$26:46:36.0&0.290&0.045&0.039&0.008&1980&4\\
\enddata
\end{deluxetable*}

\iftrue
\renewcommand{\thefigure}{A\arabic{figure}}
\newcommand{\figSize}{0.33\textwidth}
\begin{figure*}[!ht]
\caption{\textit{WISE} composite images of all 146 radio continuum VLA snapshot observations for previously ``radio quiet" candidates. Bands W2 (4.6 \microns\/), W3 (12 \microns\/), and W4 (22 \microns\/) are represented by blue, green, and red, respectively. VLA radio continuum contours at X-band in D configuration are overplotted for regions with newly detected continuum emission. Contours are placed at 33\%, 66\%, and 90\% peak radio continuum flux for sources detected in radio continuum. Peak radio continuum intensities are shown in the bottom left of each panel (in $\mu$Jy) for each detected source and infrared sizes are shown by a scale bar. The VLA beam is shown in red in the bottom right of each image. Most integrations were 4 minute snapshots, but 16 minute integrations are available for 21 sources. These 16 minute integrations are indicated by ``Deep Integration" labels. Each image is 6$^{\prime}$ on a side, and scale bars represent the regions' infrared angular diameters as cataloged in the \textit{WISE} Catalog of Galactic HII Regions \citep{anderson14}. Measured radio diameters are on average half that of infrared diameters. We smoothed radio continuum emission with a 15$^{\prime\prime}$ boxcar filter for display. Since these objects are all exceedingly faint, there are often much brighter targets in the field that are saturated. When there are multiple candidates within one frame, we only report continuum parameters for the central region. \label{fig:contSnapshotsHowMany}}
\includegraphics[width=\figSize]{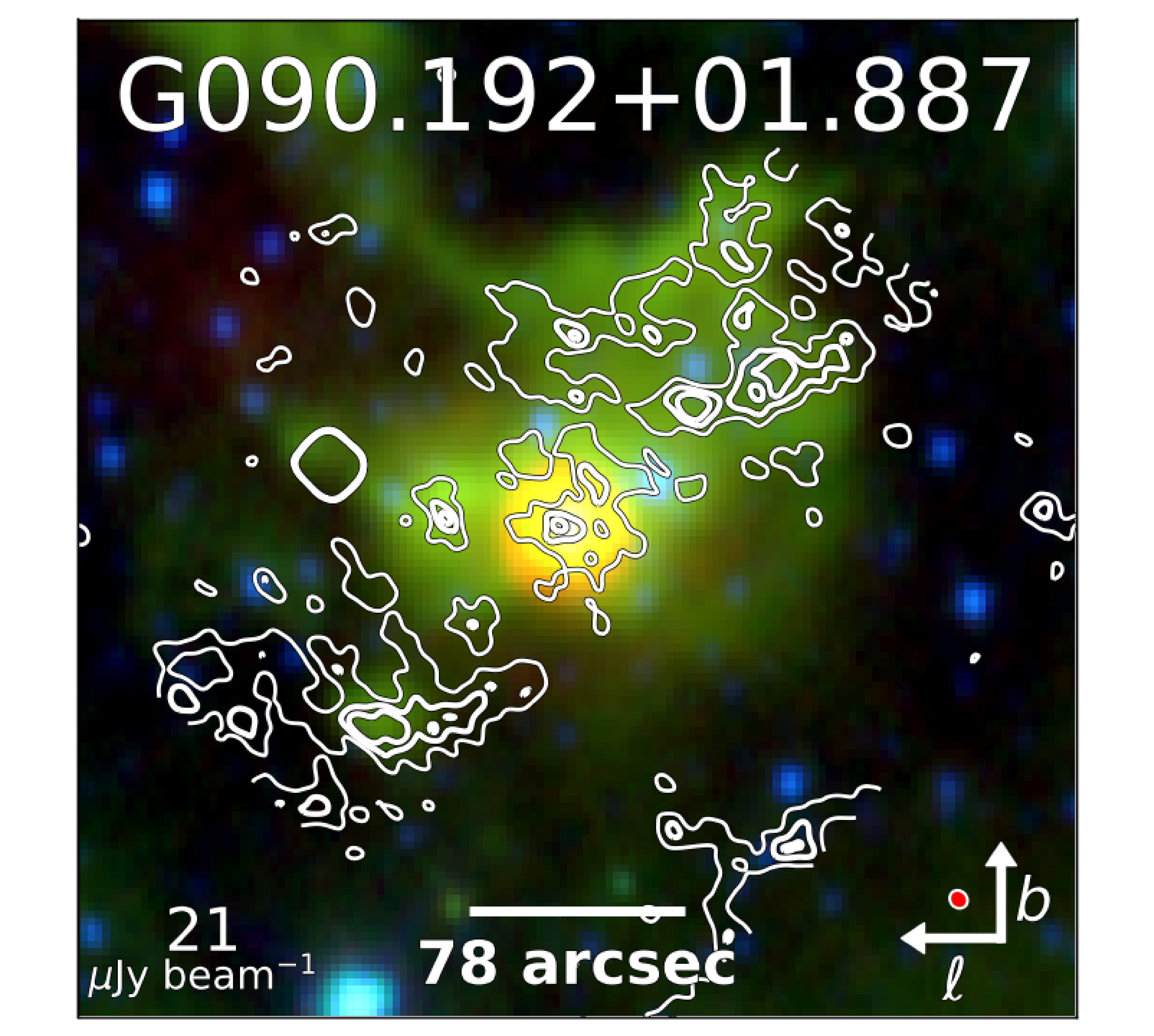}
\includegraphics[width=\figSize]{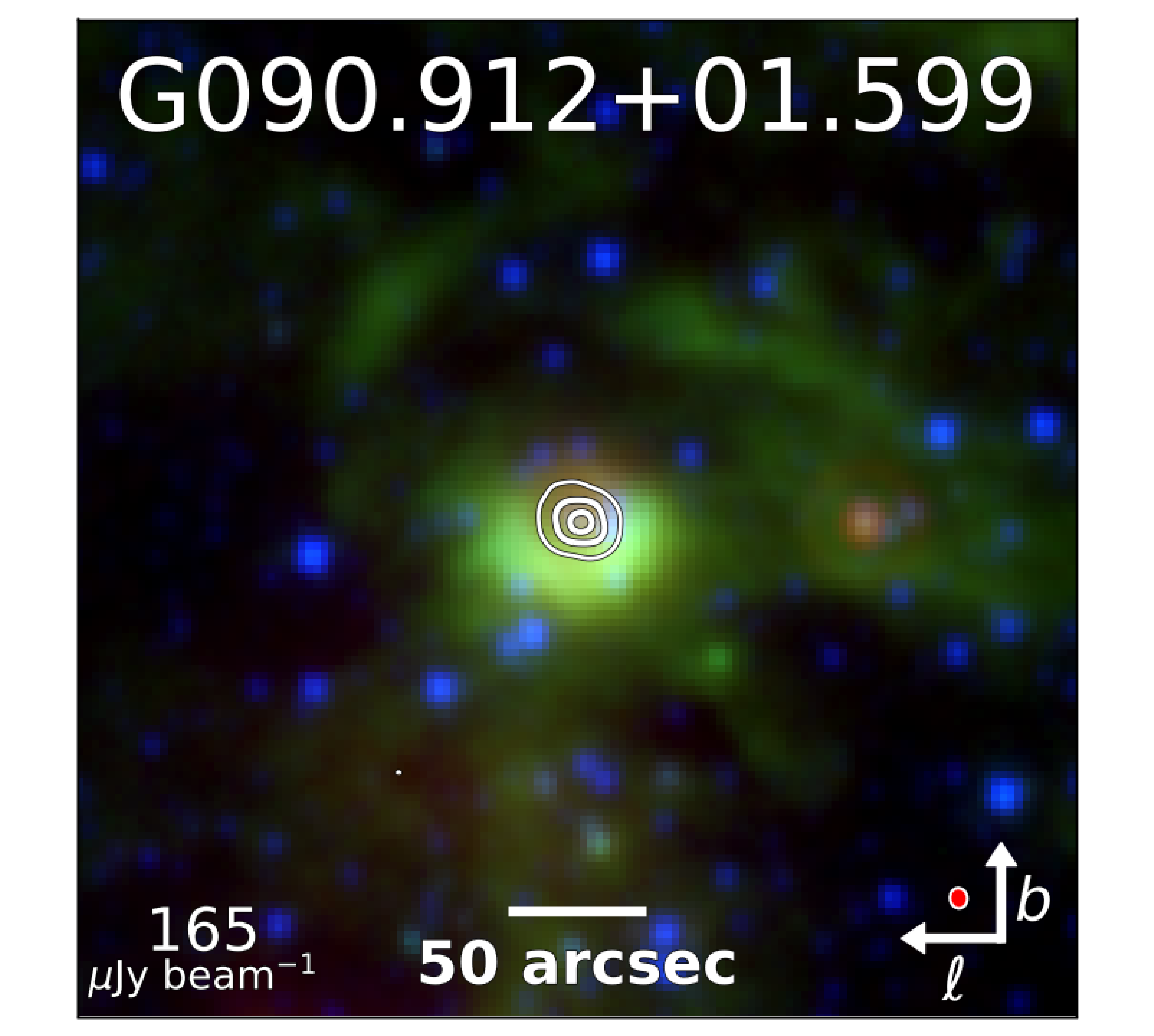}
\includegraphics[width=\figSize]{G090_919.pdf}\\
\includegraphics[width=\figSize]{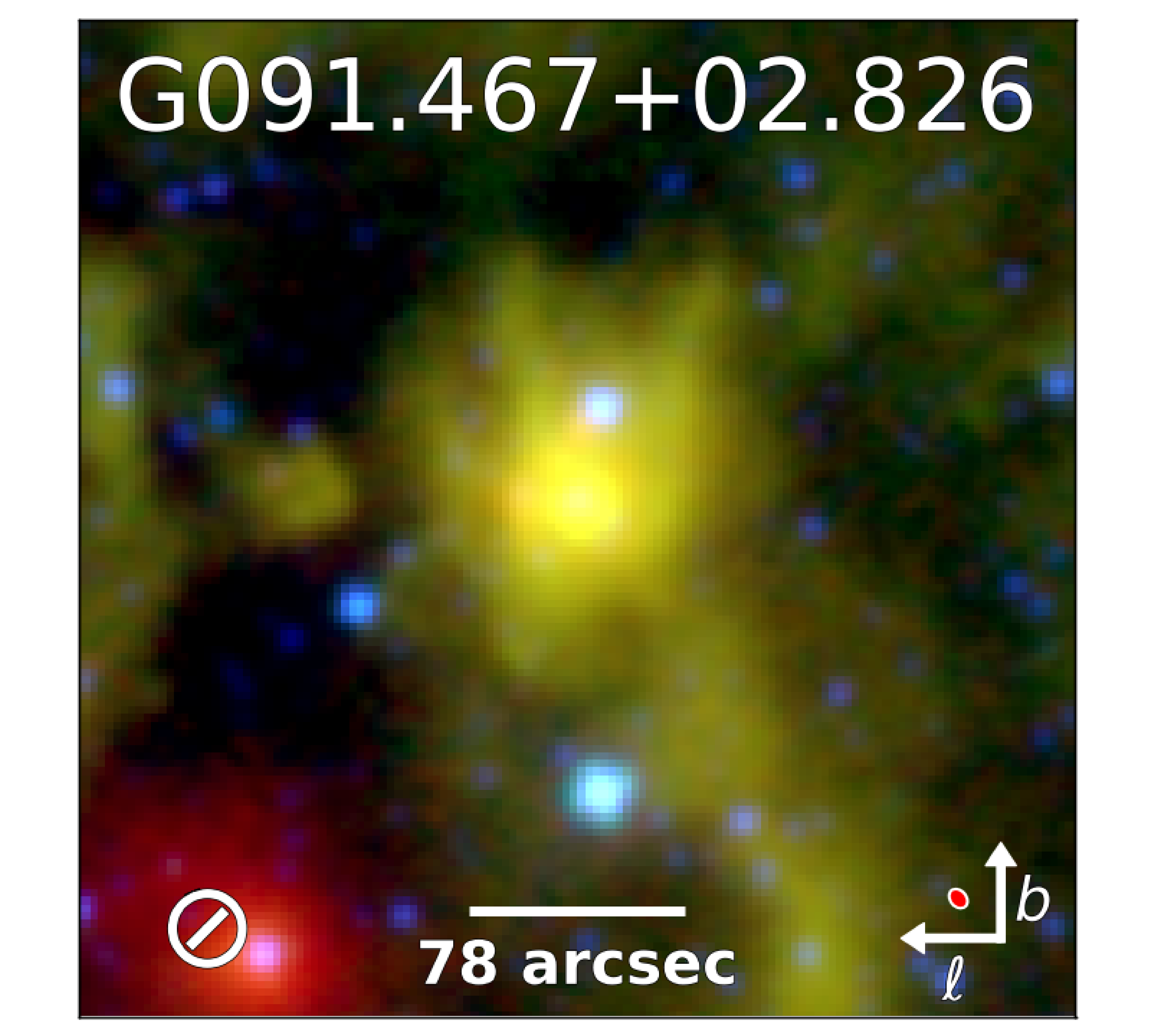}
\includegraphics[width=\figSize]{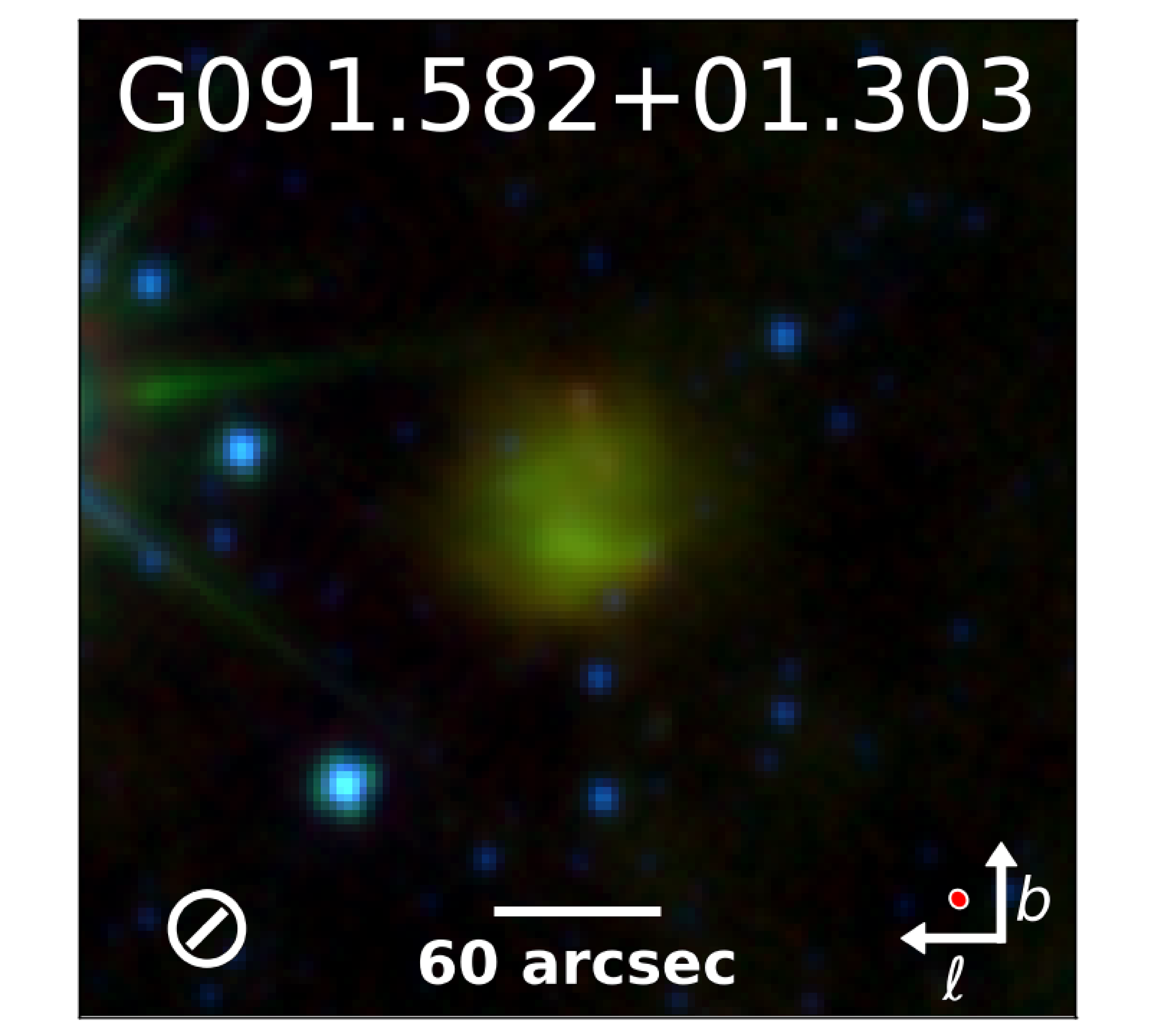}
\includegraphics[width=\figSize]{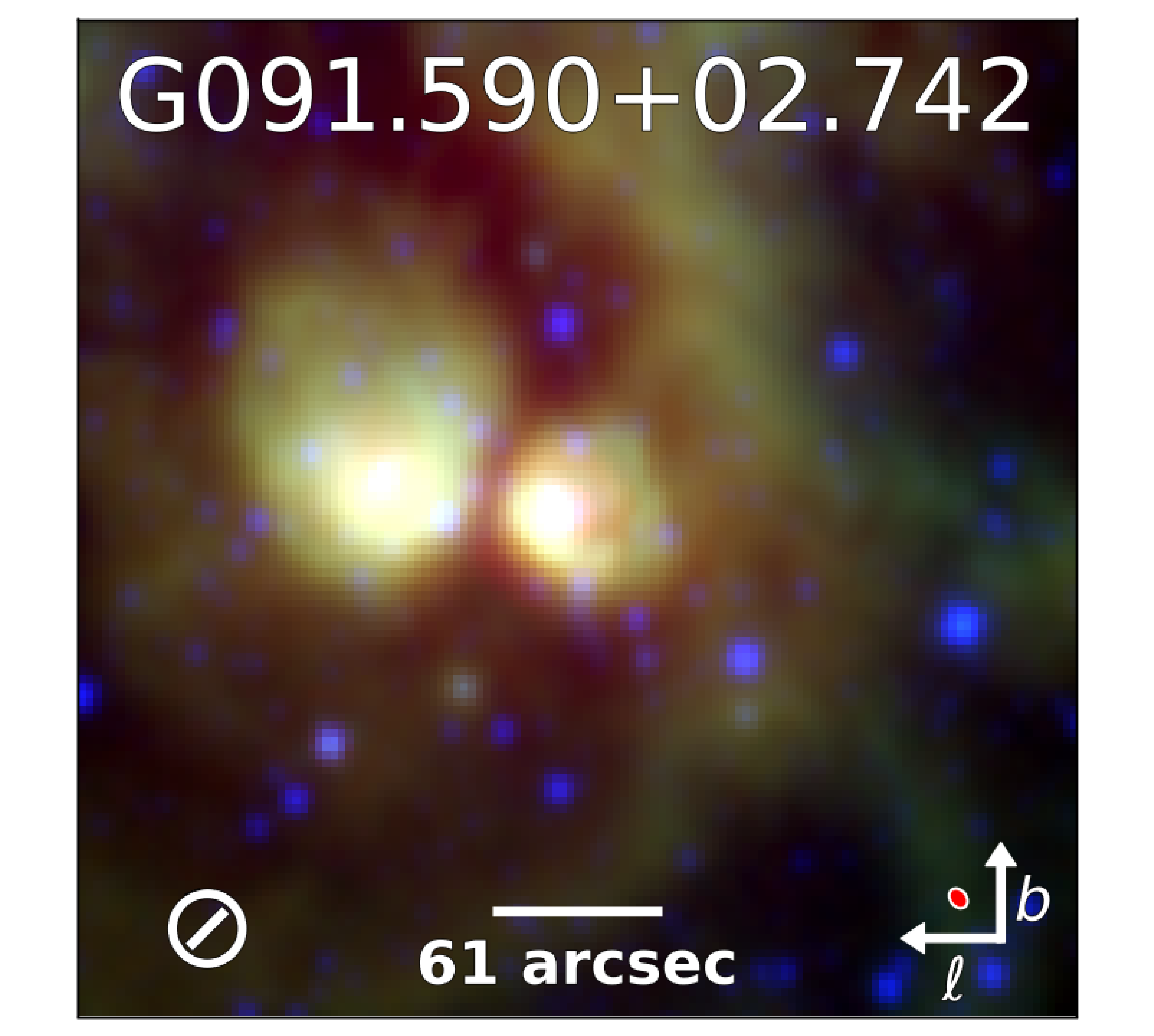}\\
\includegraphics[width=\figSize]{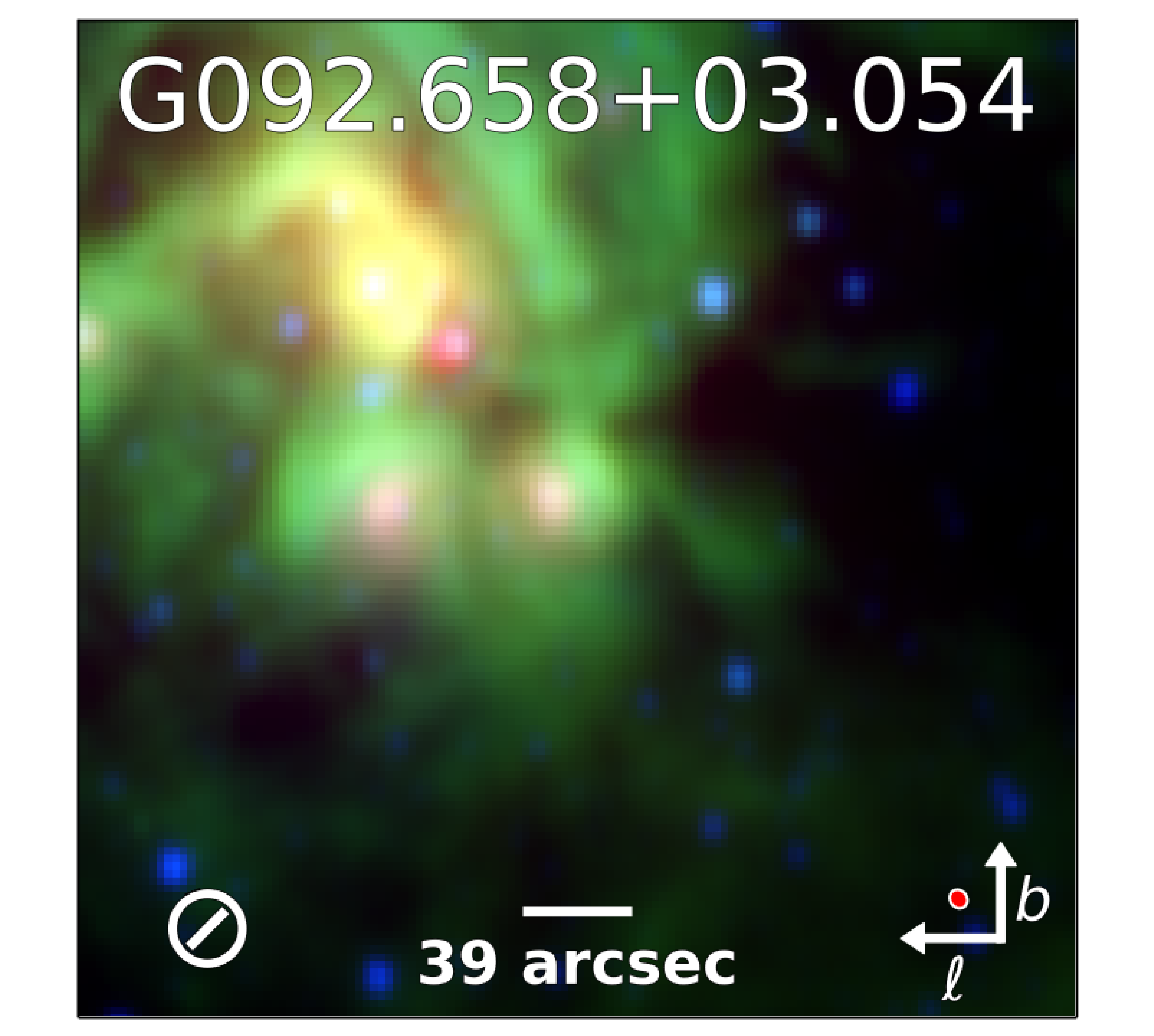} 
\includegraphics[width=\figSize]{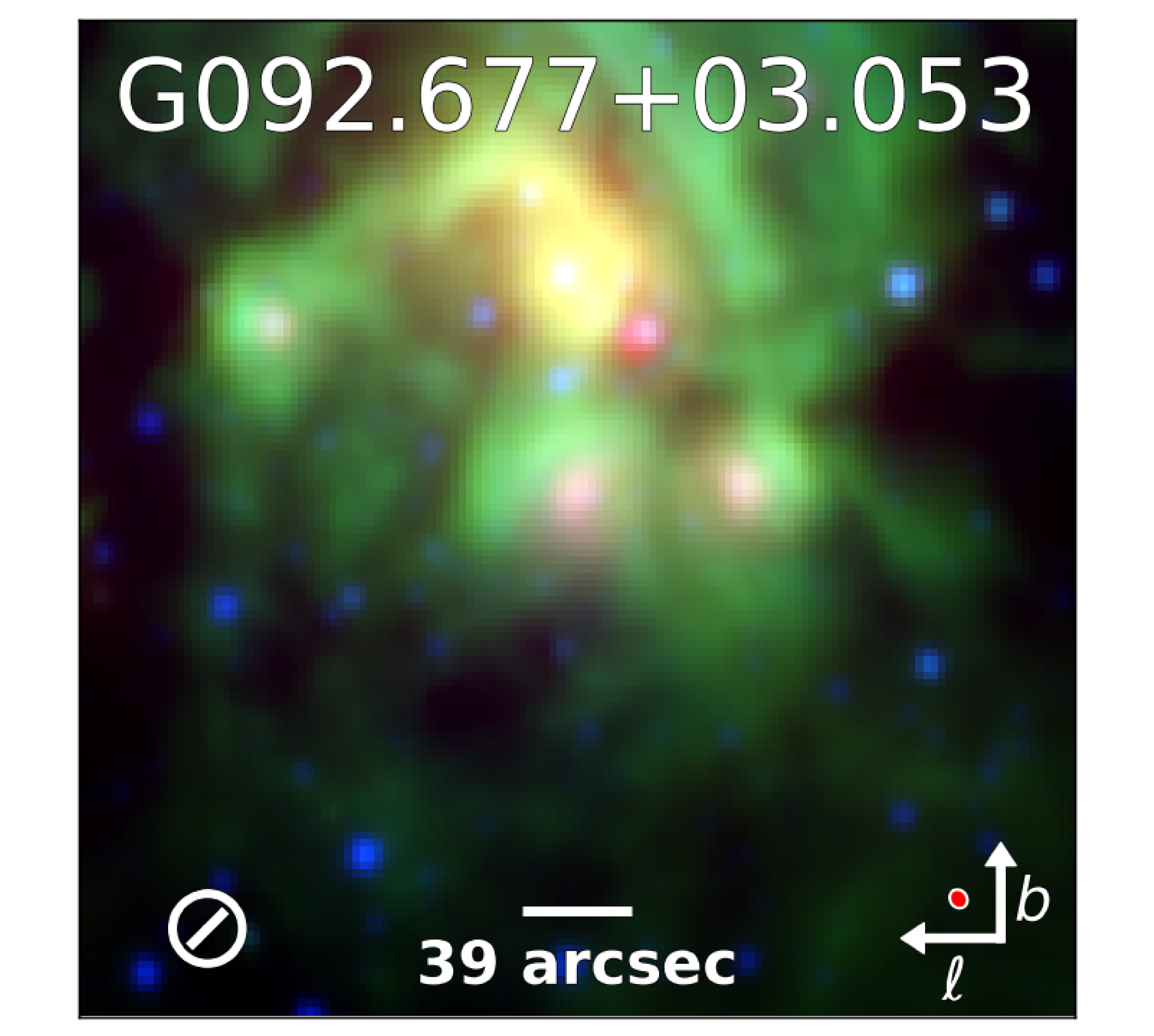}
\includegraphics[width=\figSize]{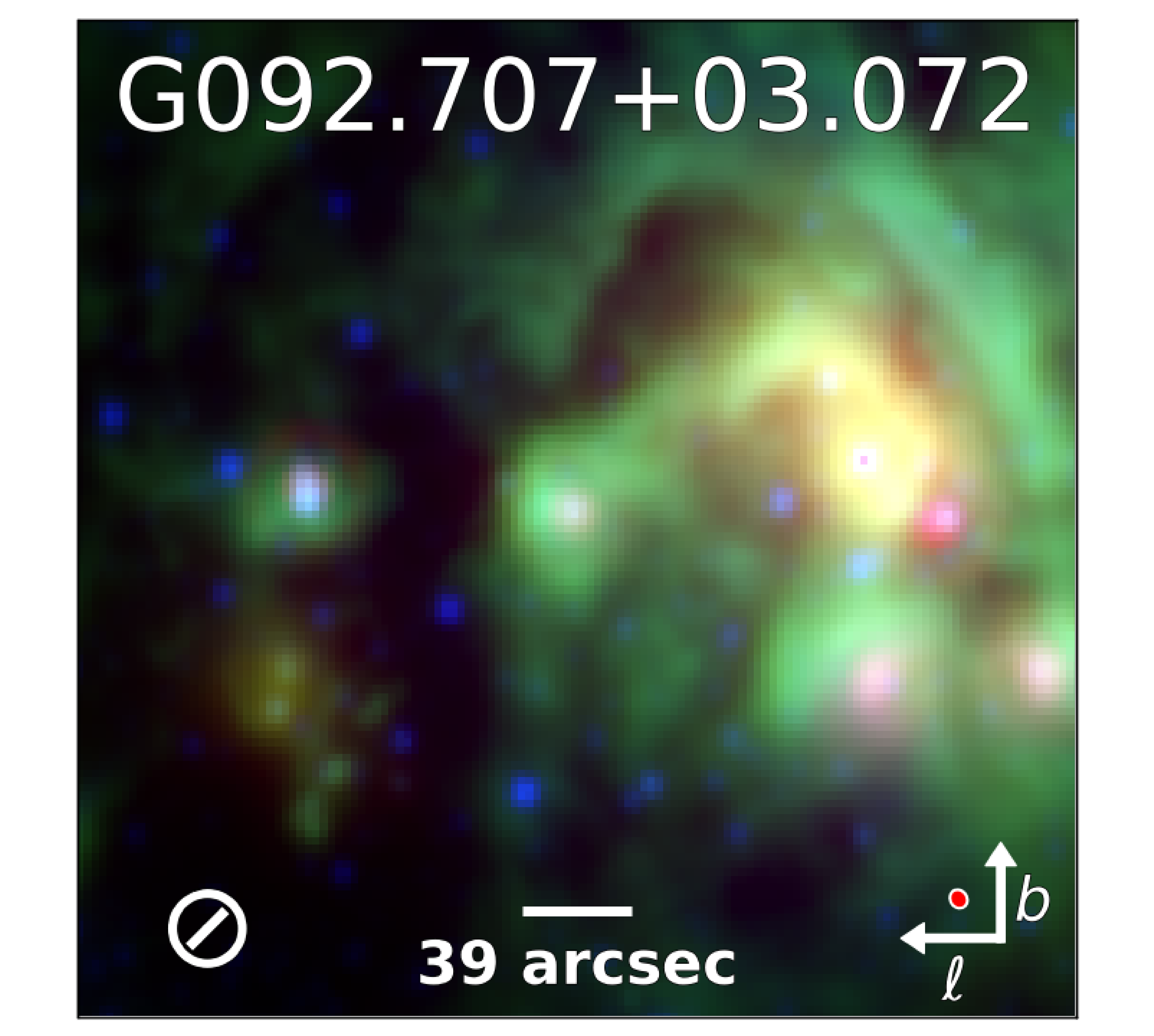} 
\end{figure*}
\begin{figure*}[!htb]
\includegraphics[width=\figSize]{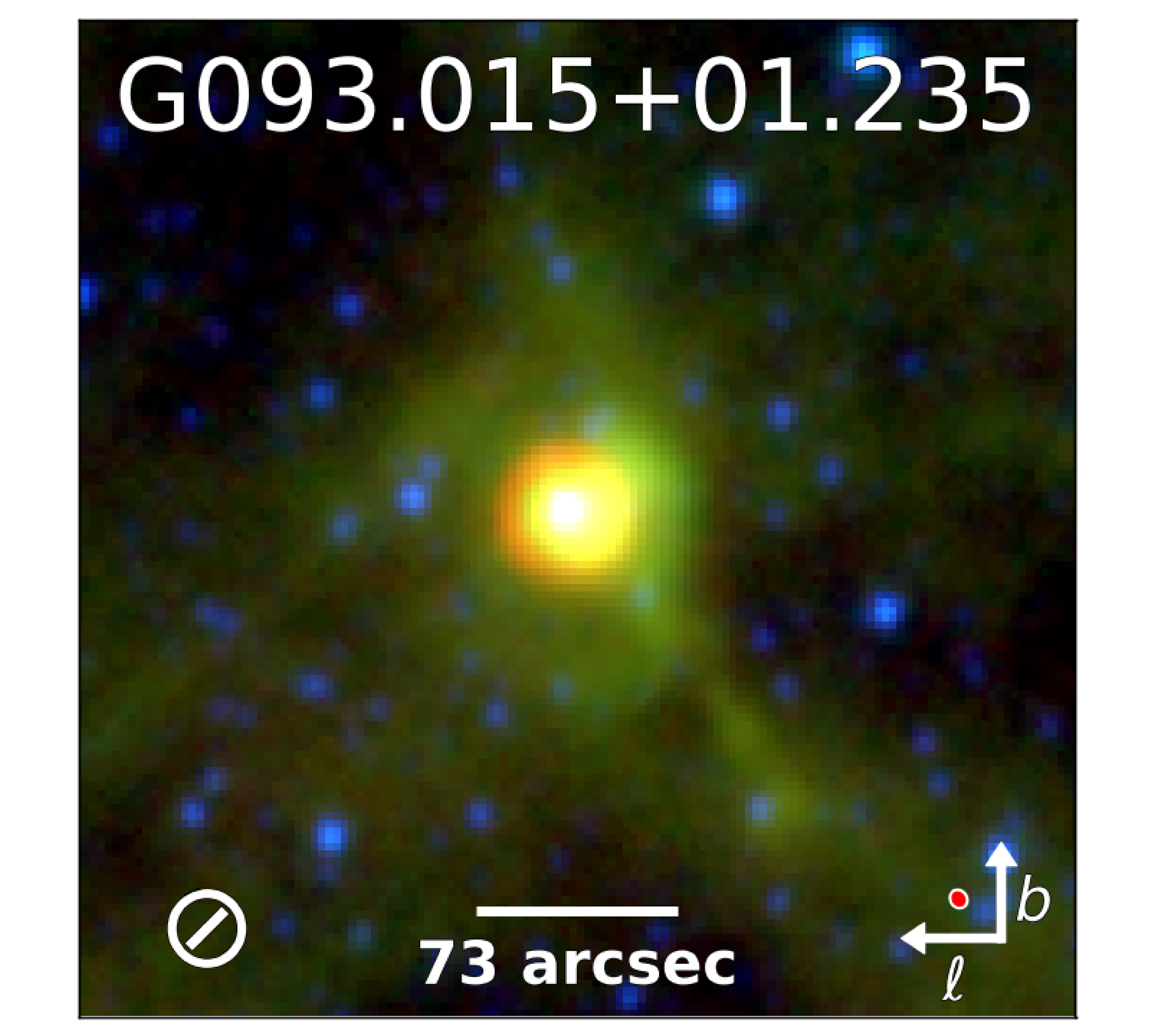}
\includegraphics[width=\figSize]{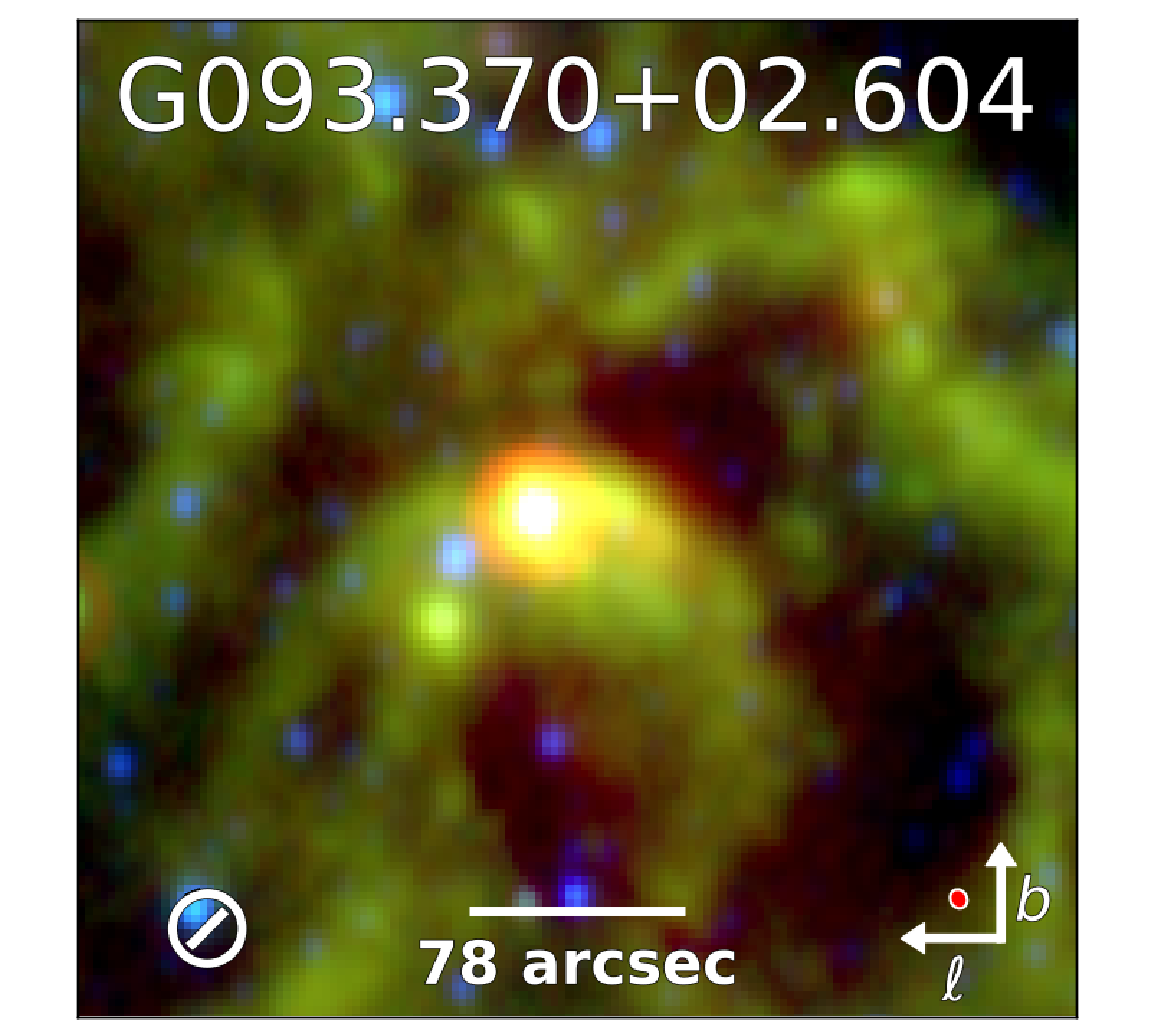}
\includegraphics[width=\figSize]{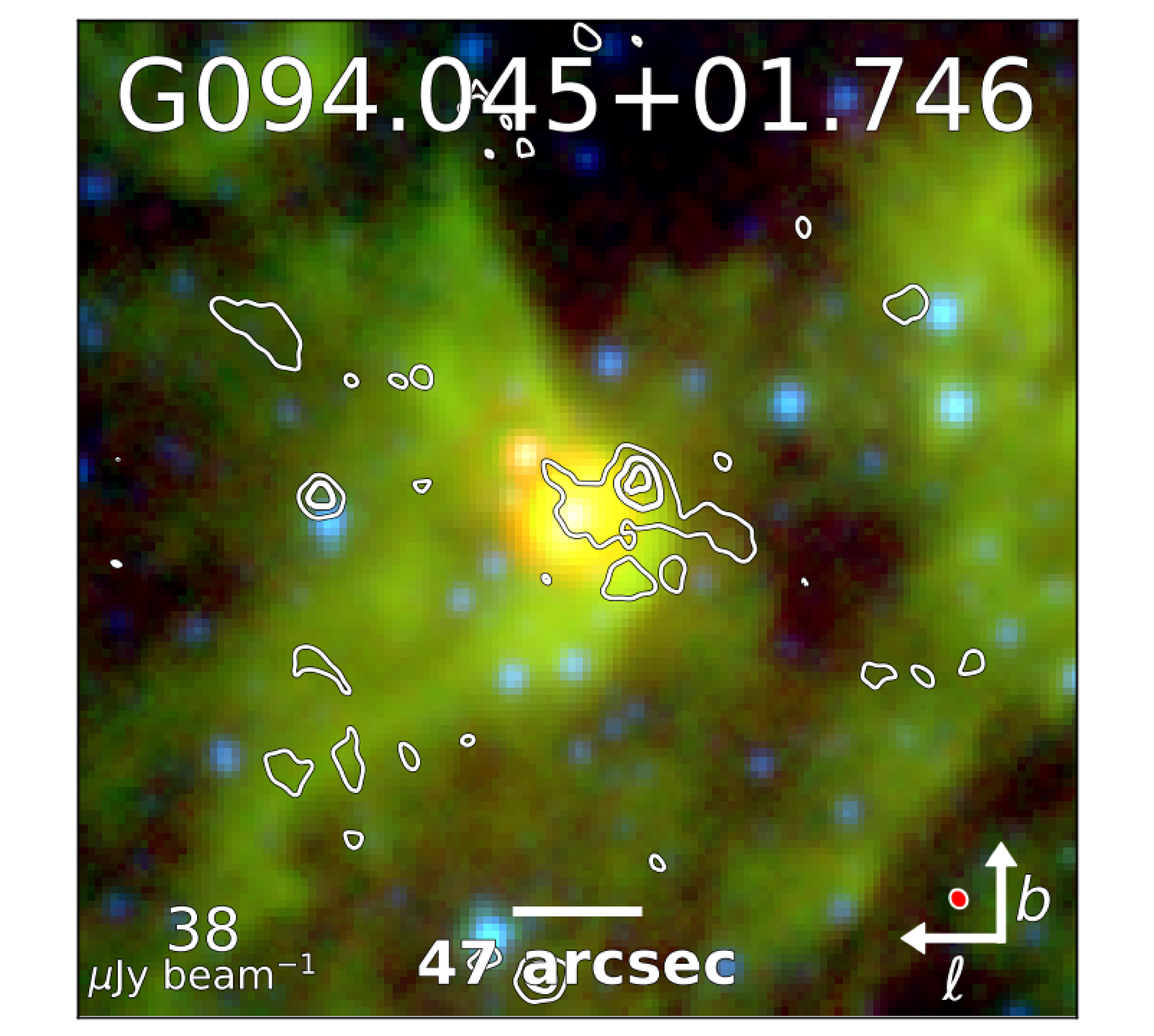}\\
\includegraphics[width=\figSize]{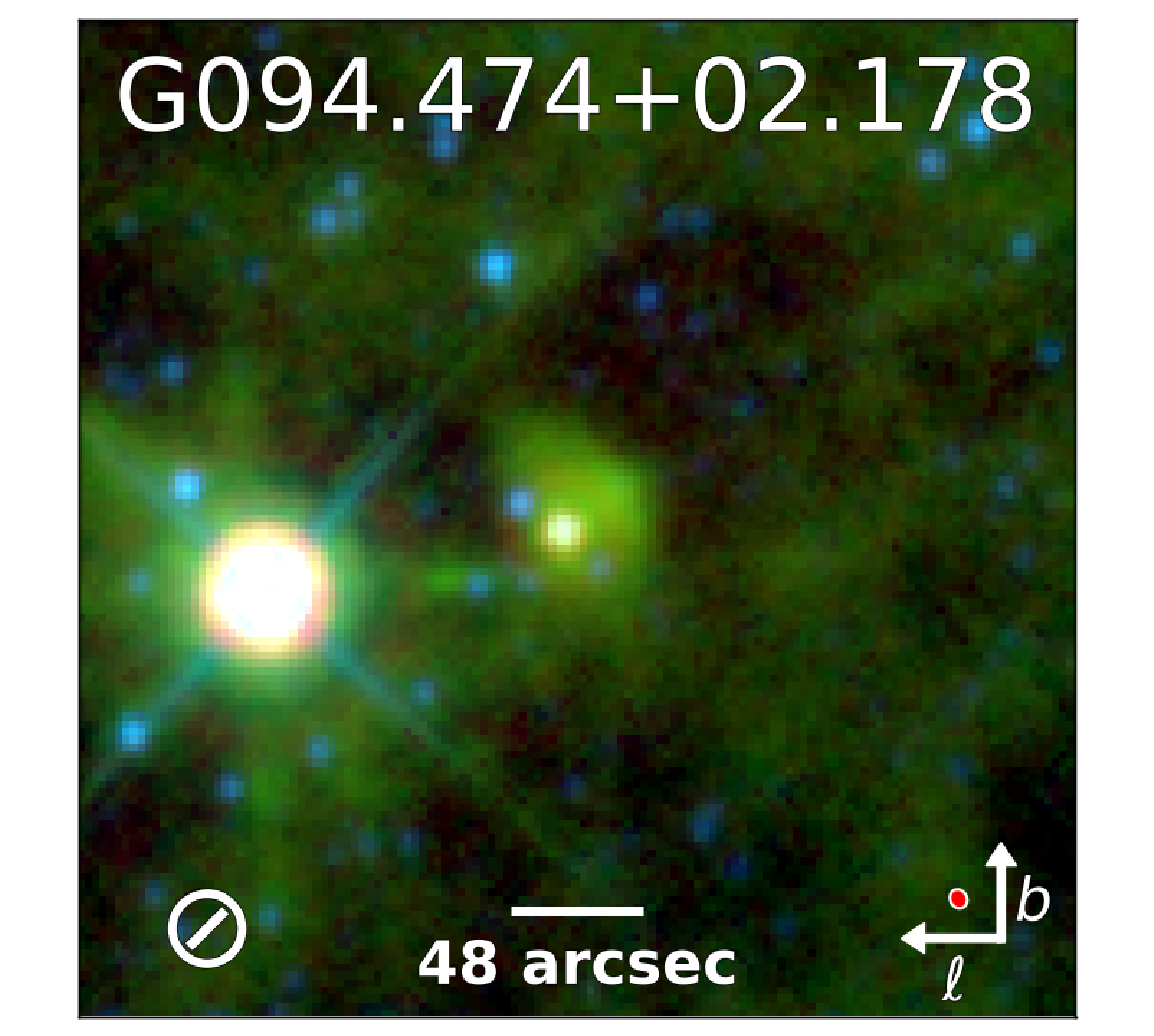}
\includegraphics[width=\figSize]{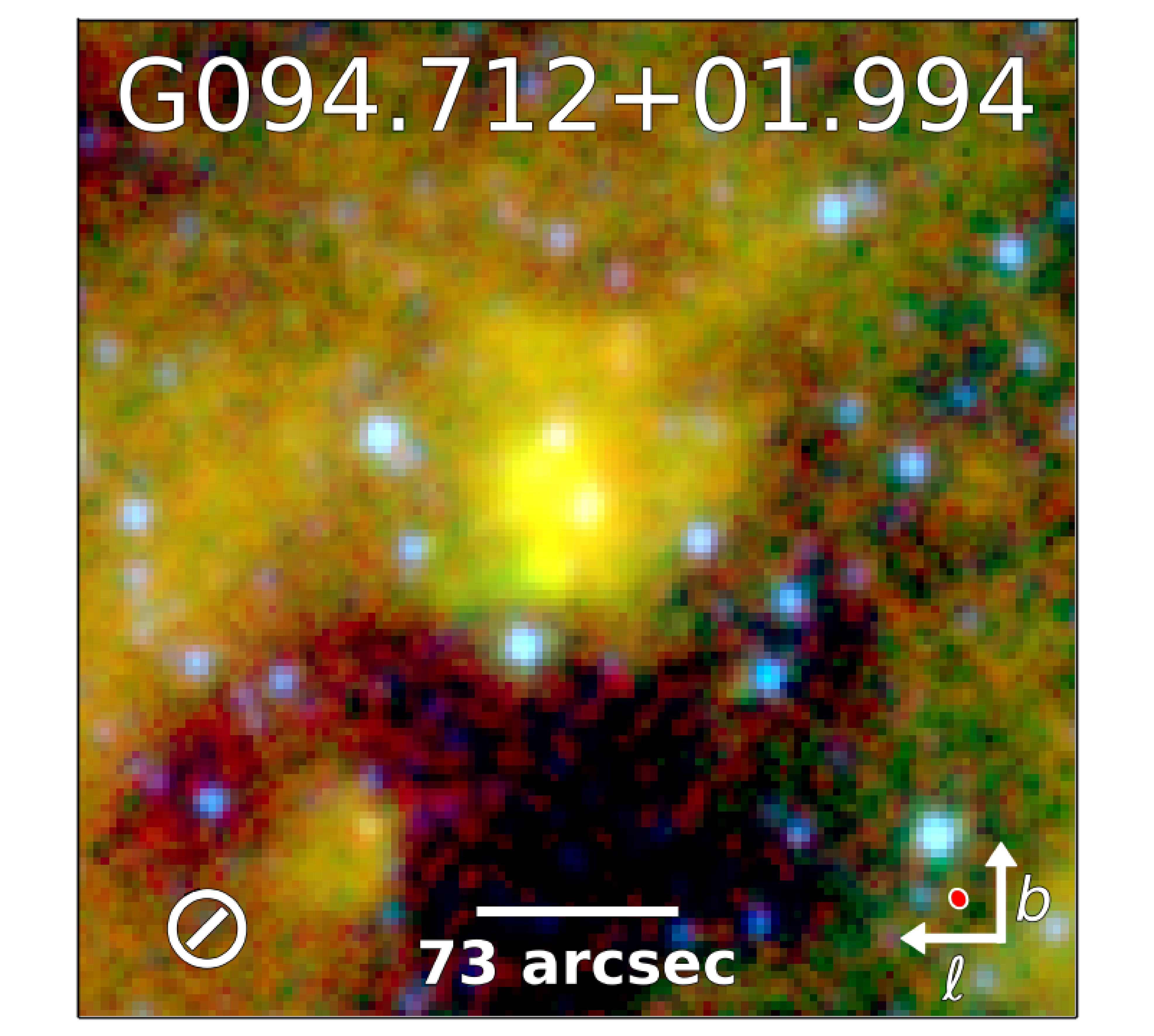}
\includegraphics[width=\figSize]{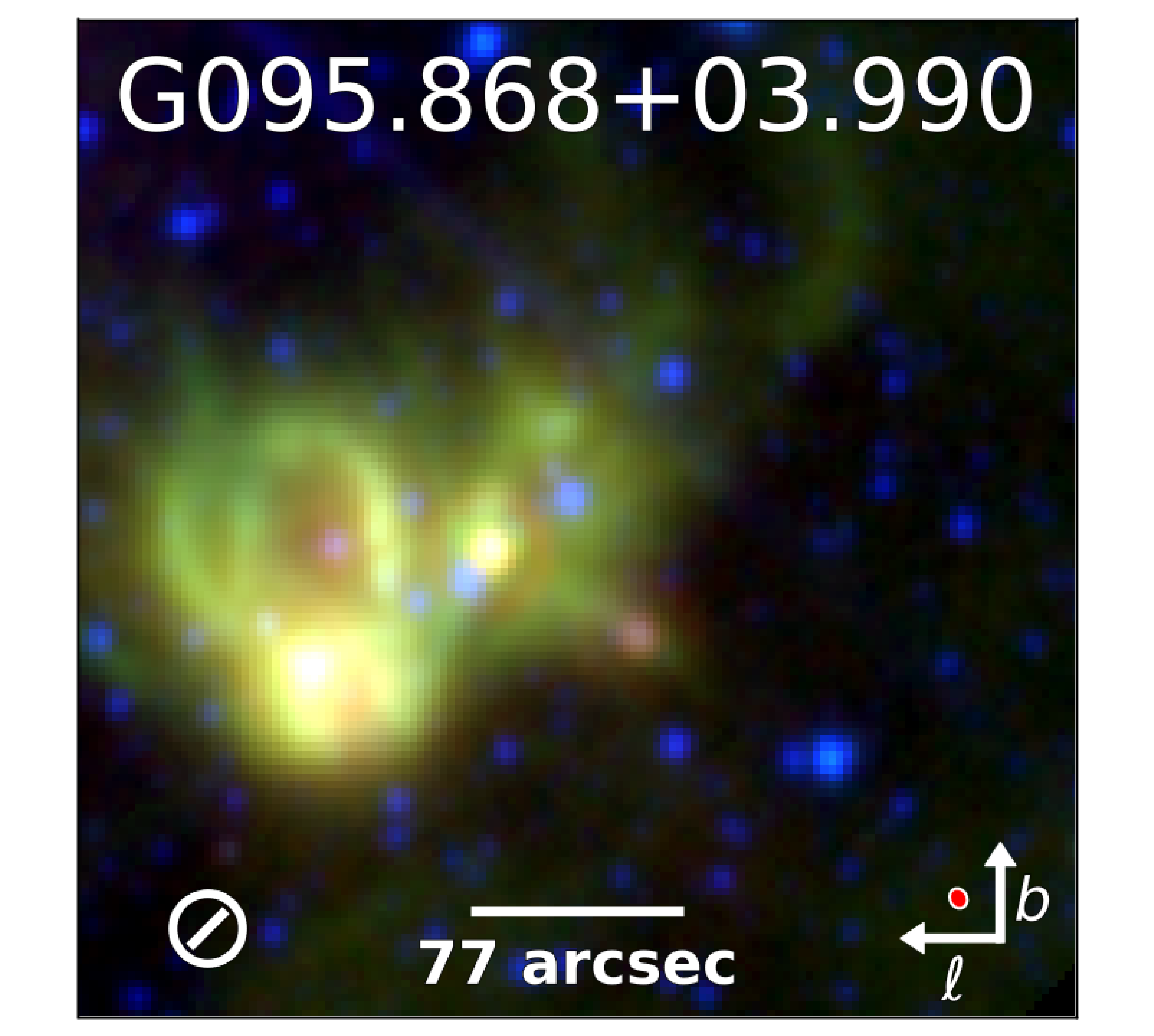}\\
\includegraphics[width=\figSize]{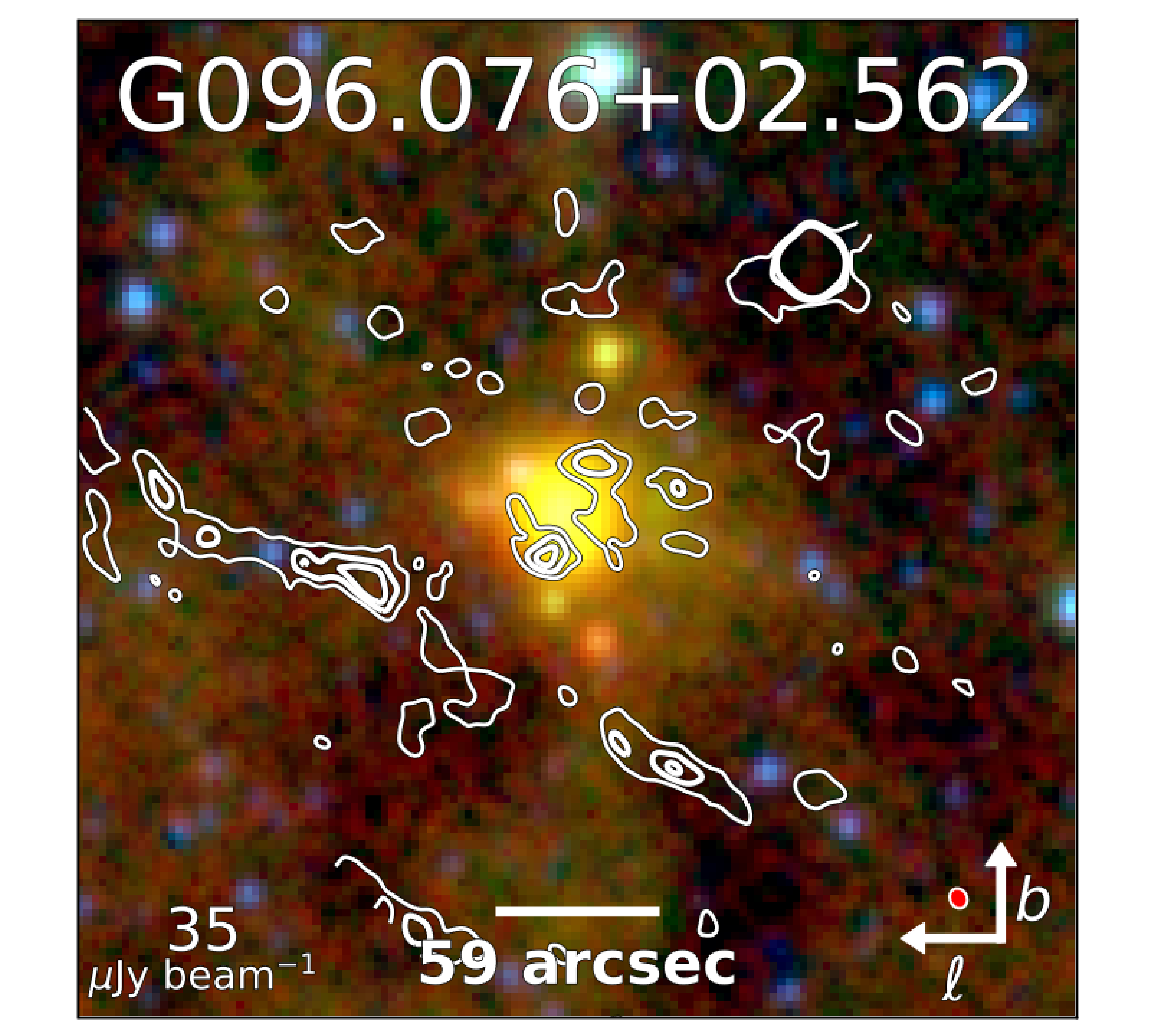}
\includegraphics[width=\figSize]{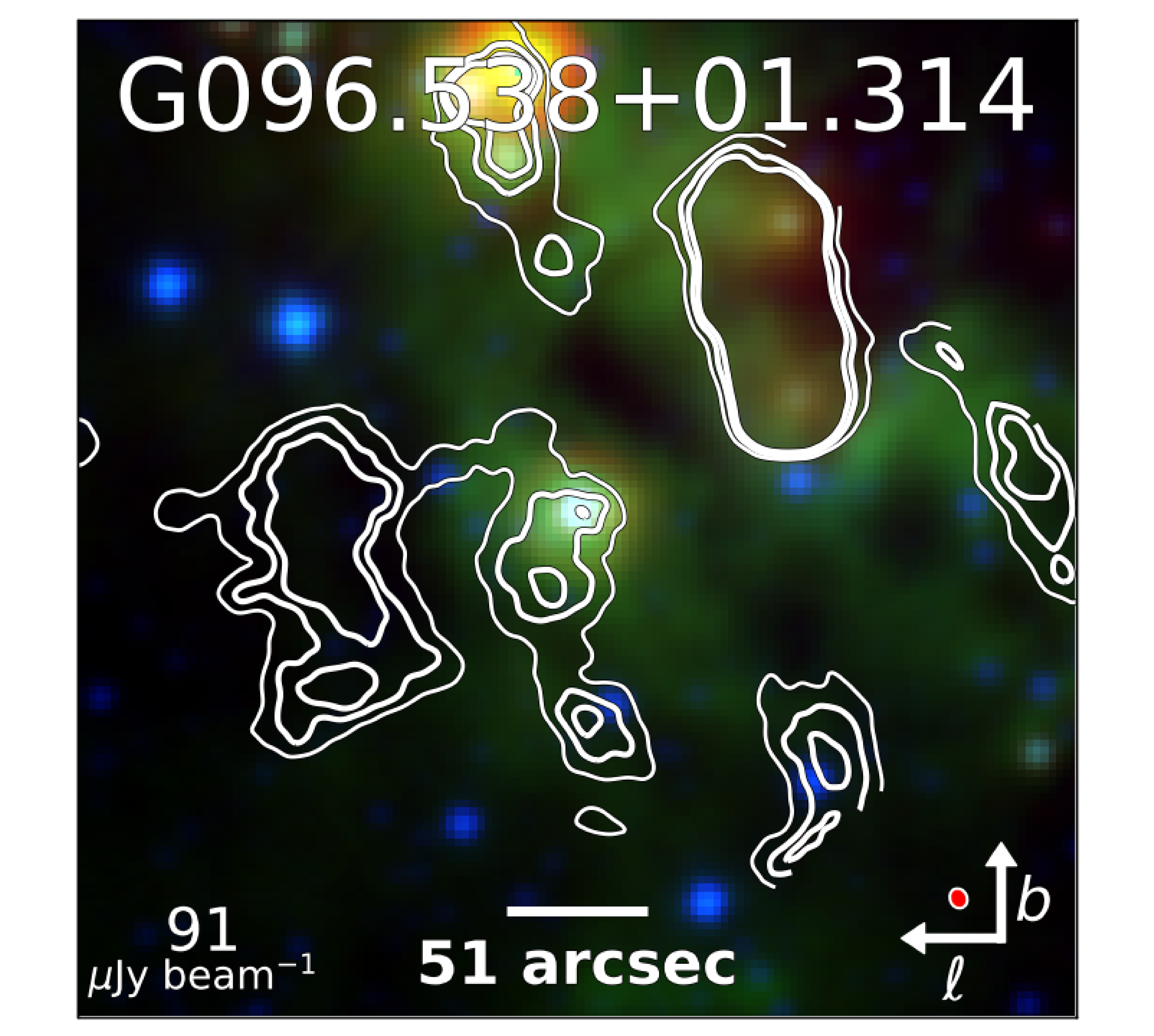}
\includegraphics[width=\figSize]{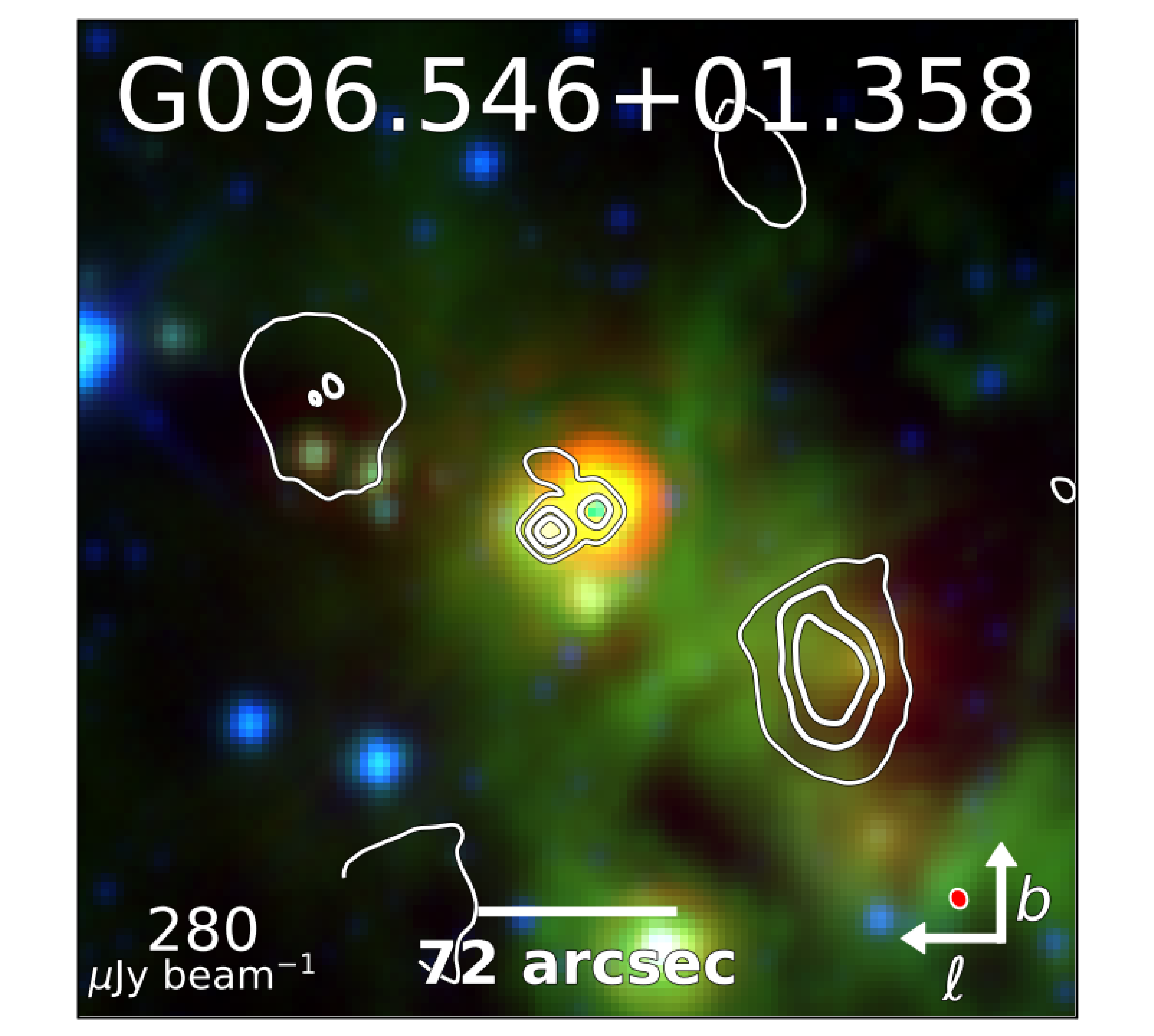}\\
\includegraphics[width=\figSize]{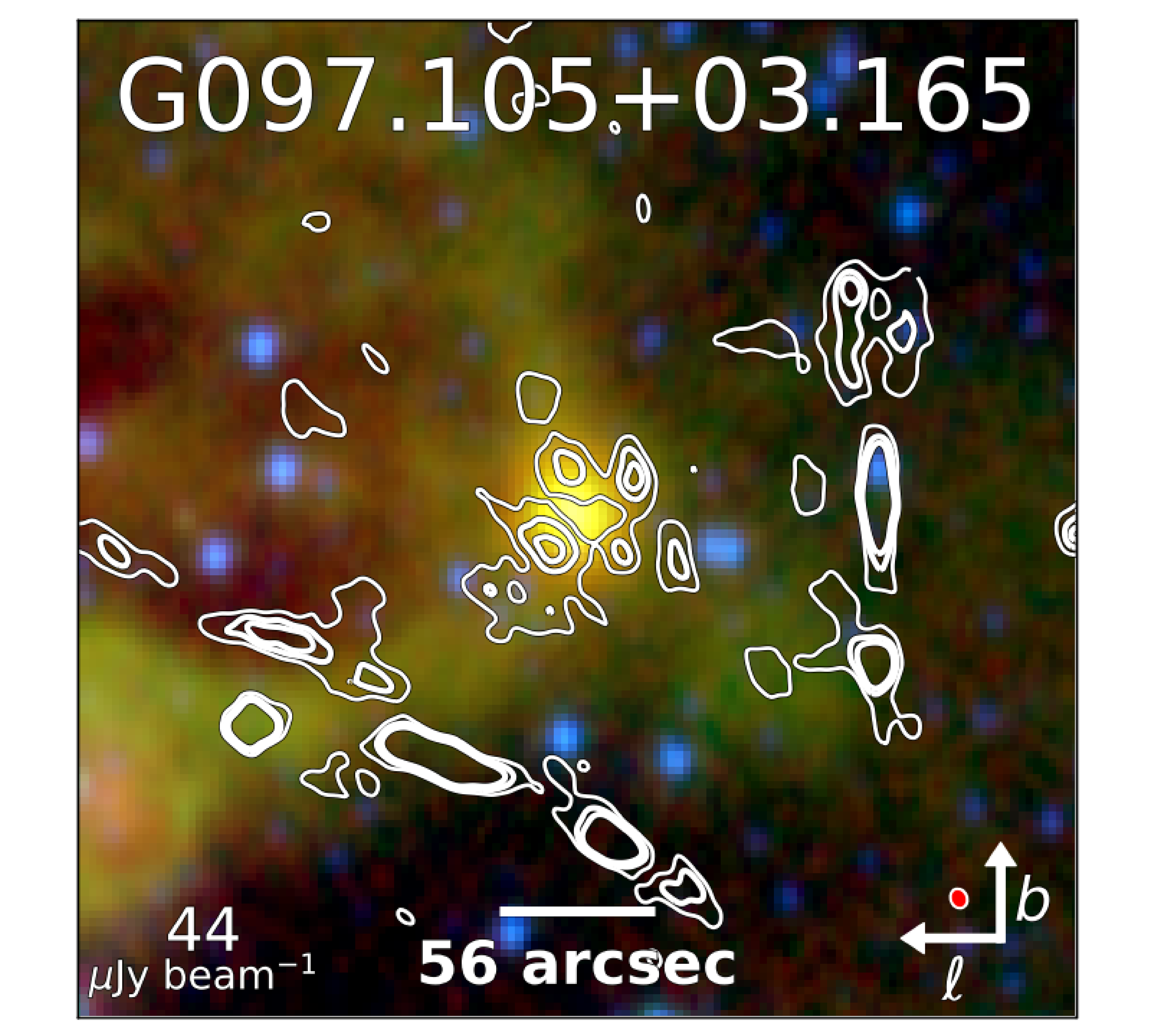}
\includegraphics[width=\figSize]{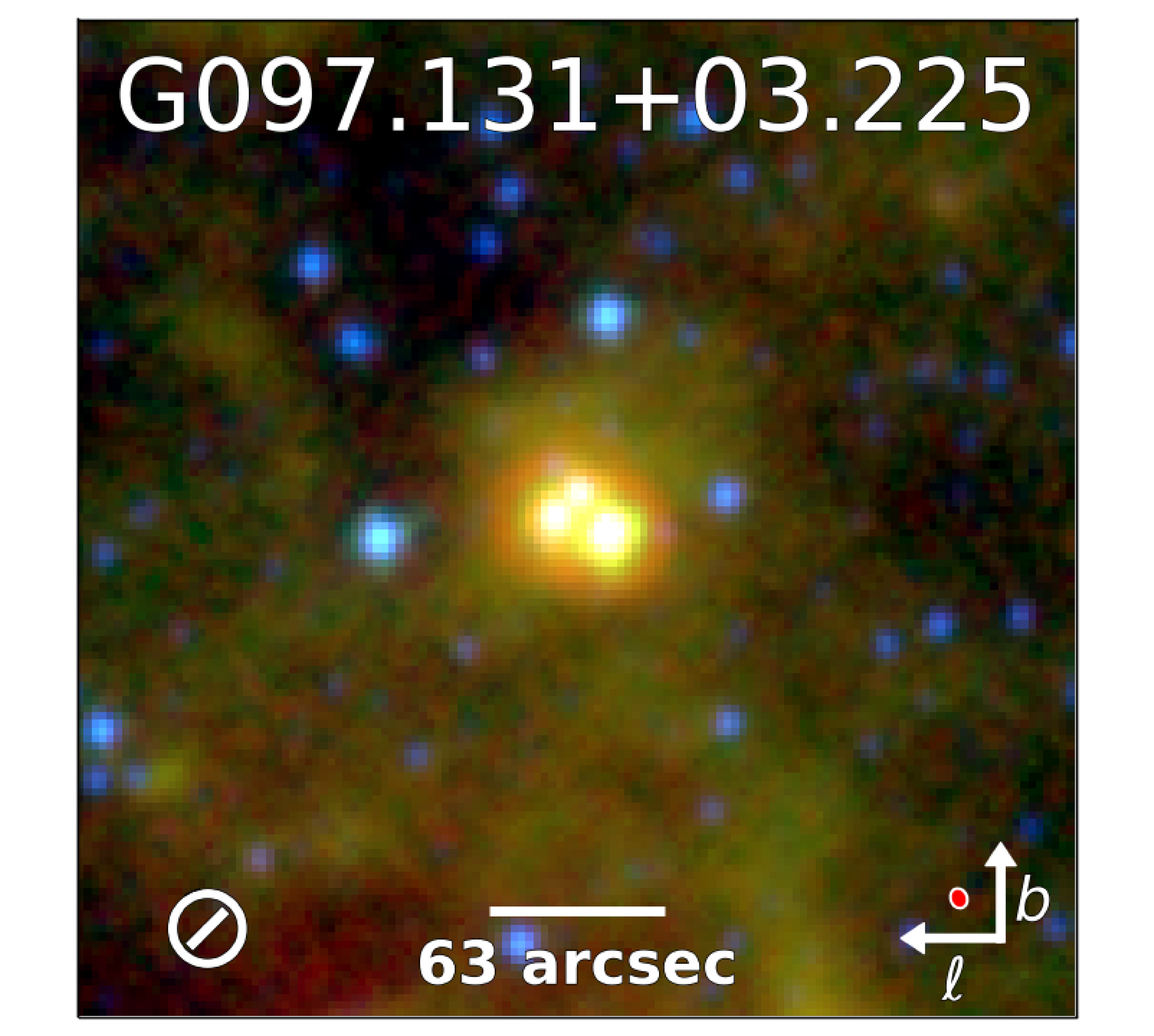}
\includegraphics[width=\figSize]{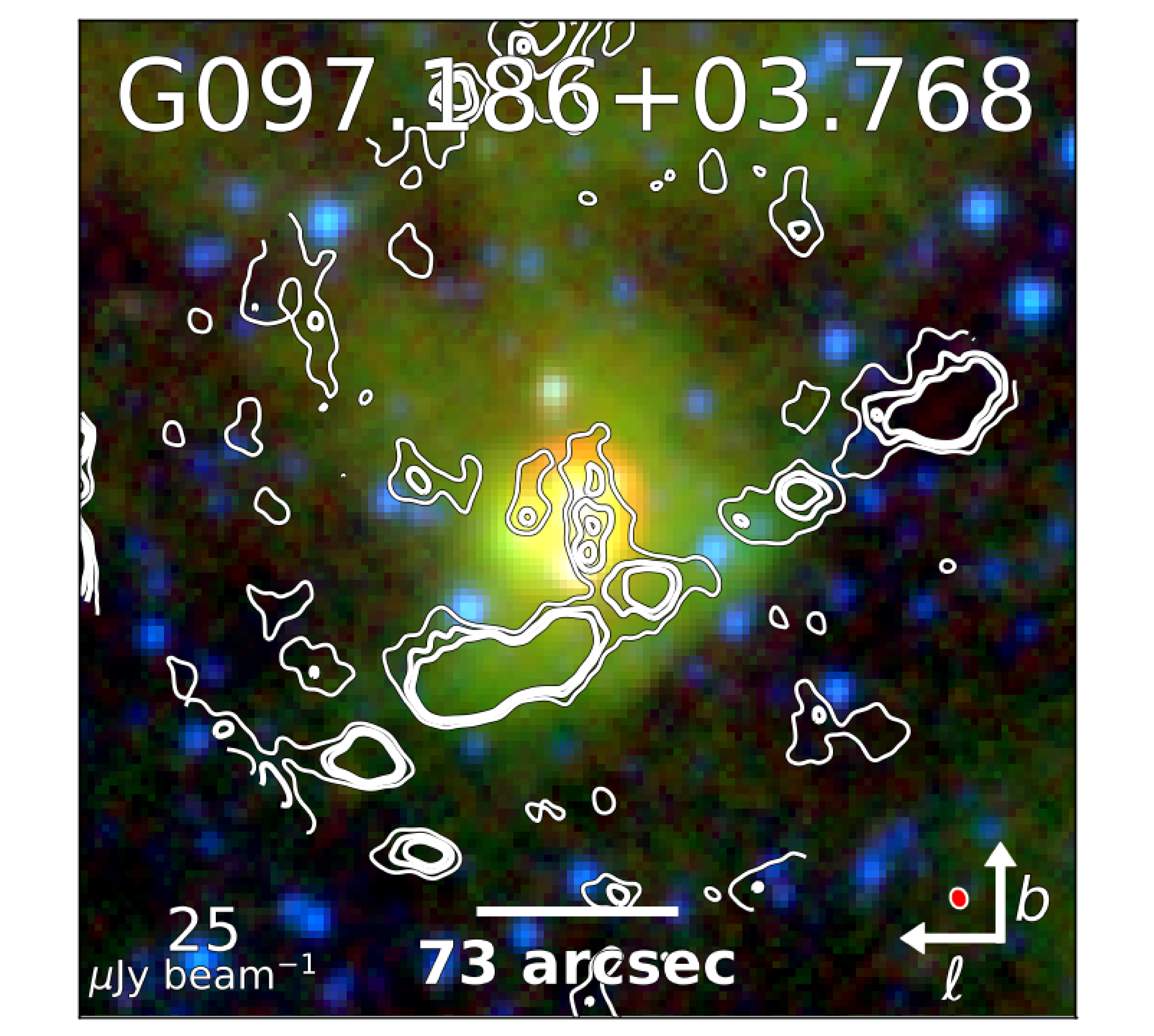}
\end{figure*}
\begin{figure*}[!htb]
\includegraphics[width=\figSize]{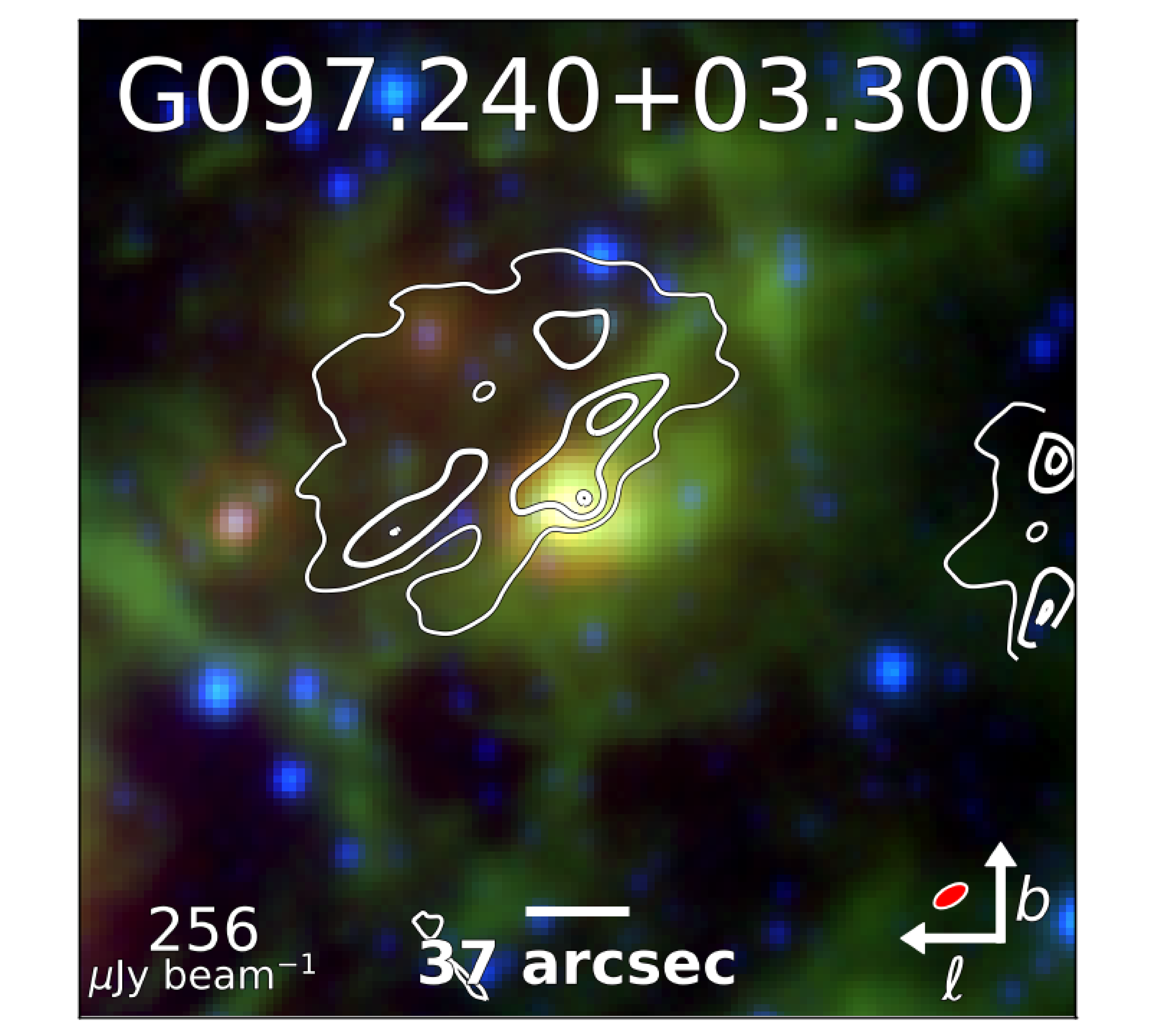}
\includegraphics[width=\figSize]{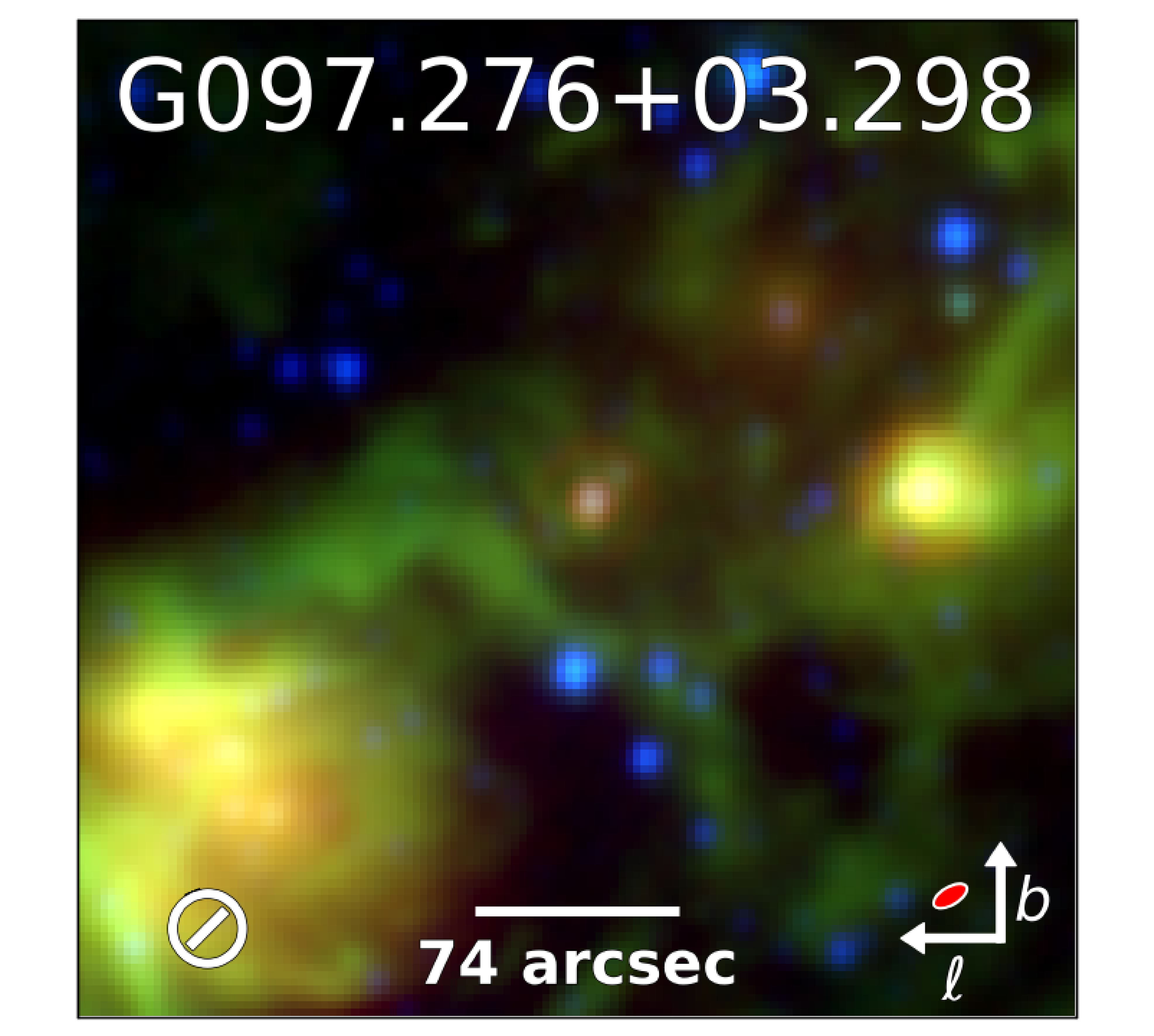}
\includegraphics[width=\figSize]{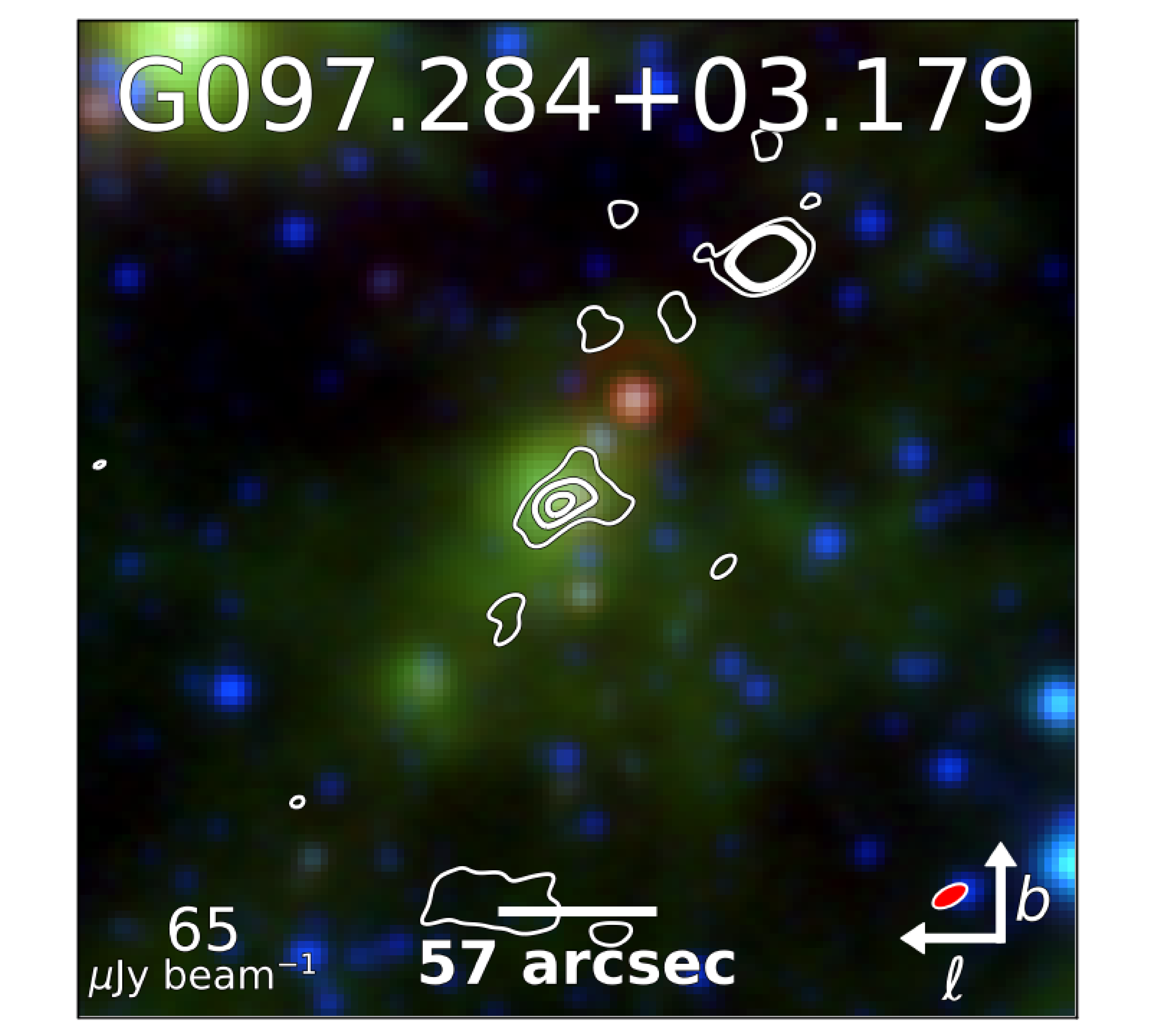}\\
\includegraphics[width=\figSize]{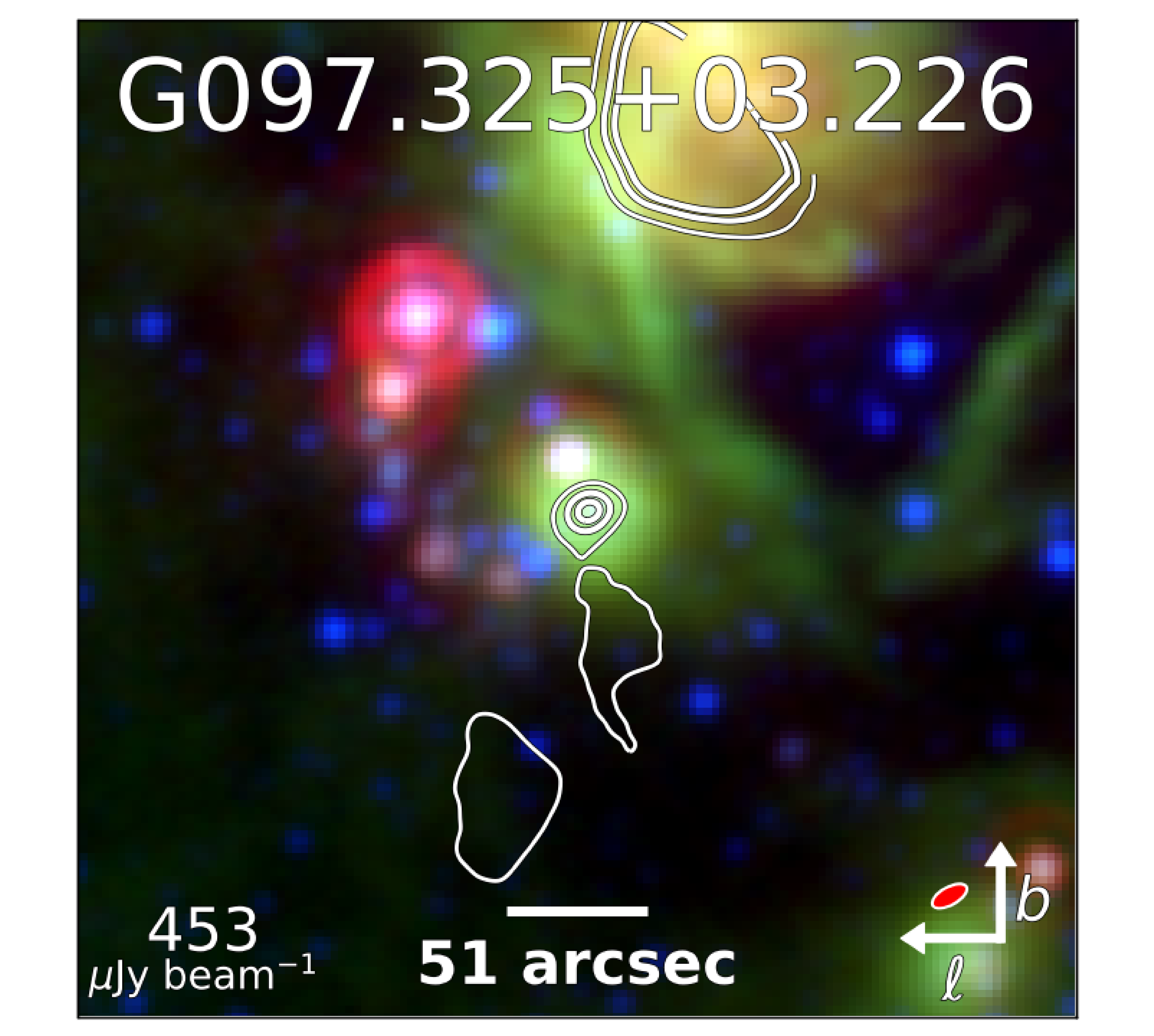}
\includegraphics[width=\figSize]{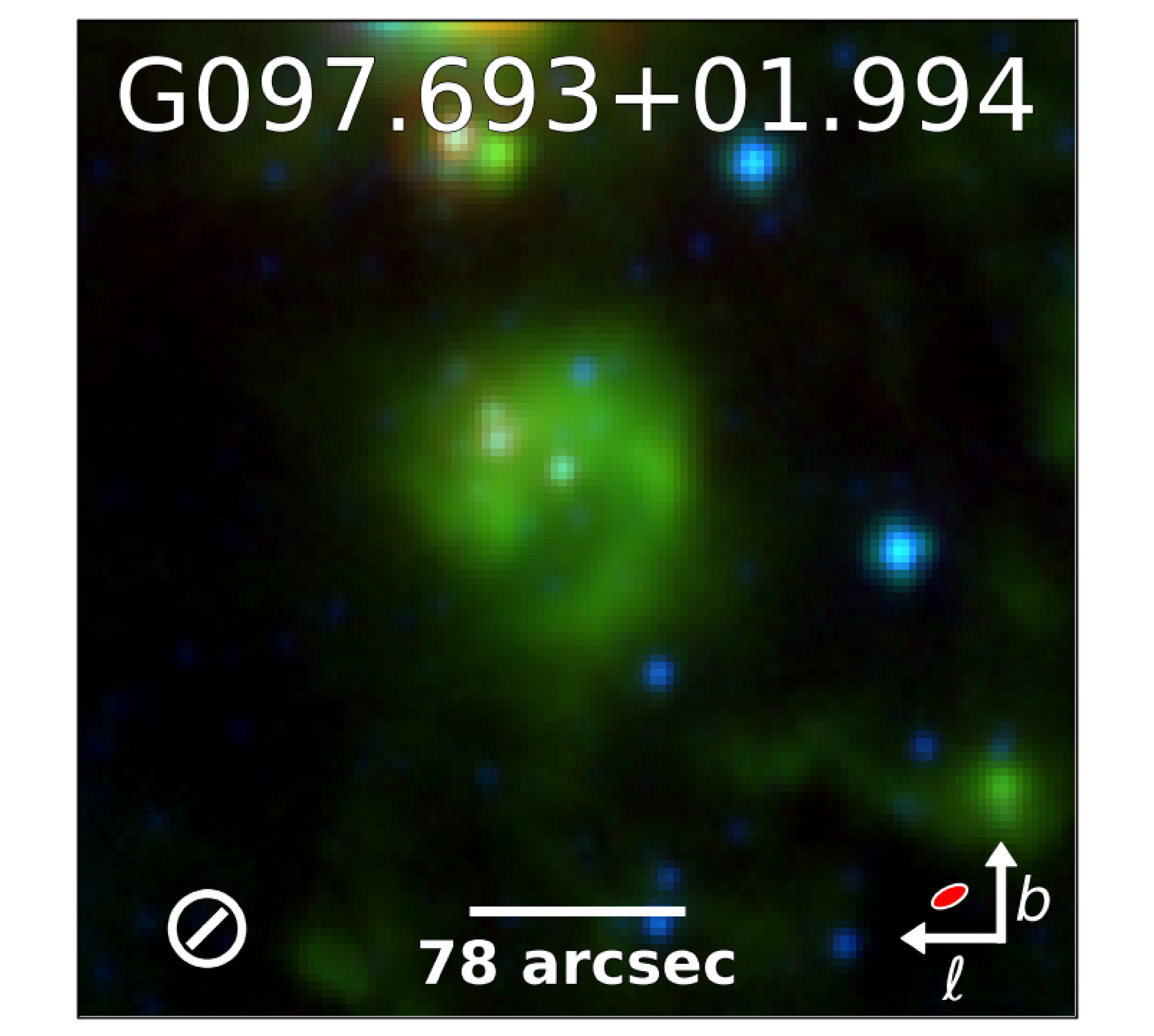}
\includegraphics[width=\figSize]{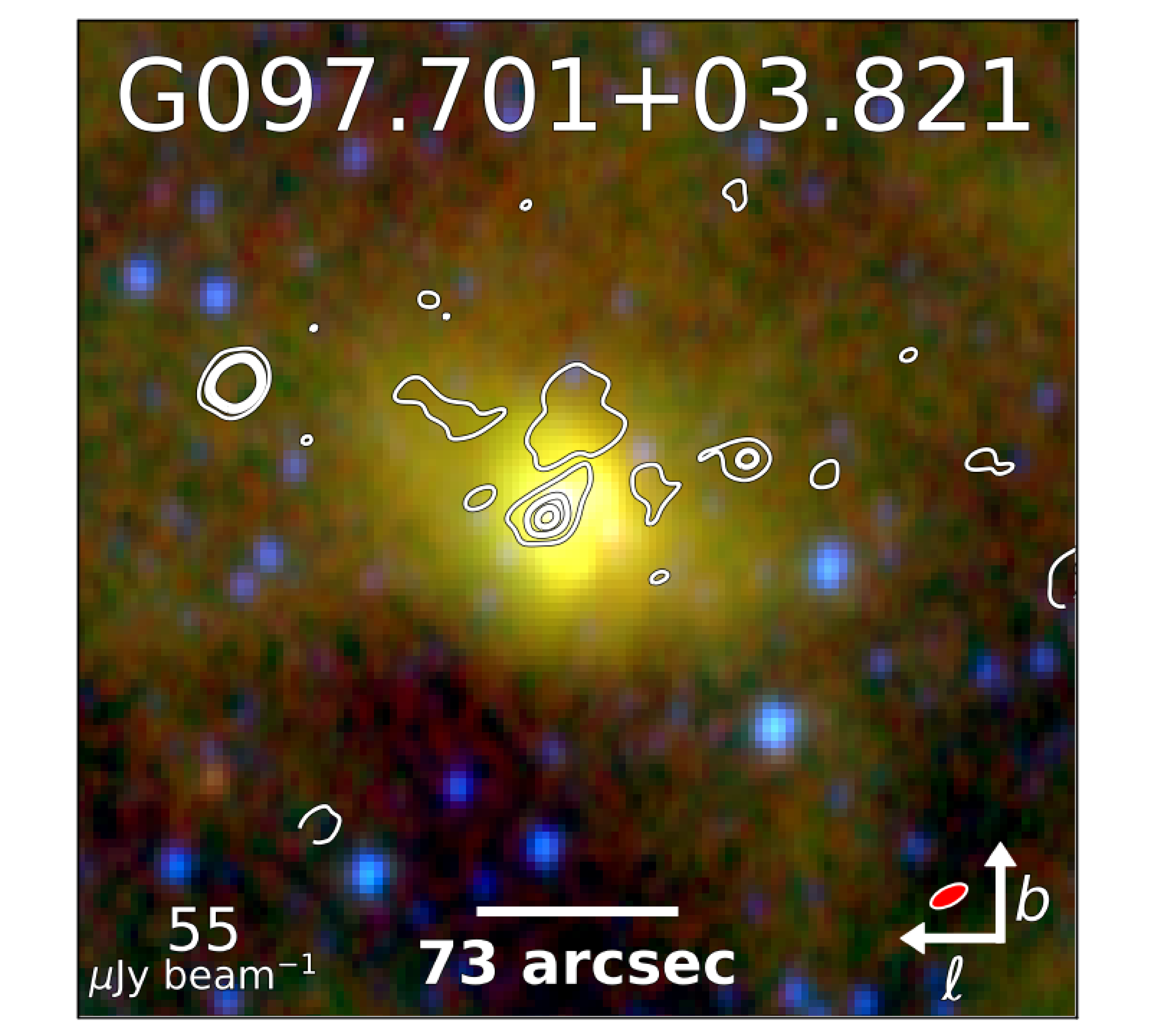}\\
\includegraphics[width=\figSize]{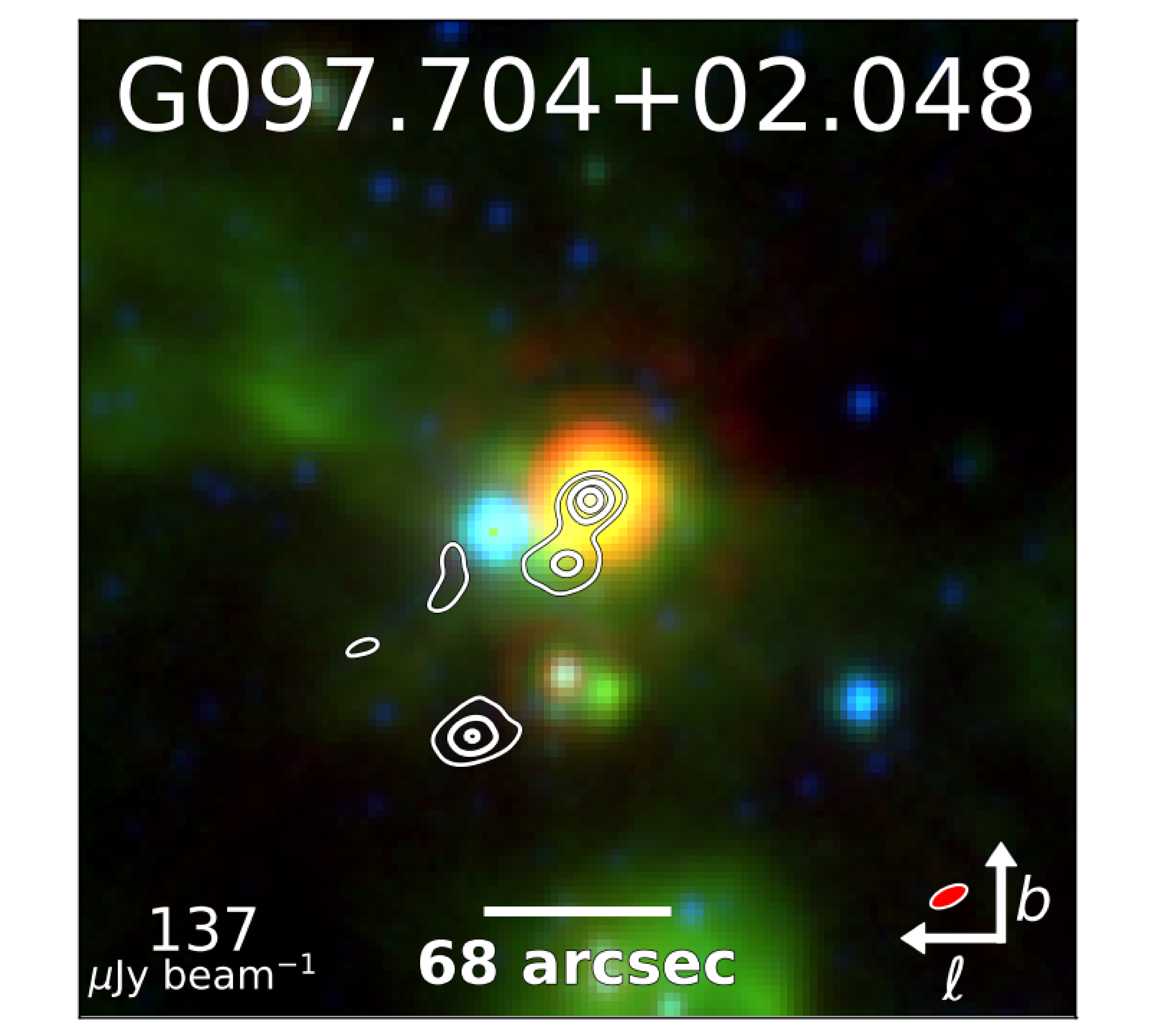}
\includegraphics[width=\figSize]{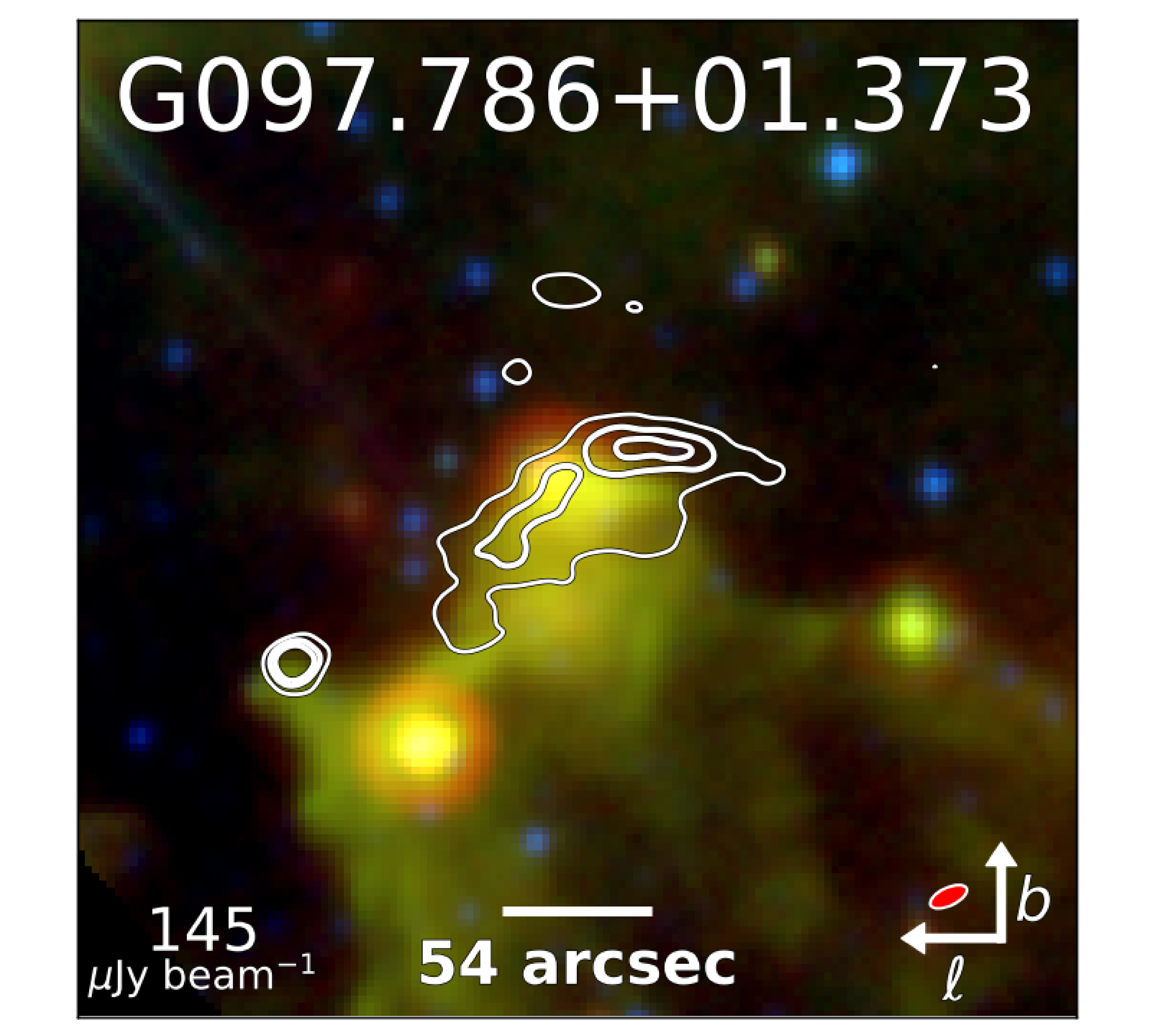}
\includegraphics[width=\figSize]{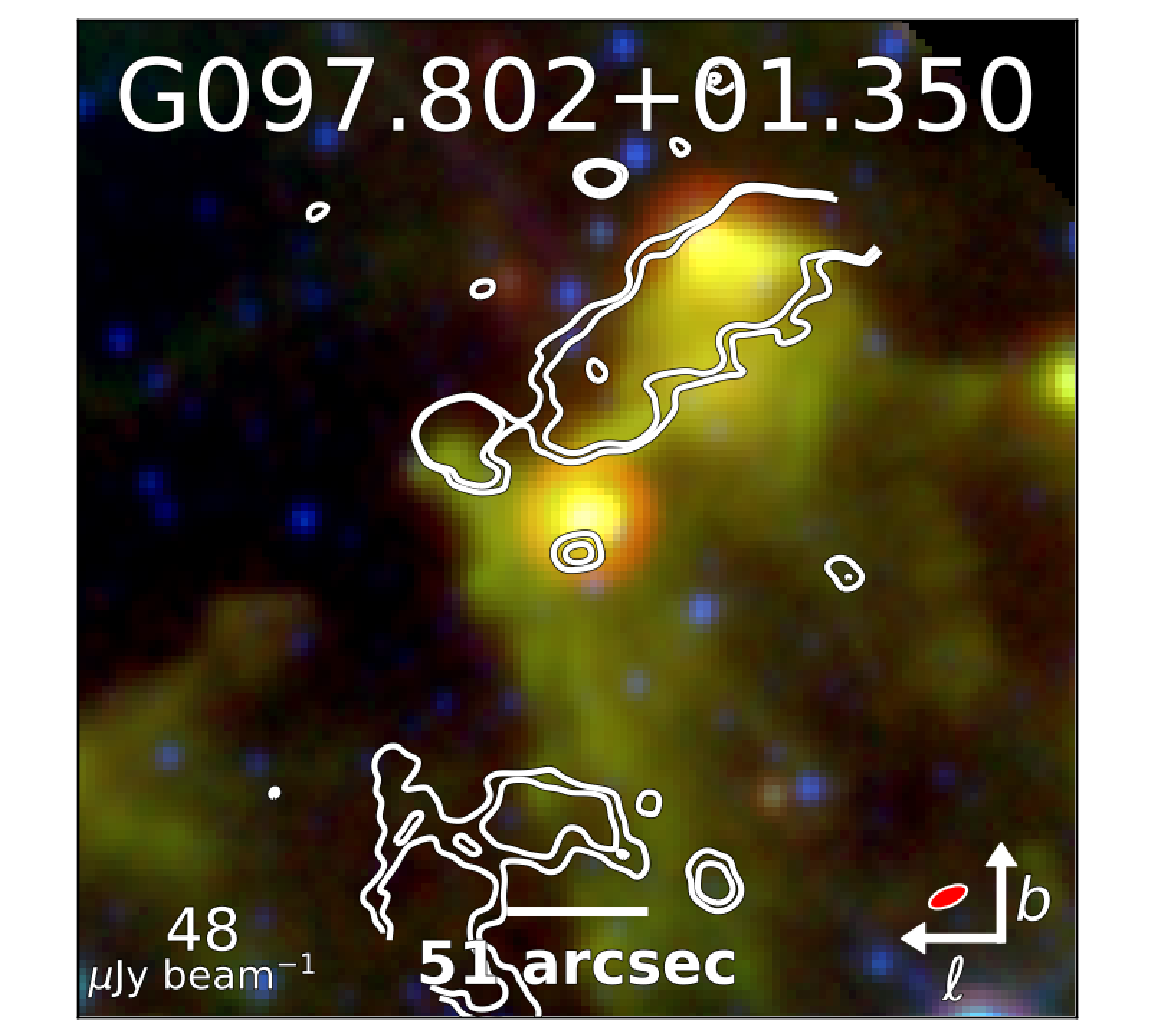}\\
\includegraphics[width=\figSize]{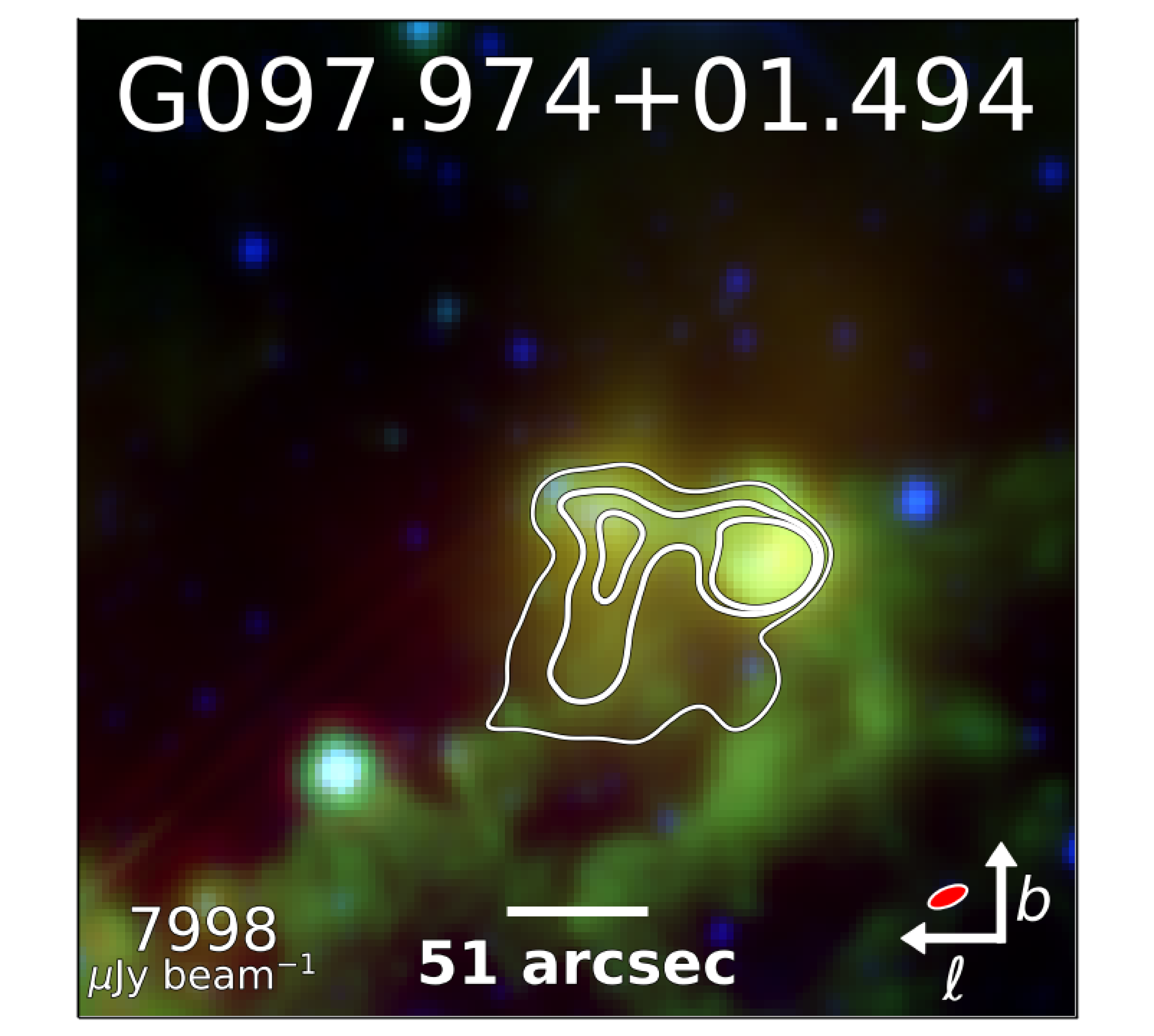}
\includegraphics[width=\figSize]{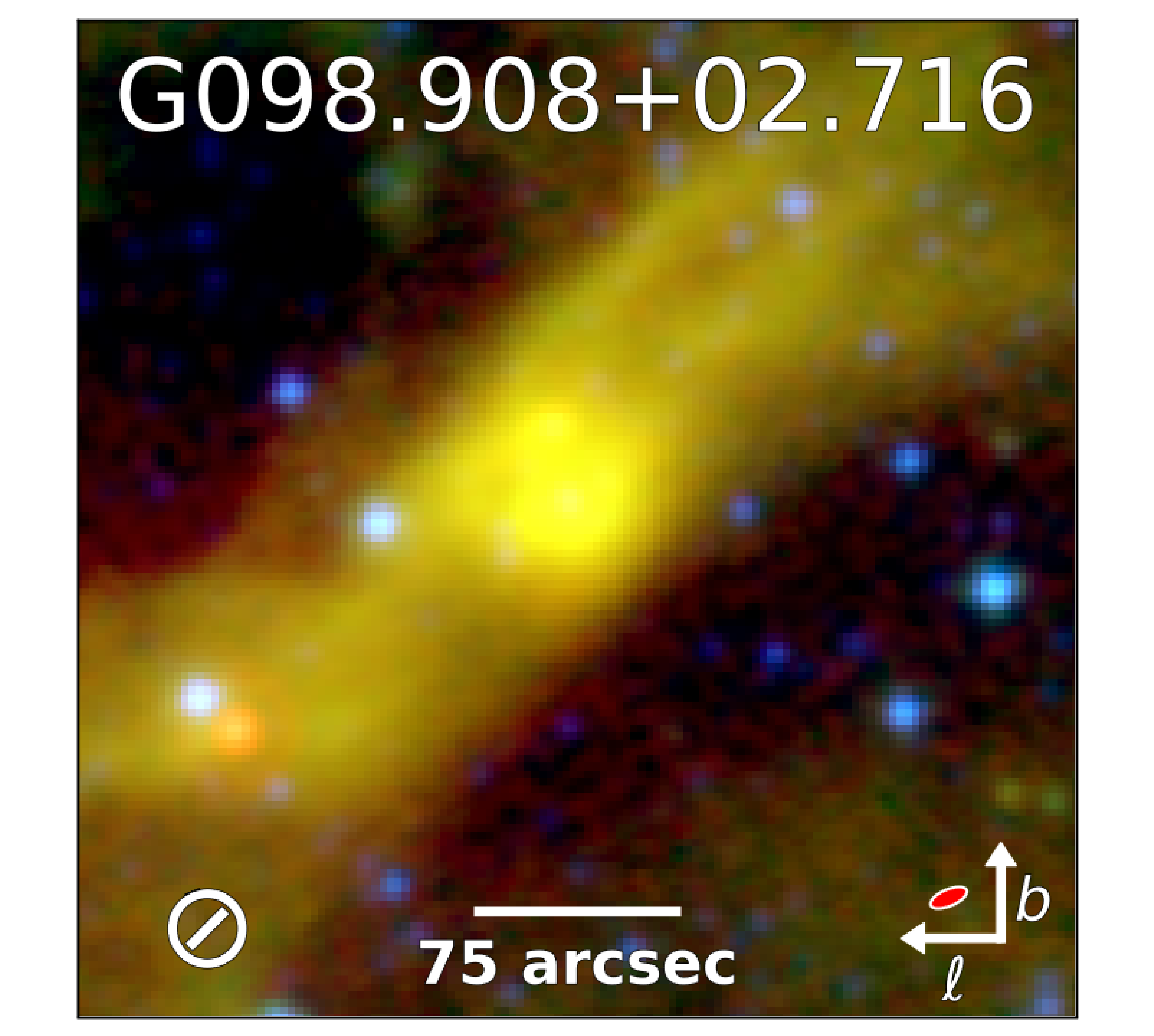}
\includegraphics[width=\figSize]{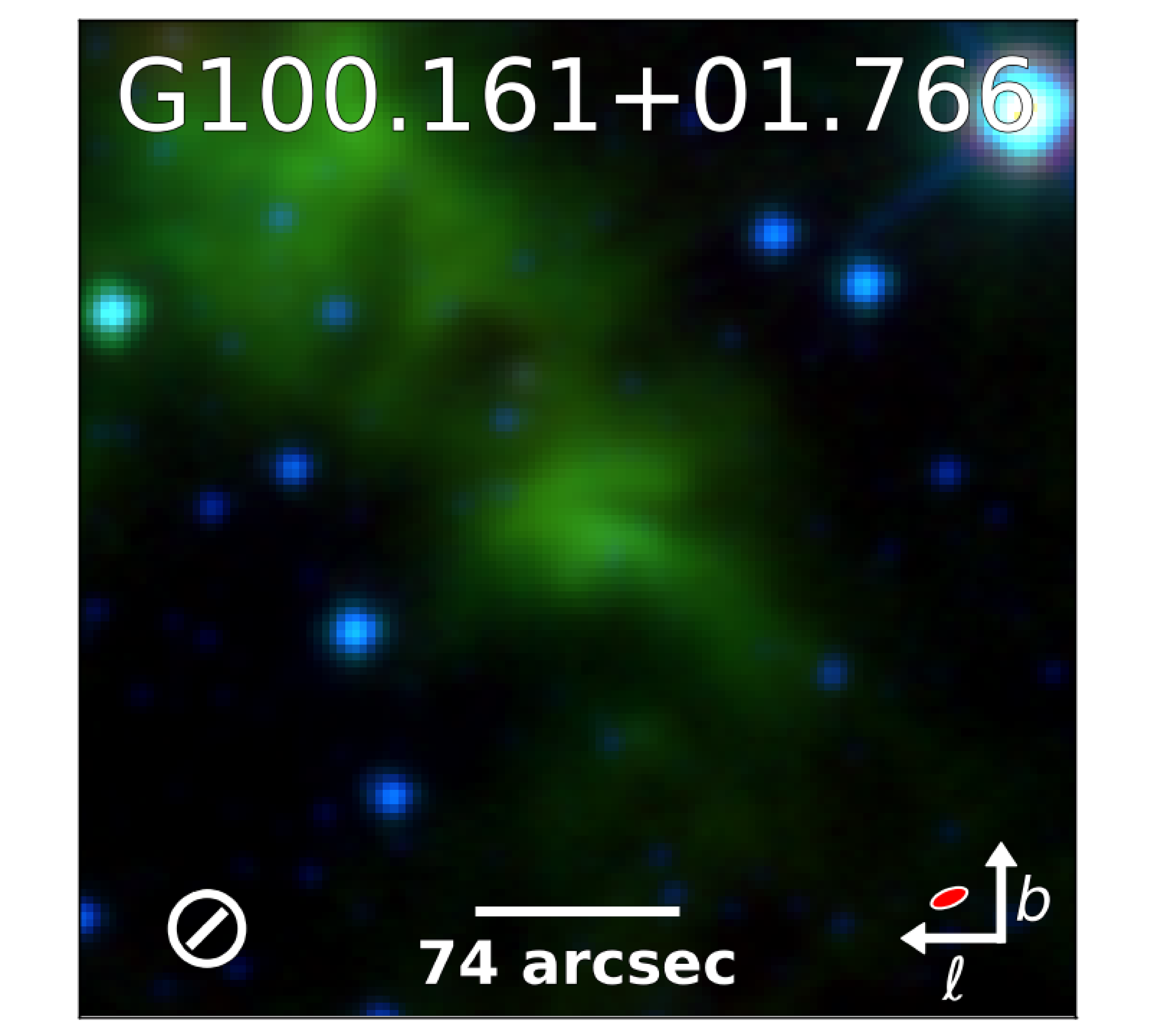}
\end{figure*}
\begin{figure*}[!htb]
\includegraphics[width=\figSize]{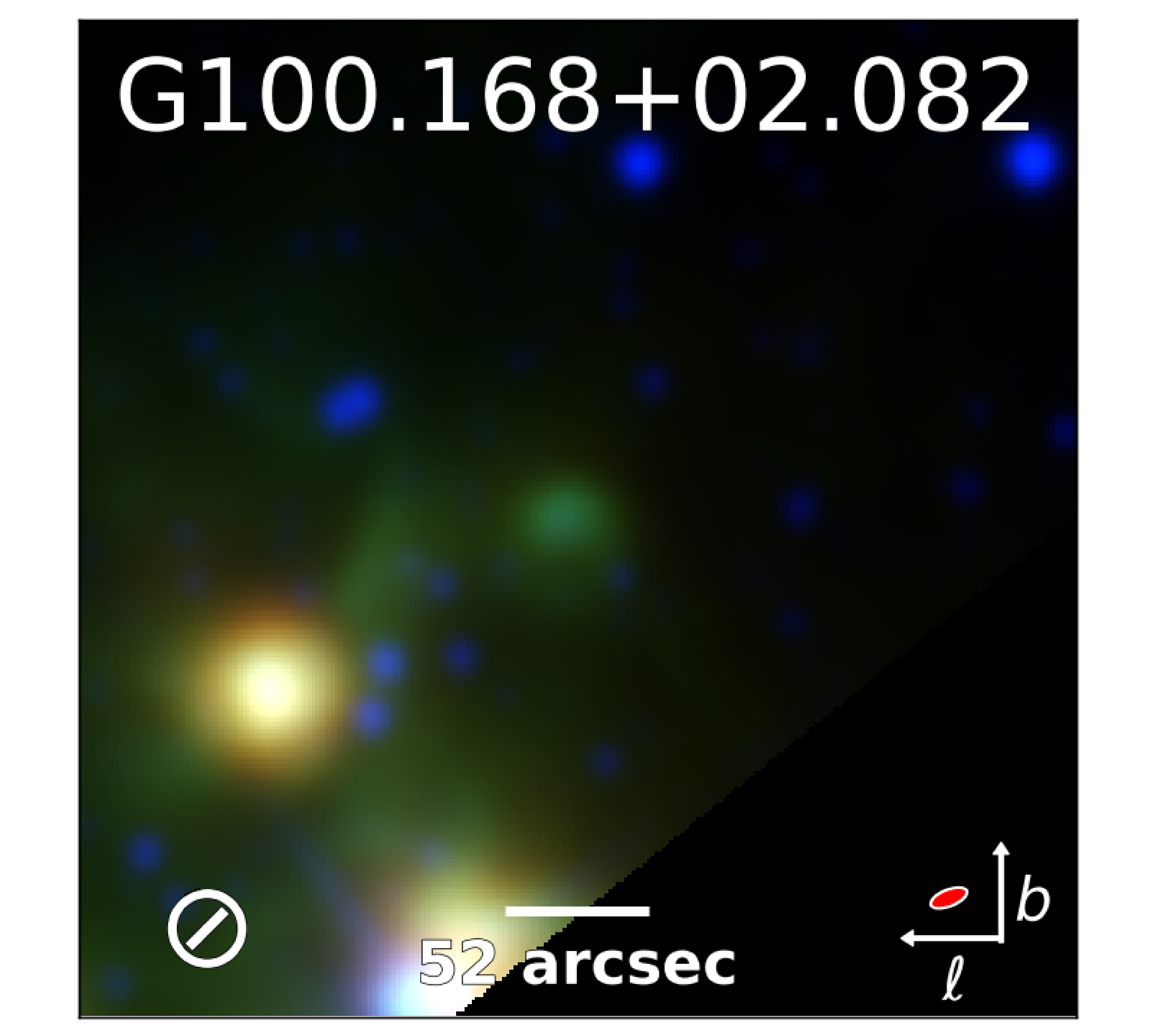}
\includegraphics[width=\figSize]{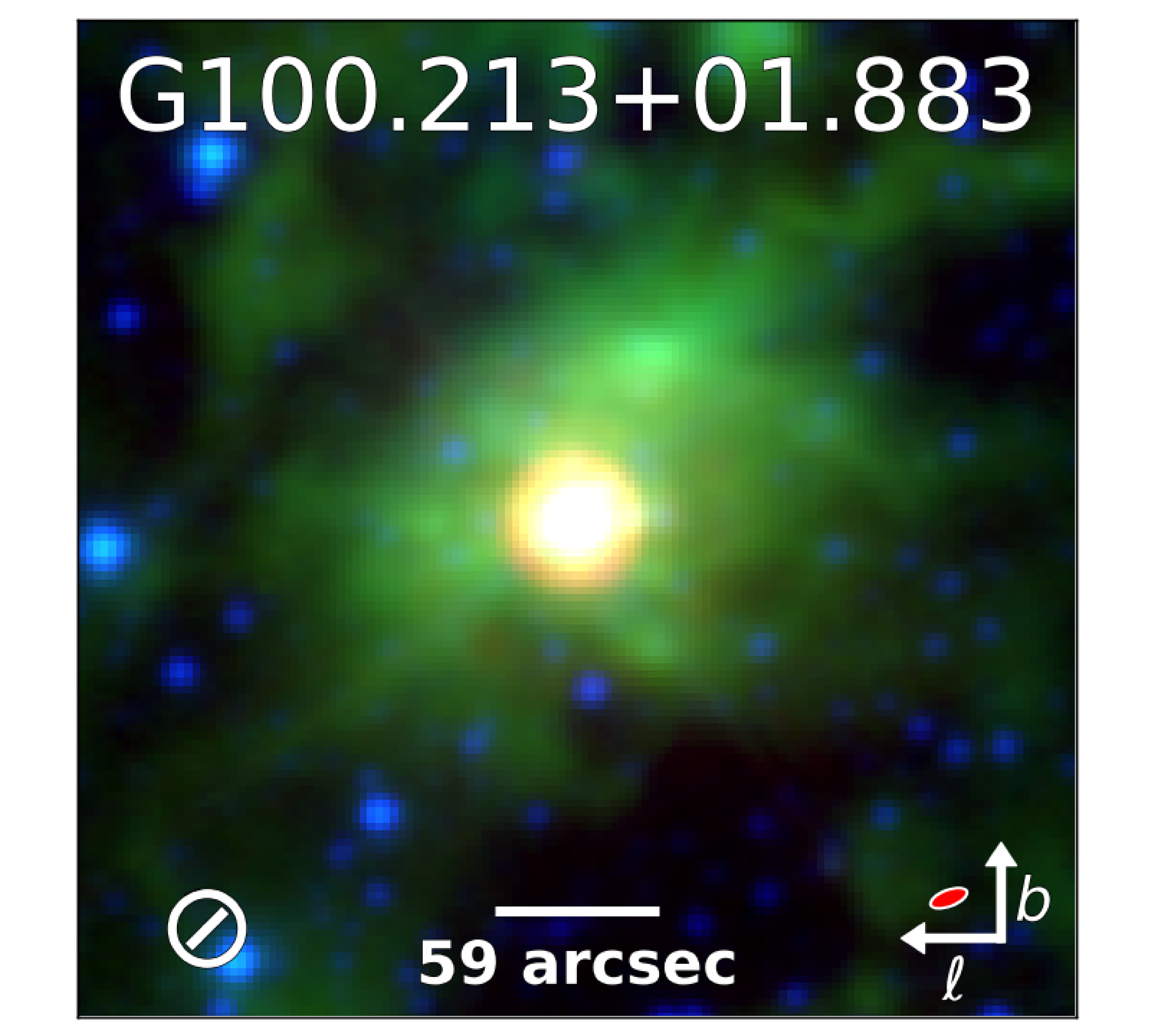}
\includegraphics[width=\figSize]{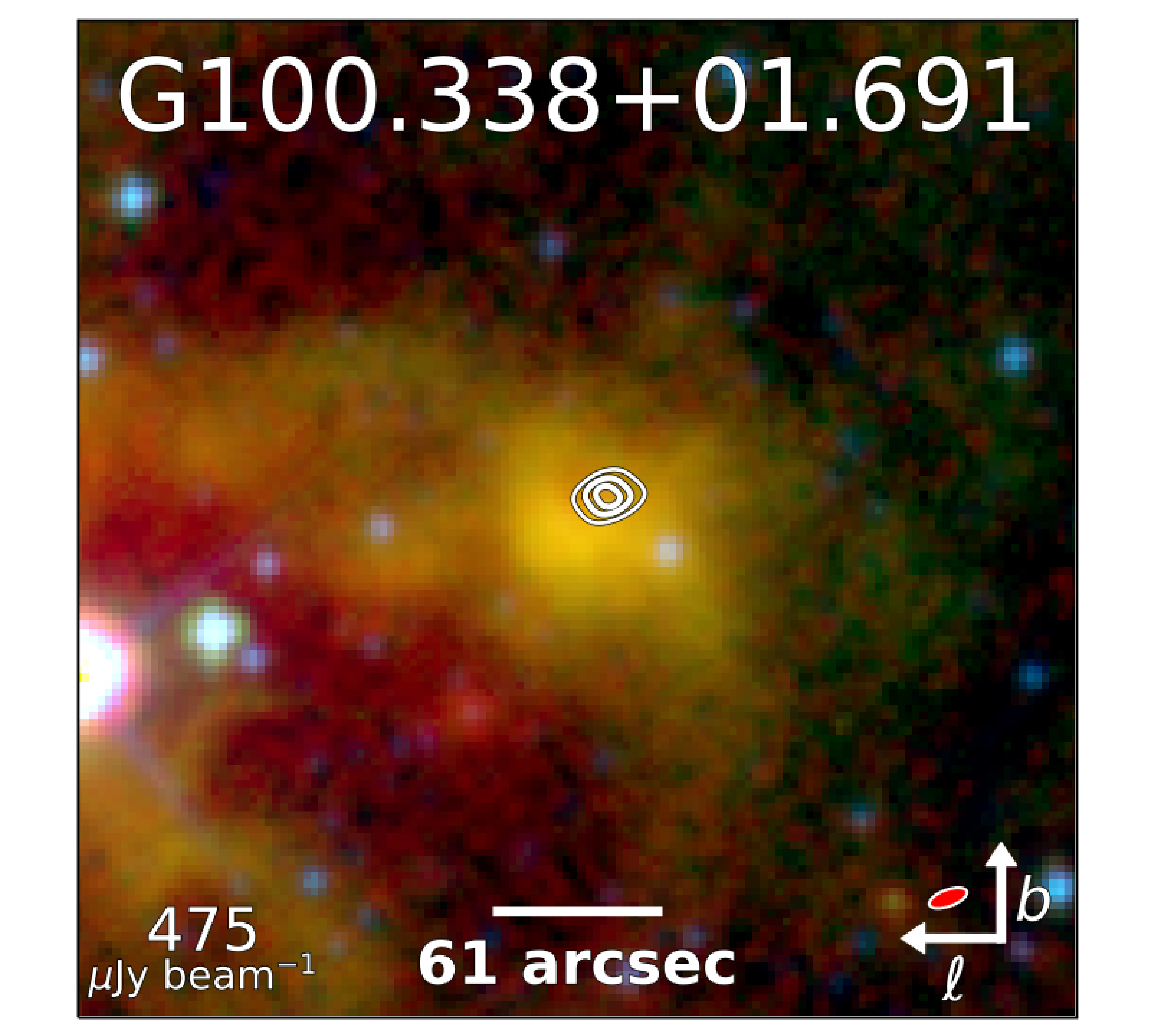}\\
\includegraphics[width=\figSize]{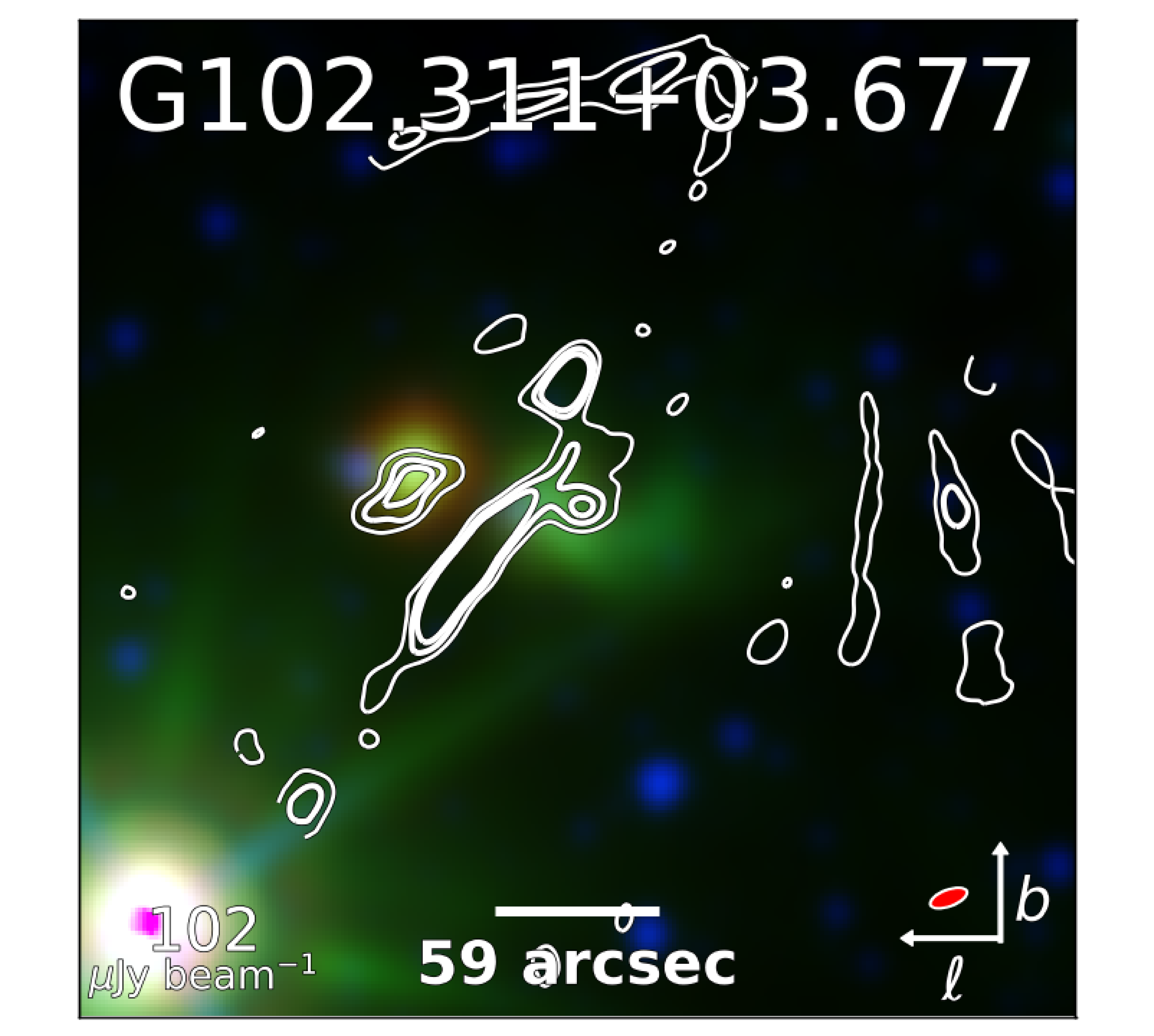}
\includegraphics[width=\figSize]{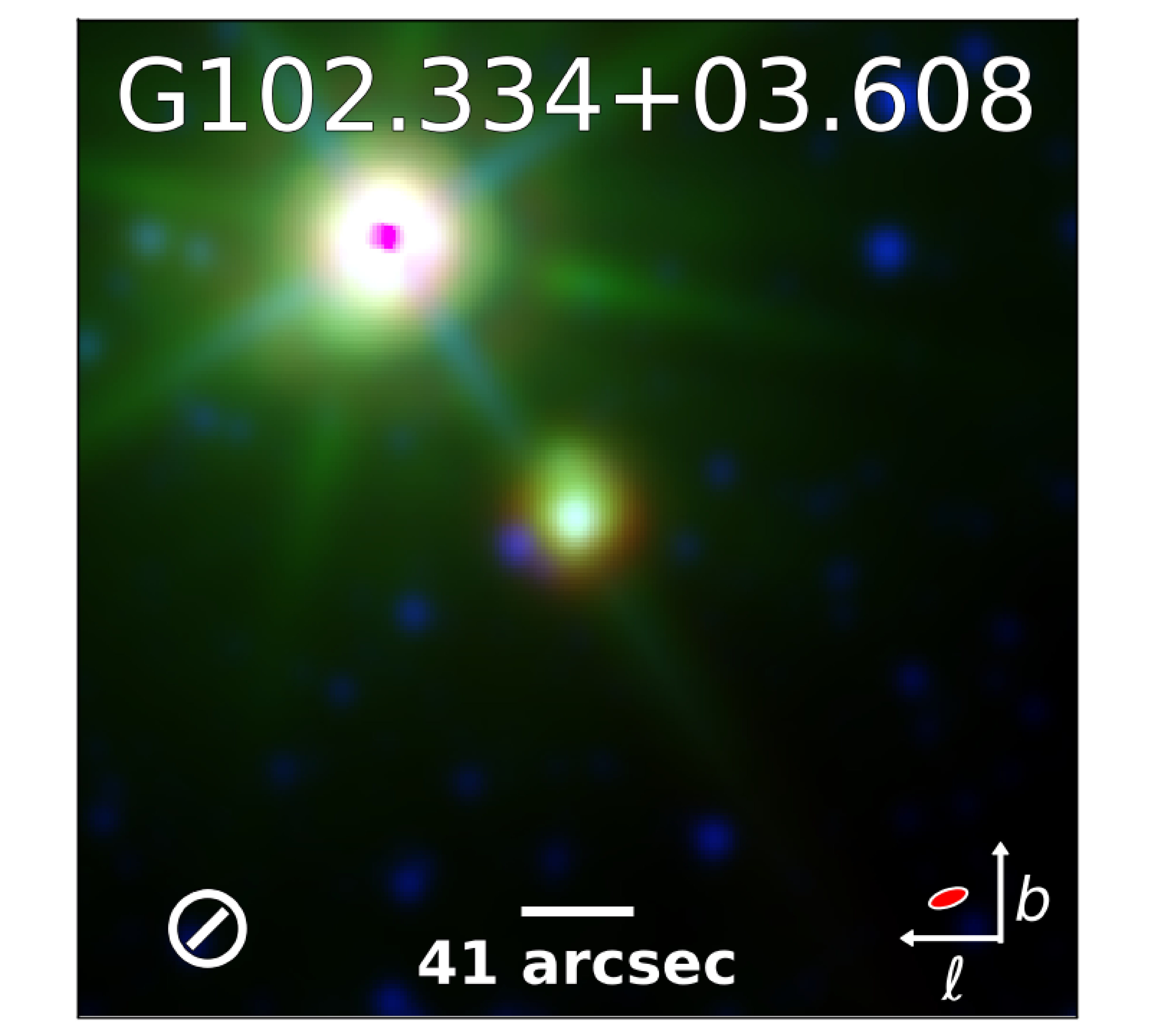}
\includegraphics[width=\figSize]{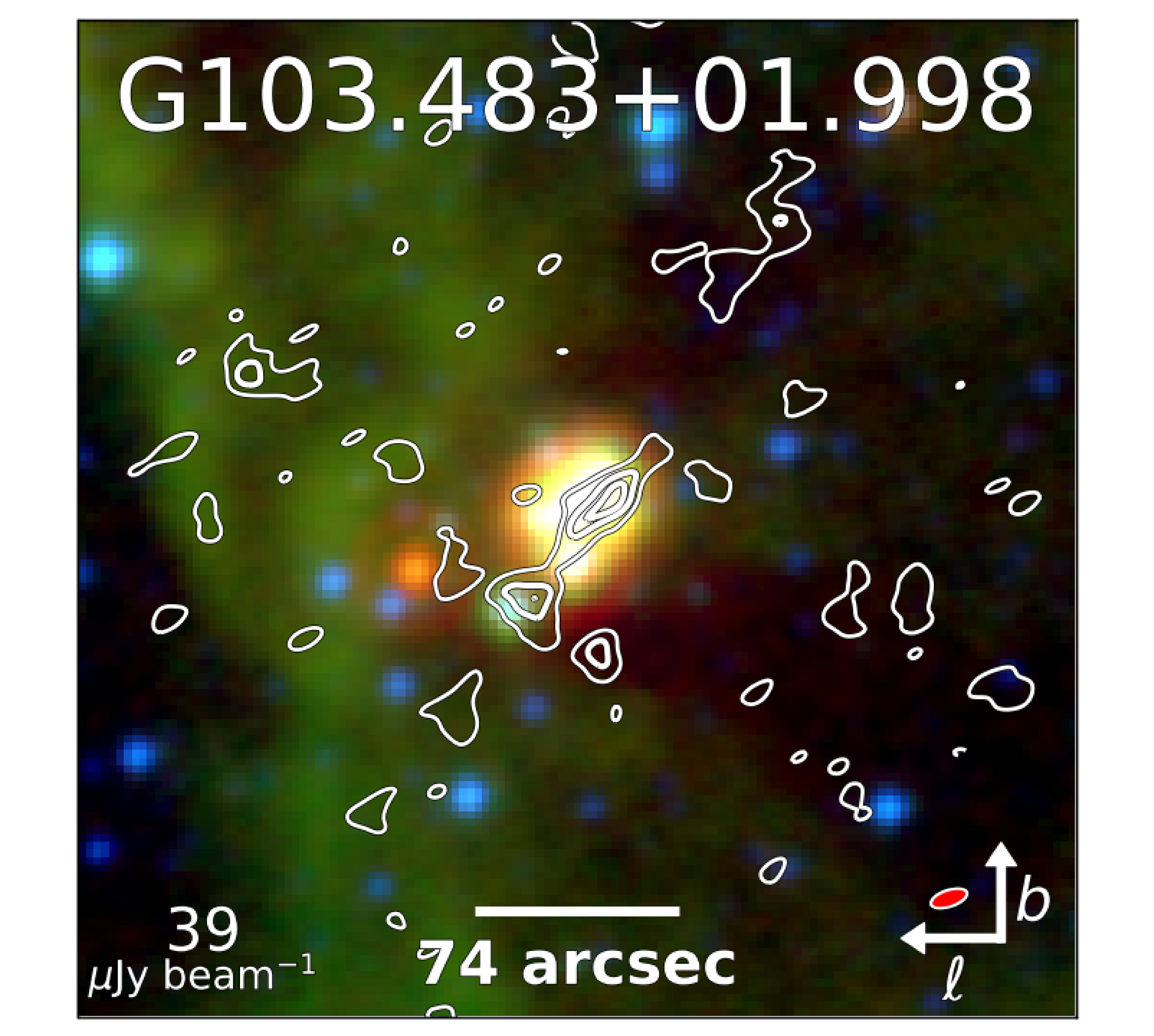}\\
\includegraphics[width=\figSize]{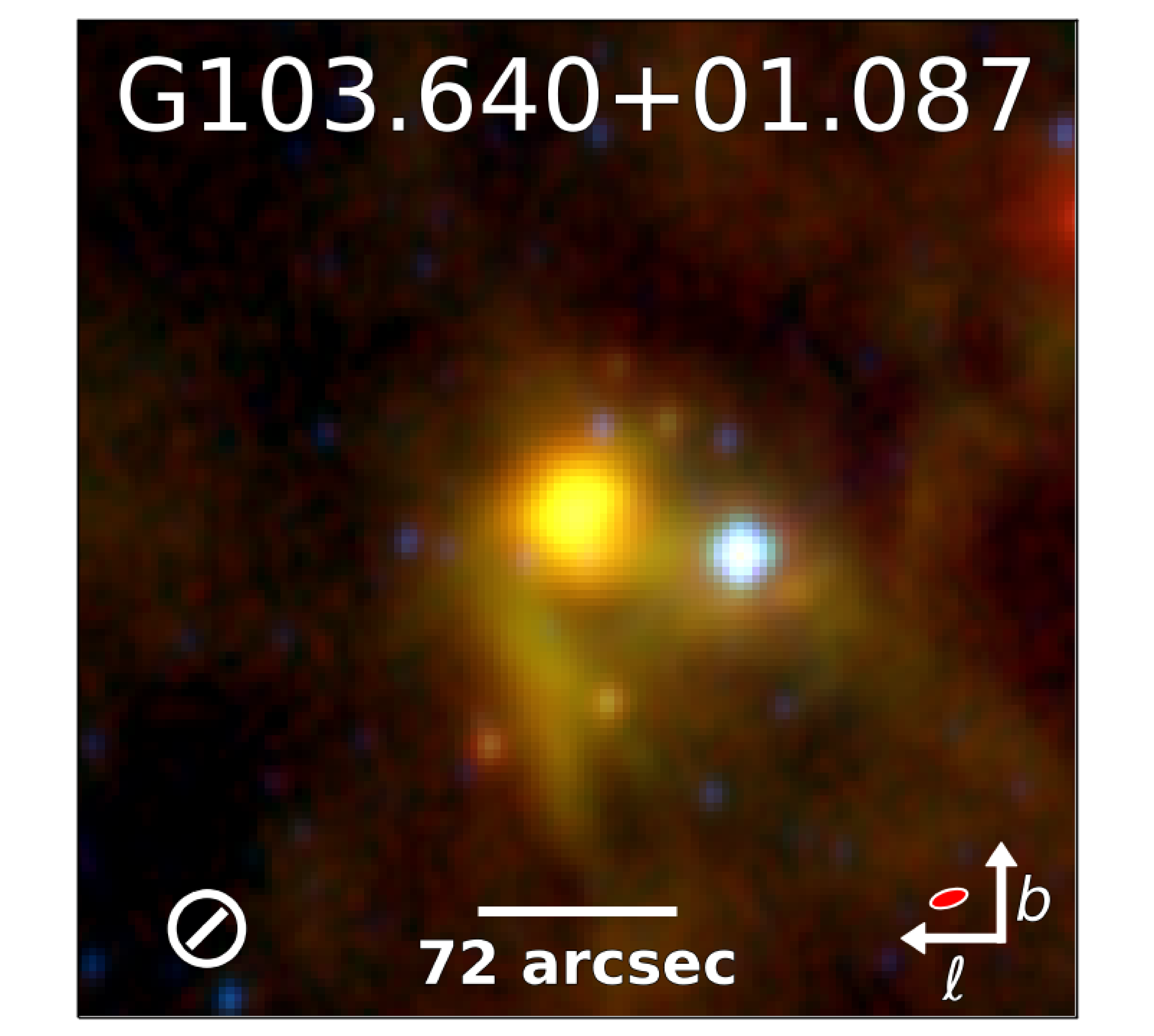}
\includegraphics[width=\figSize]{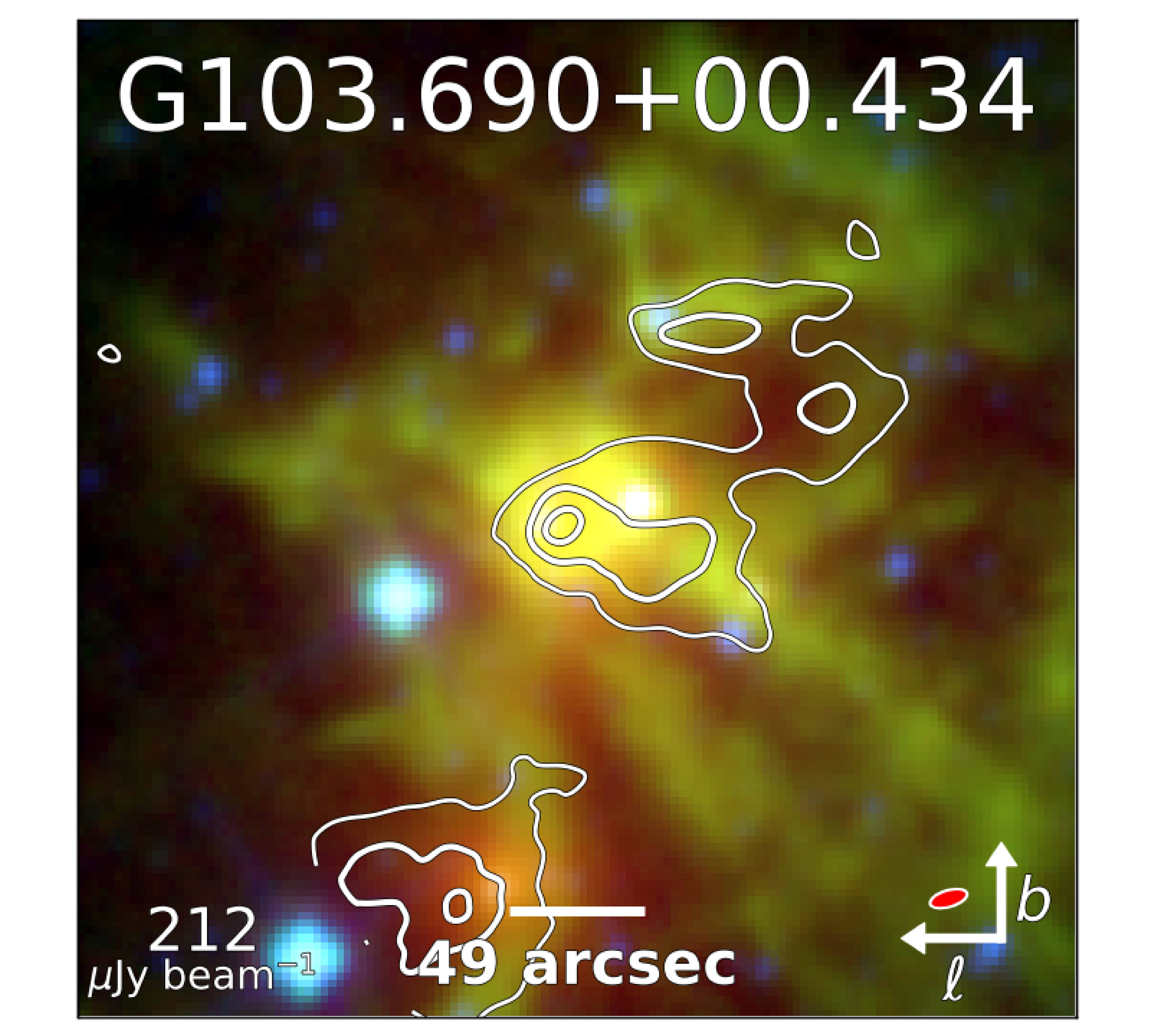}
\includegraphics[width=\figSize]{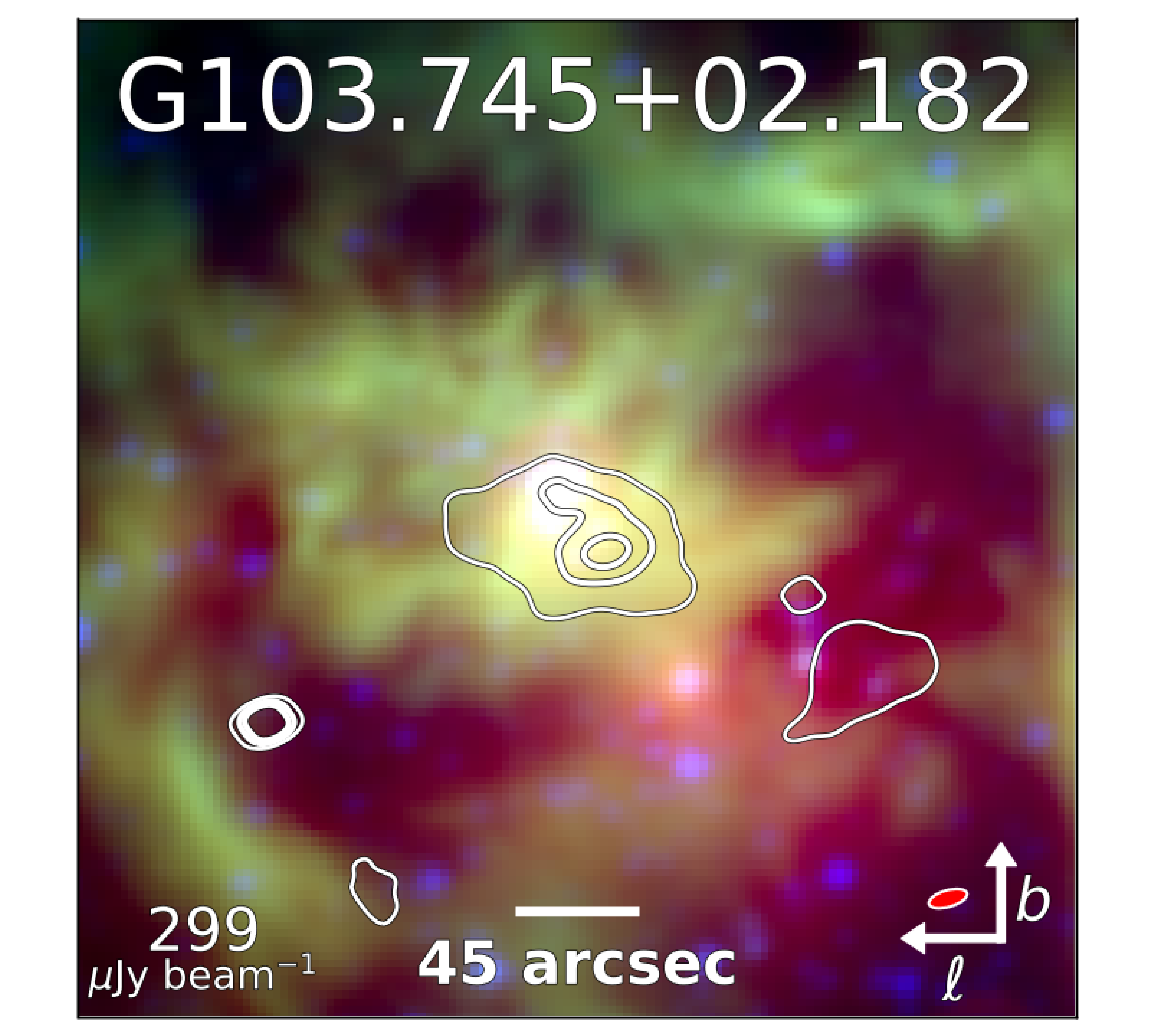}\\
\includegraphics[width=\figSize]{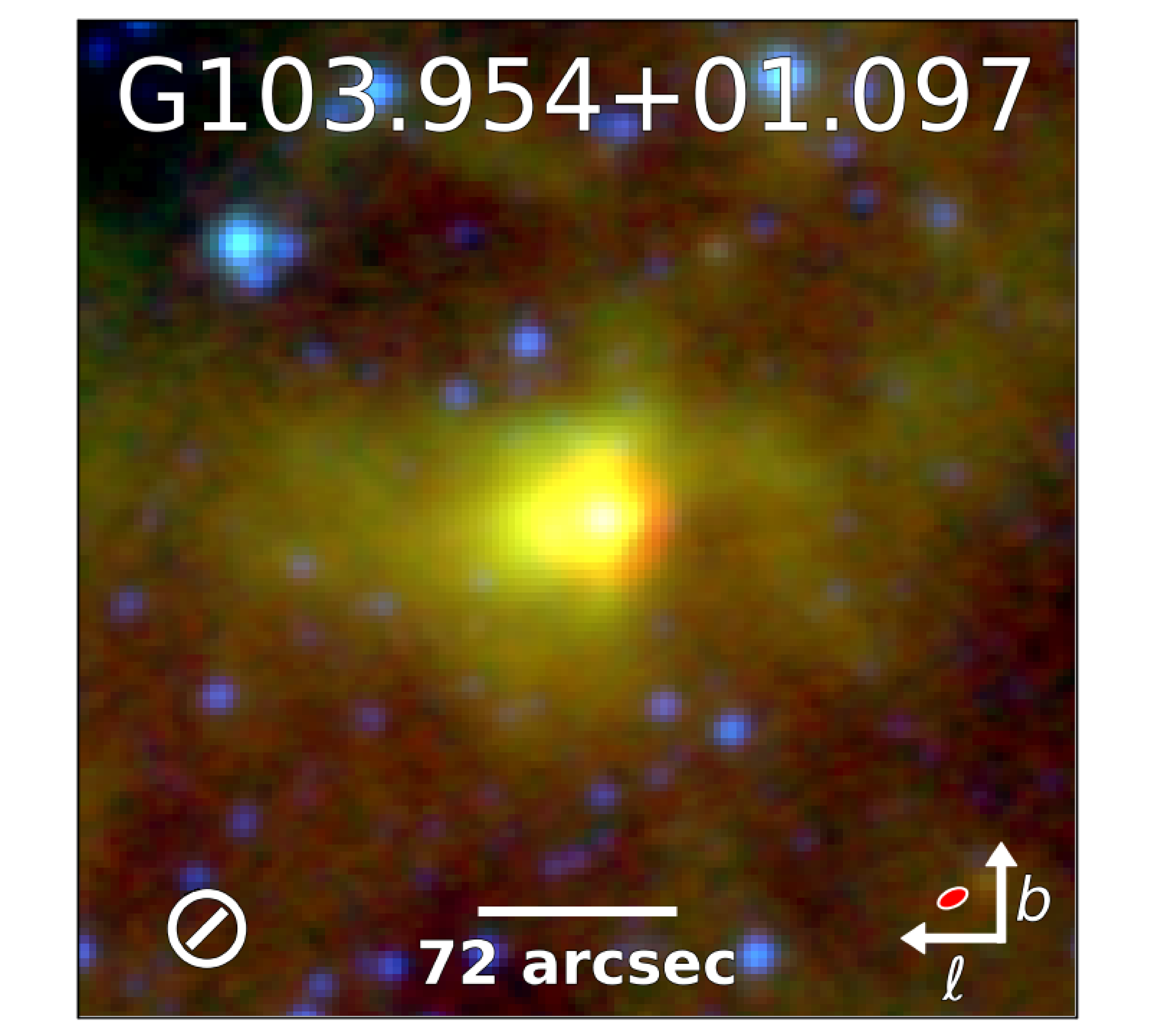}
\includegraphics[width=\figSize]{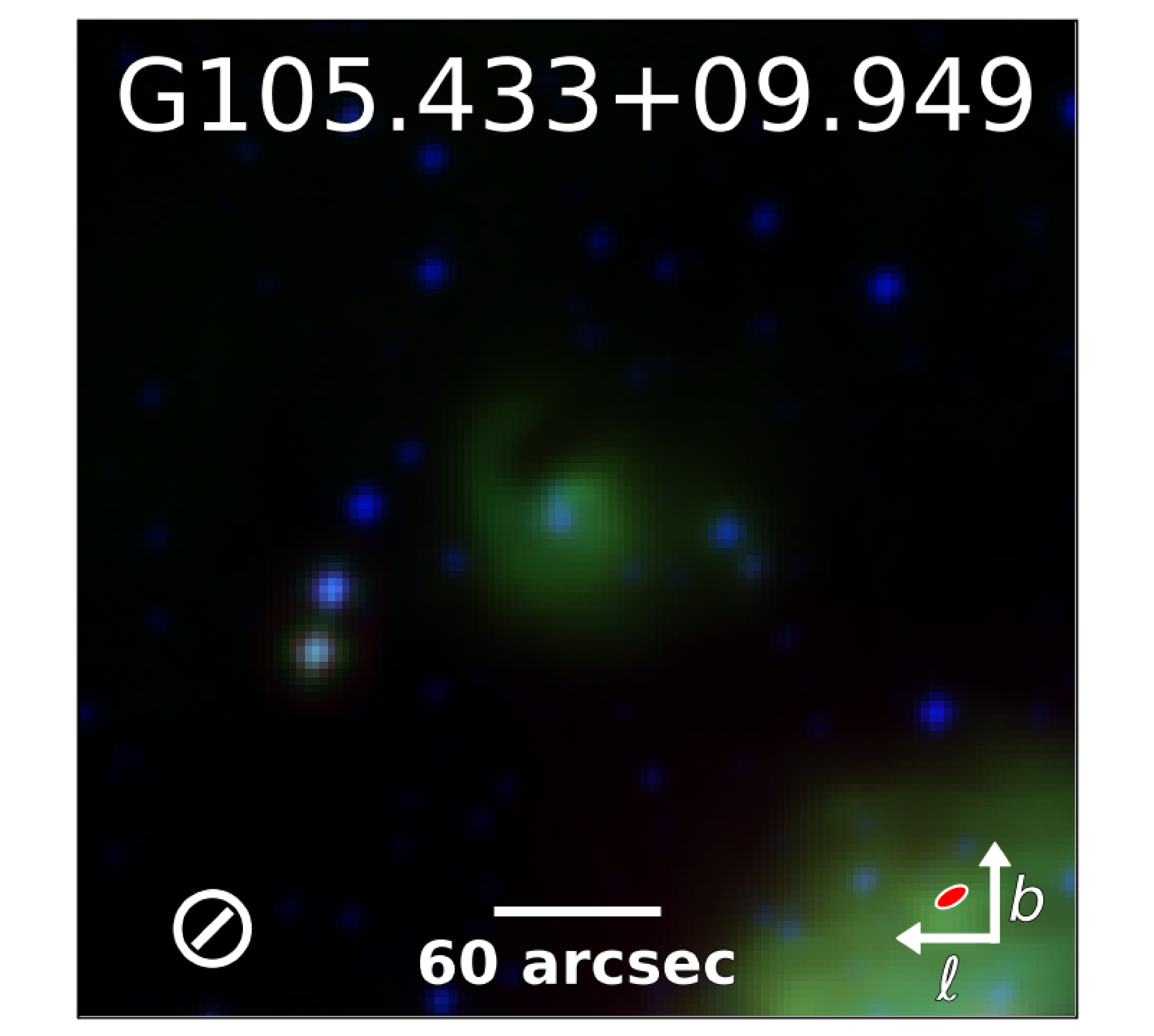}
\includegraphics[width=\figSize]{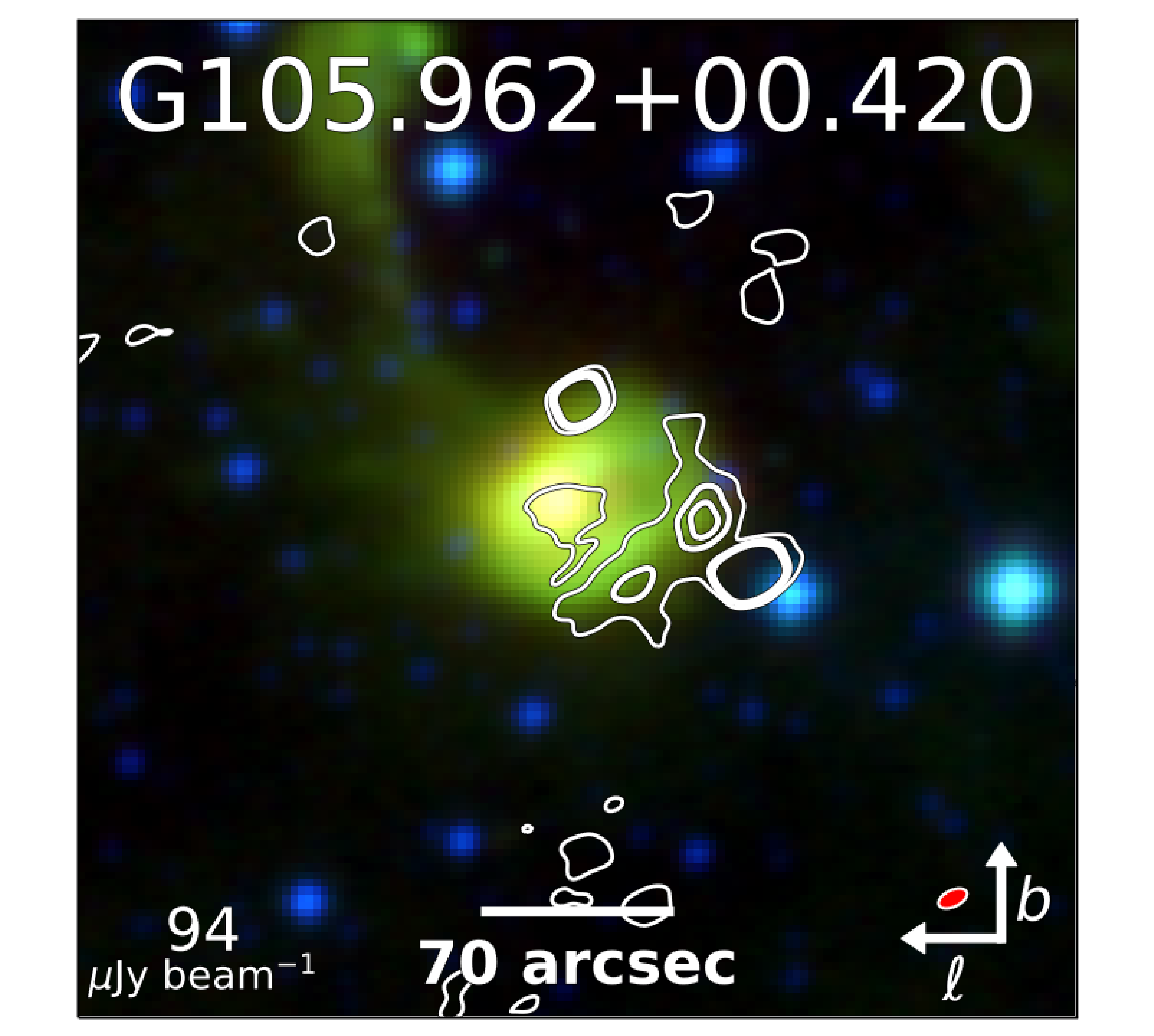}
\end{figure*}
\begin{figure*}[!htb]
\includegraphics[width=\figSize]{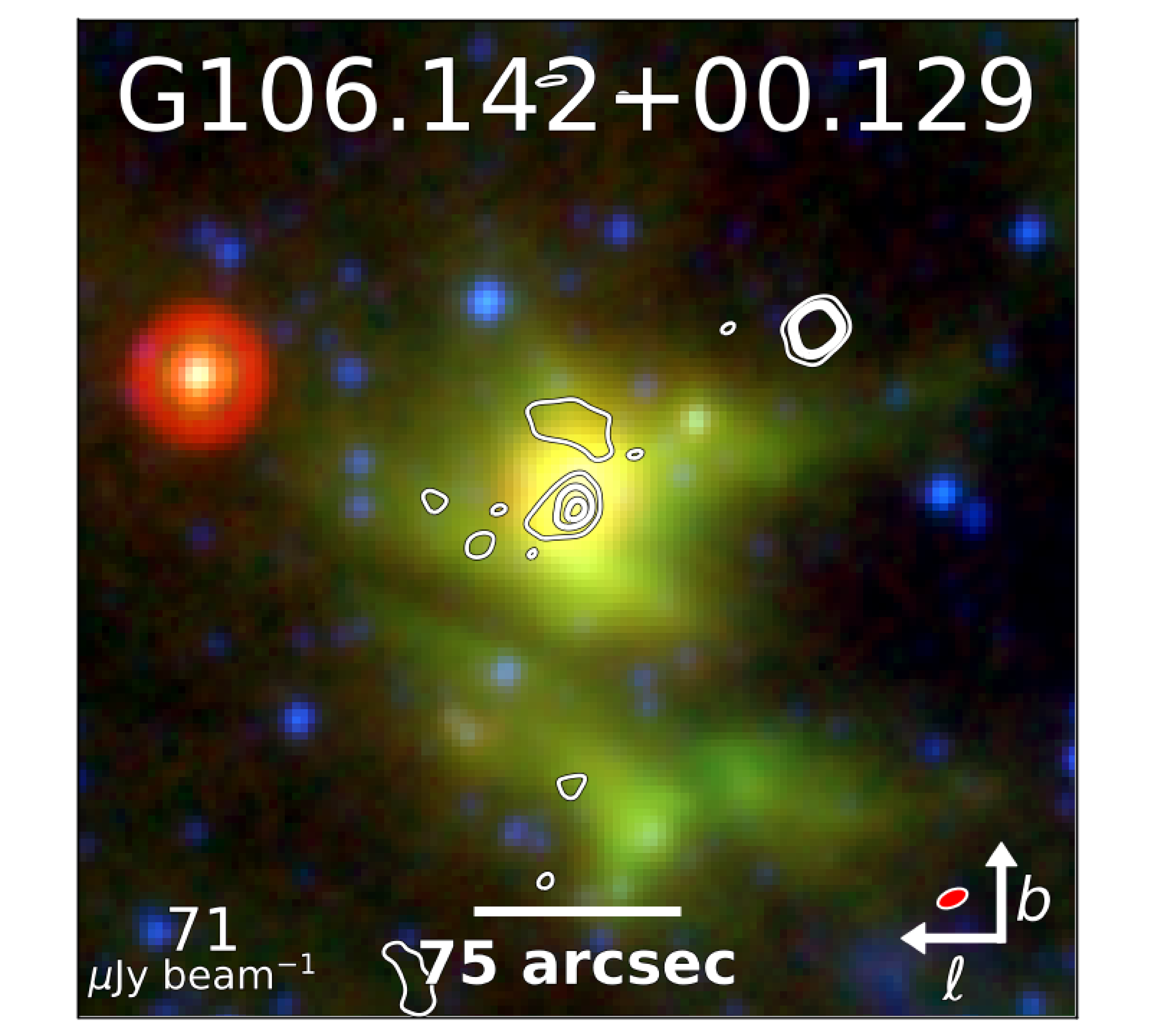}
\includegraphics[width=\figSize]{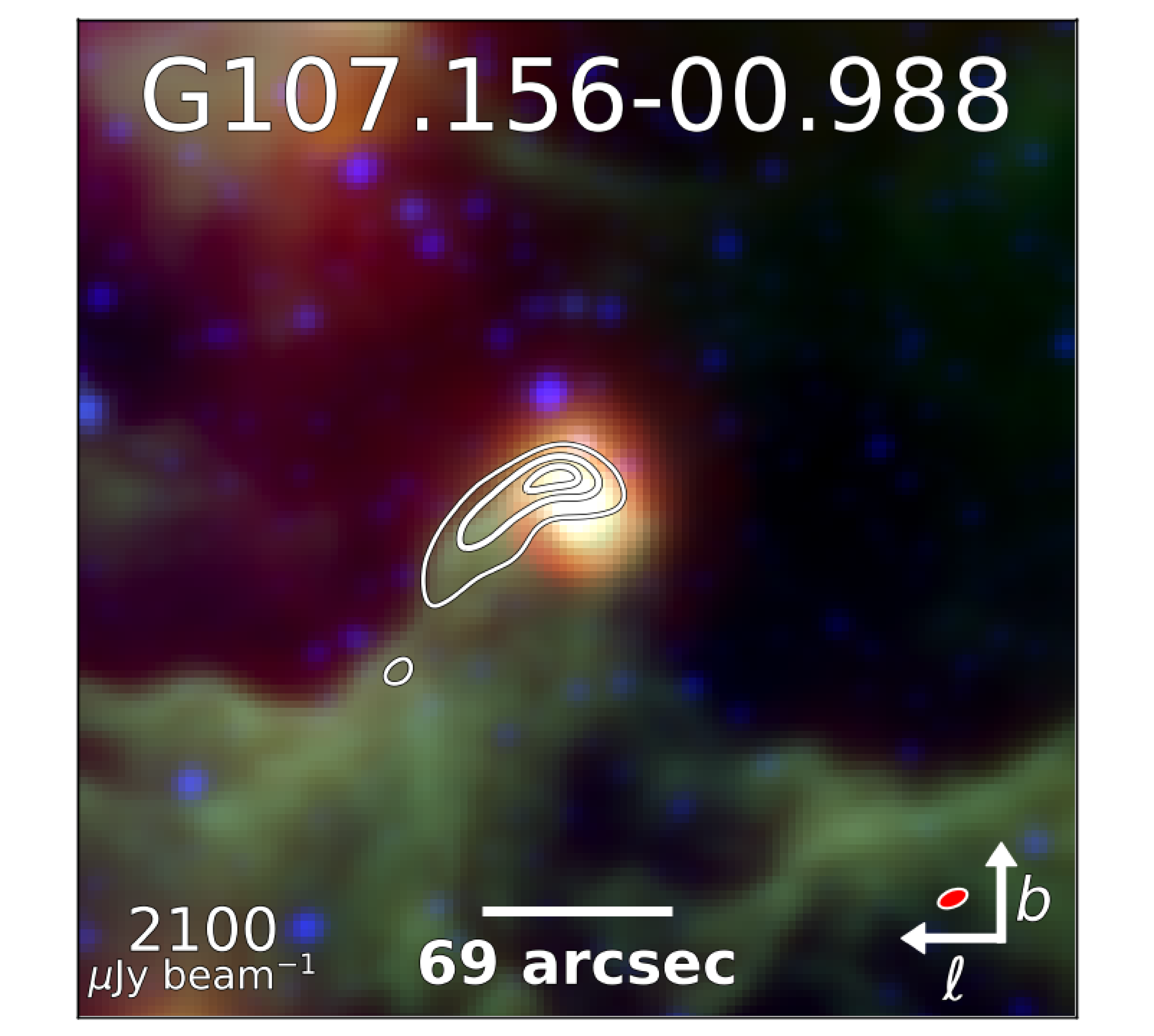}
\includegraphics[width=\figSize]{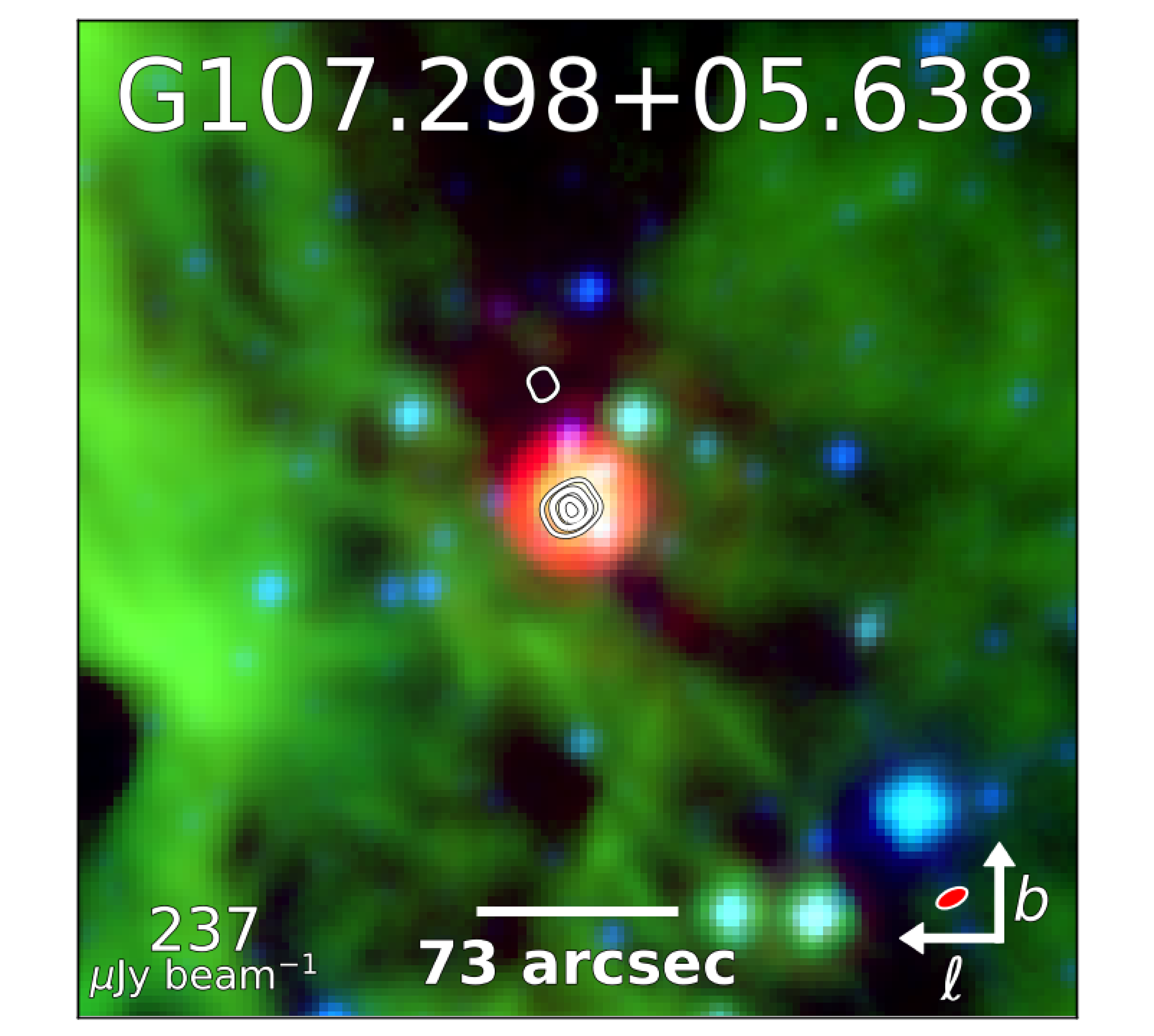}\\
\includegraphics[width=\figSize]{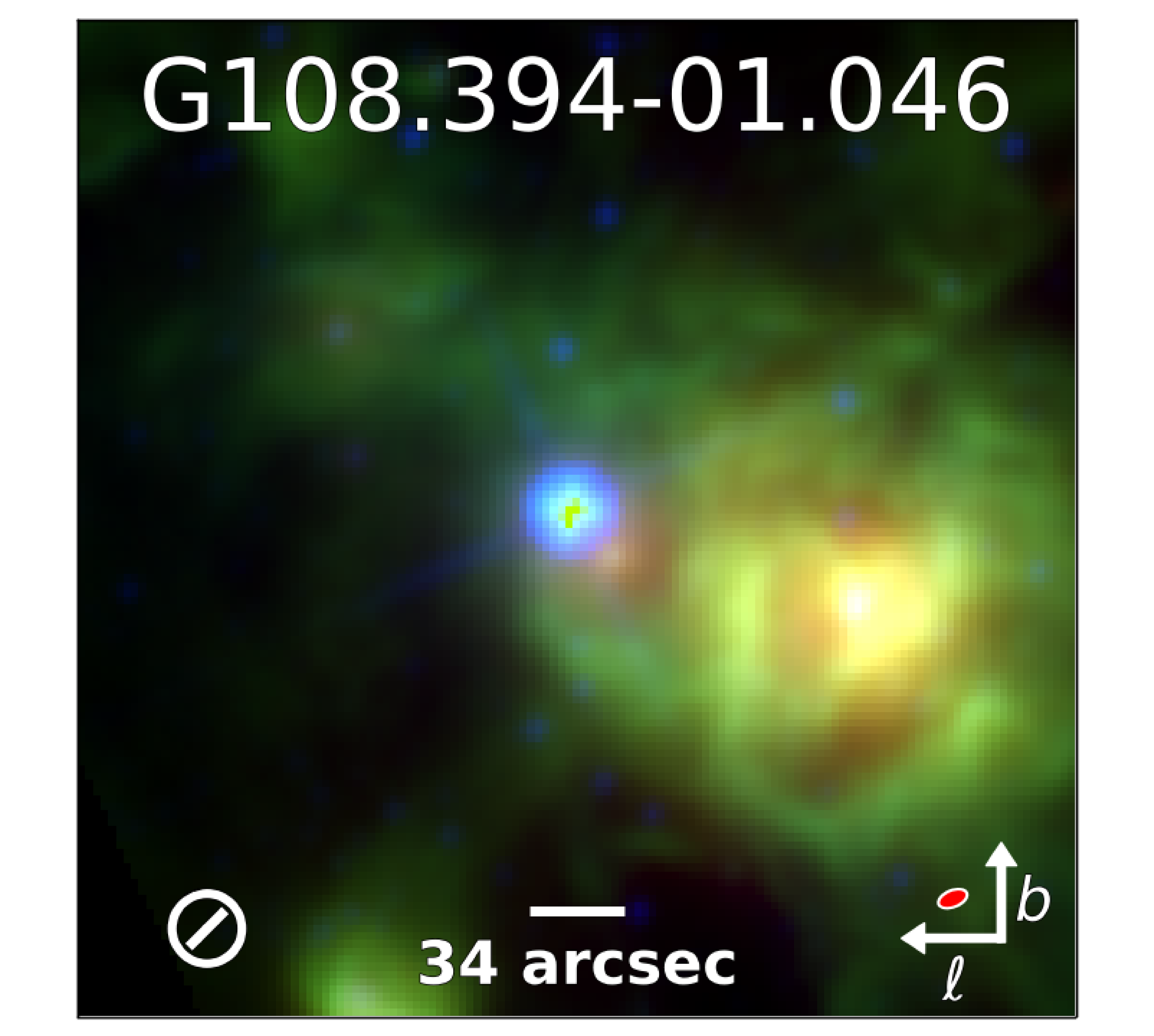}
\includegraphics[width=\figSize]{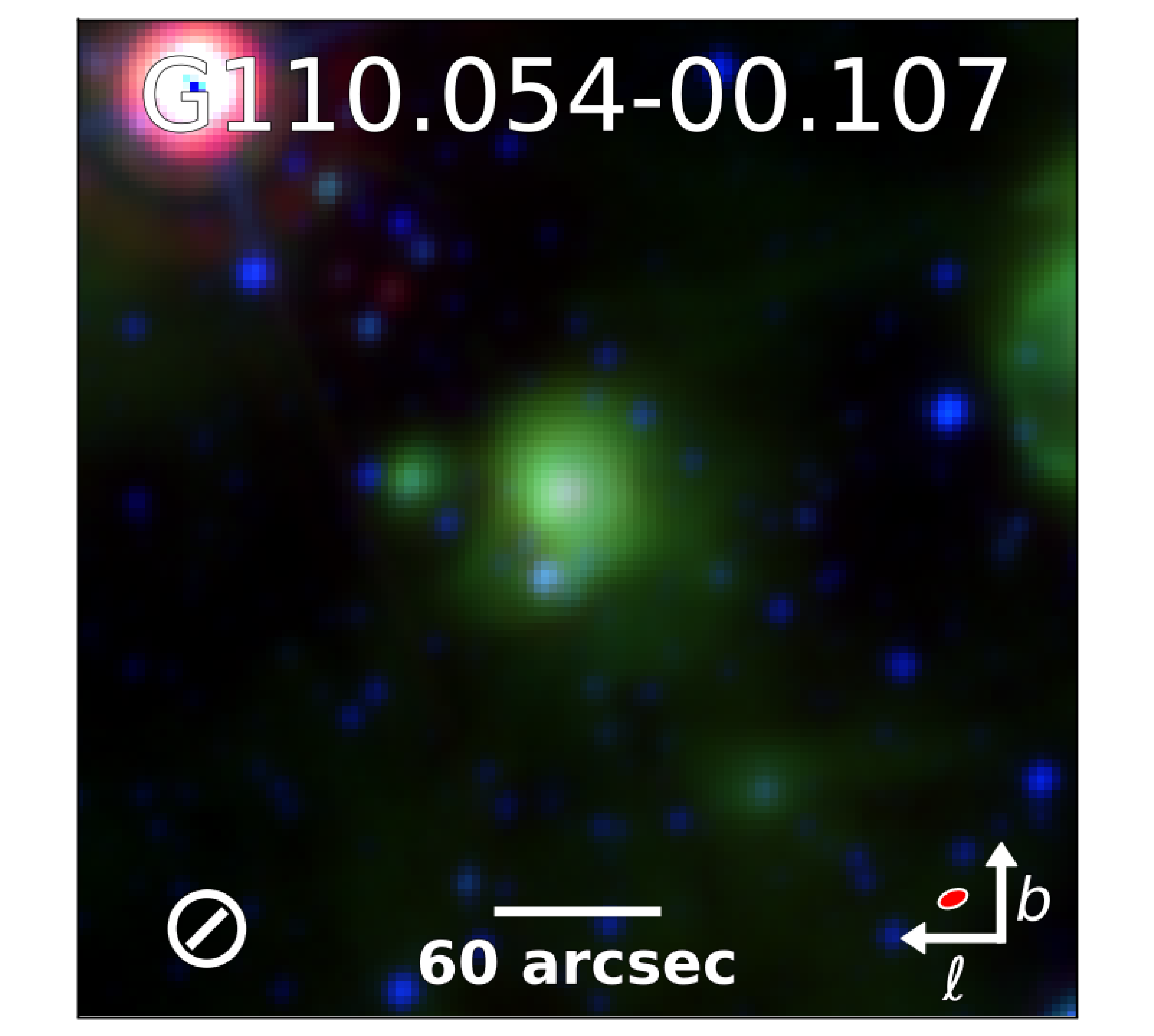}
\includegraphics[width=\figSize]{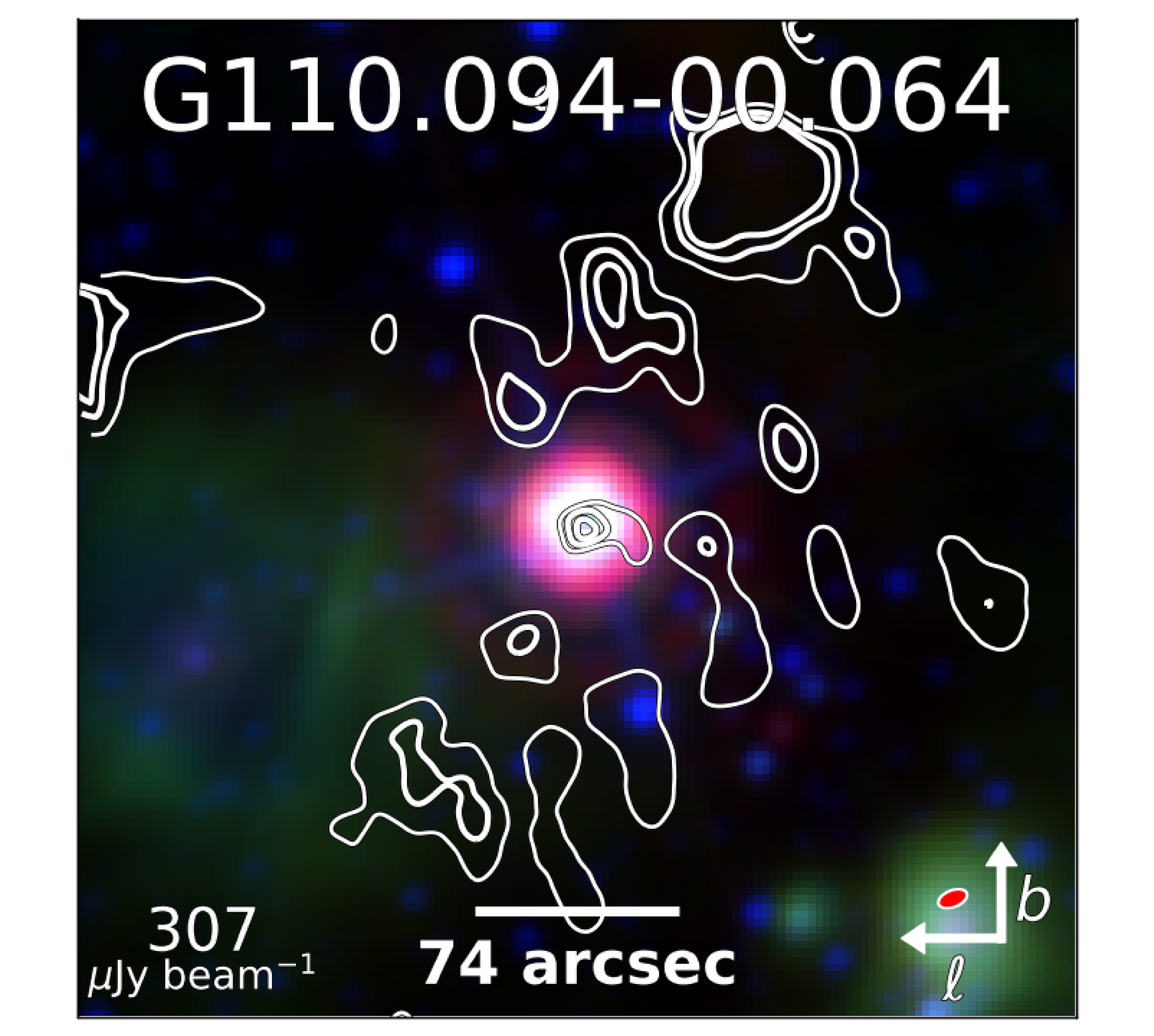}\\
\includegraphics[width=\figSize]{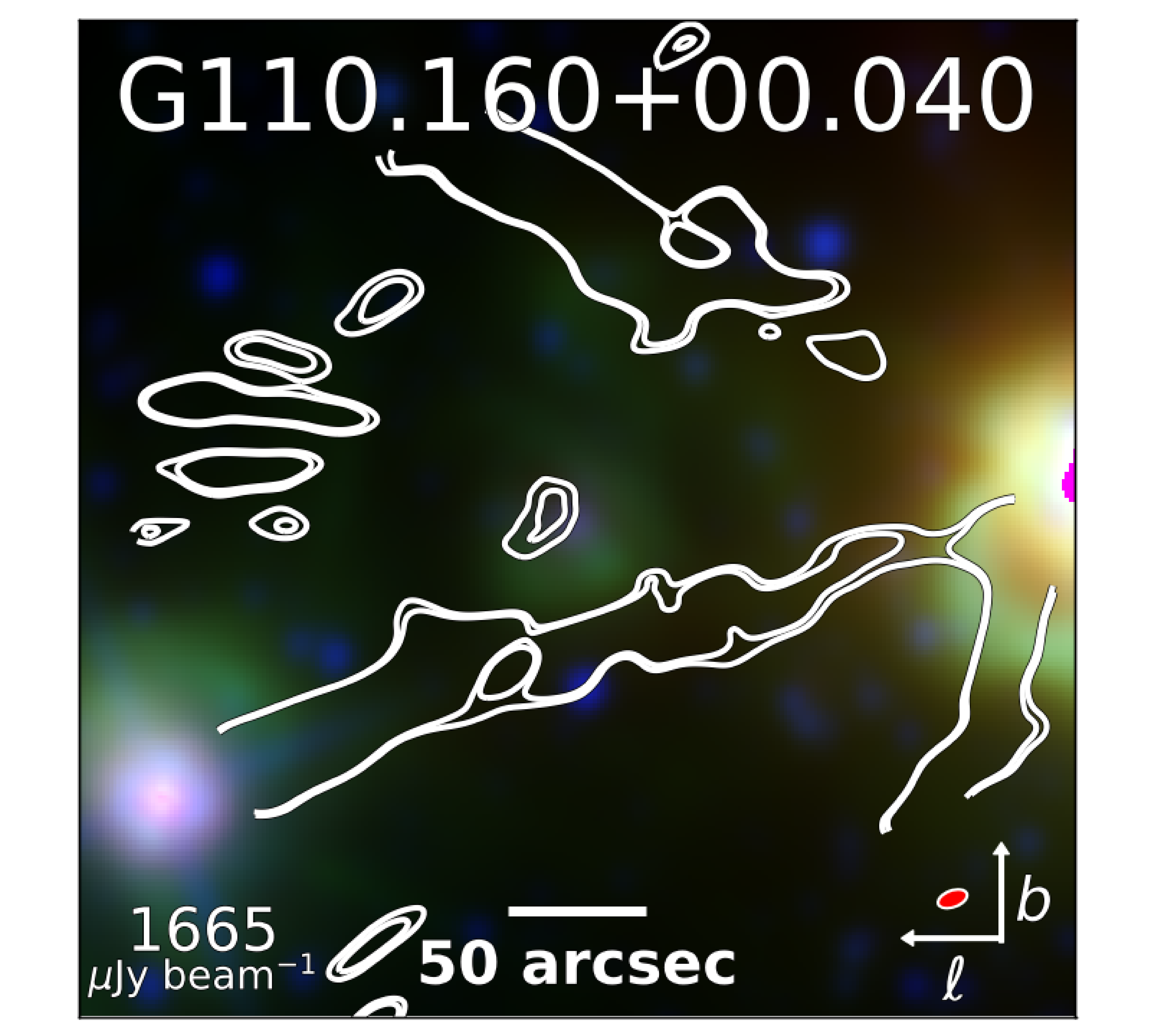}
\includegraphics[width=\figSize]{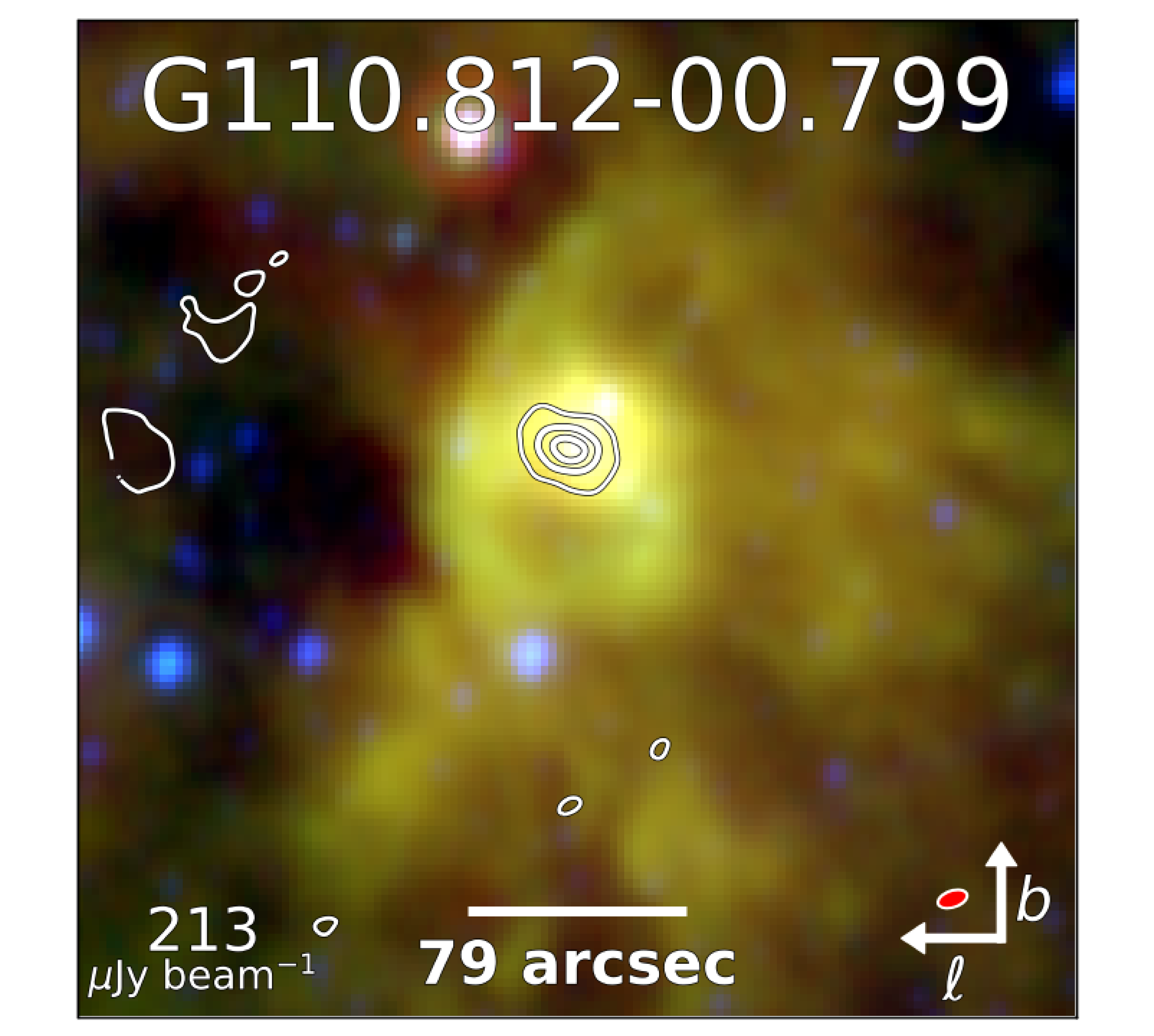}
\includegraphics[width=\figSize]{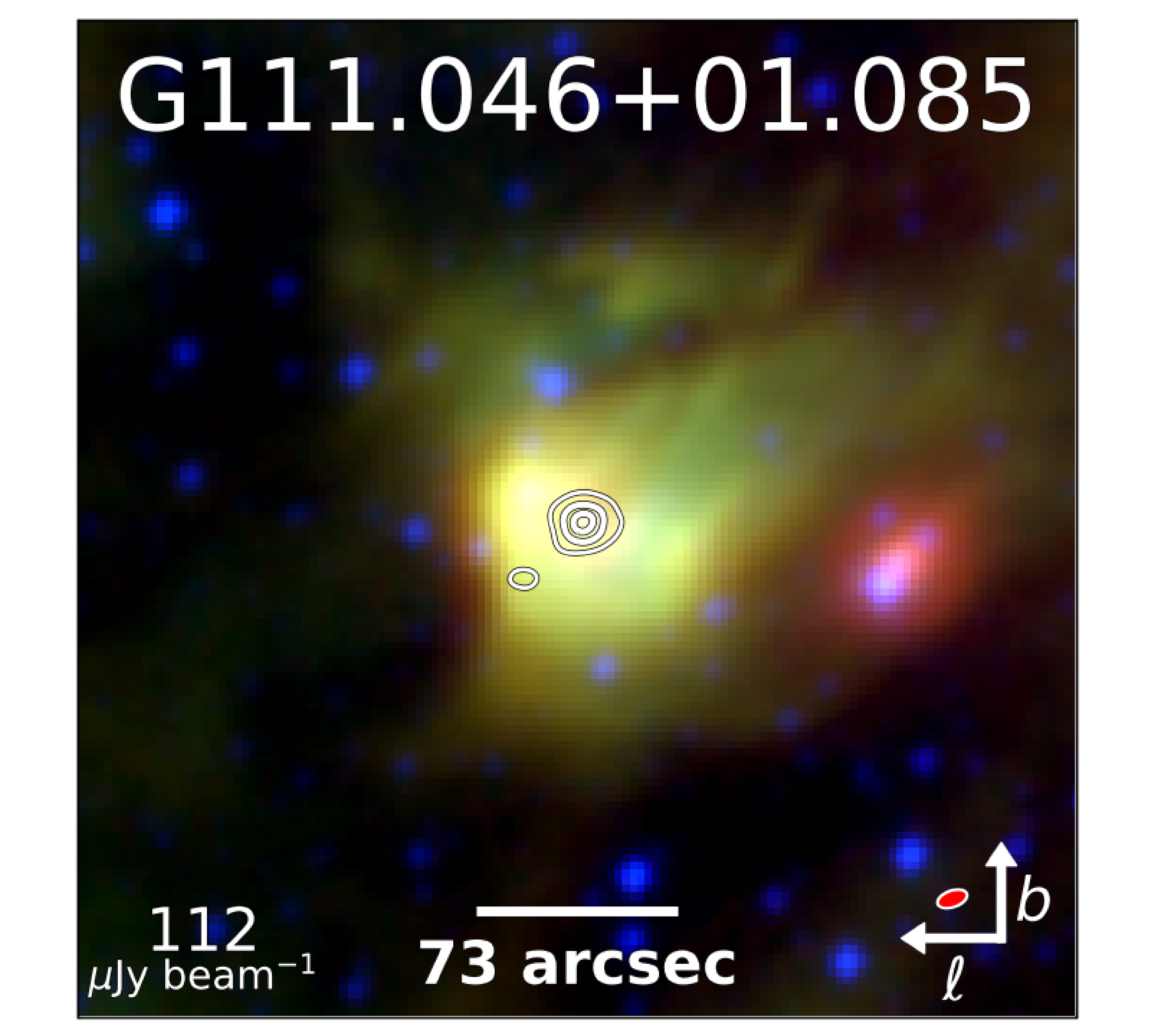}\\
\includegraphics[width=\figSize]{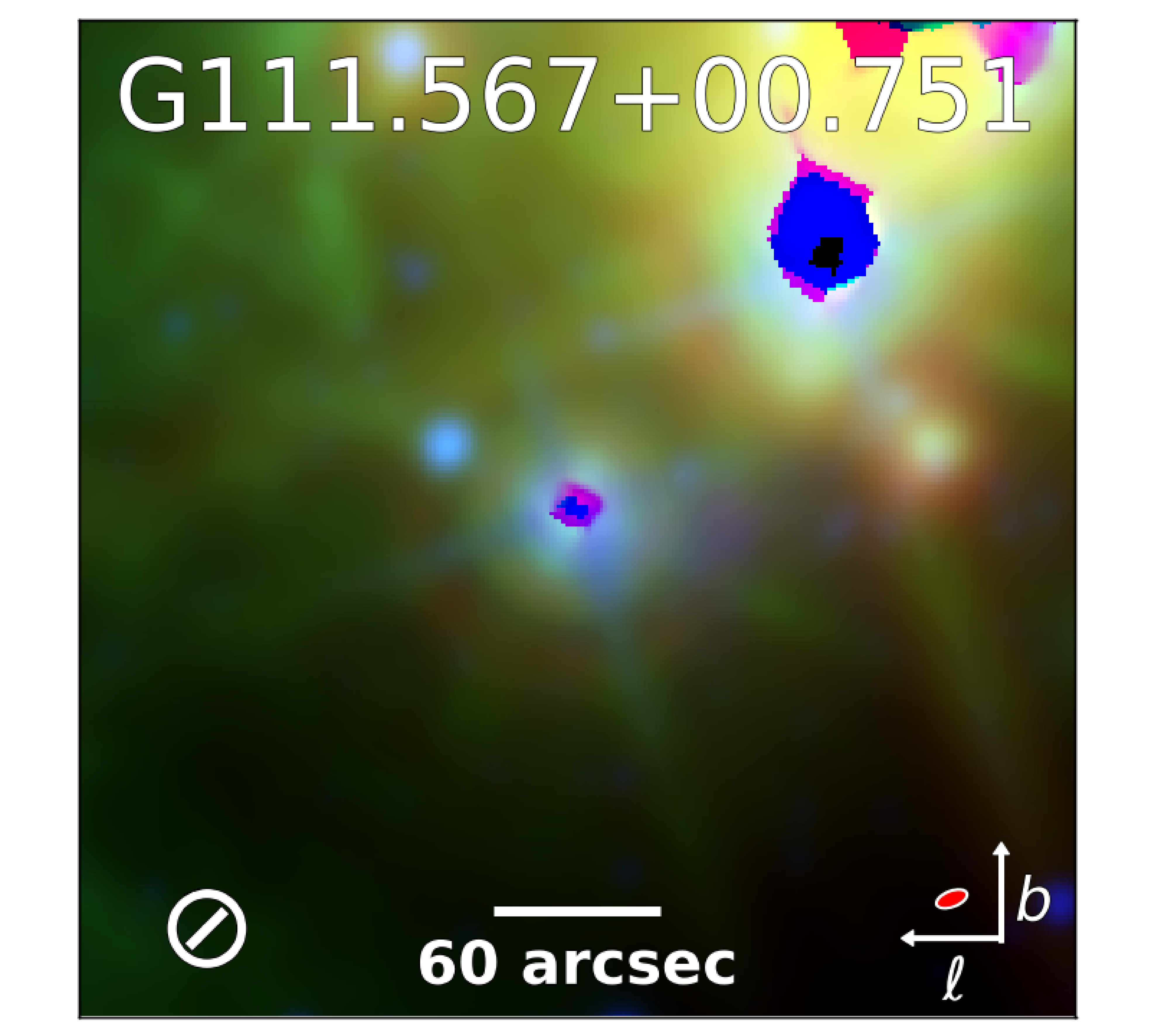}
\includegraphics[width=\figSize]{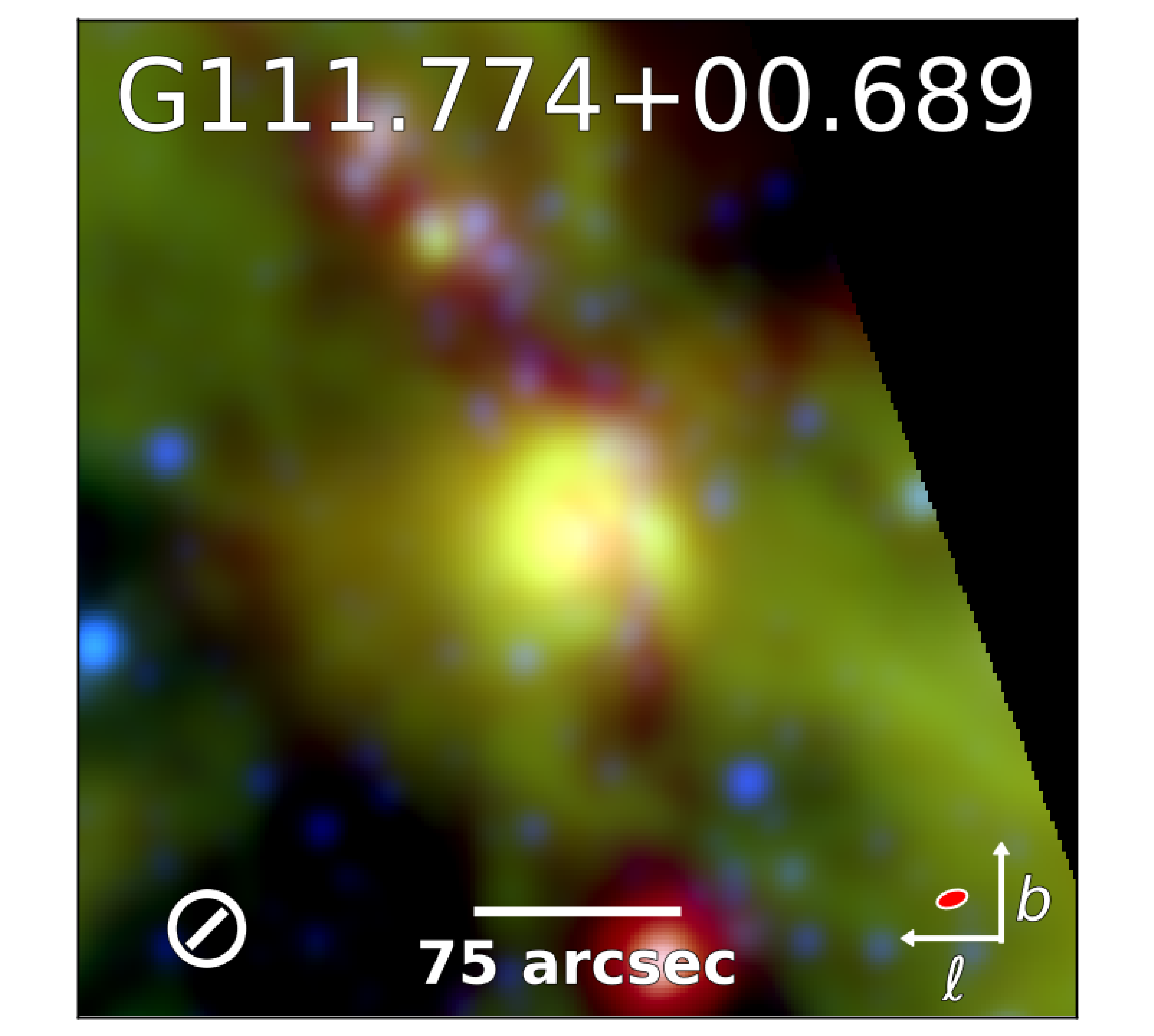}
\includegraphics[width=\figSize]{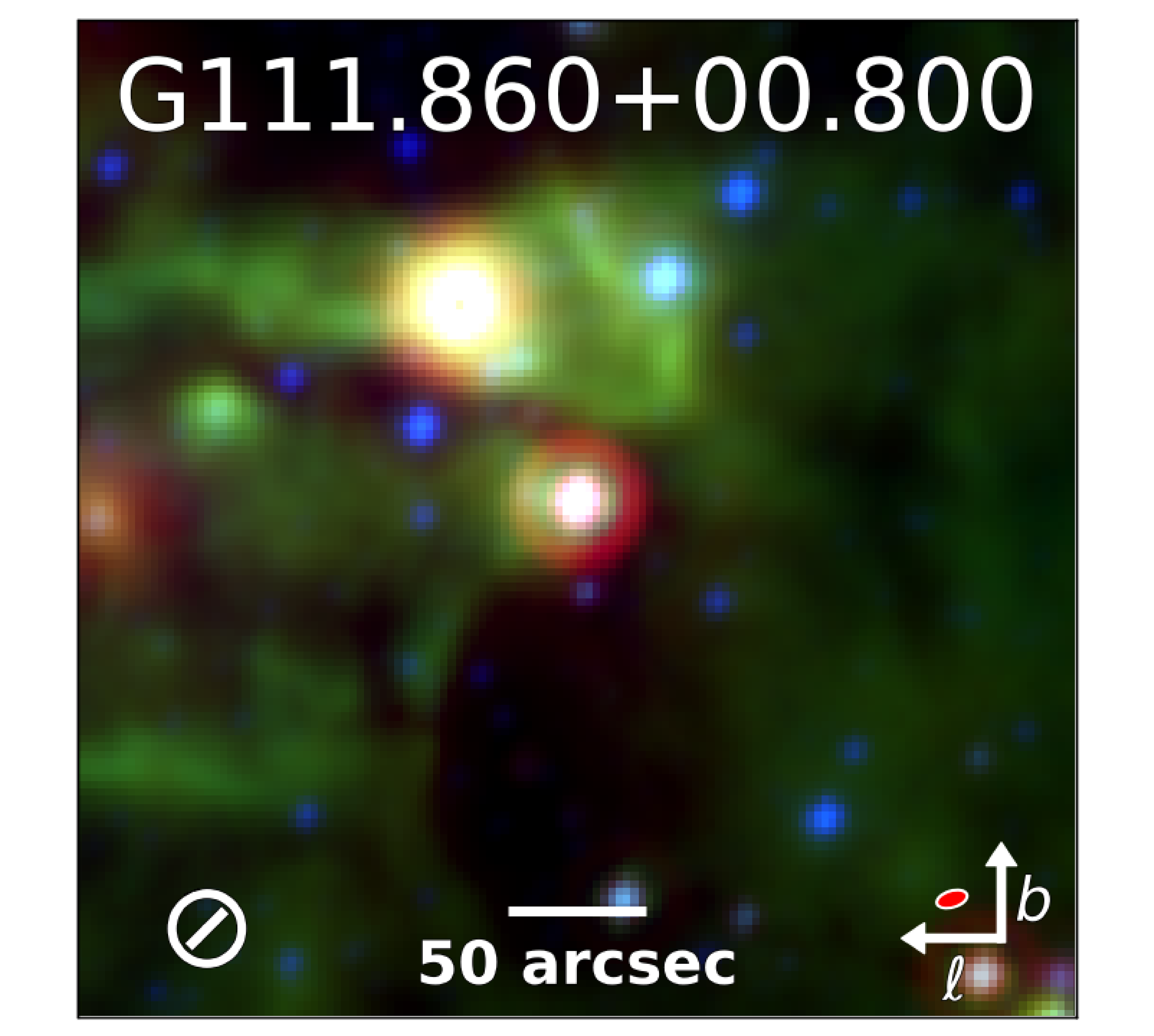}
\end{figure*}
\begin{figure*}[!htb]
\includegraphics[width=\figSize]{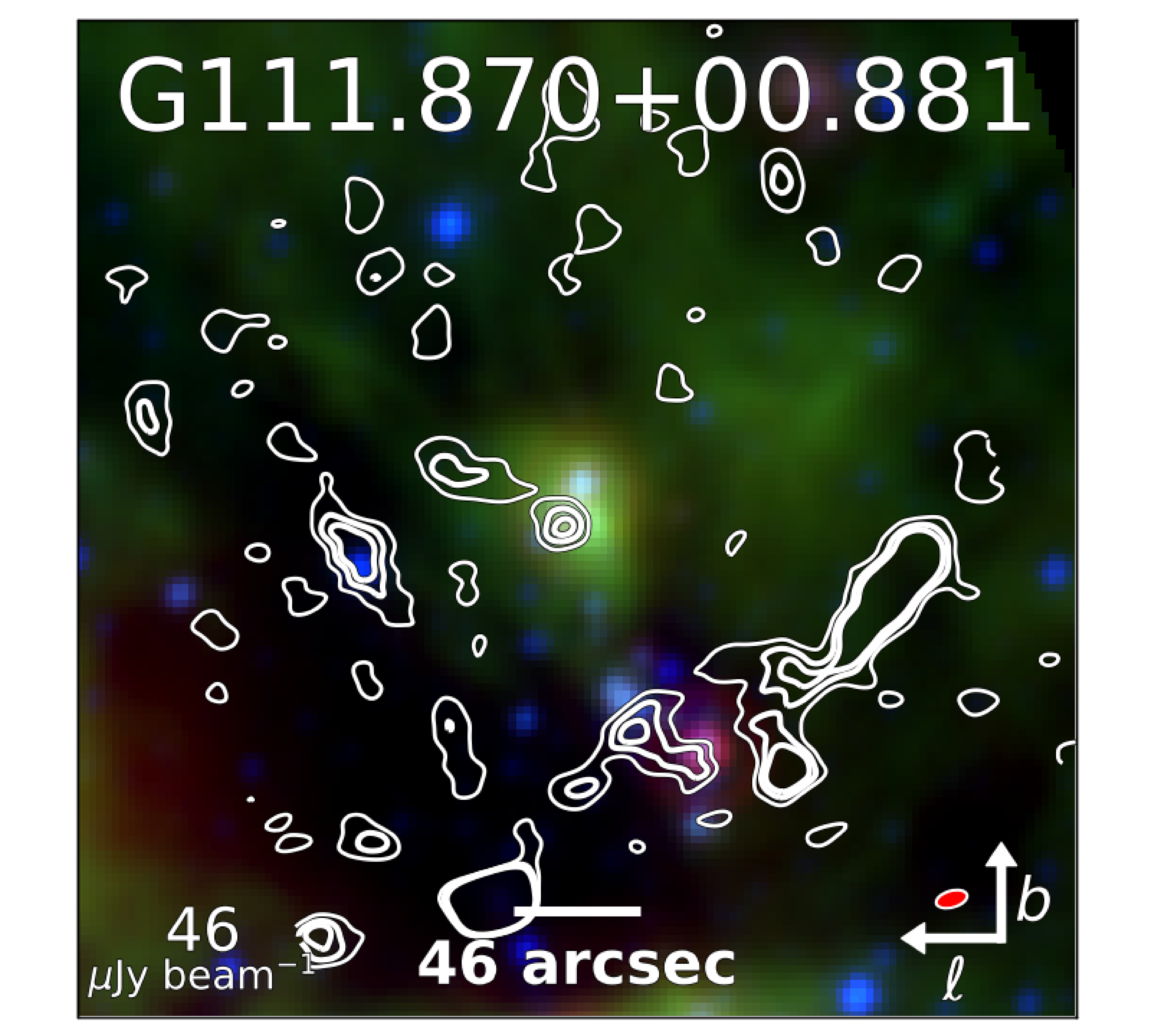}
\includegraphics[width=\figSize]{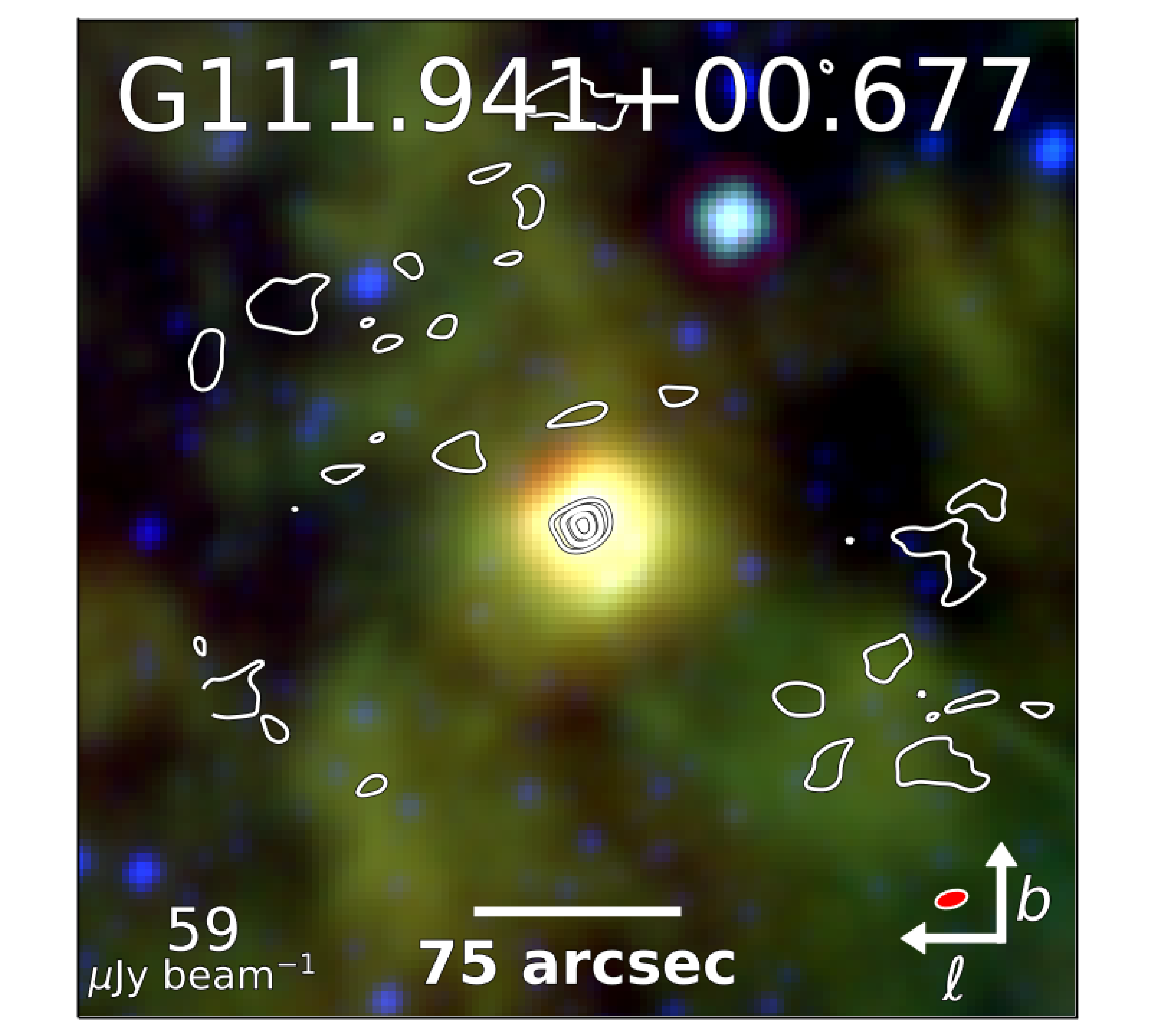}
\includegraphics[width=\figSize]{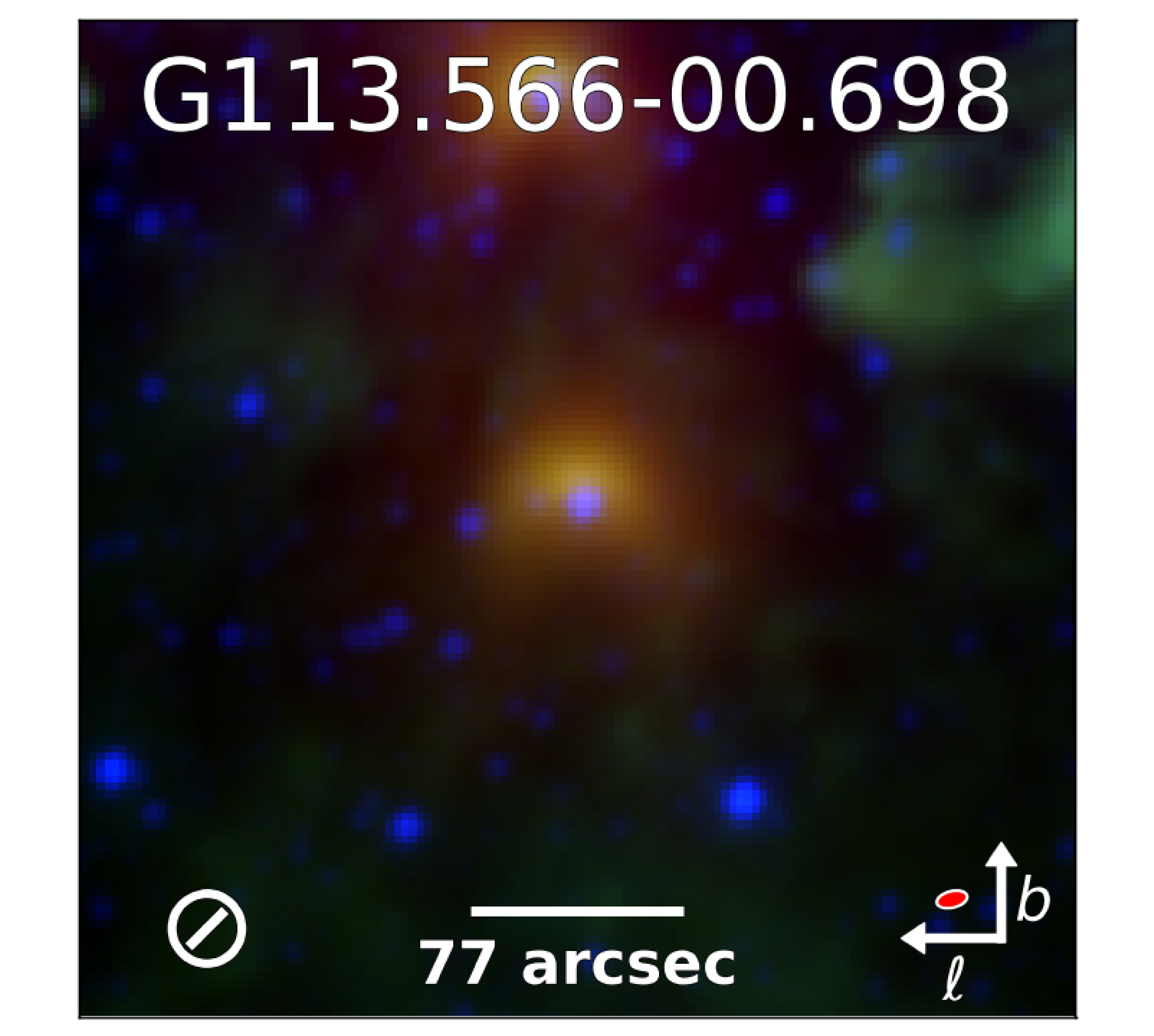}\\
\includegraphics[width=\figSize]{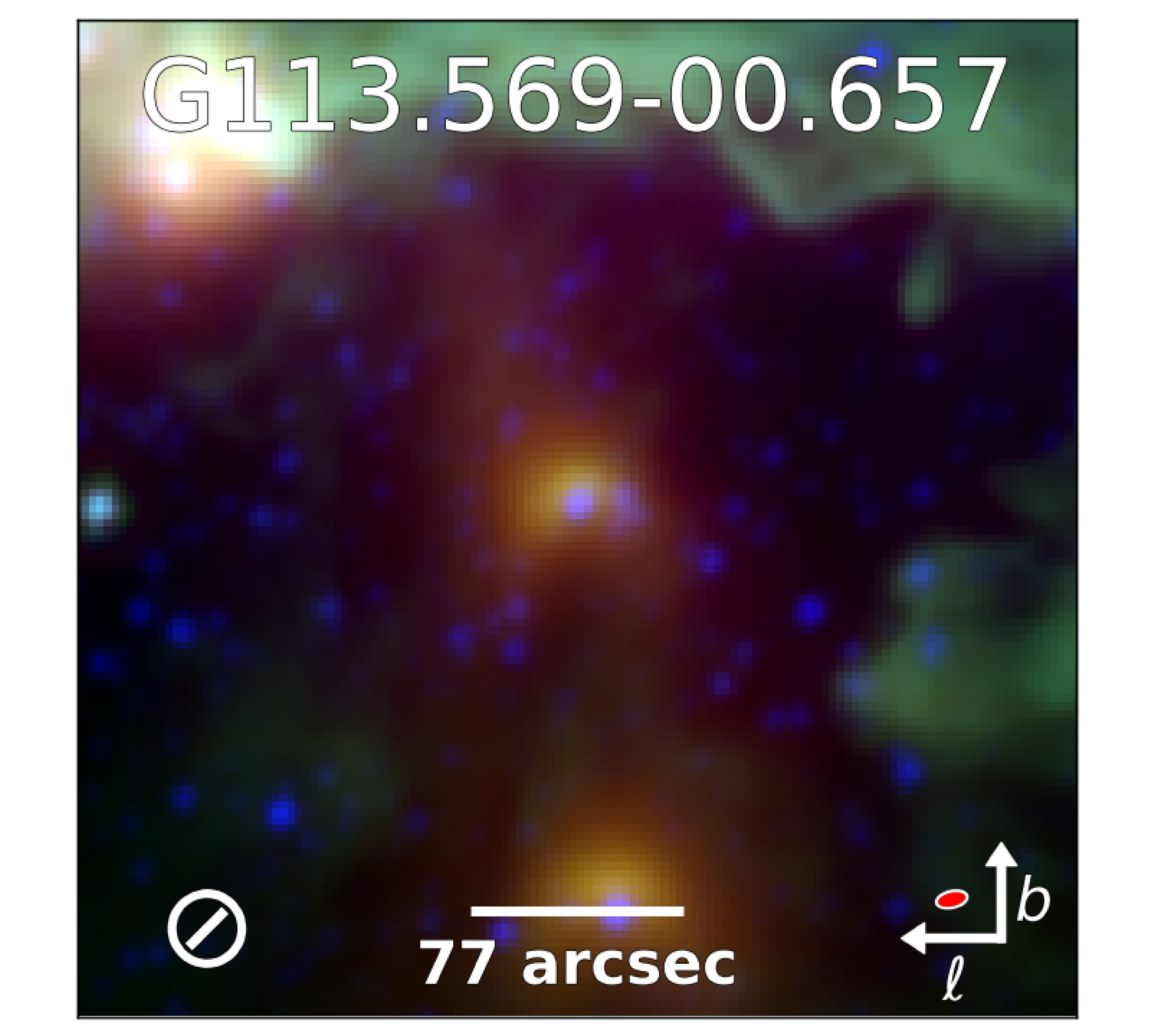}
\includegraphics[width=\figSize]{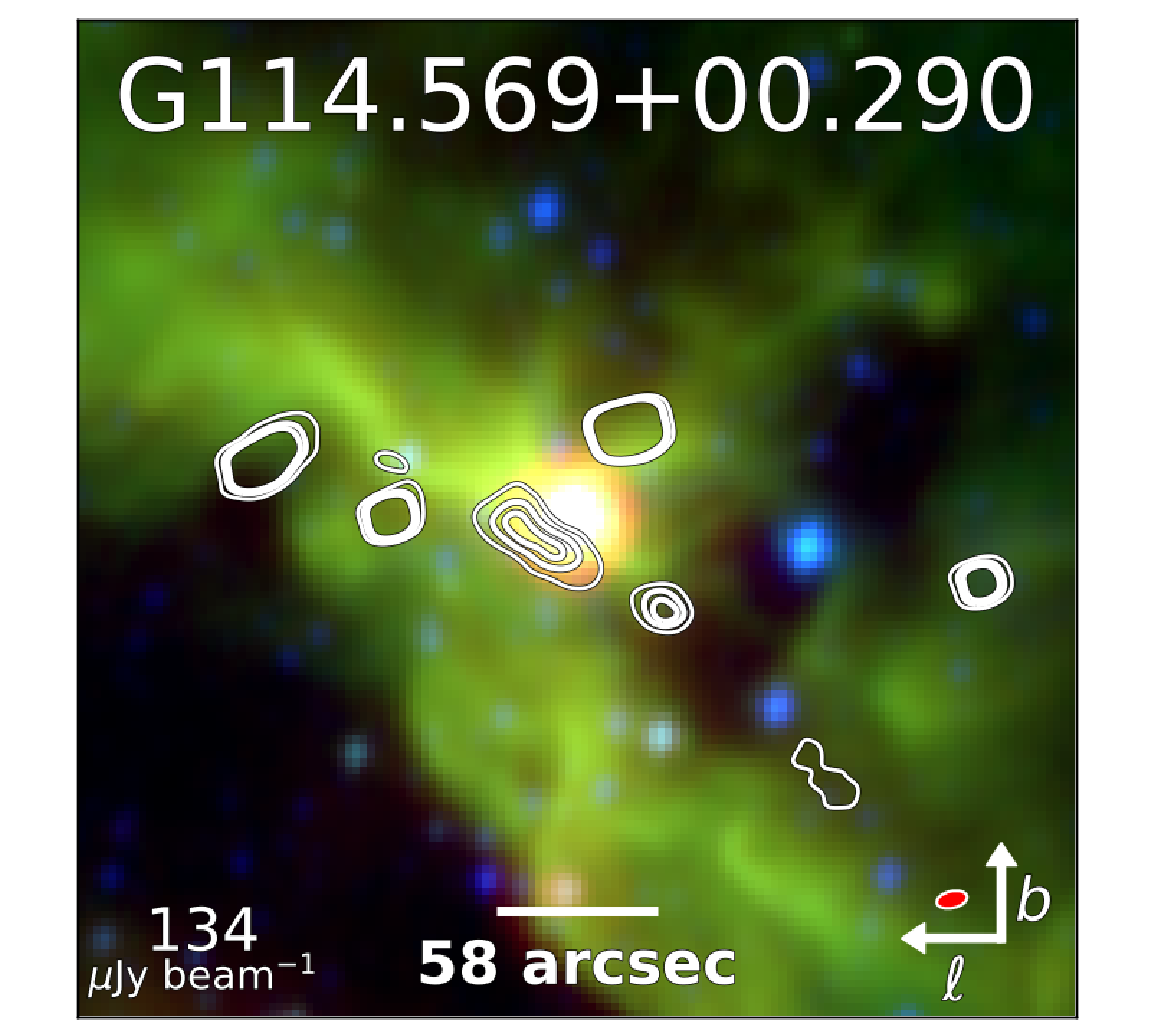}
\includegraphics[width=\figSize]{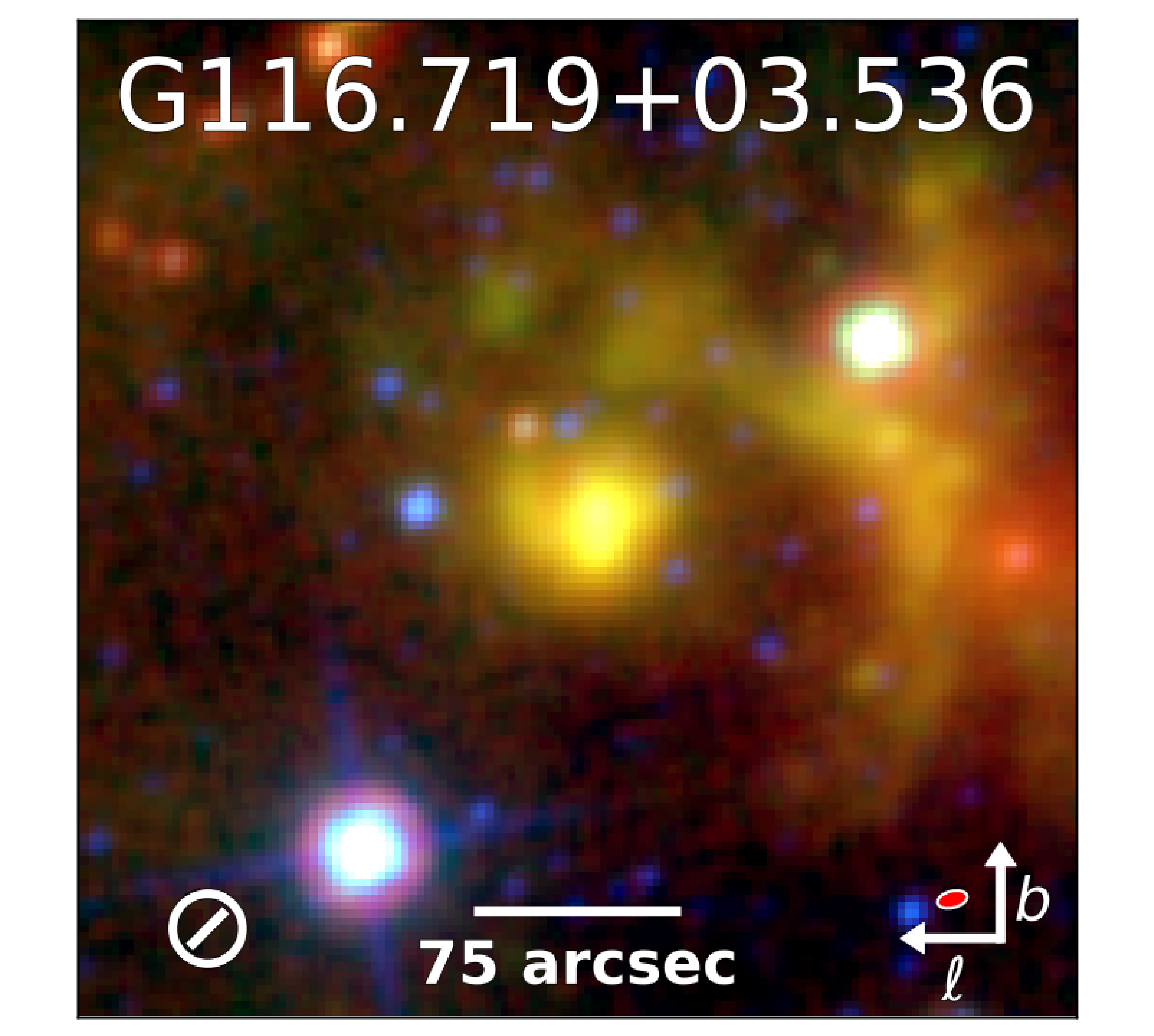}\\
\includegraphics[width=\figSize]{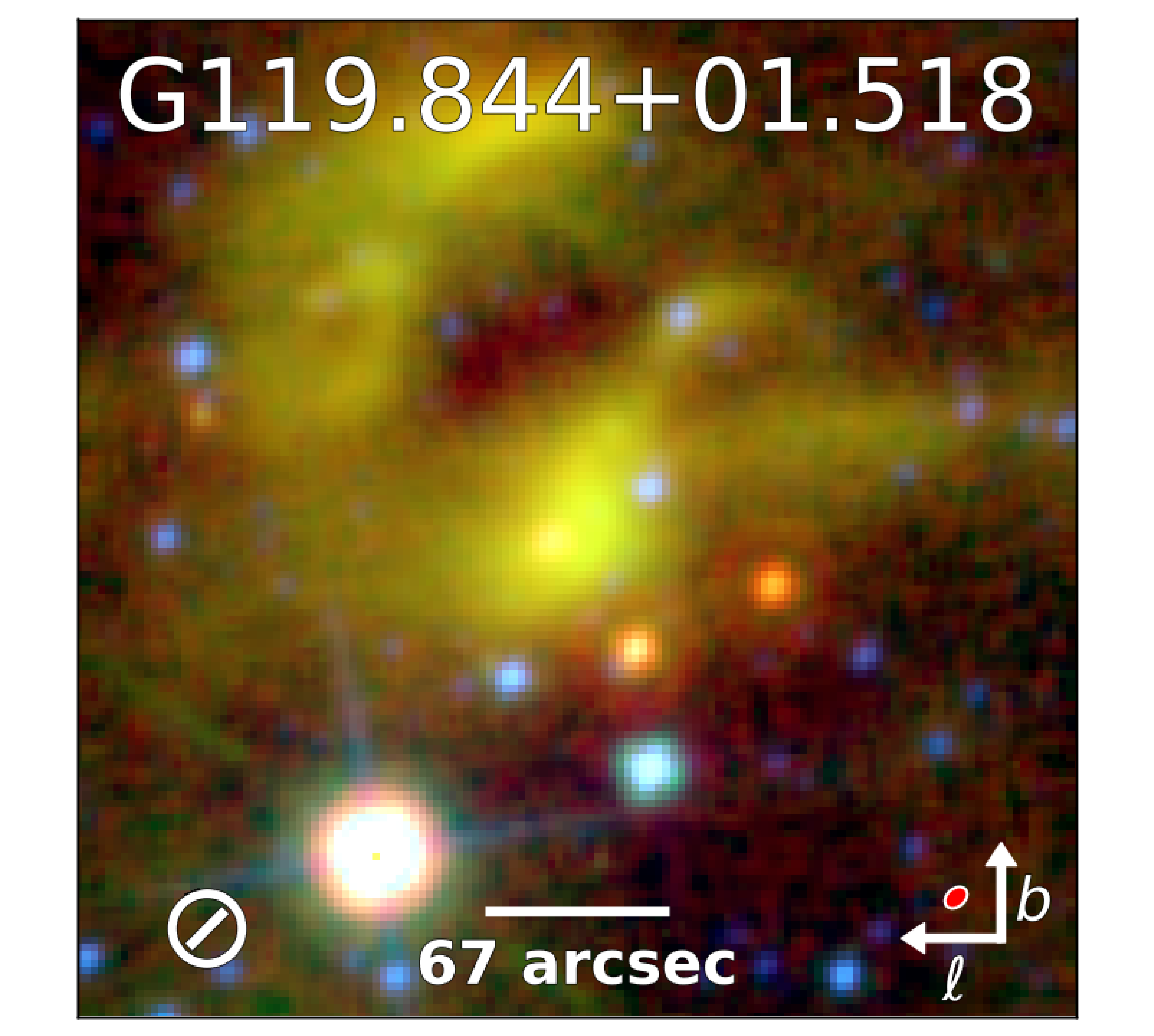}
\includegraphics[width=\figSize]{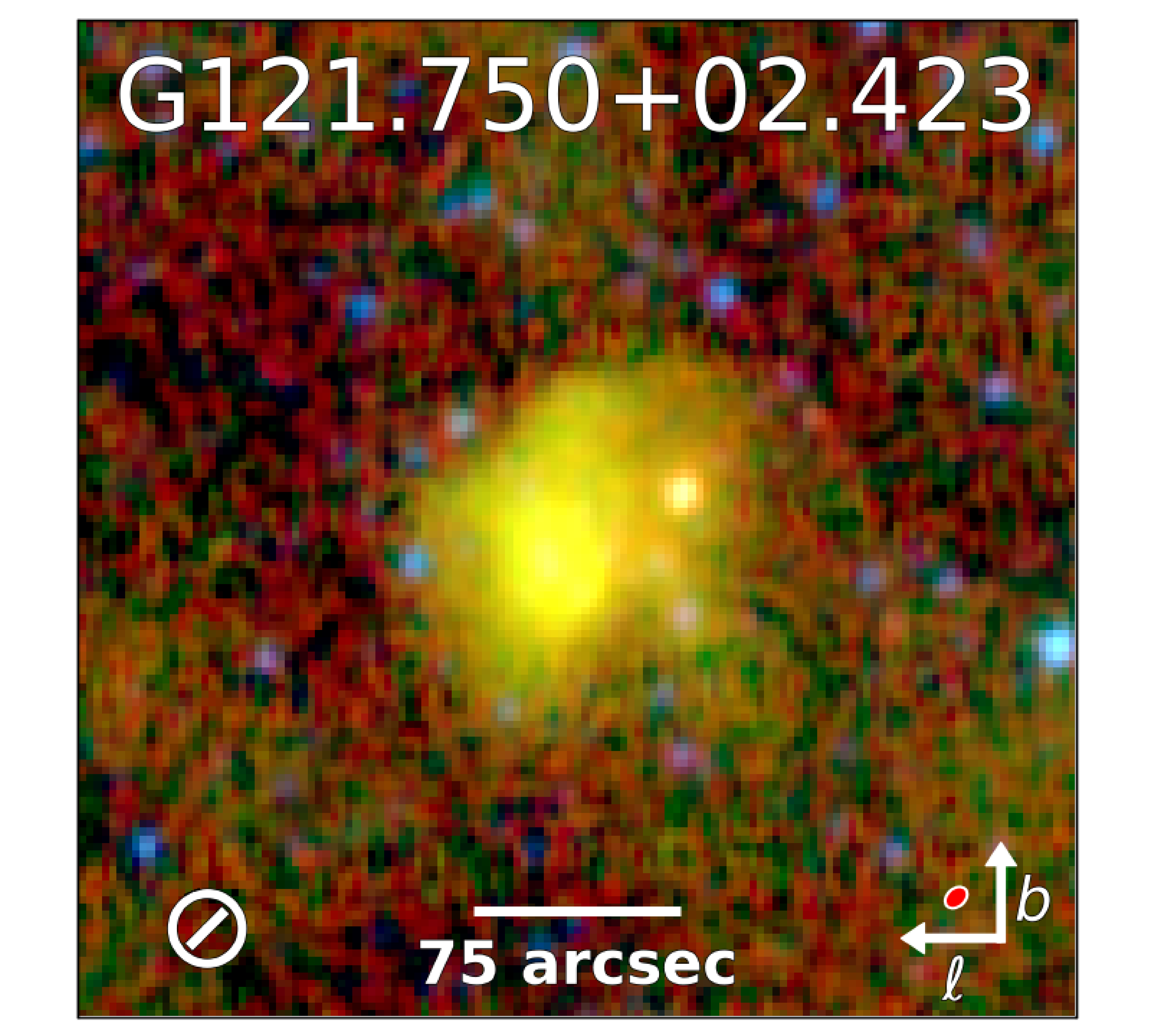}
\includegraphics[width=\figSize]{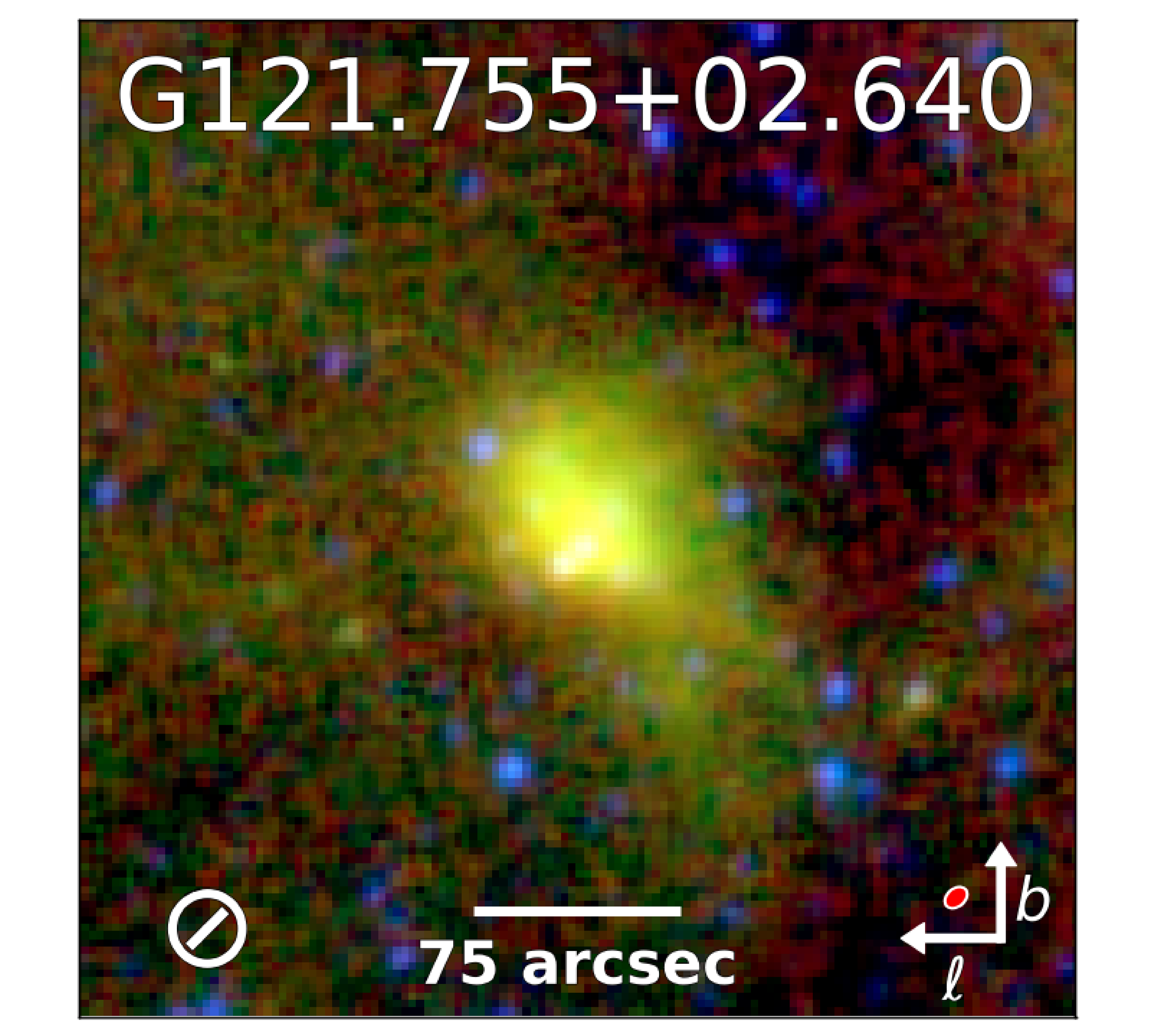}\\
\includegraphics[width=\figSize]{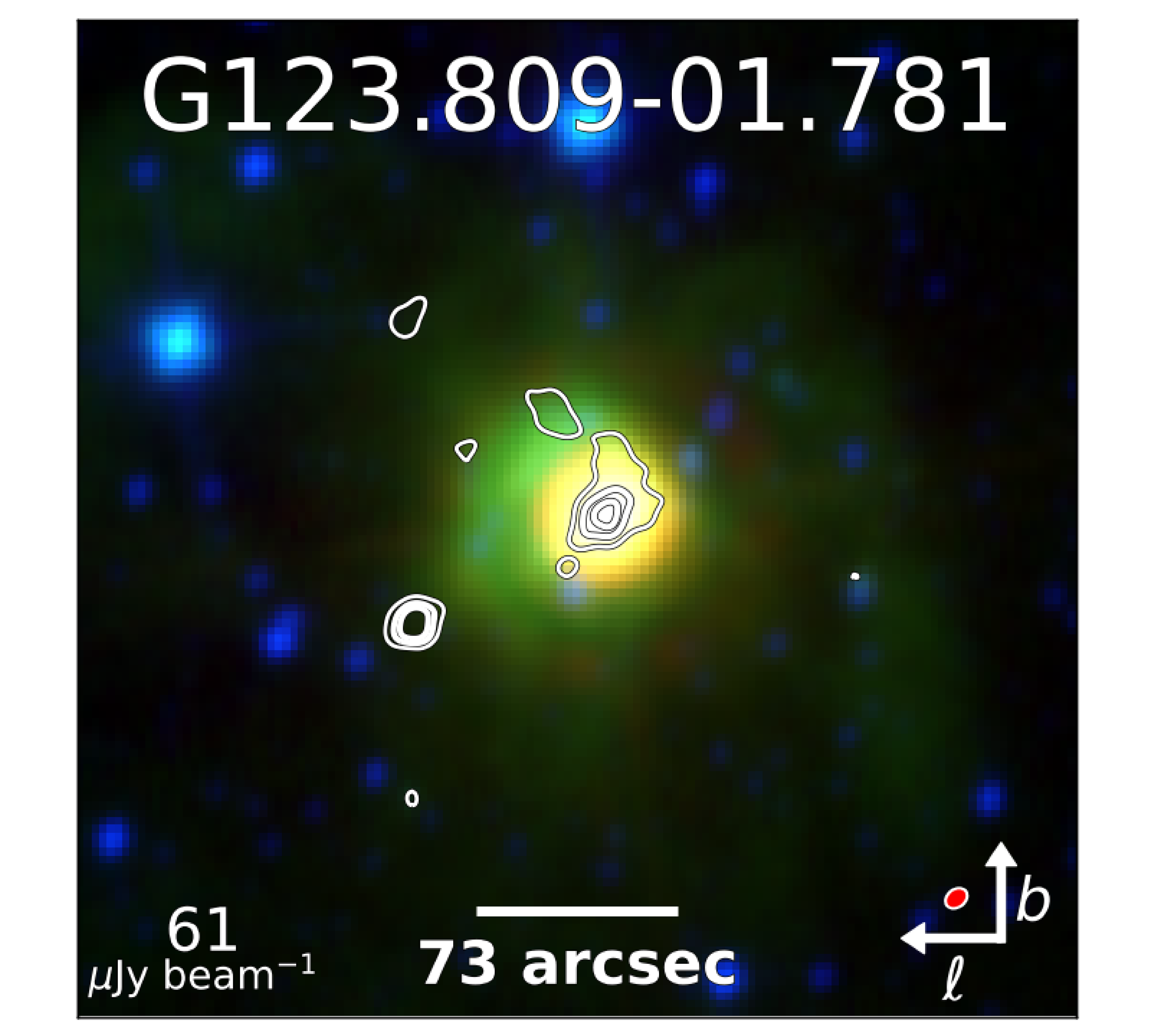}
\includegraphics[width=\figSize]{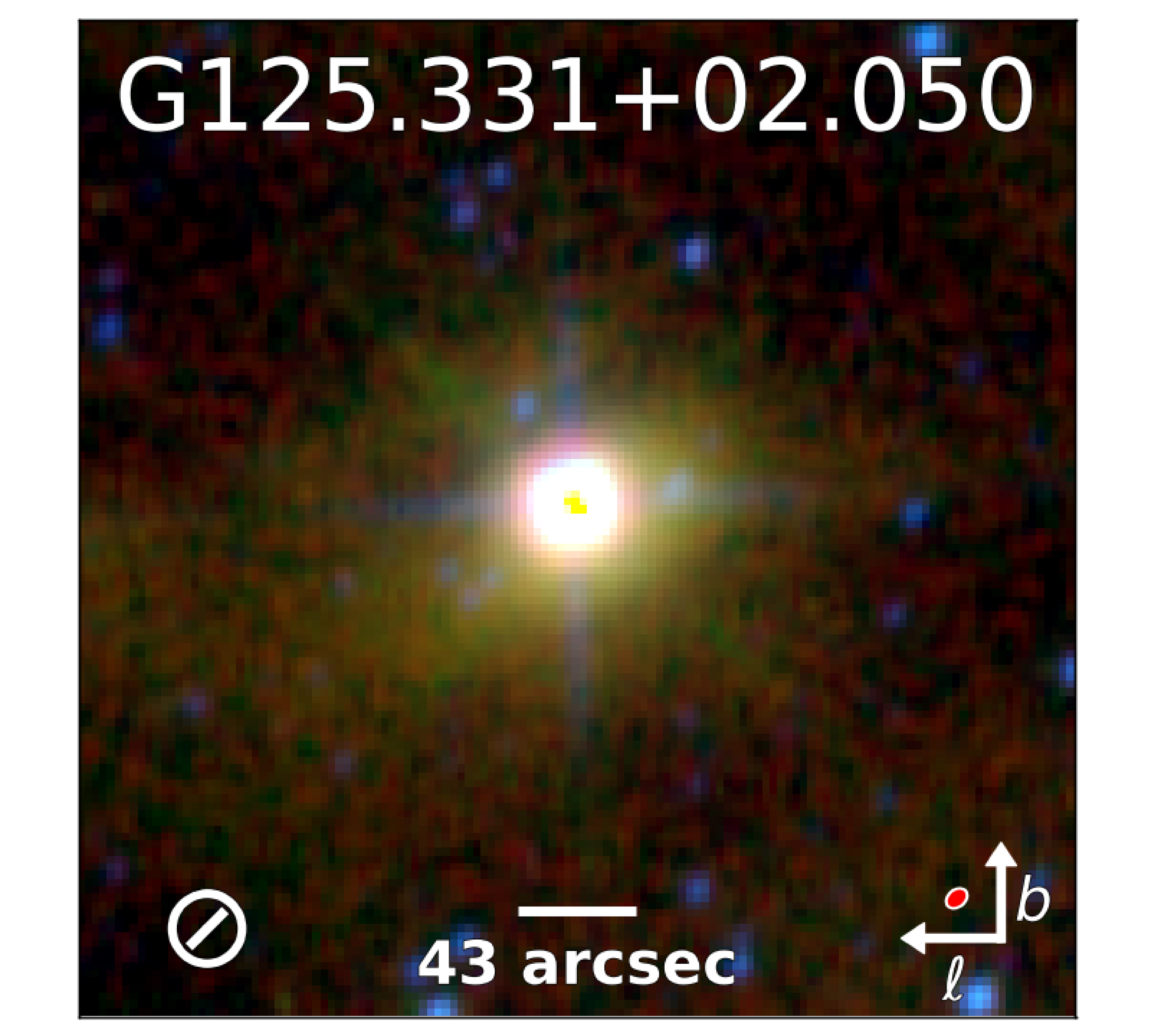}
\includegraphics[width=\figSize]{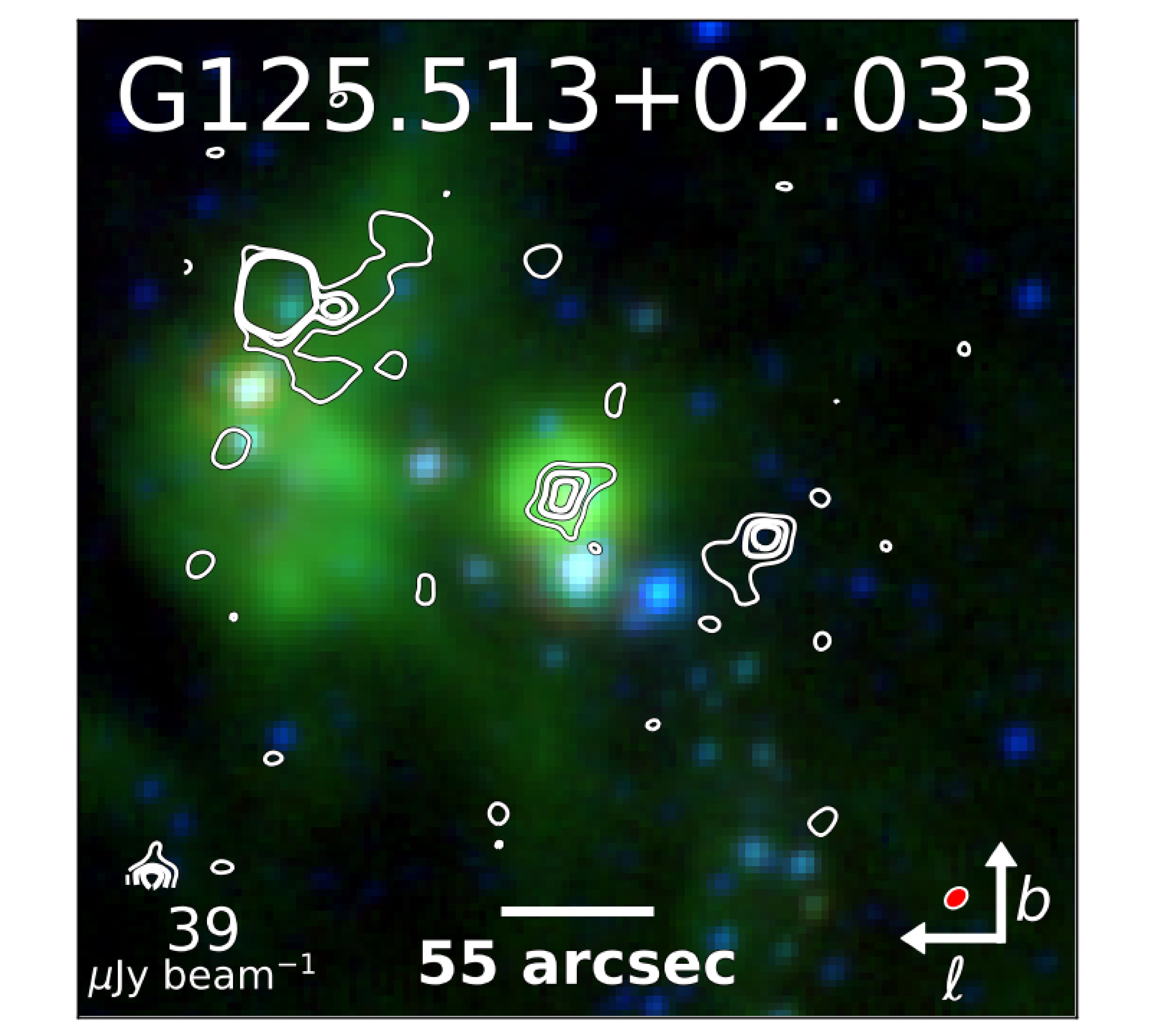}
\end{figure*}
\begin{figure*}[!htb]
\includegraphics[width=\figSize]{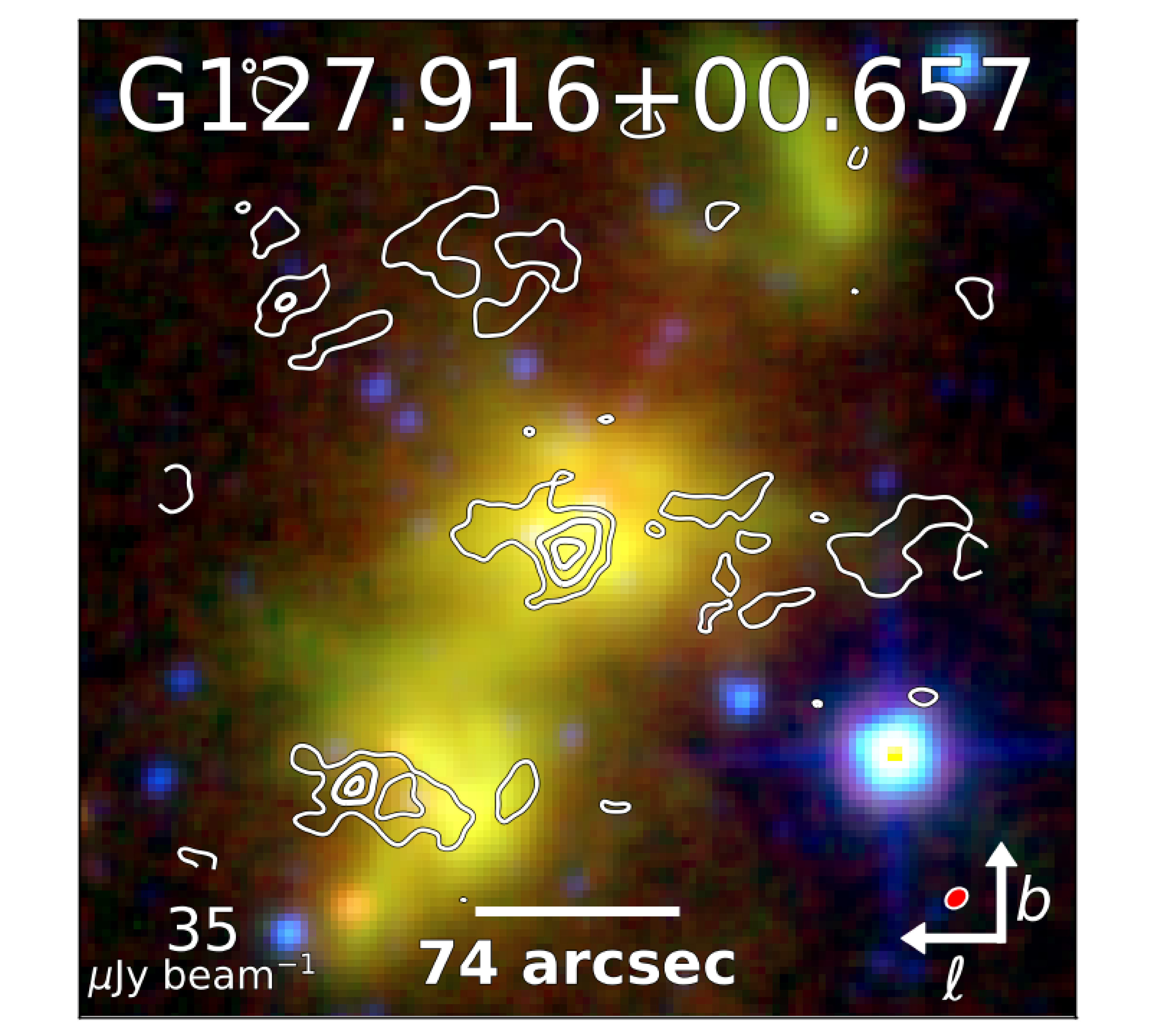}
\includegraphics[width=\figSize]{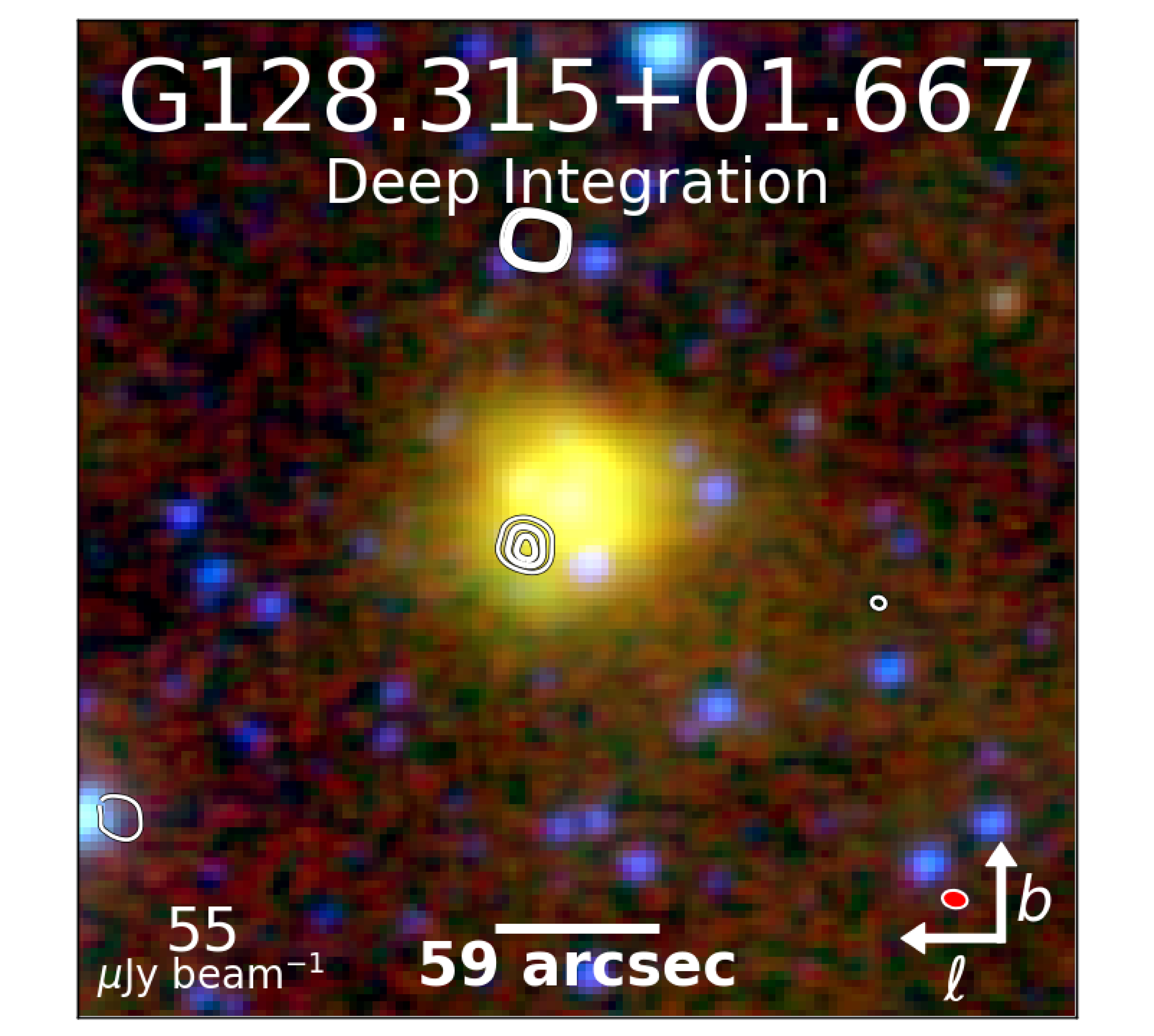}
\includegraphics[width=\figSize]{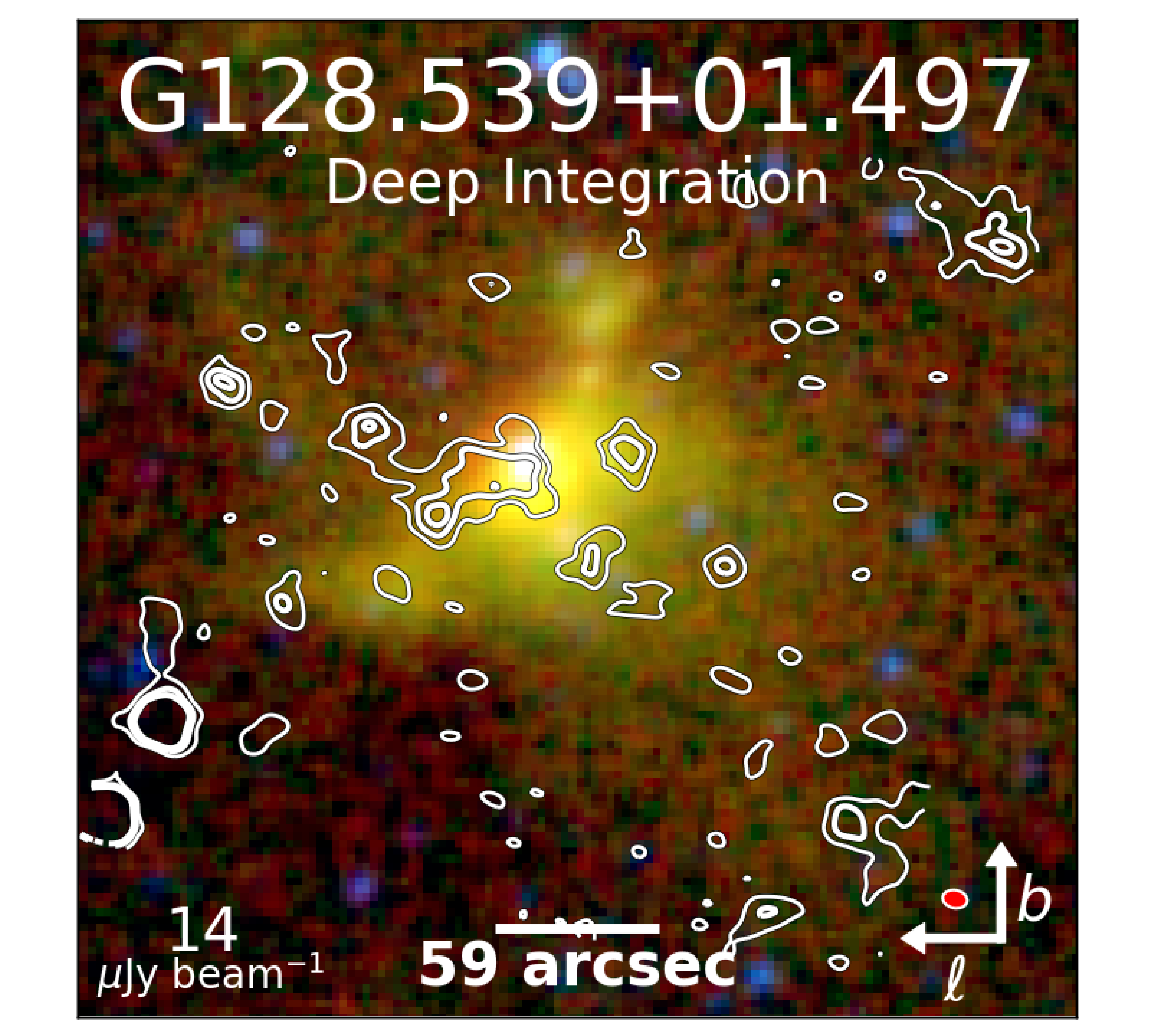}\\
\includegraphics[width=\figSize]{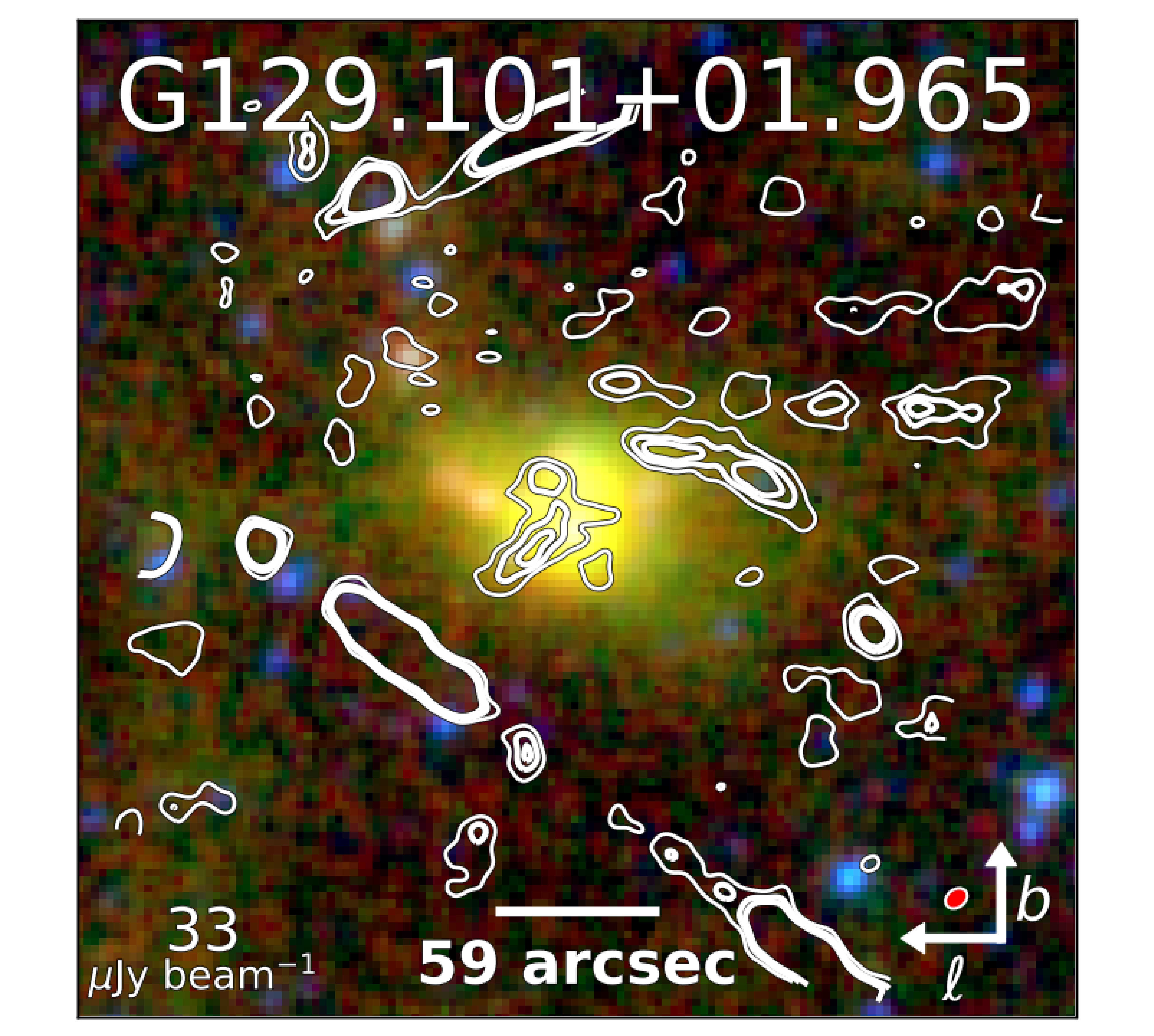}
\includegraphics[width=\figSize]{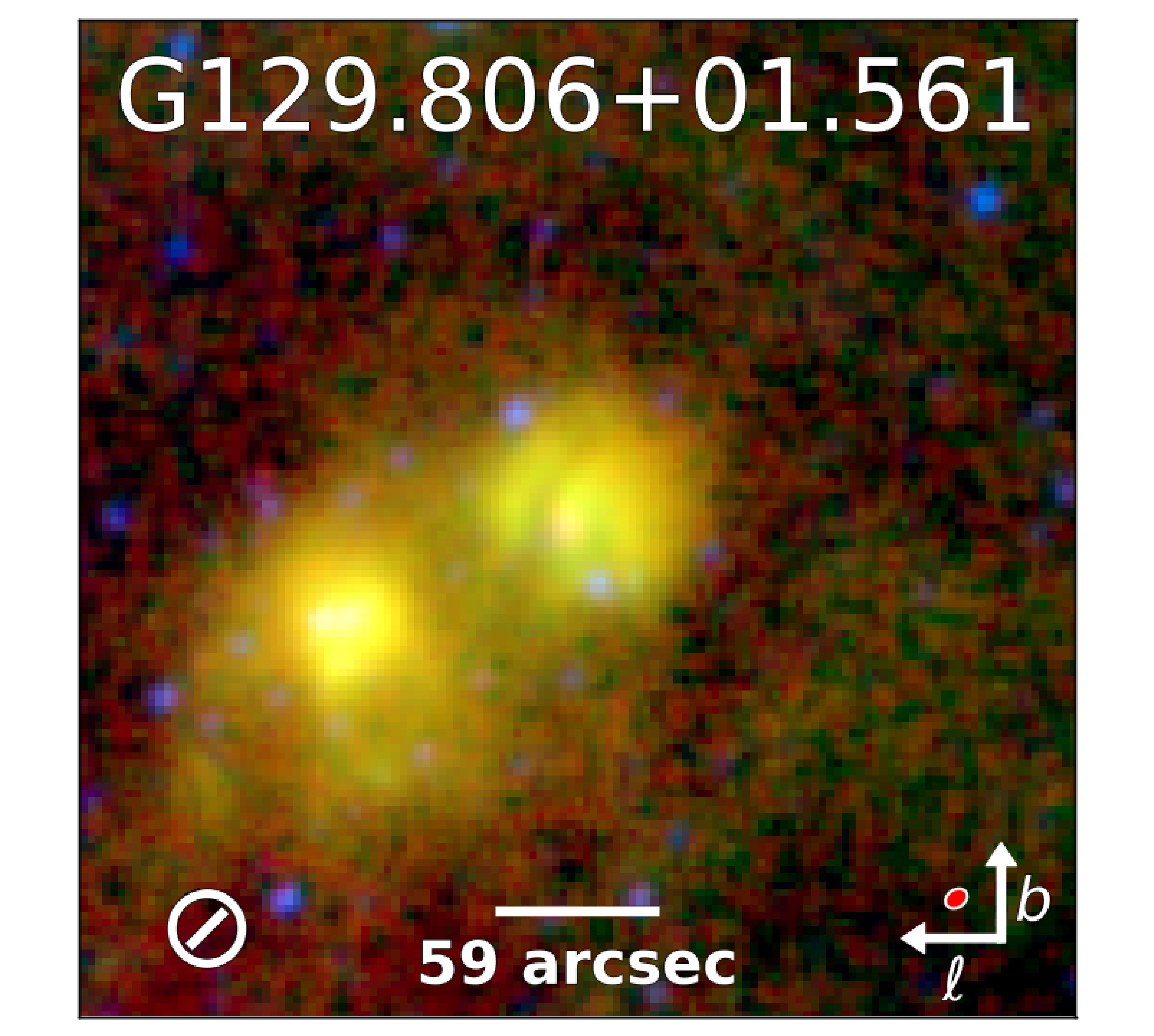}
\includegraphics[width=\figSize]{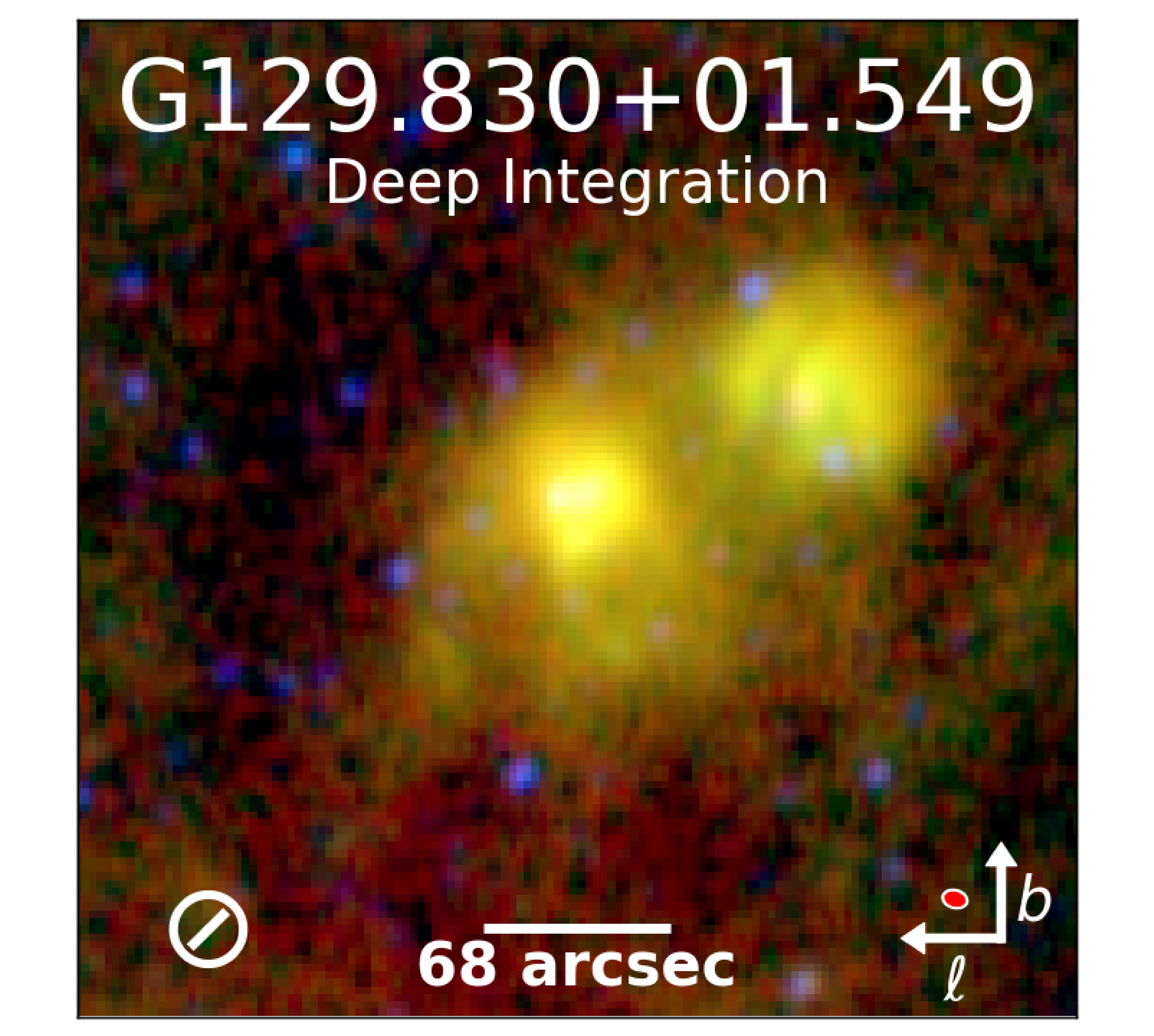}\\
\includegraphics[width=\figSize]{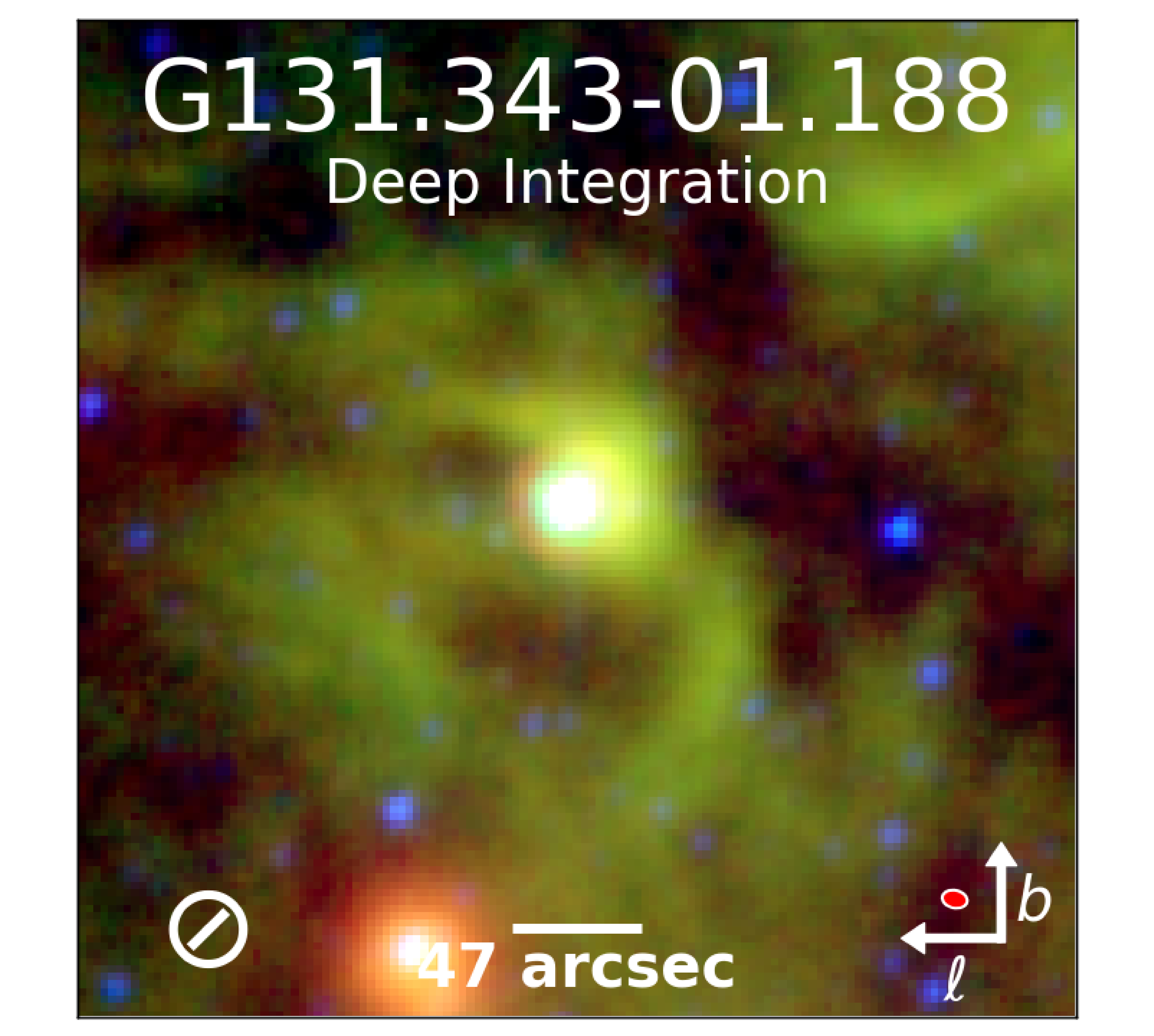}
\includegraphics[width=\figSize]{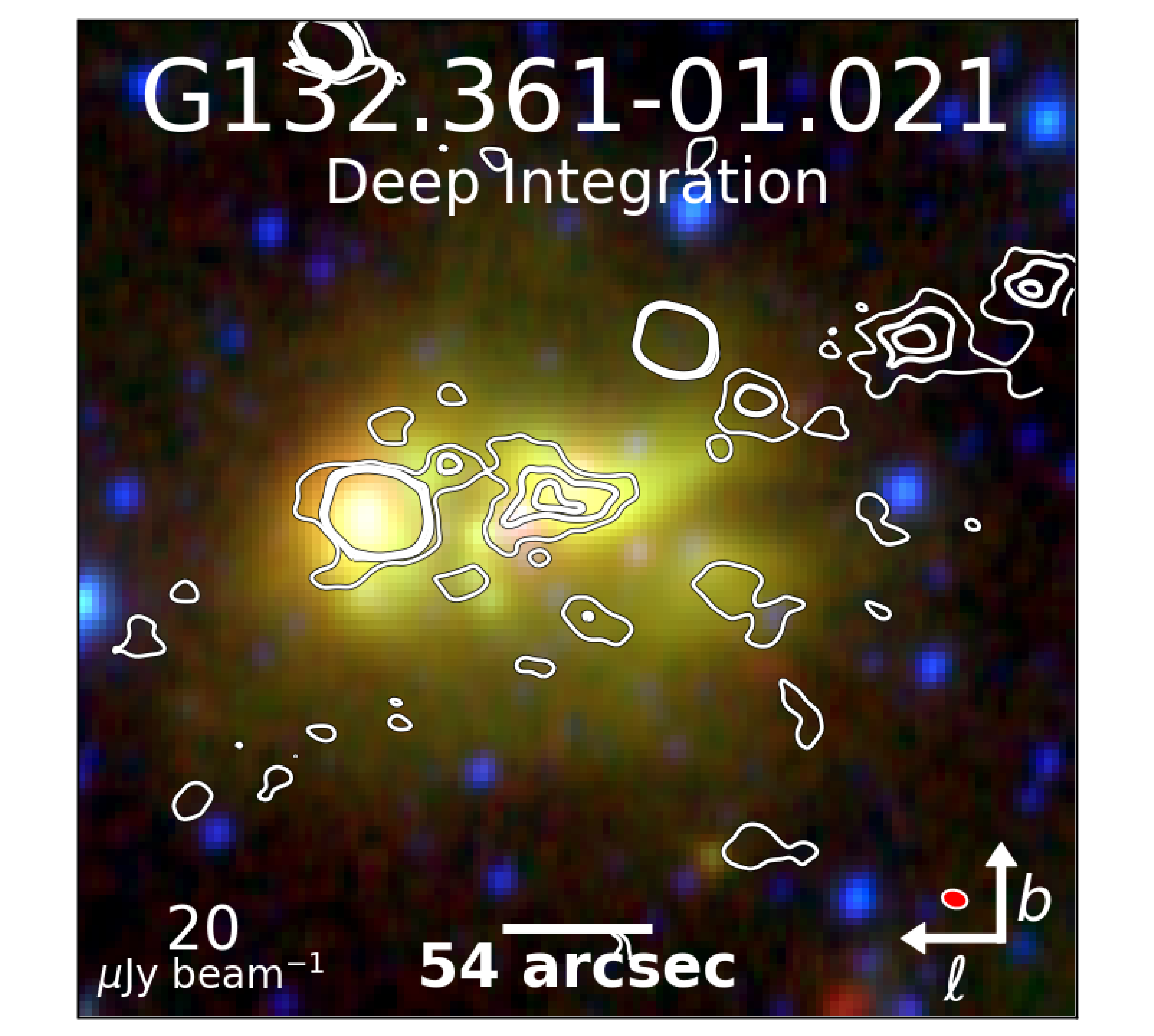}
\includegraphics[width=\figSize]{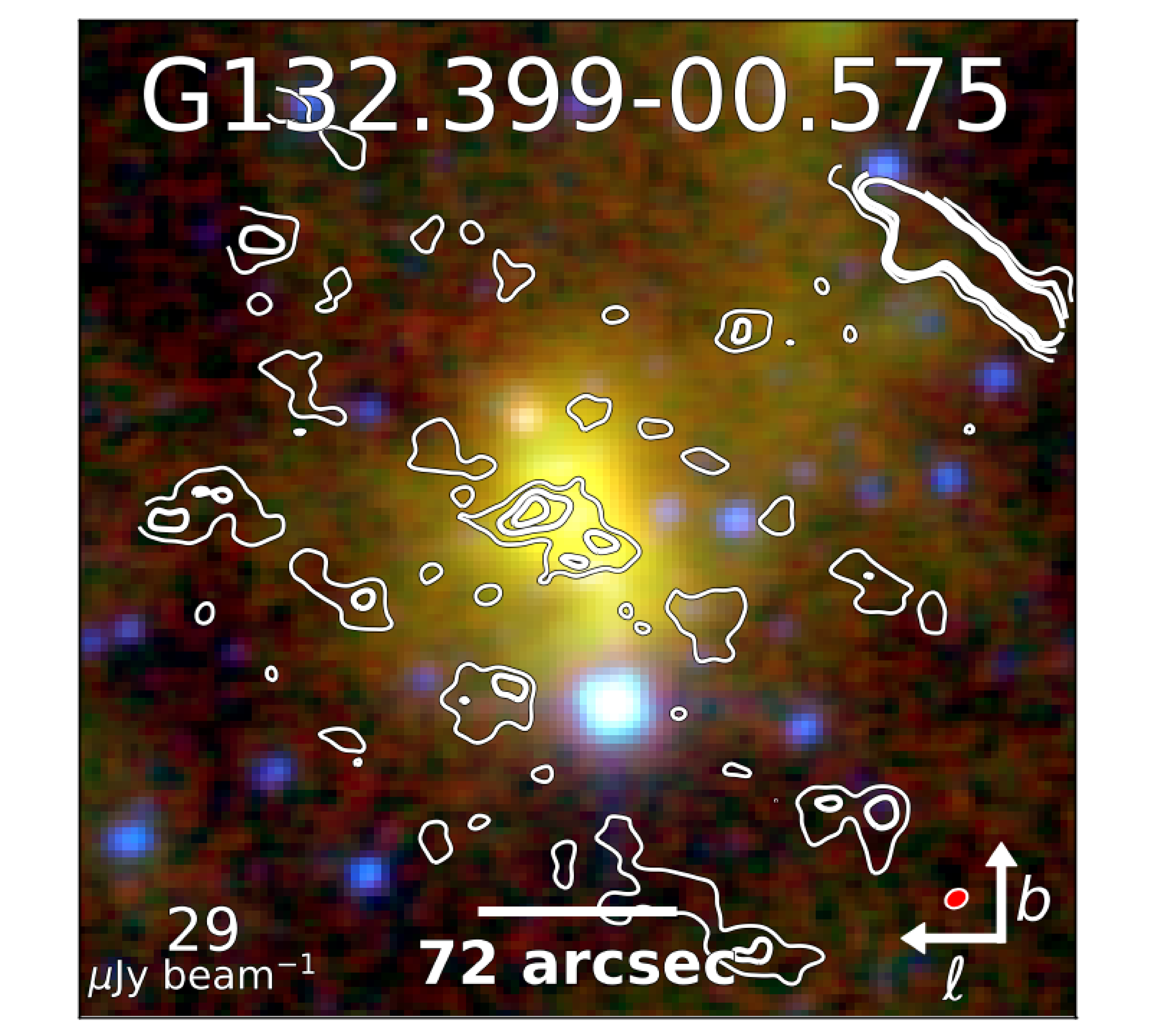}\\
\includegraphics[width=\figSize]{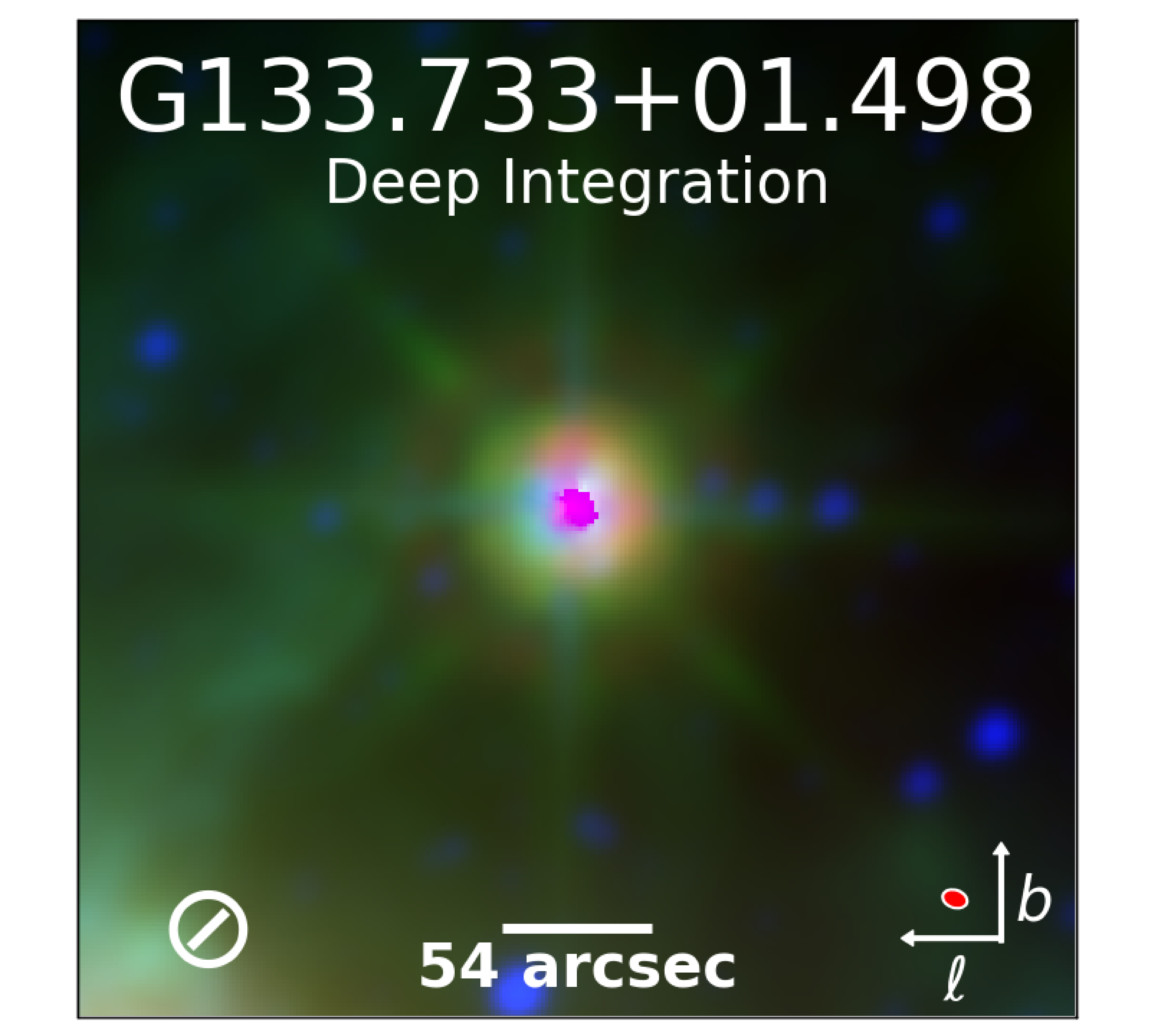}
\includegraphics[width=\figSize]{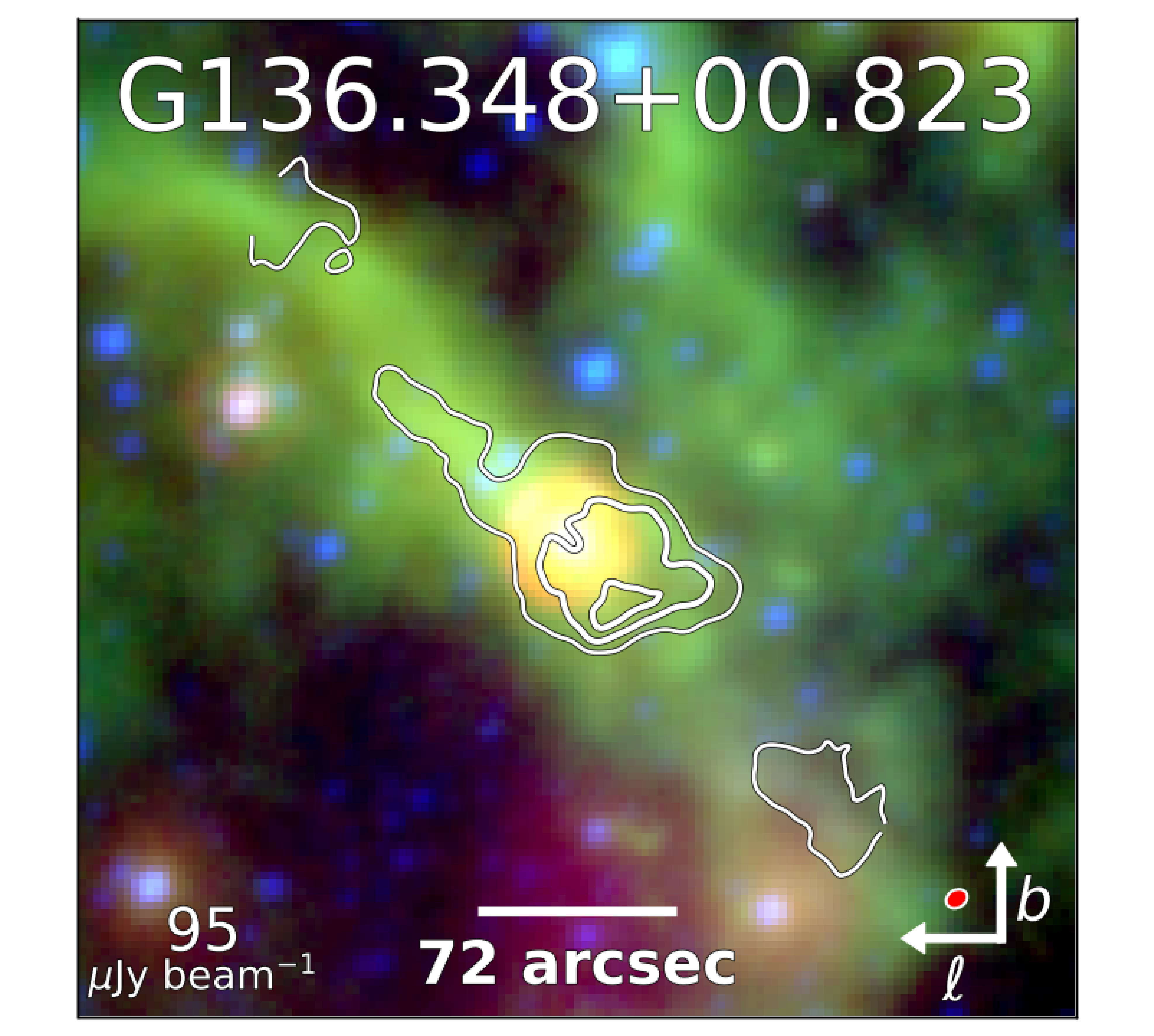}
\includegraphics[width=\figSize]{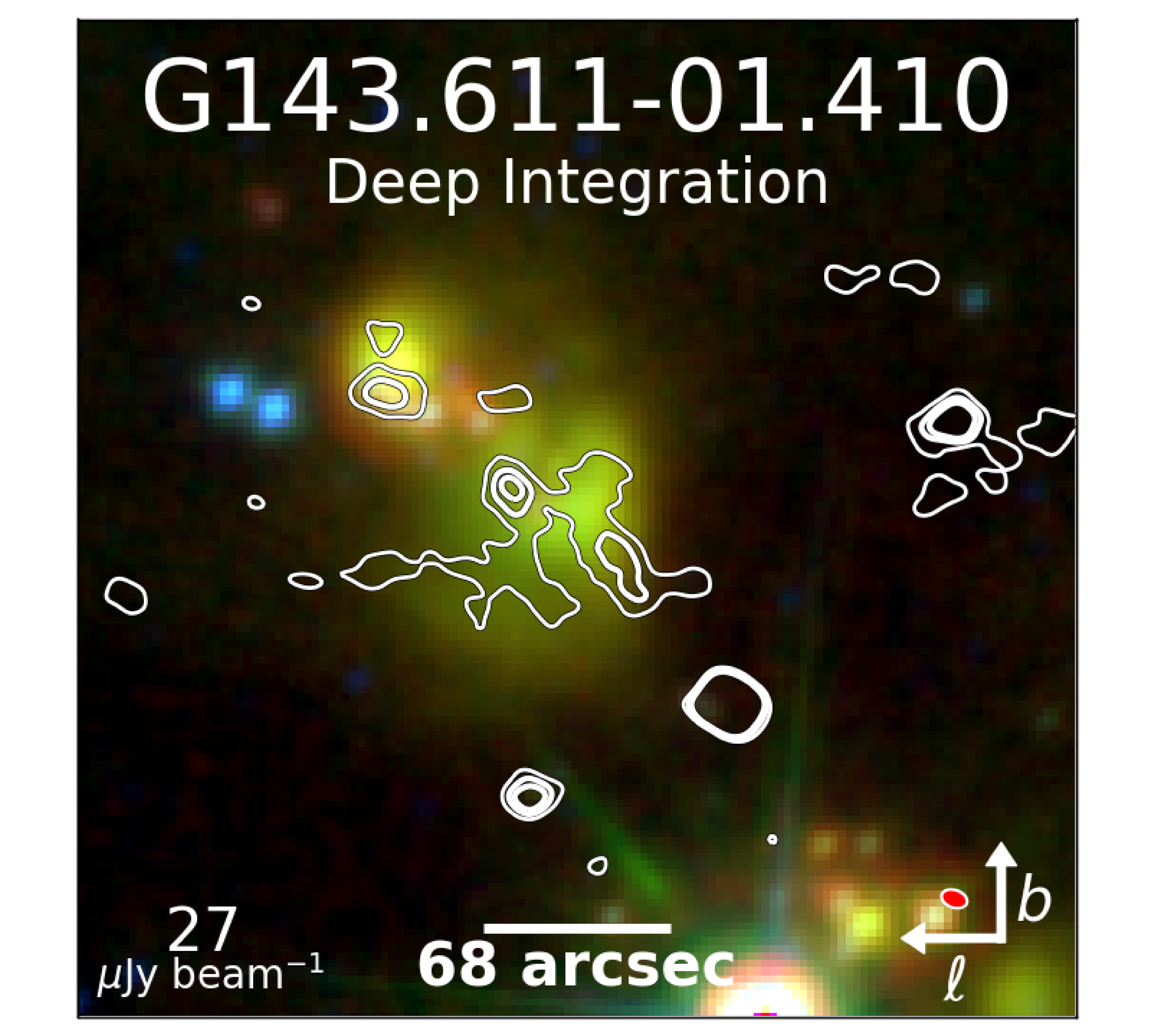}
\end{figure*}
\begin{figure*}[!htb]
\includegraphics[width=\figSize]{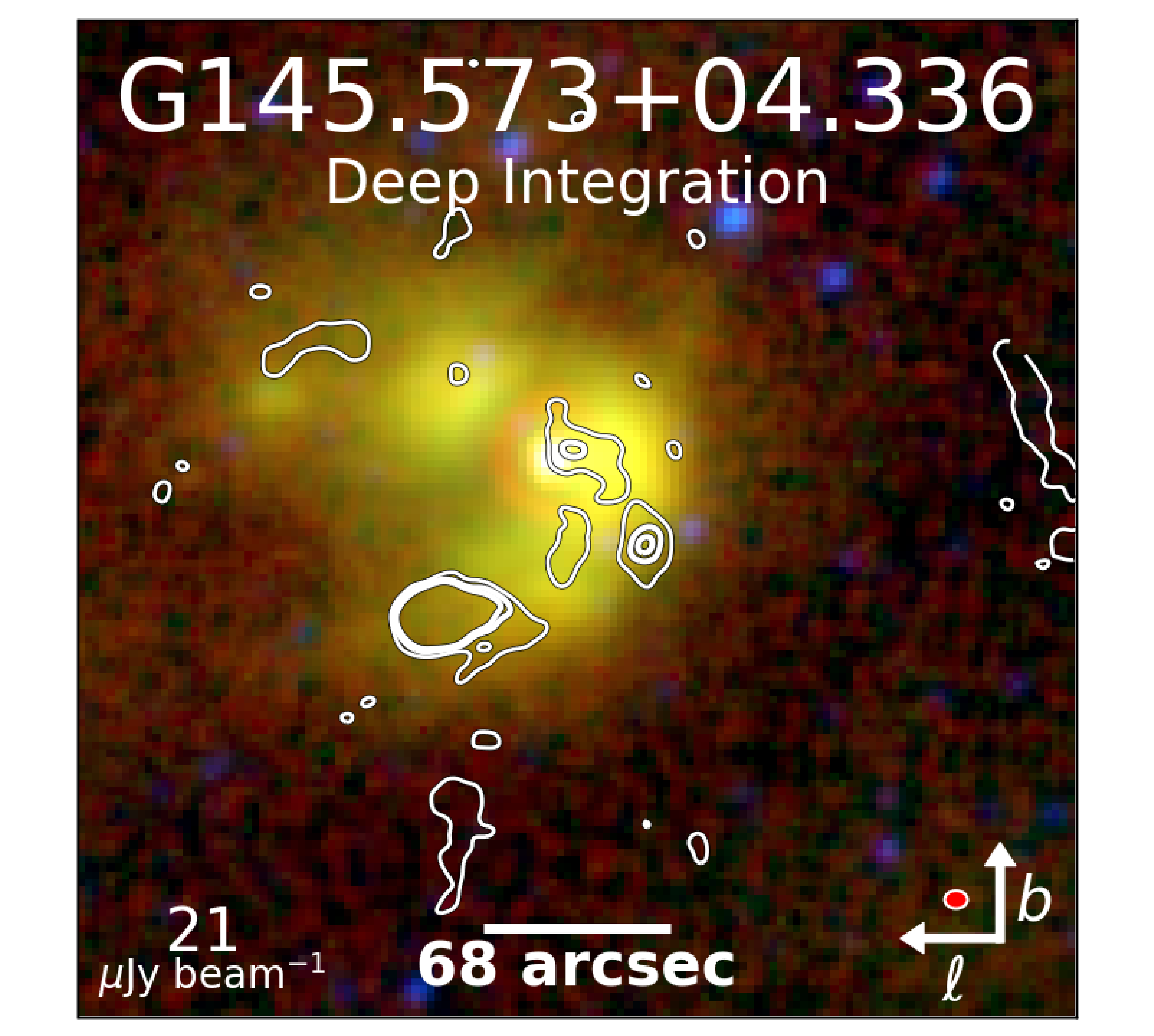}
\includegraphics[width=\figSize]{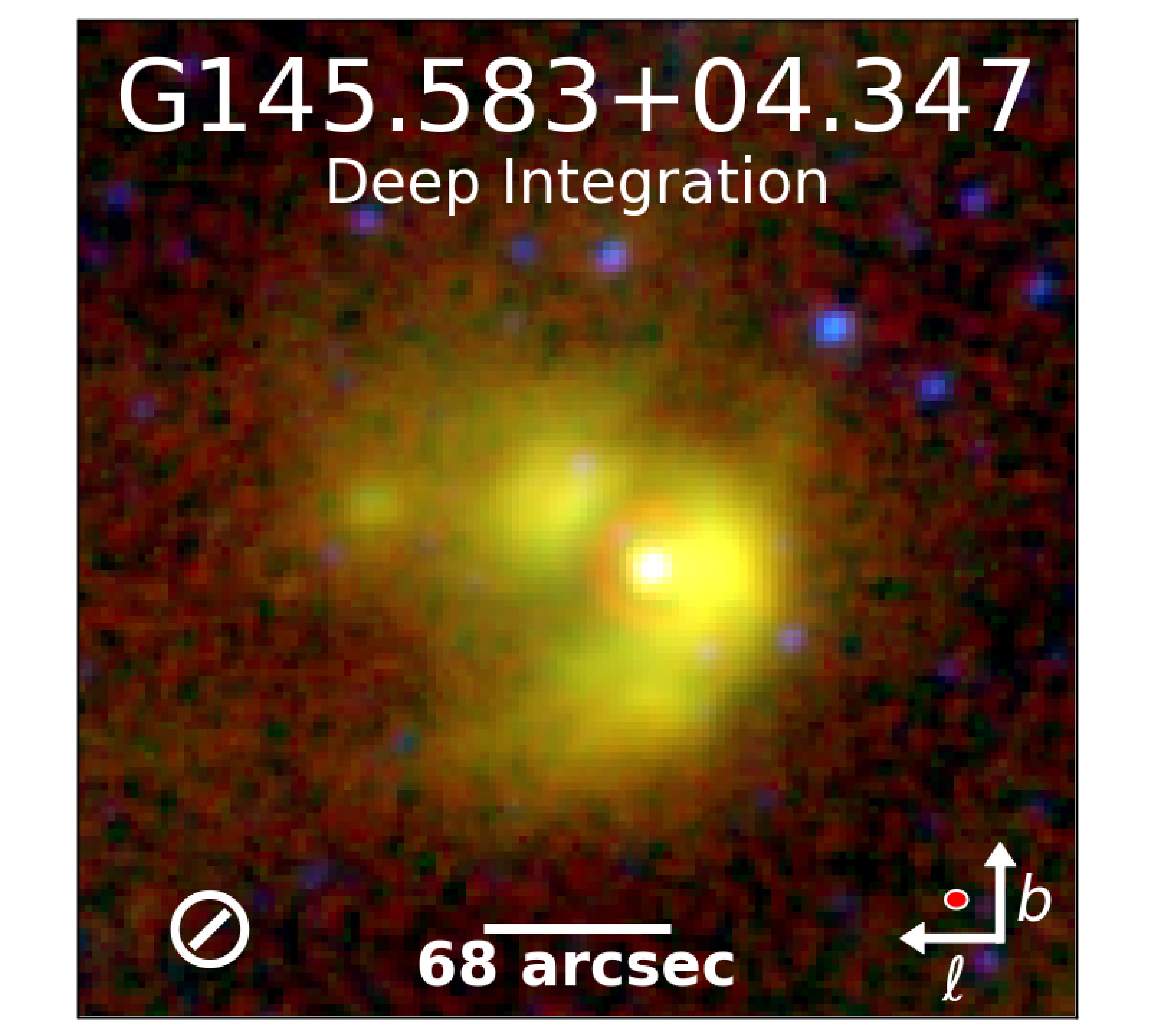}
\includegraphics[width=\figSize]{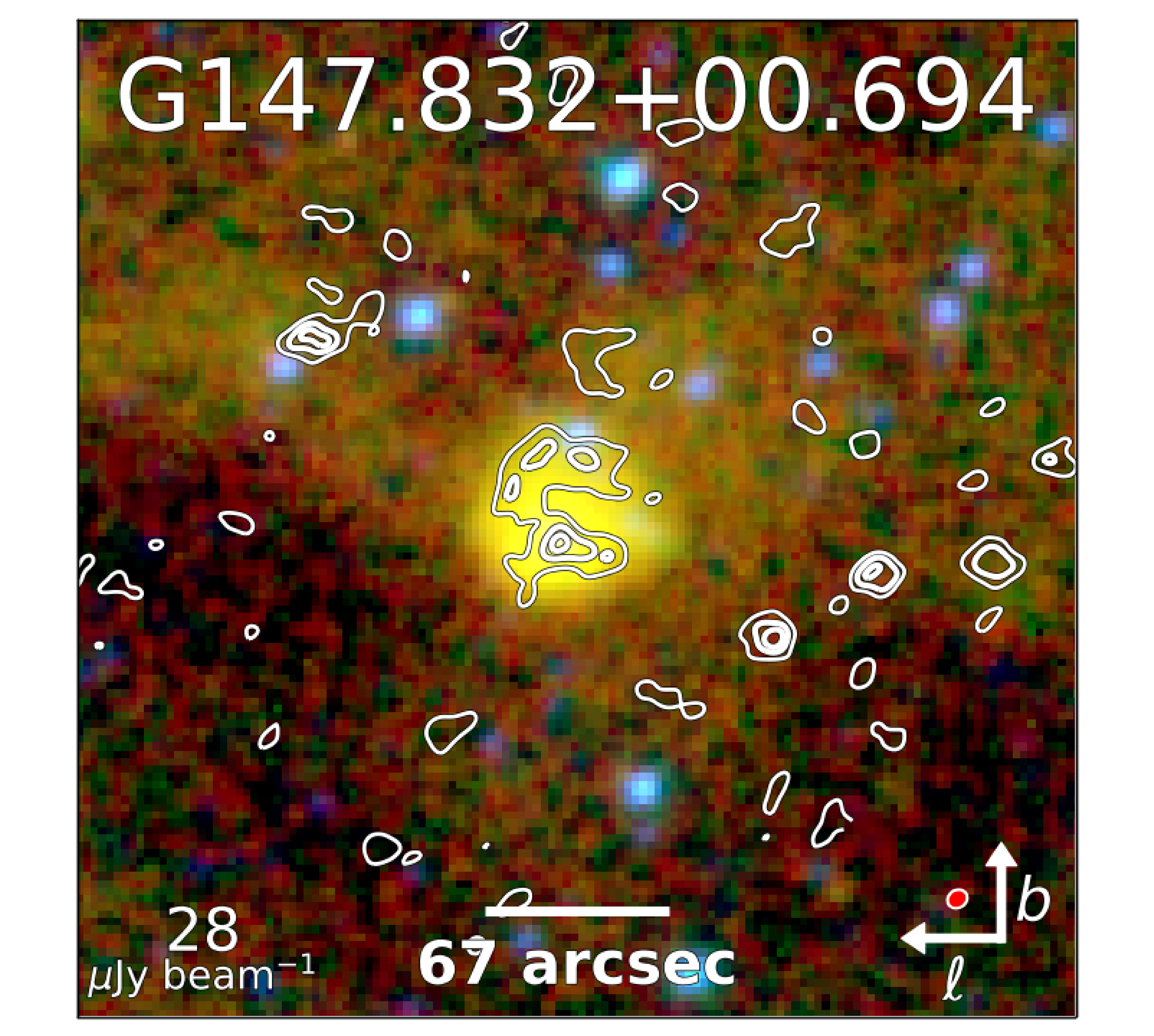}\\
\includegraphics[width=\figSize]{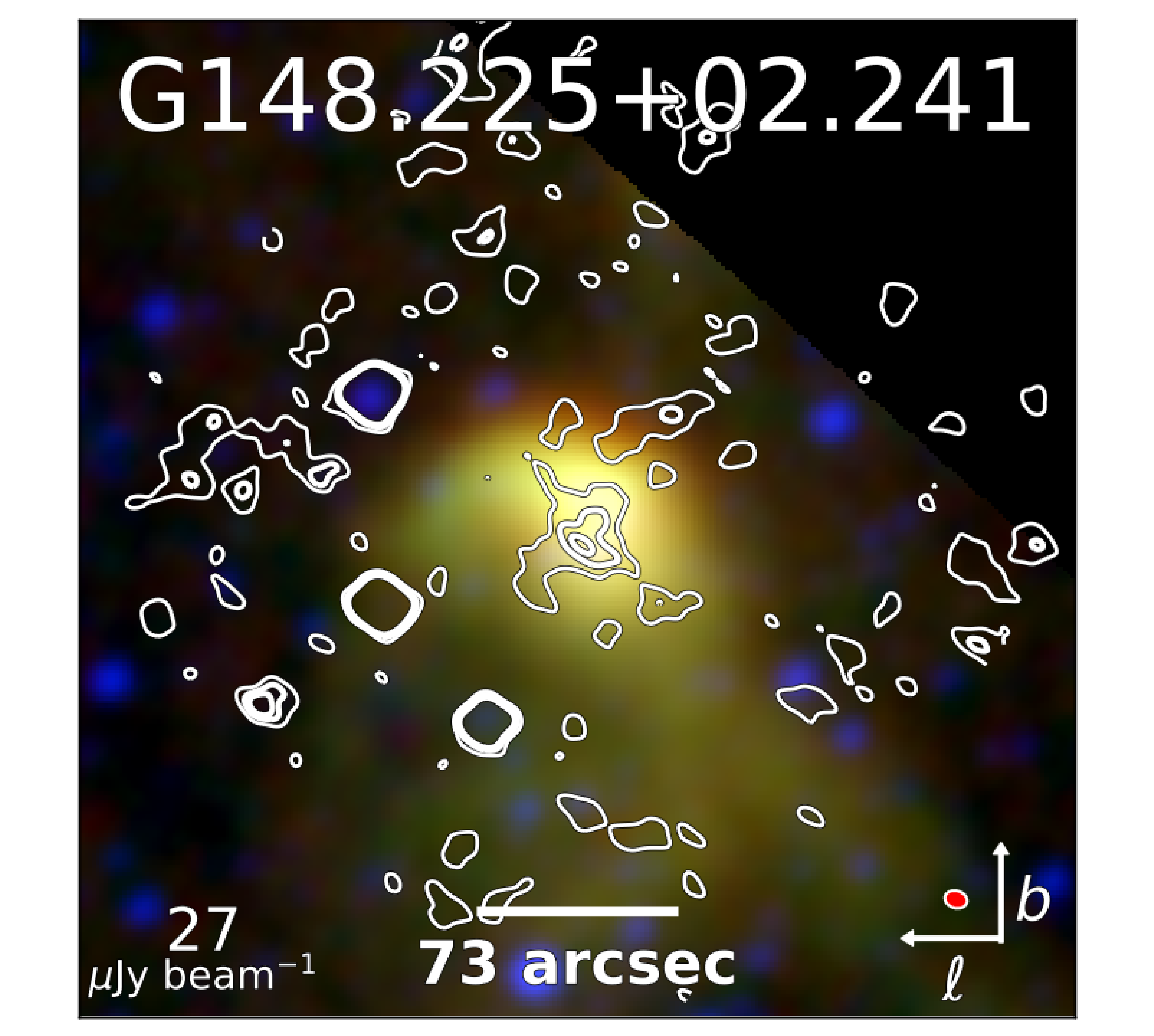}
\includegraphics[width=\figSize]{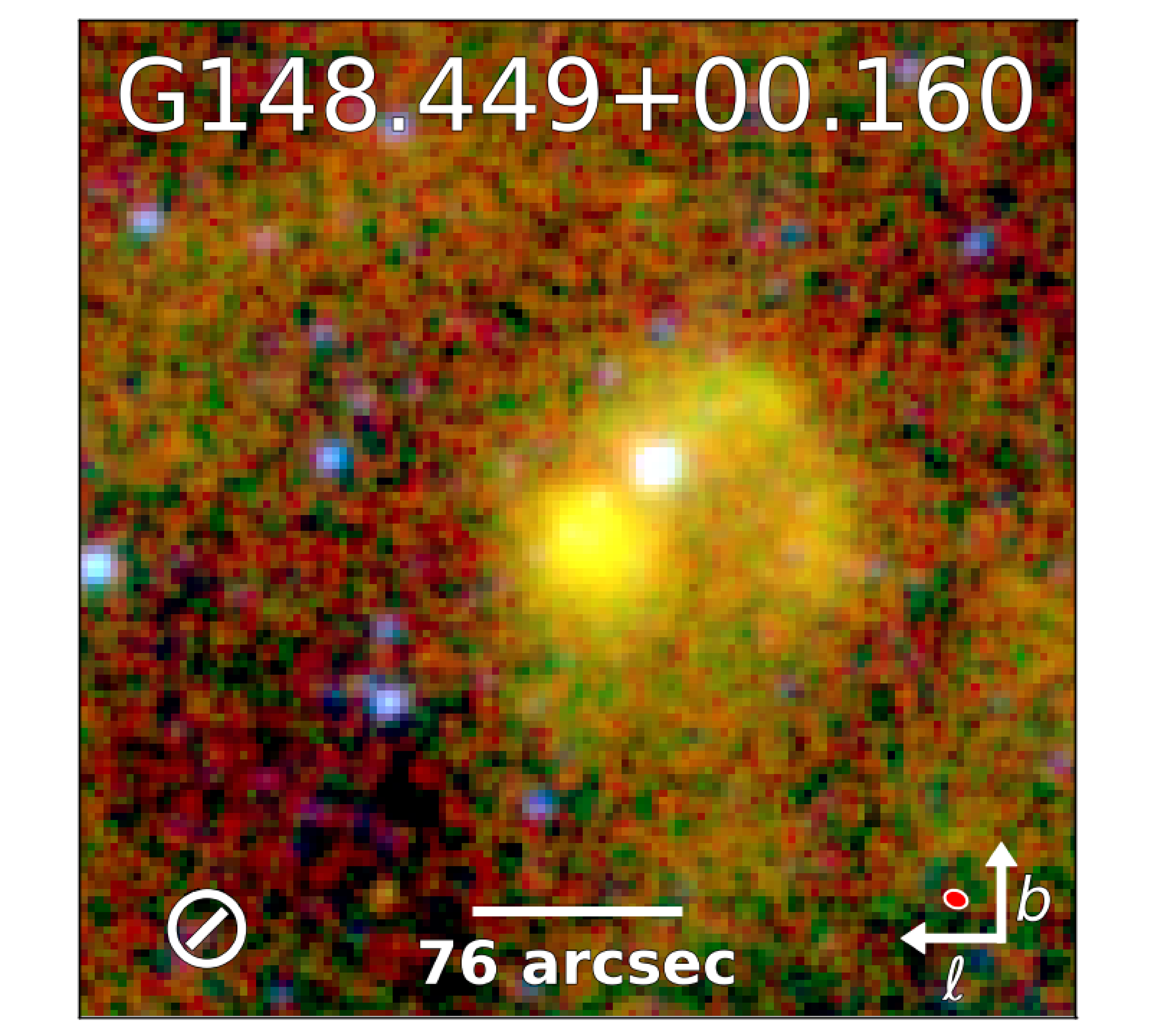}
\includegraphics[width=\figSize]{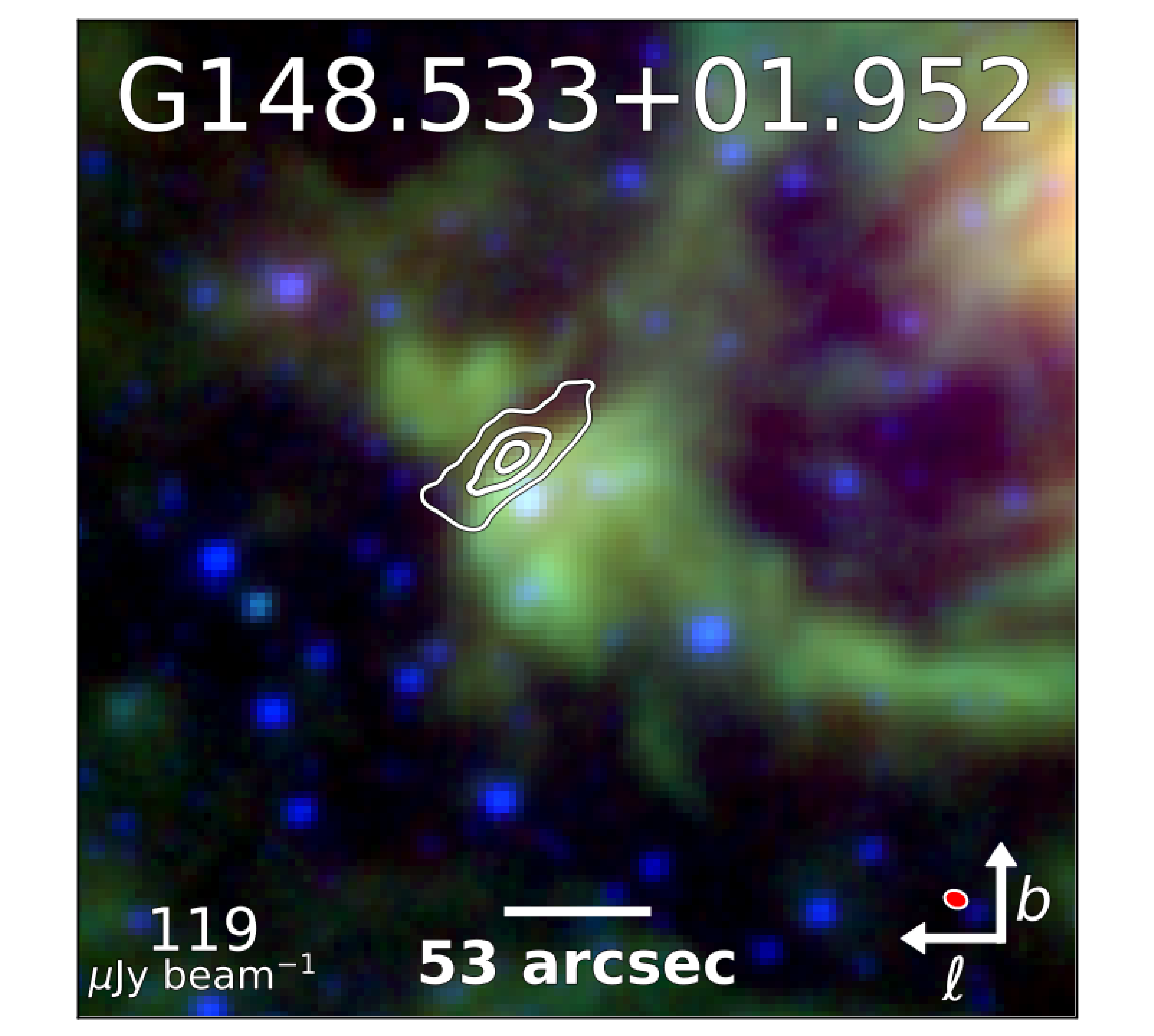}\\
\includegraphics[width=\figSize]{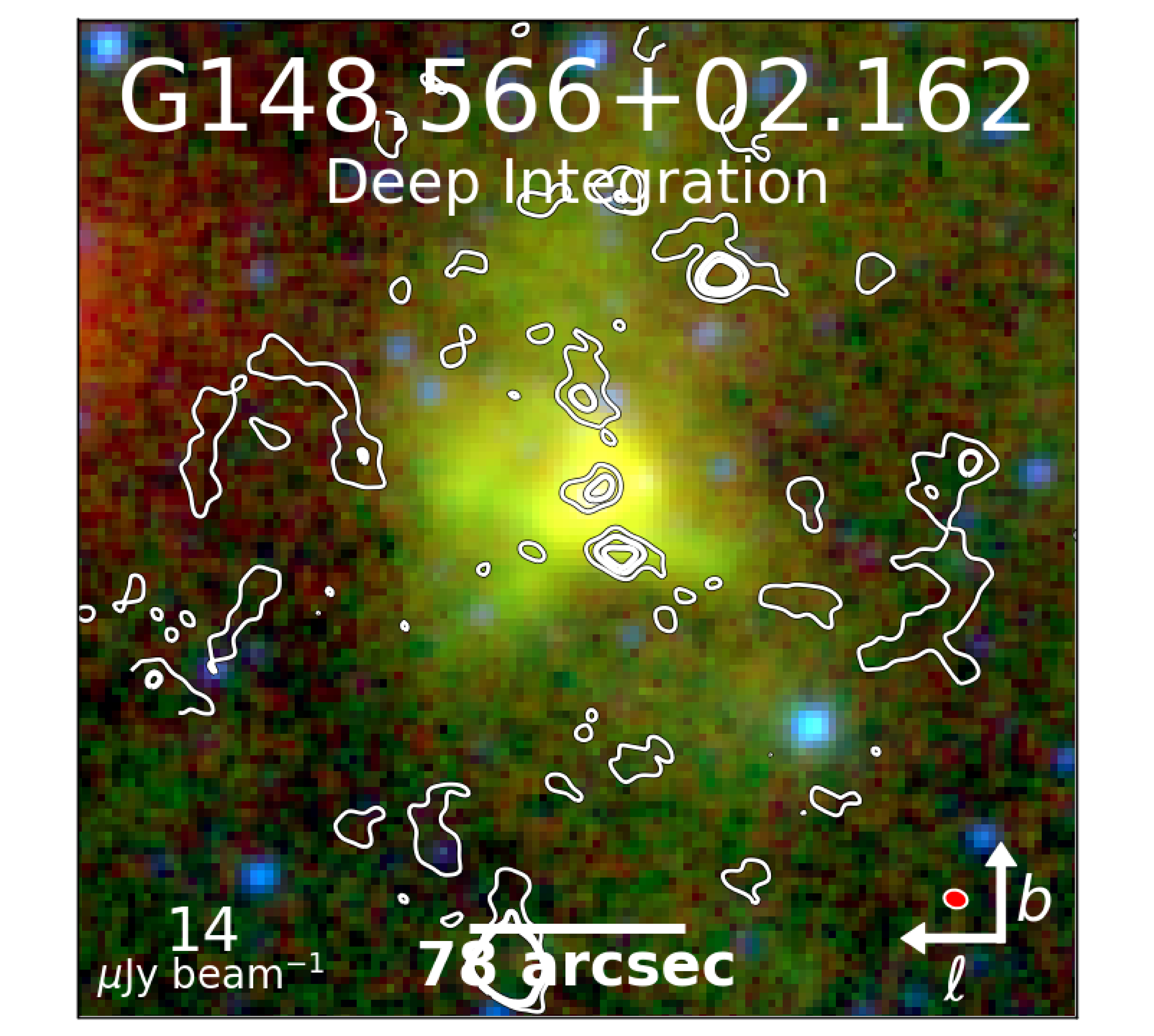}
\includegraphics[width=\figSize]{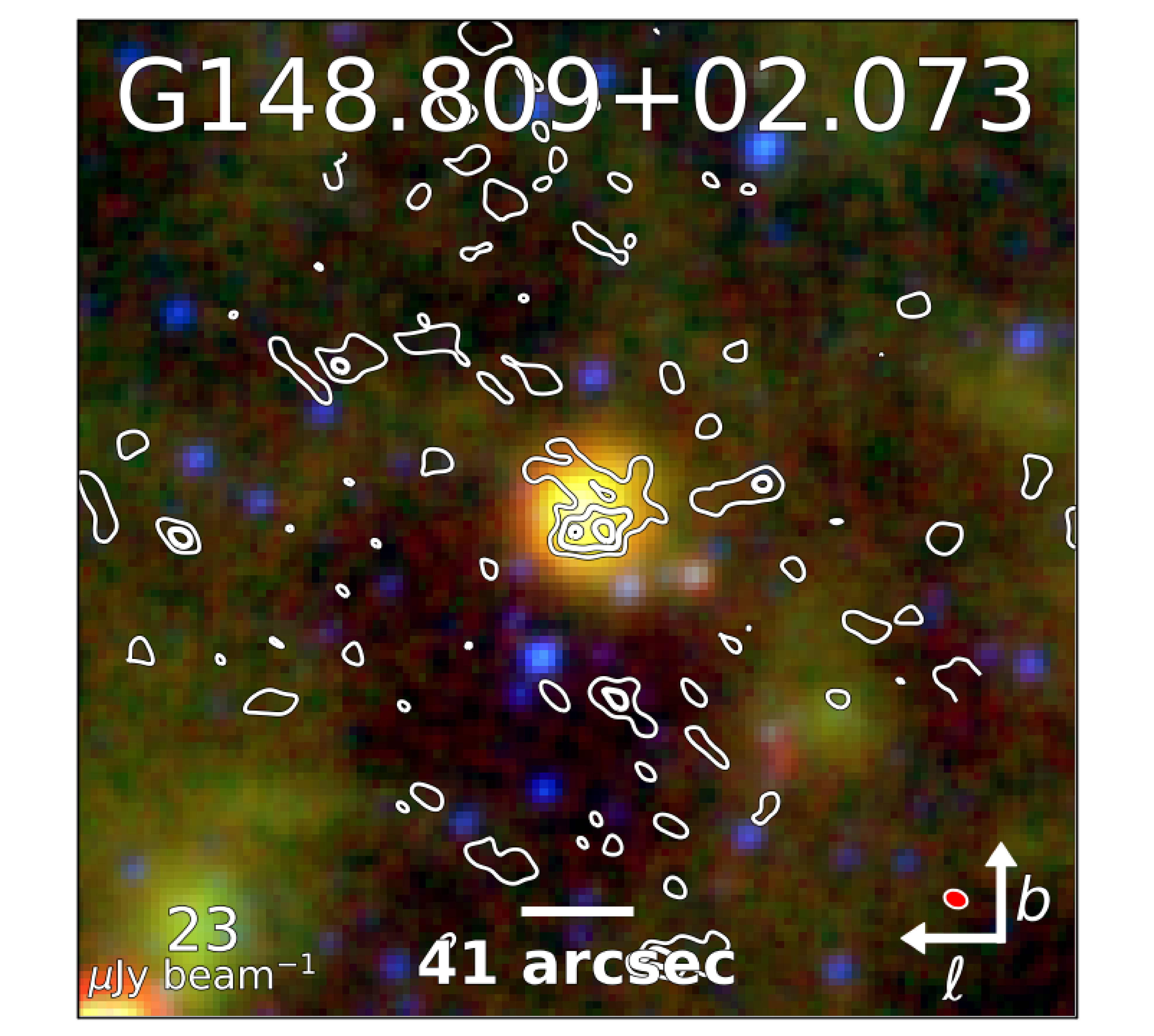}
\includegraphics[width=\figSize]{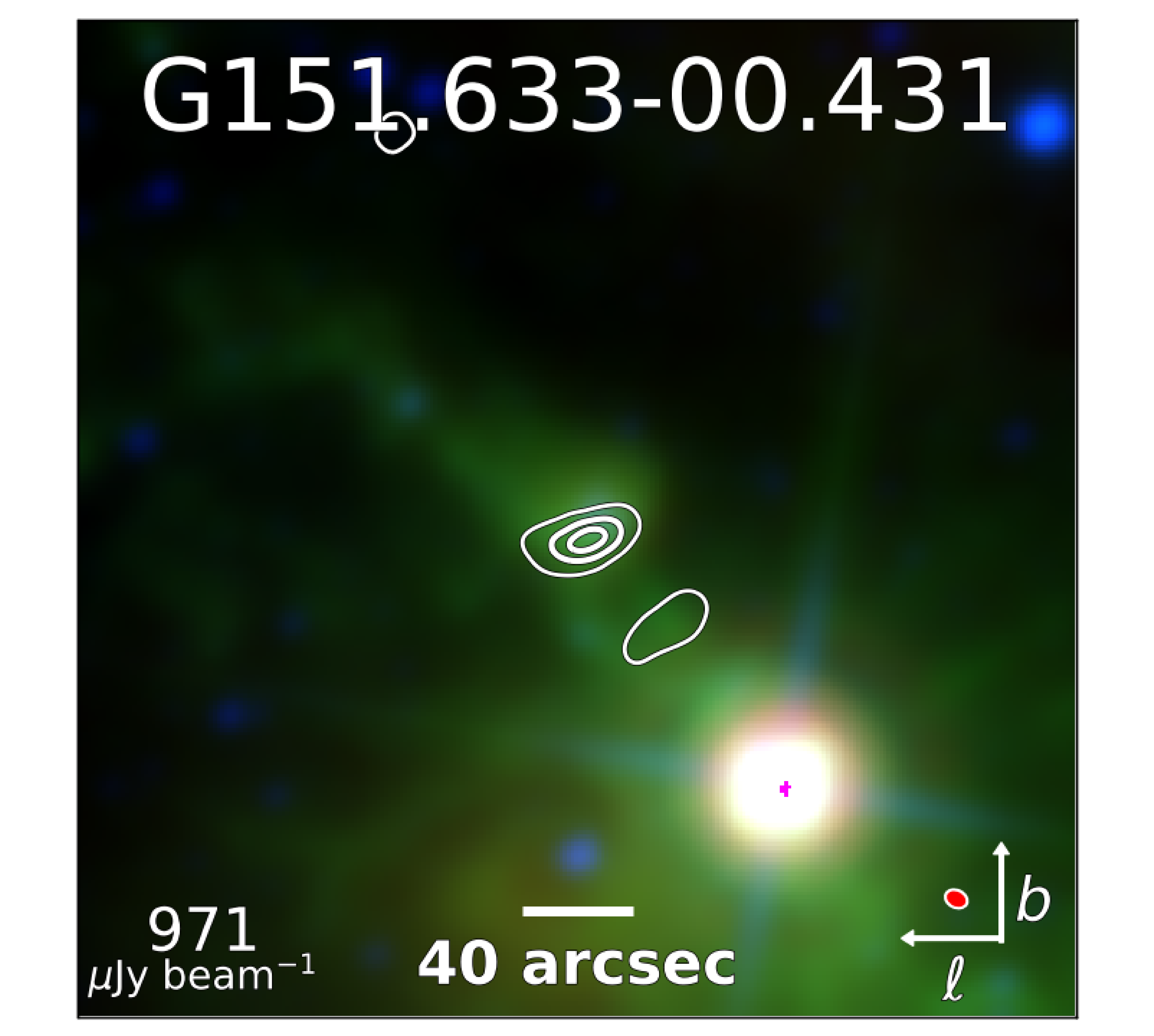}\\
\includegraphics[width=\figSize]{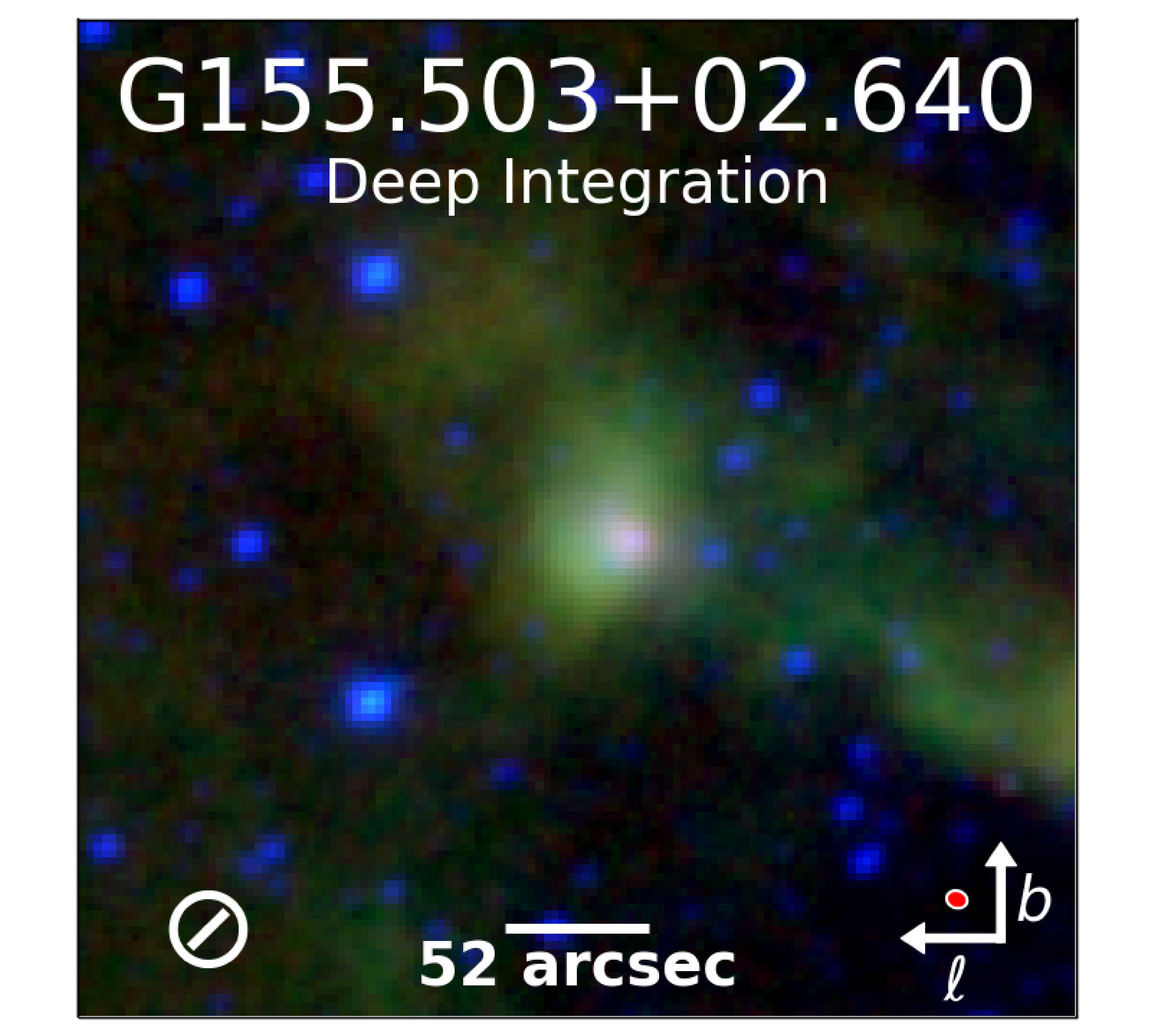}
\includegraphics[width=\figSize]{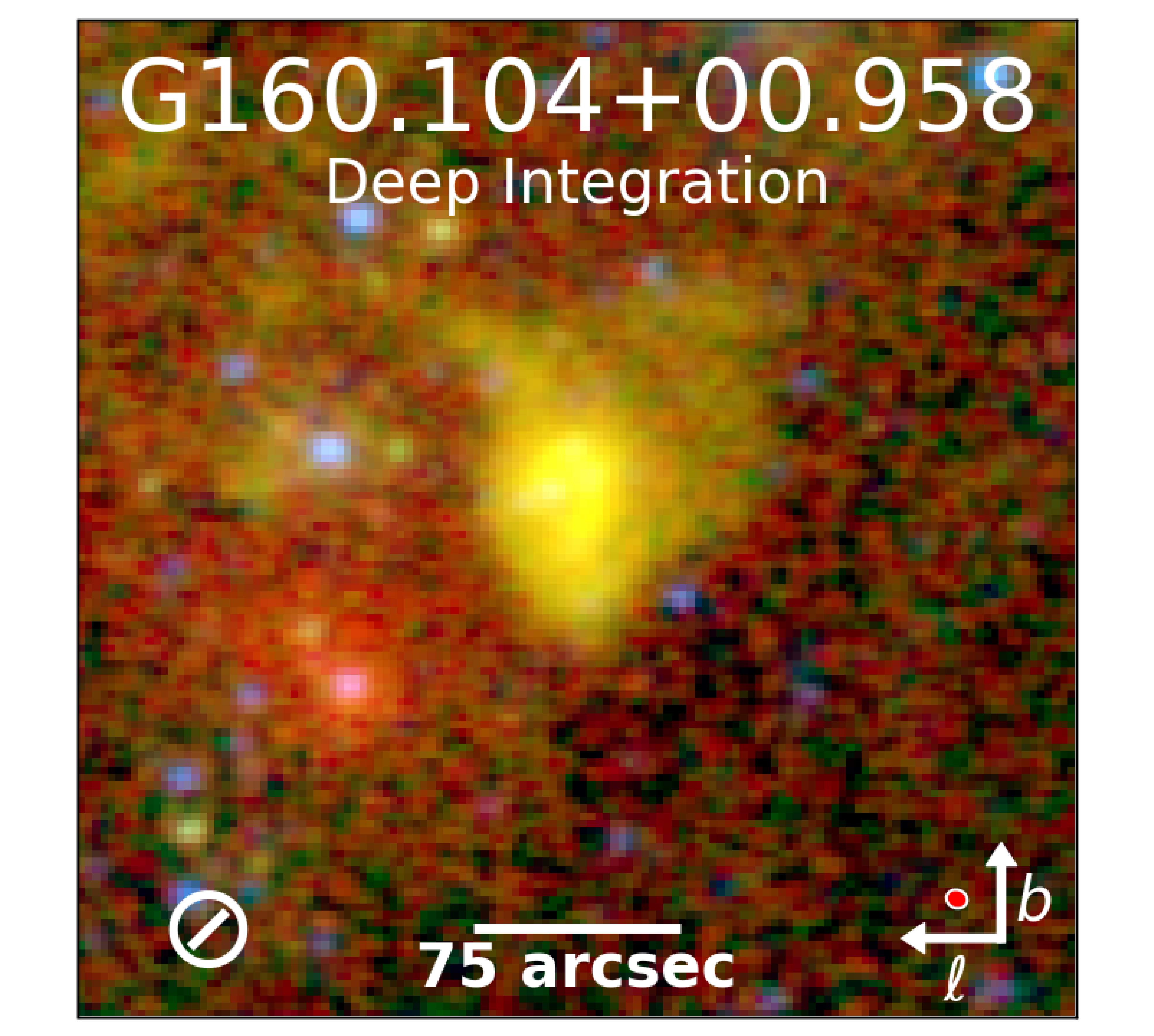}
\includegraphics[width=\figSize]{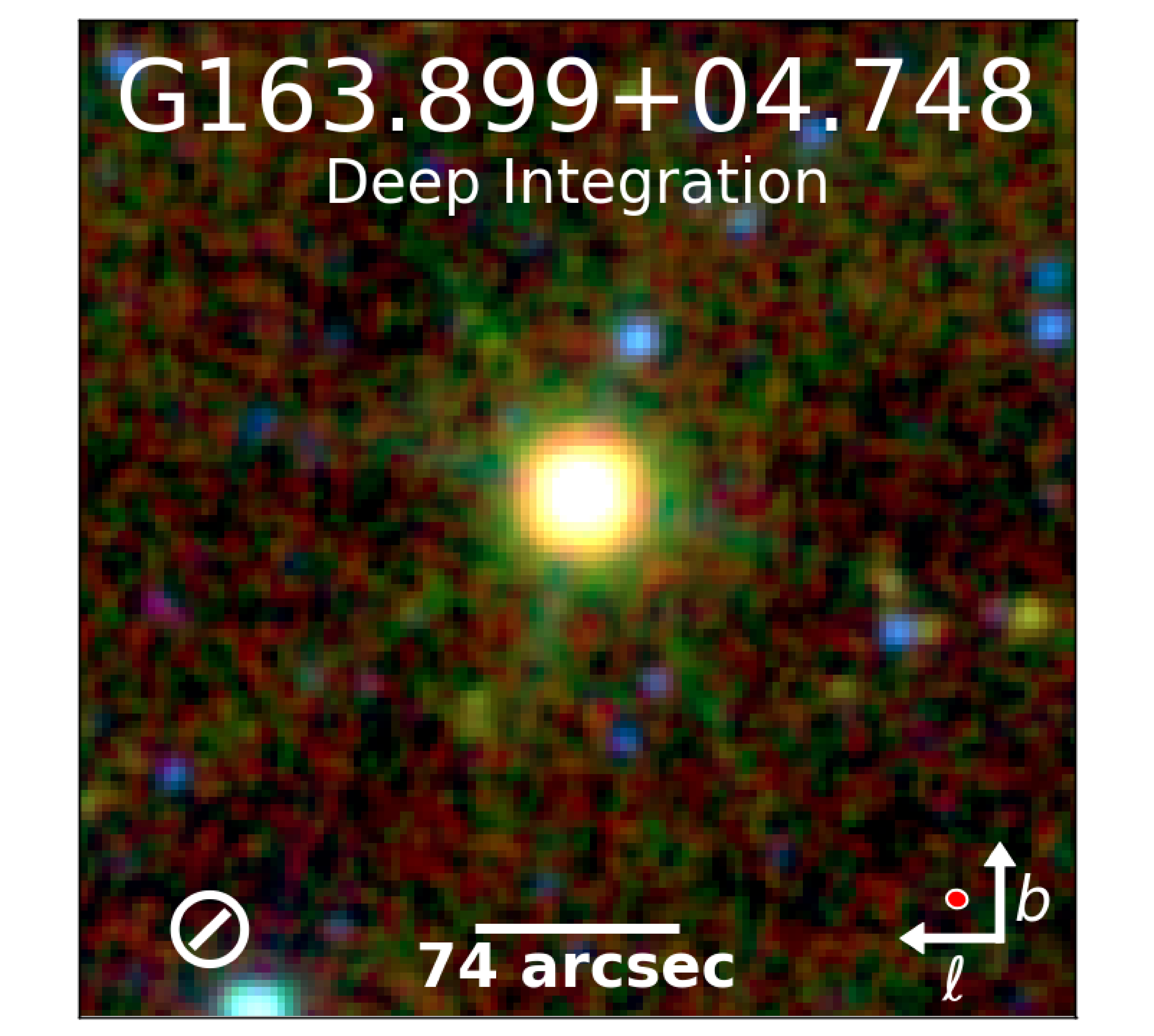}
\end{figure*}
\begin{figure*}[!htb]
\includegraphics[width=\figSize]{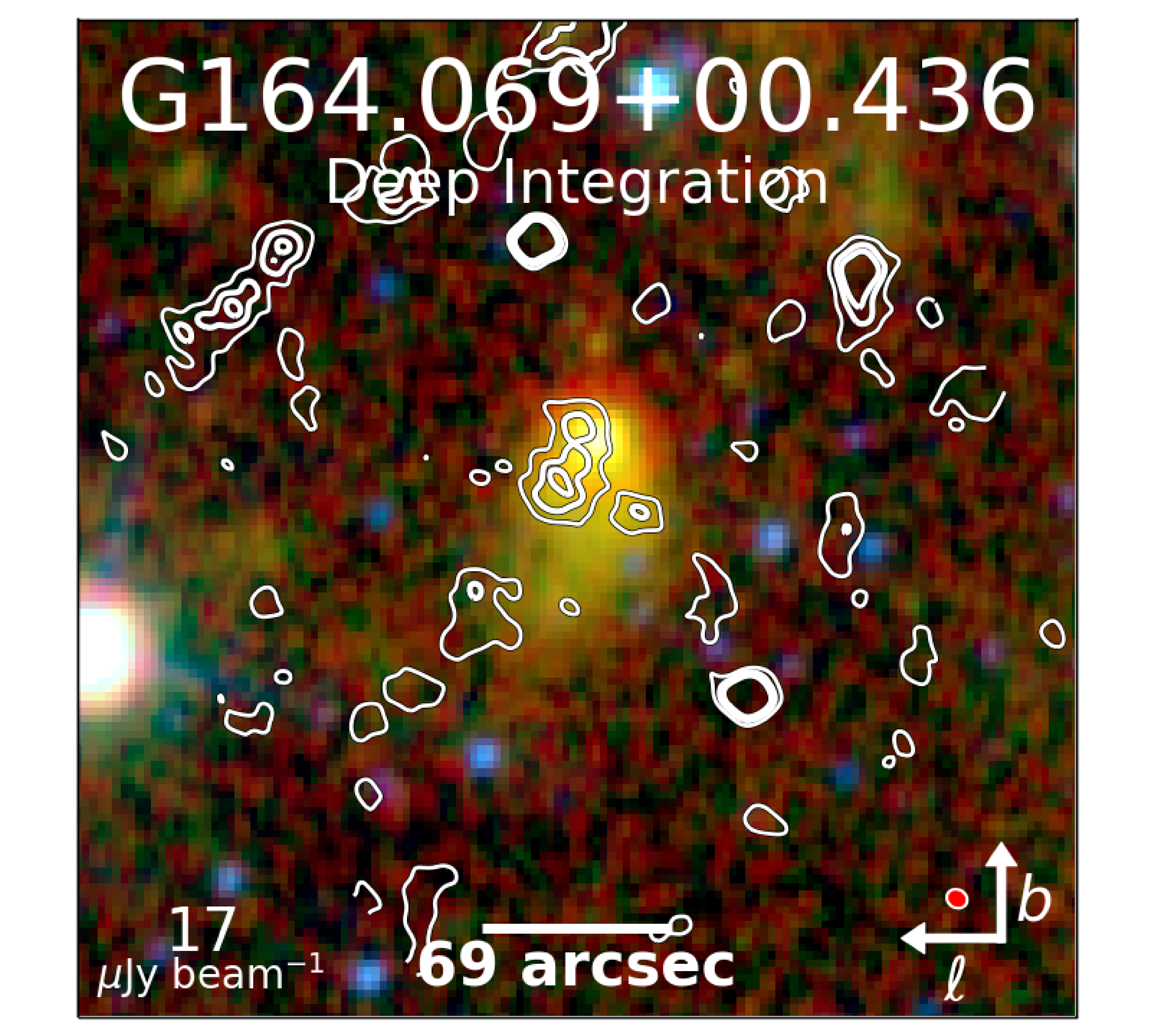}
\includegraphics[width=\figSize]{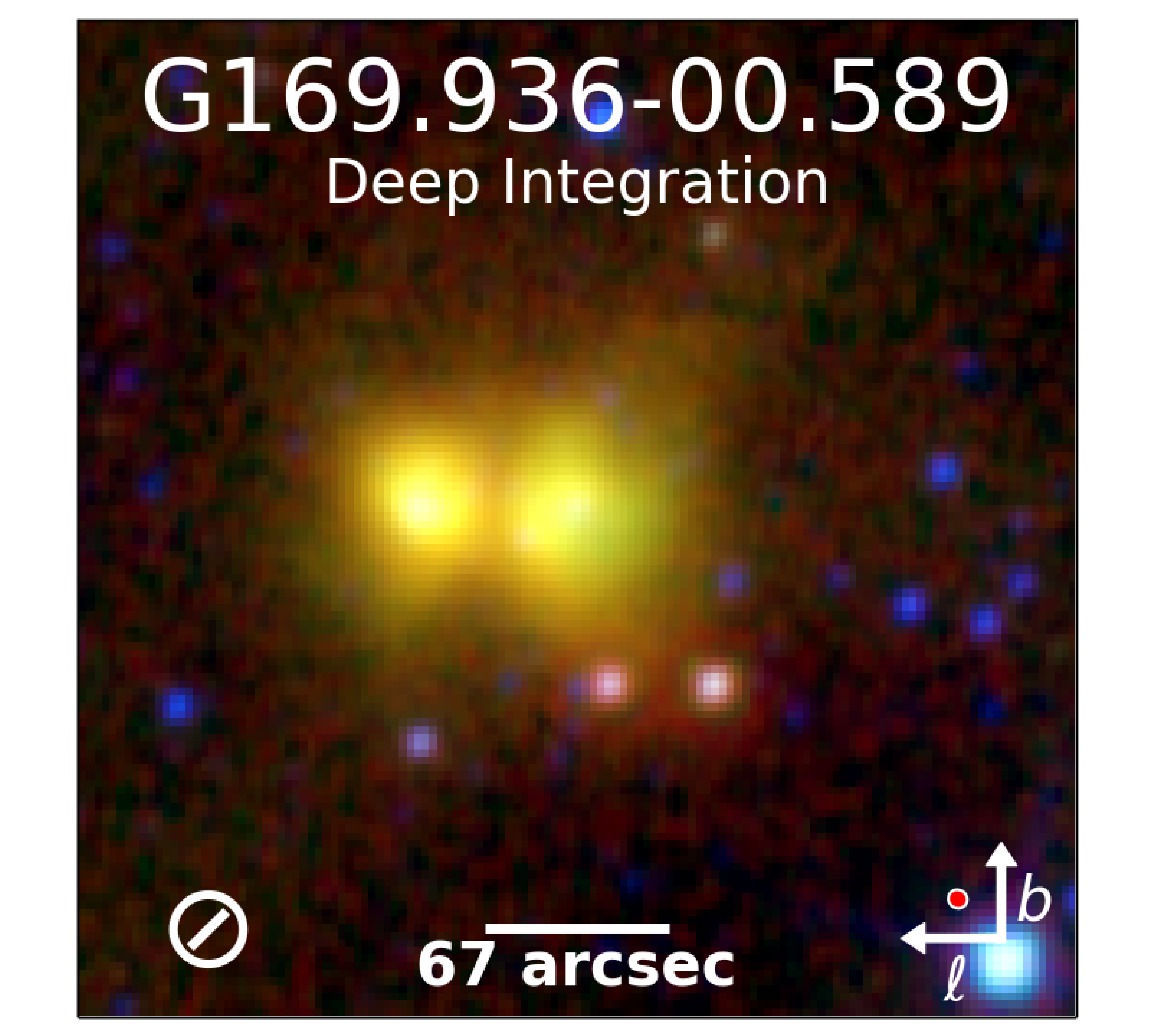}
\includegraphics[width=\figSize]{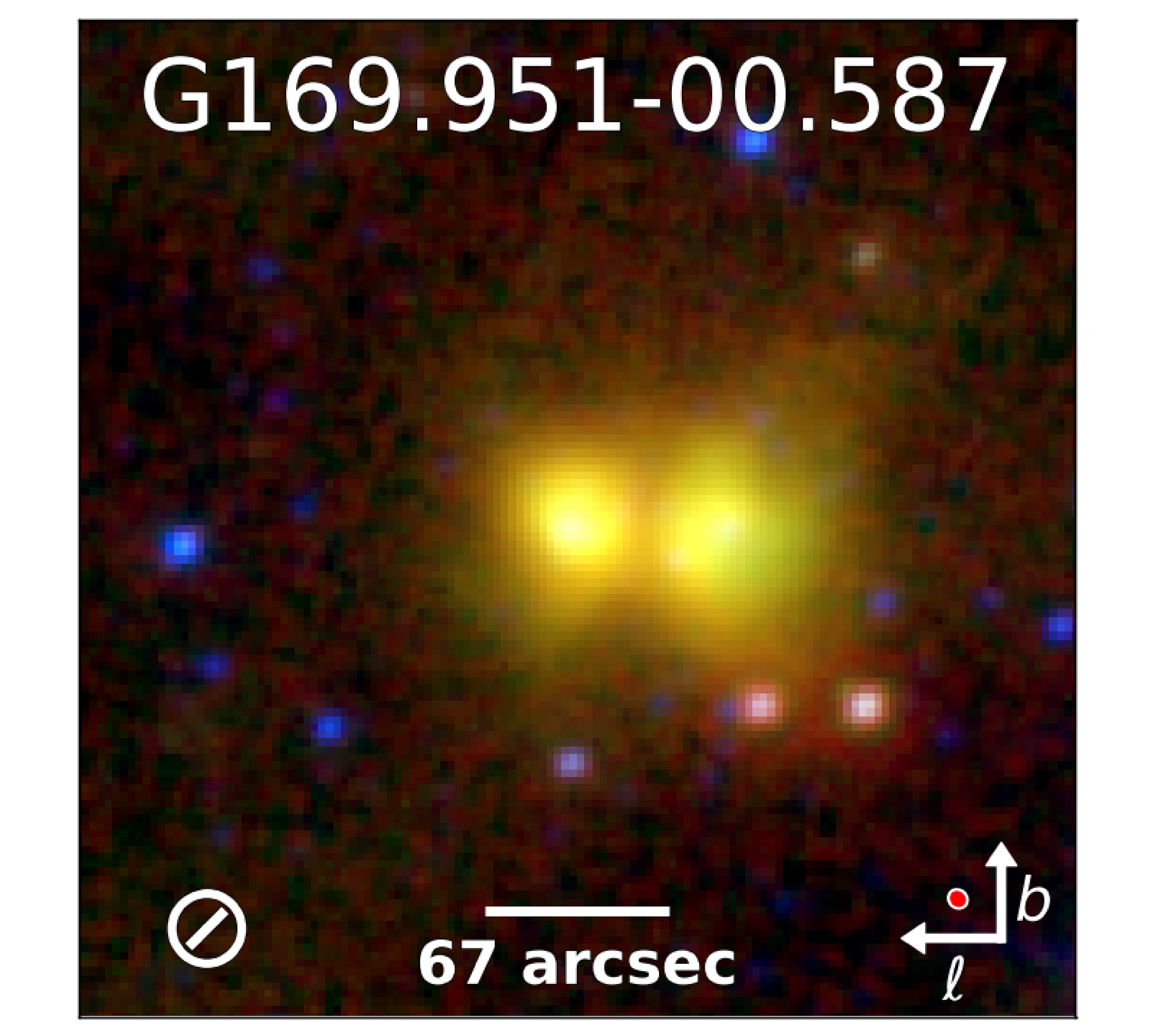}\\
\includegraphics[width=\figSize]{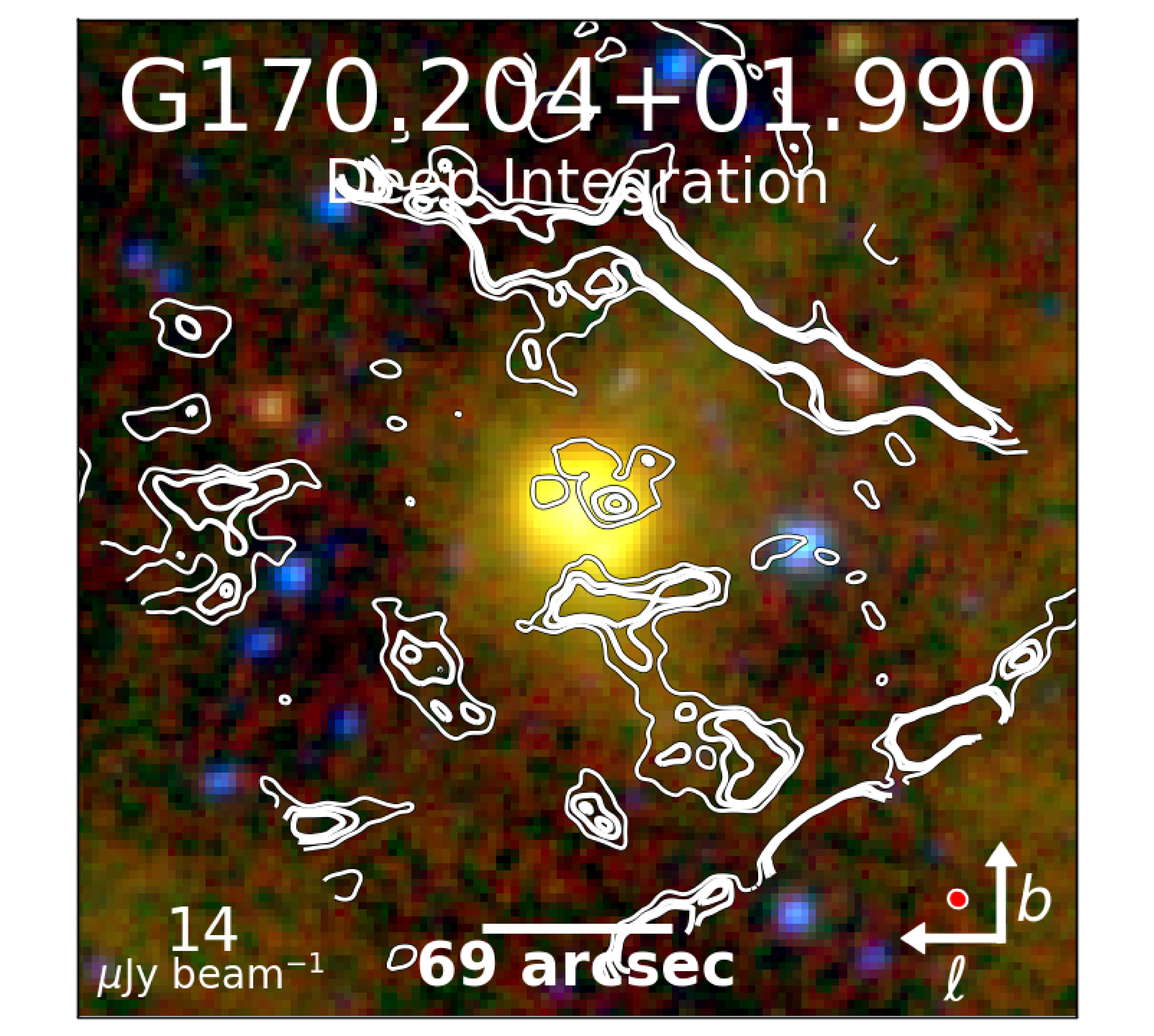}
\includegraphics[width=\figSize]{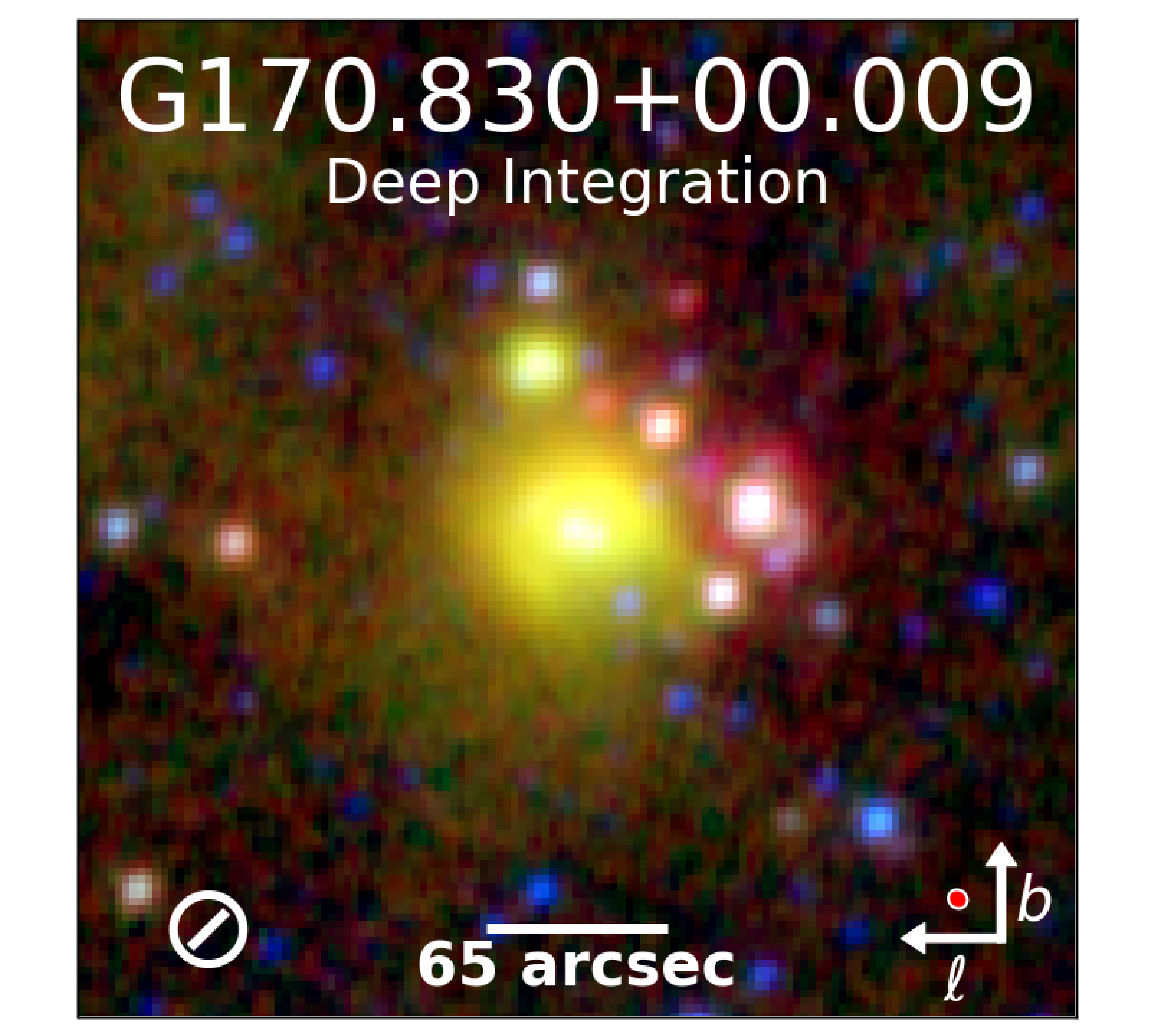}
\includegraphics[width=\figSize]{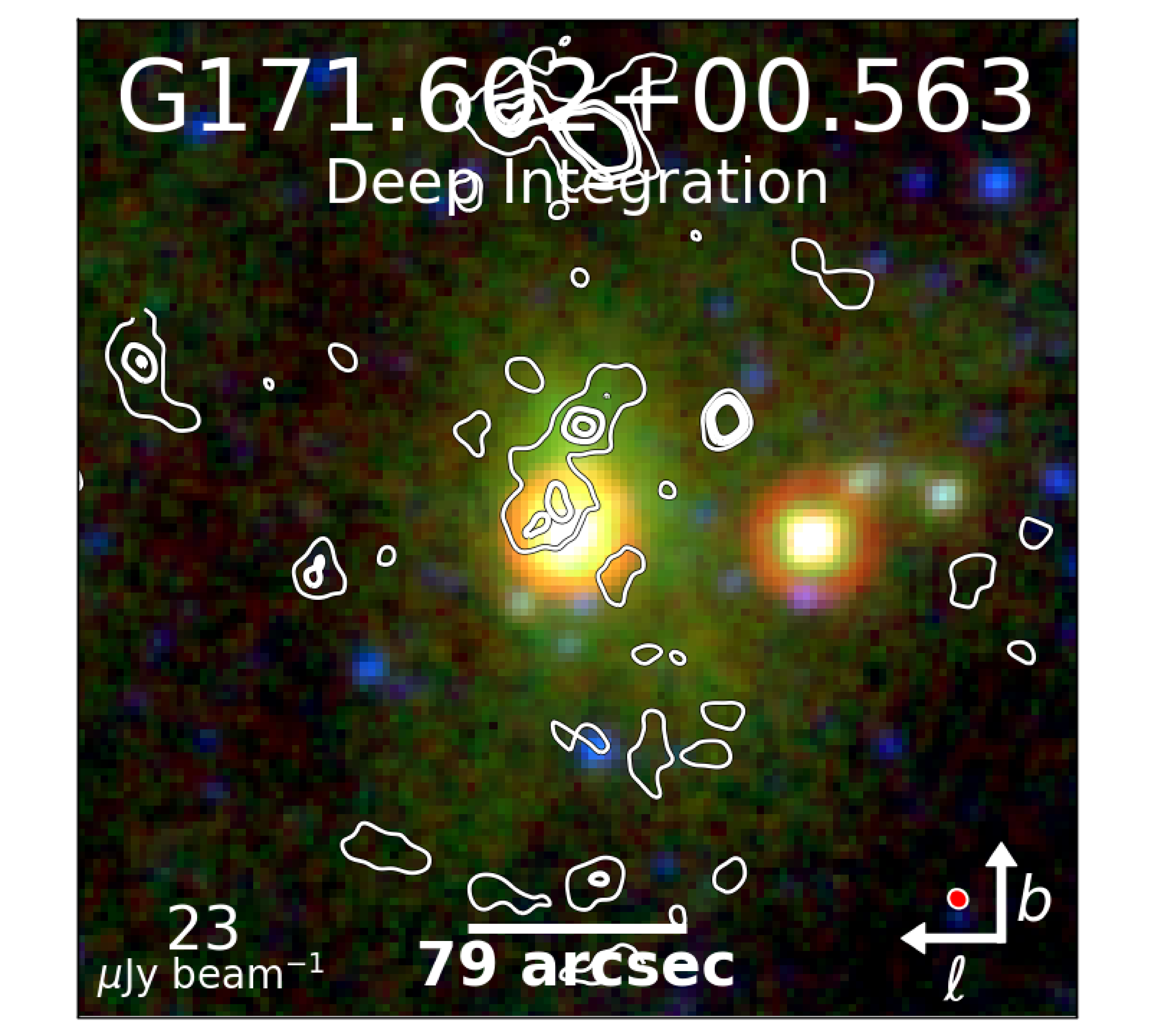}\\
\includegraphics[width=\figSize]{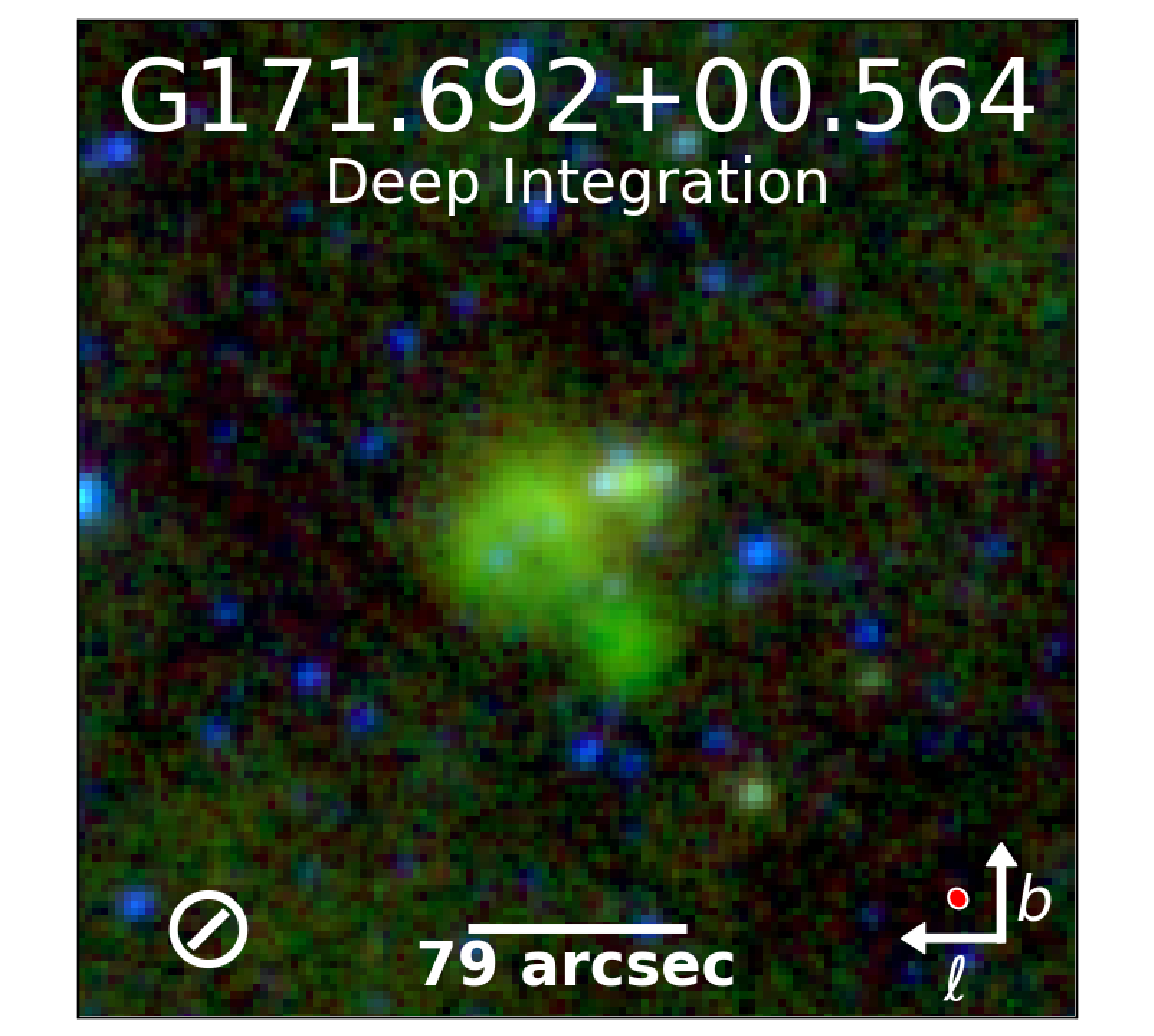}
\includegraphics[width=\figSize]{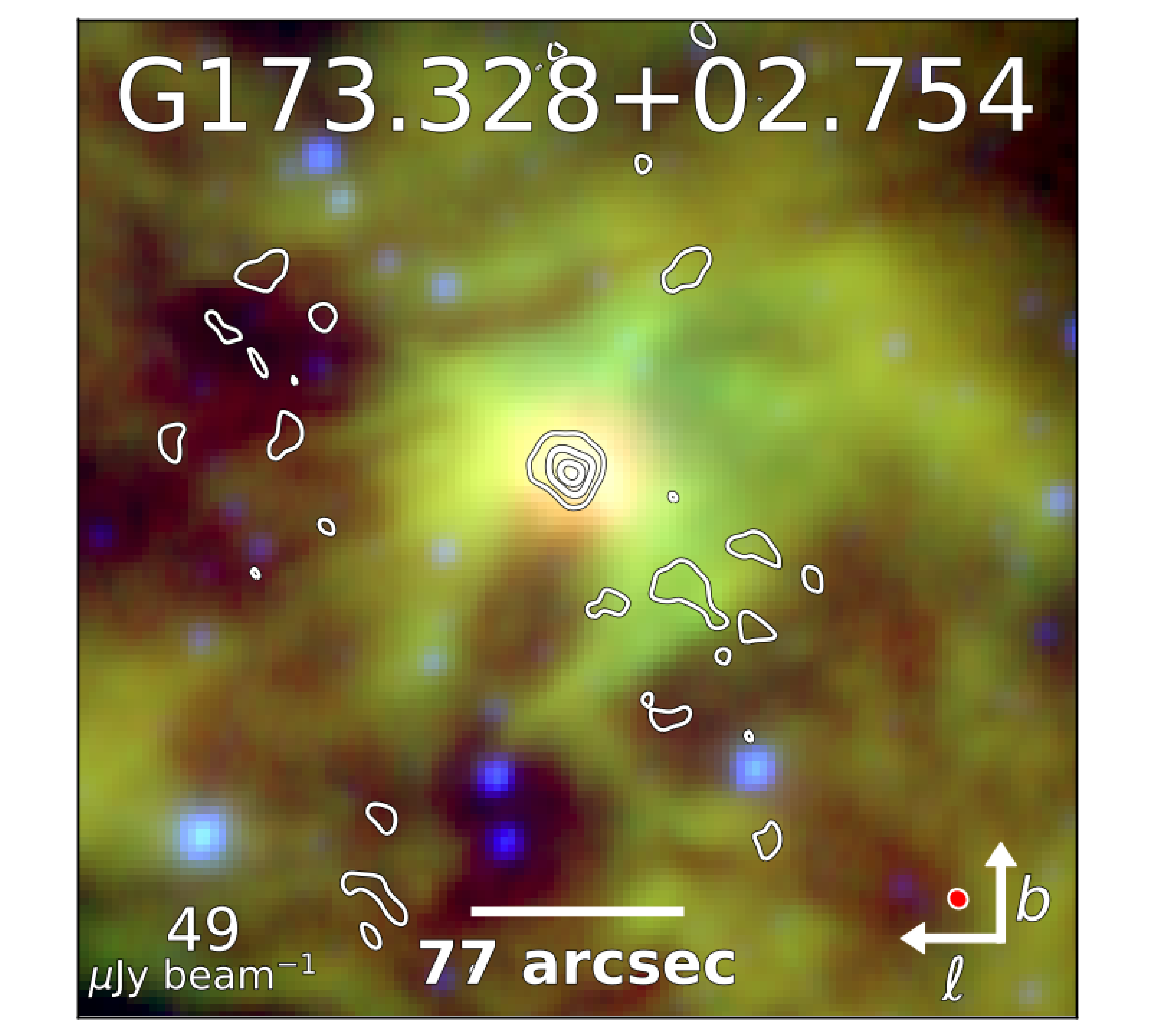}
\includegraphics[width=\figSize]{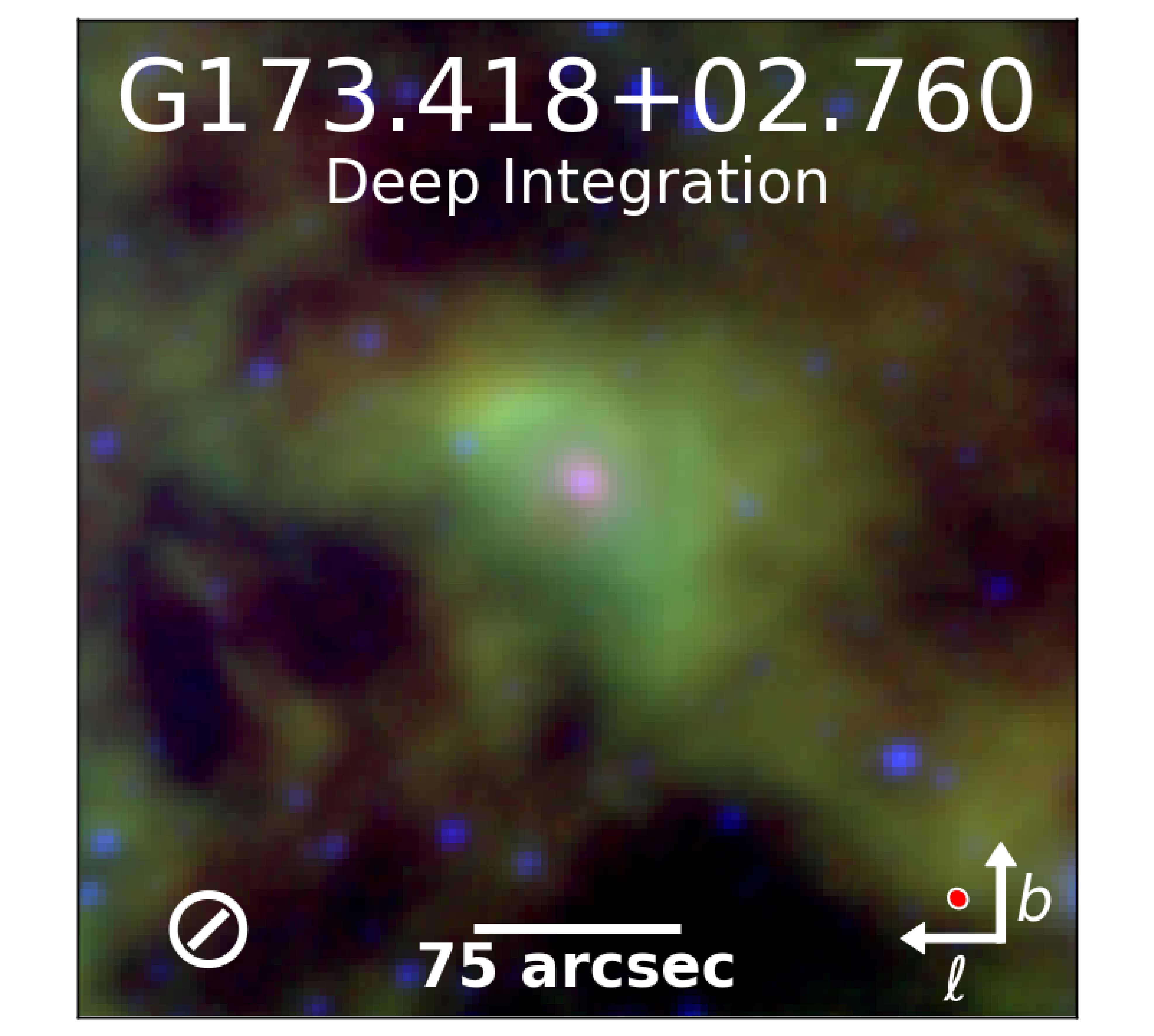}\\
\includegraphics[width=\figSize]{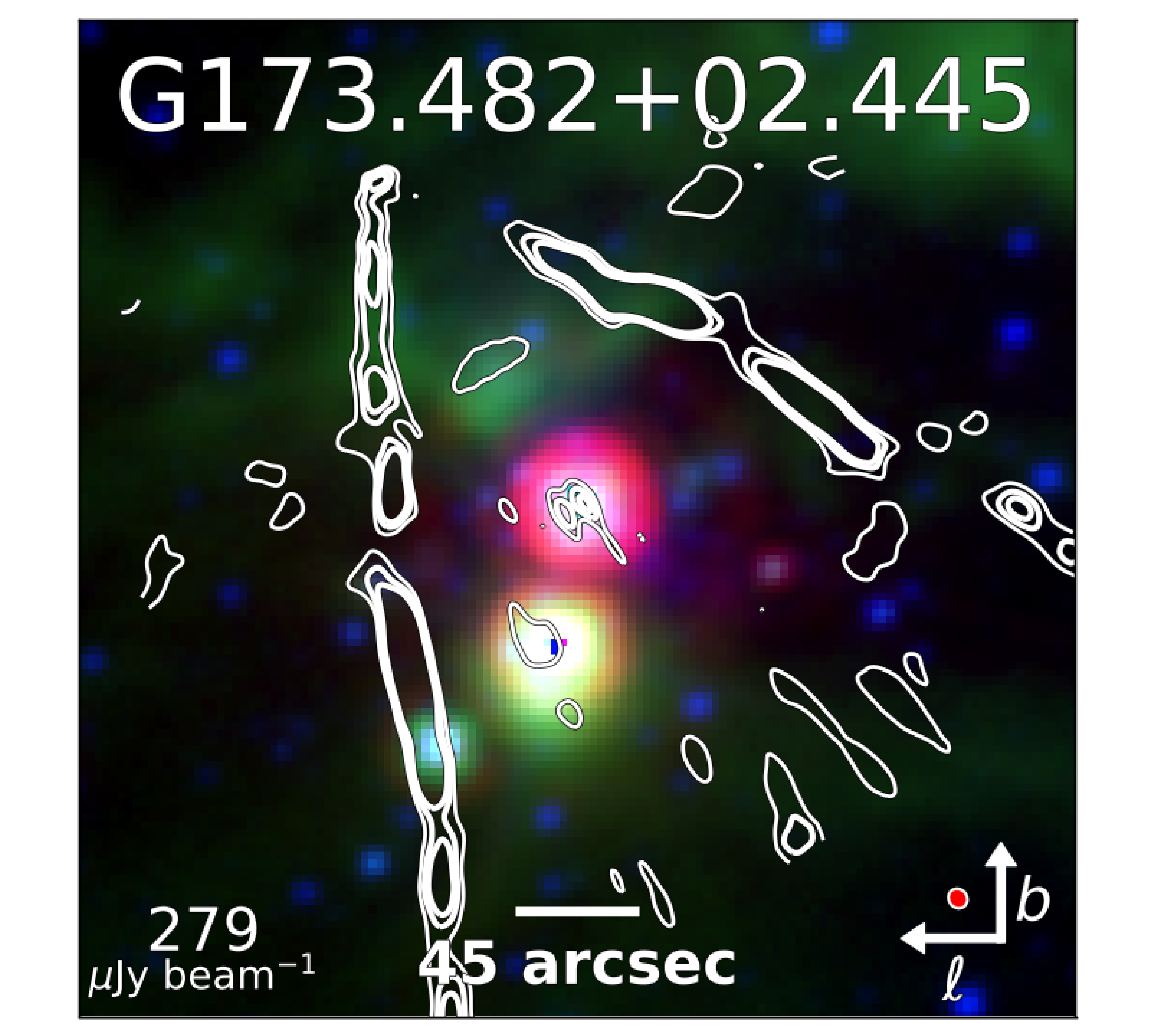}
\includegraphics[width=\figSize]{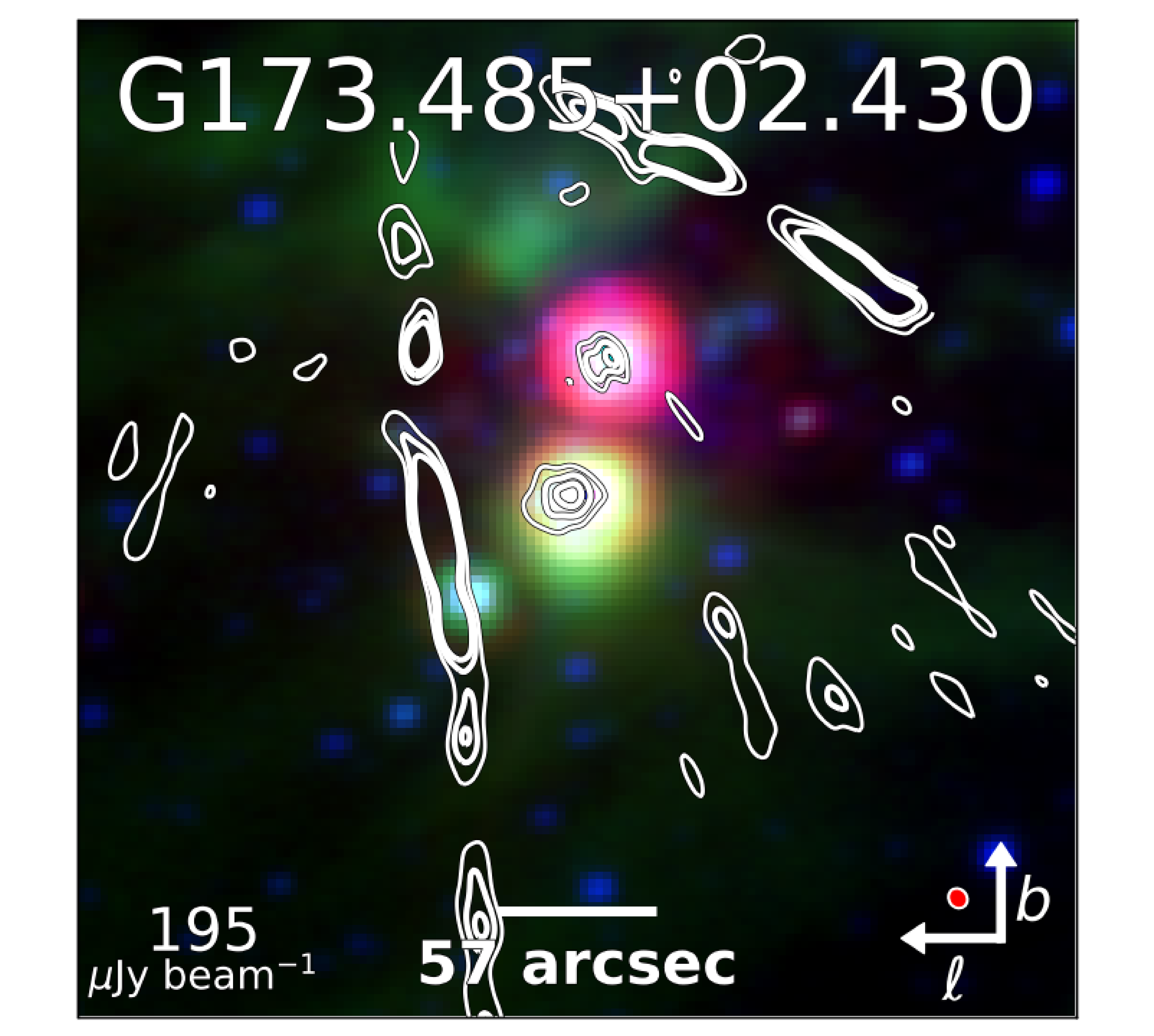}
\includegraphics[width=\figSize]{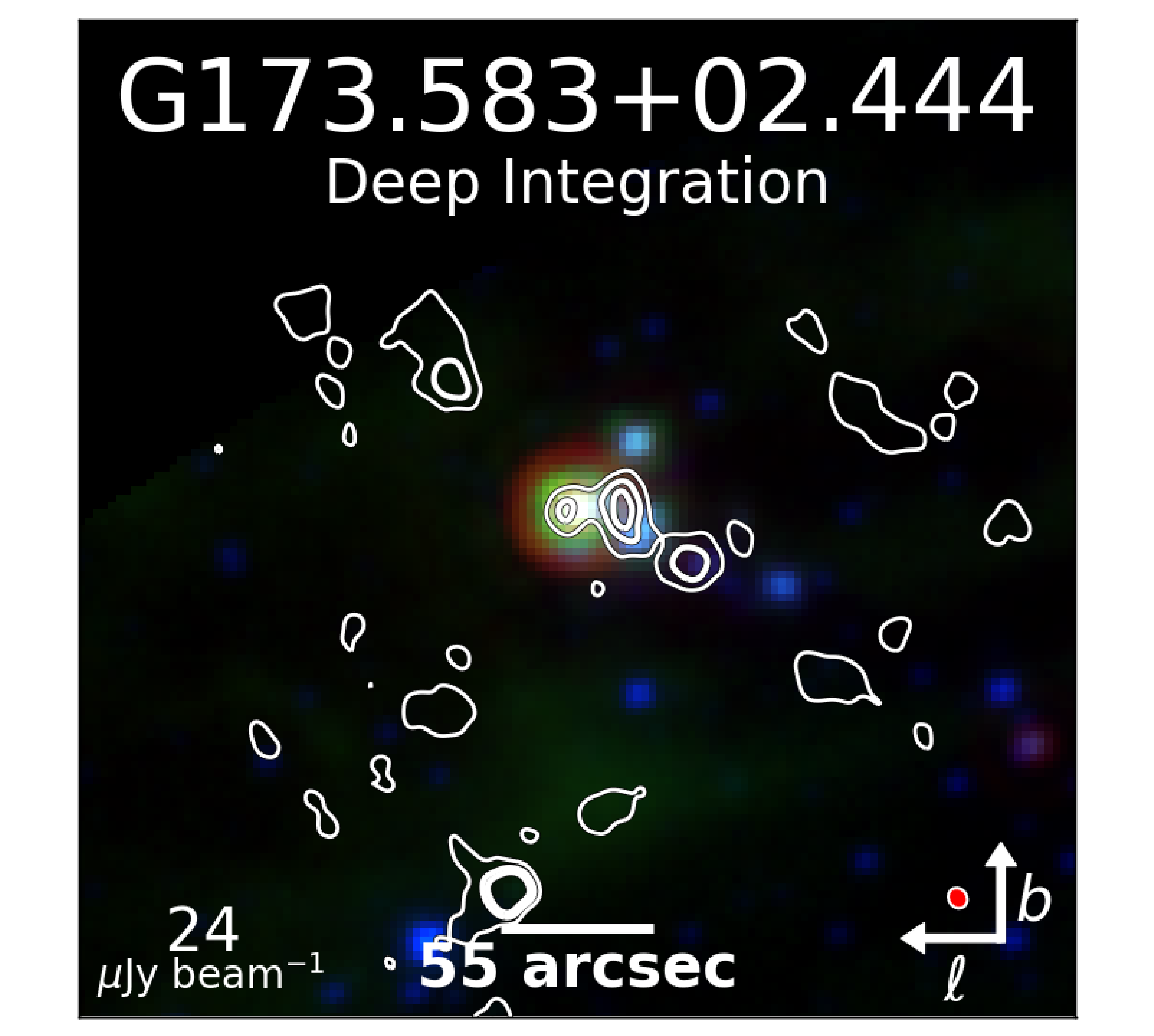}
\end{figure*}
\begin{figure*}[!htb]
\includegraphics[width=\figSize]{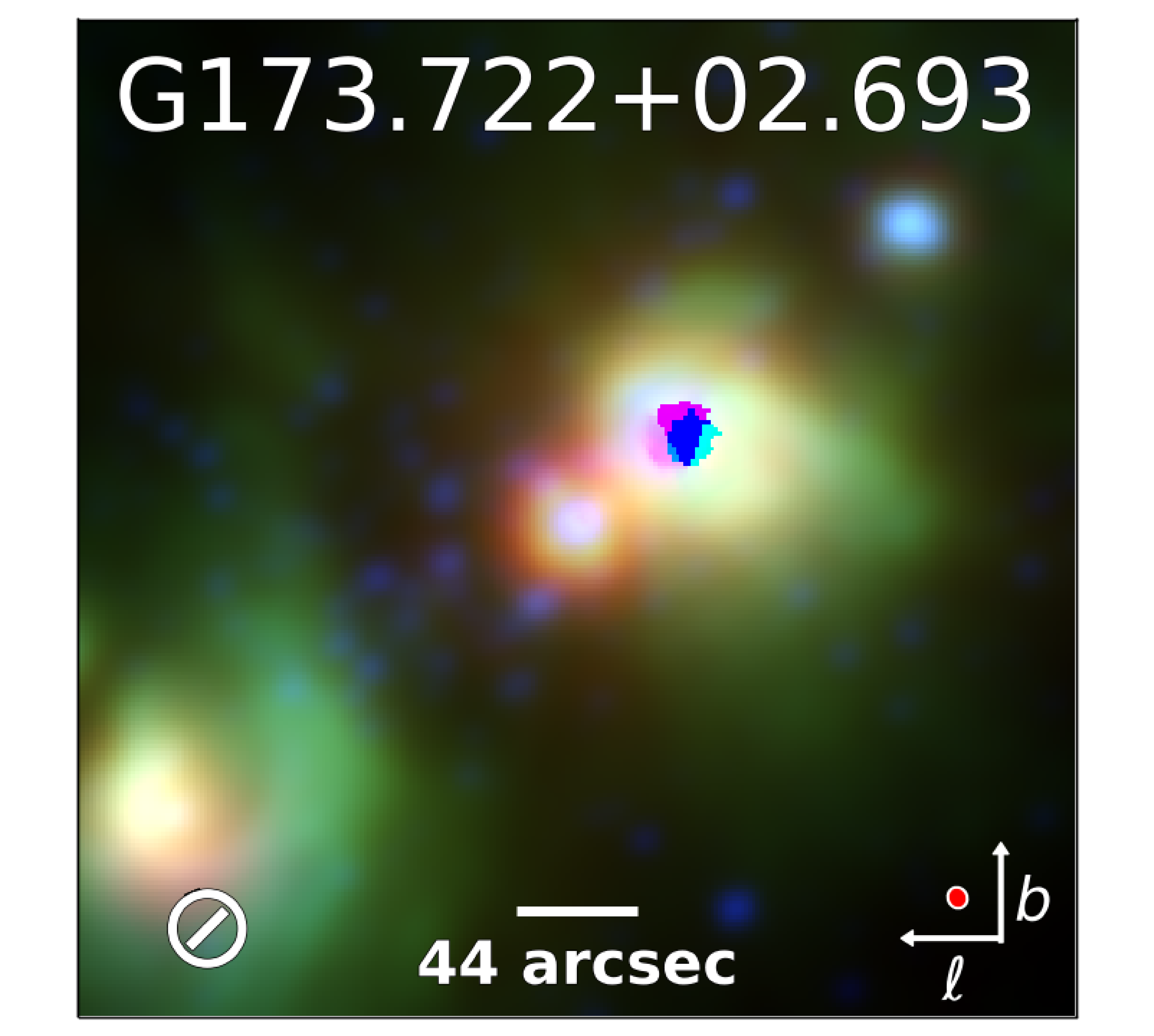}
\includegraphics[width=\figSize]{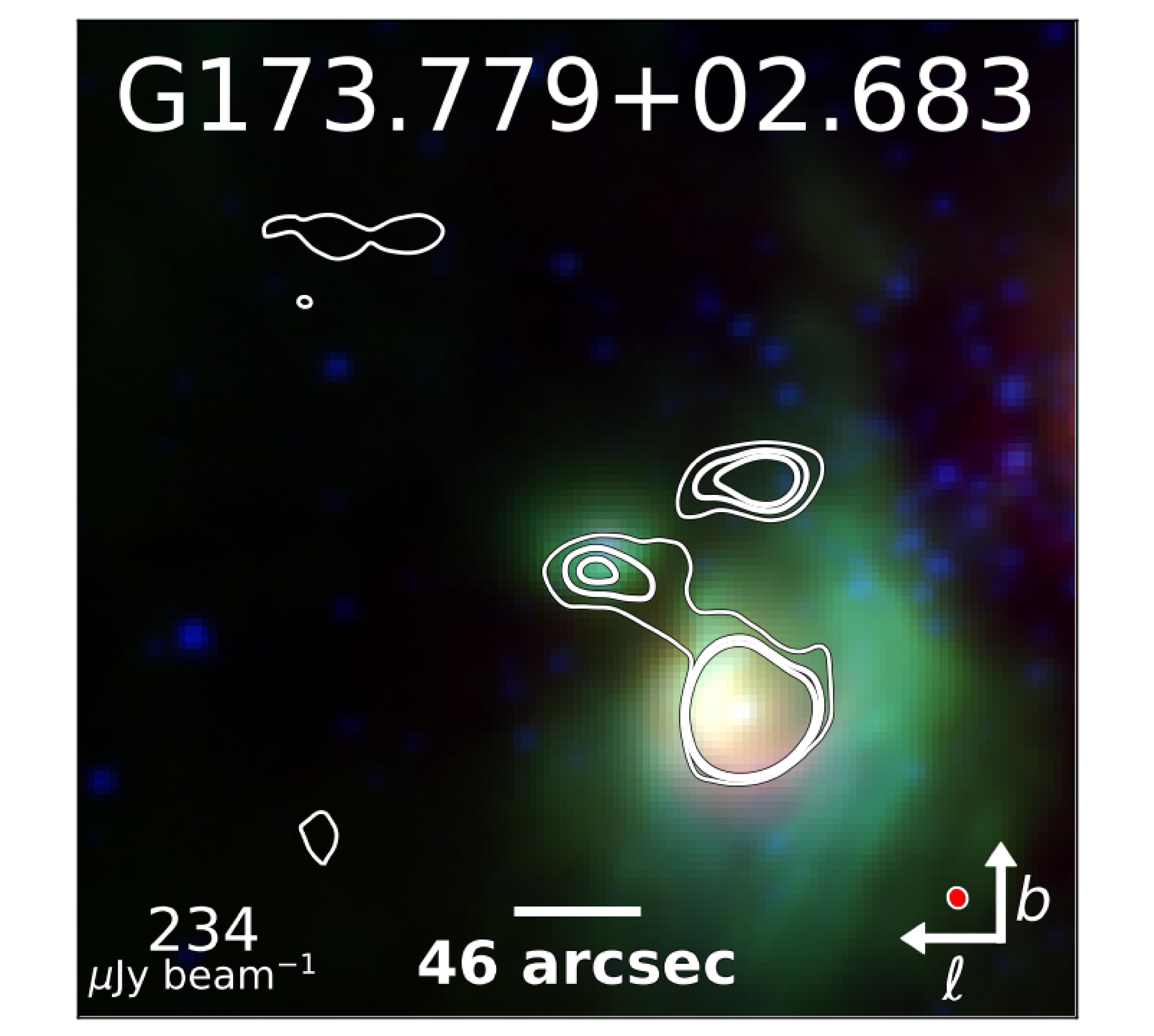}
\includegraphics[width=\figSize]{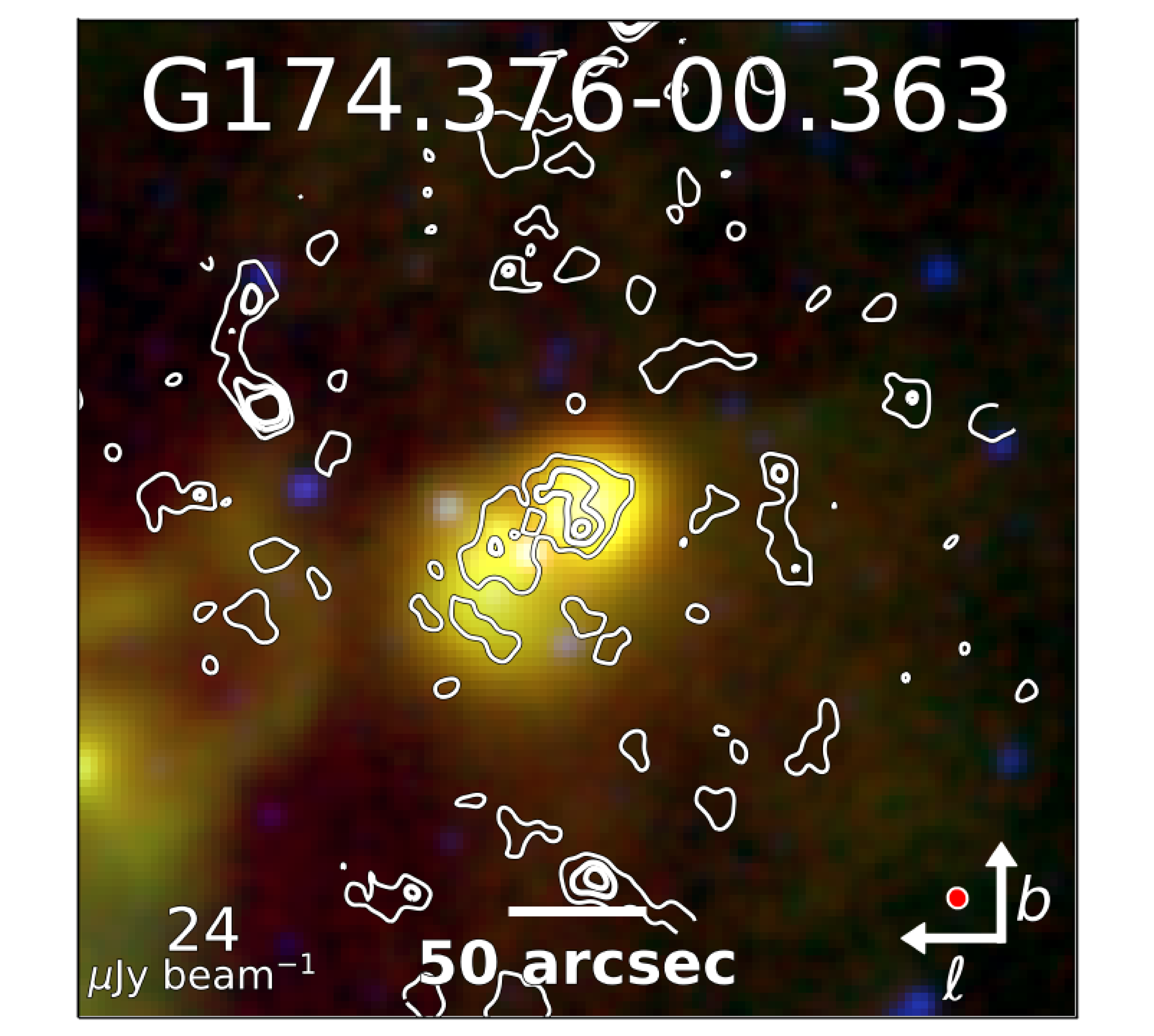}\\
\includegraphics[width=\figSize]{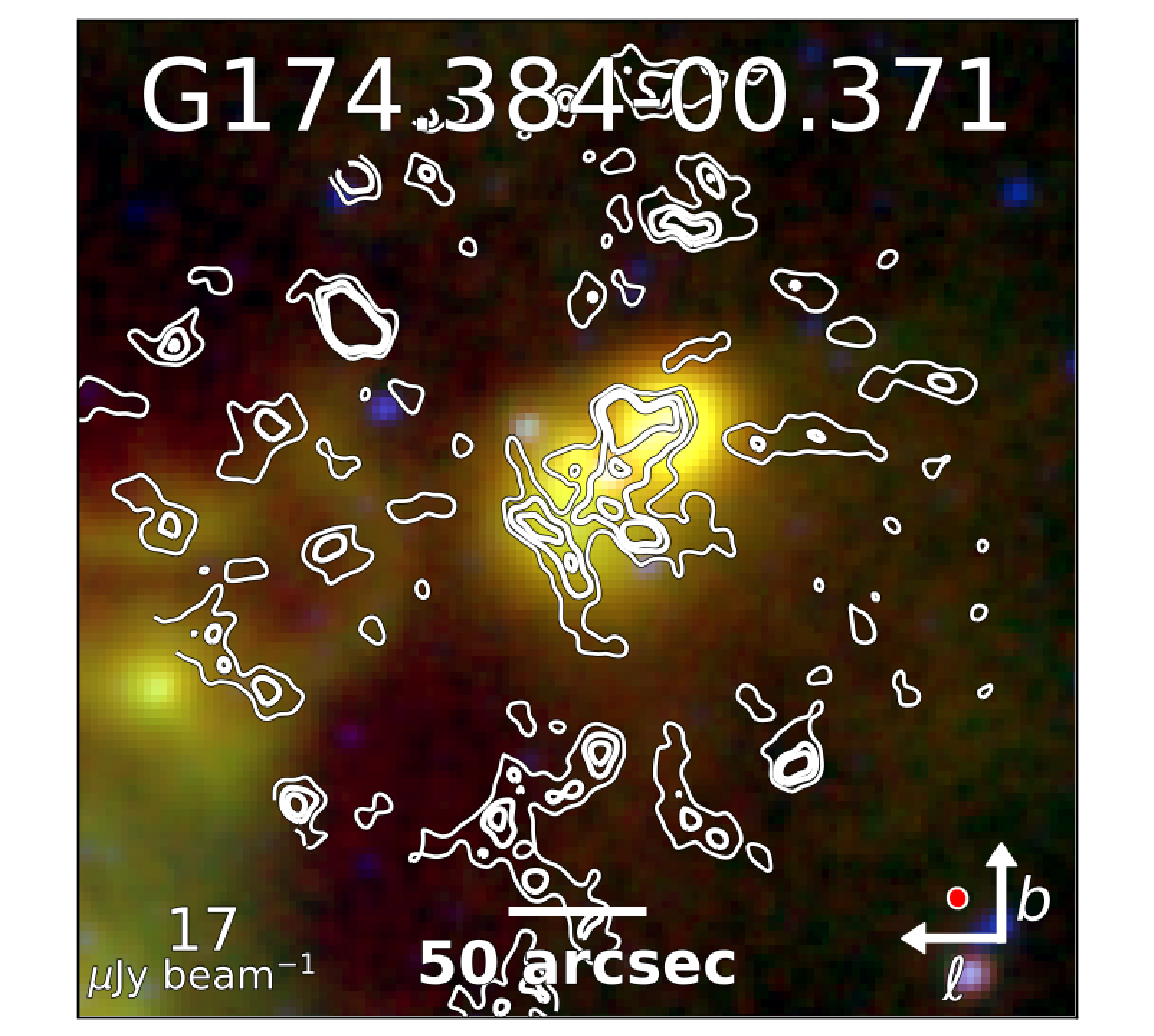}
\includegraphics[width=\figSize]{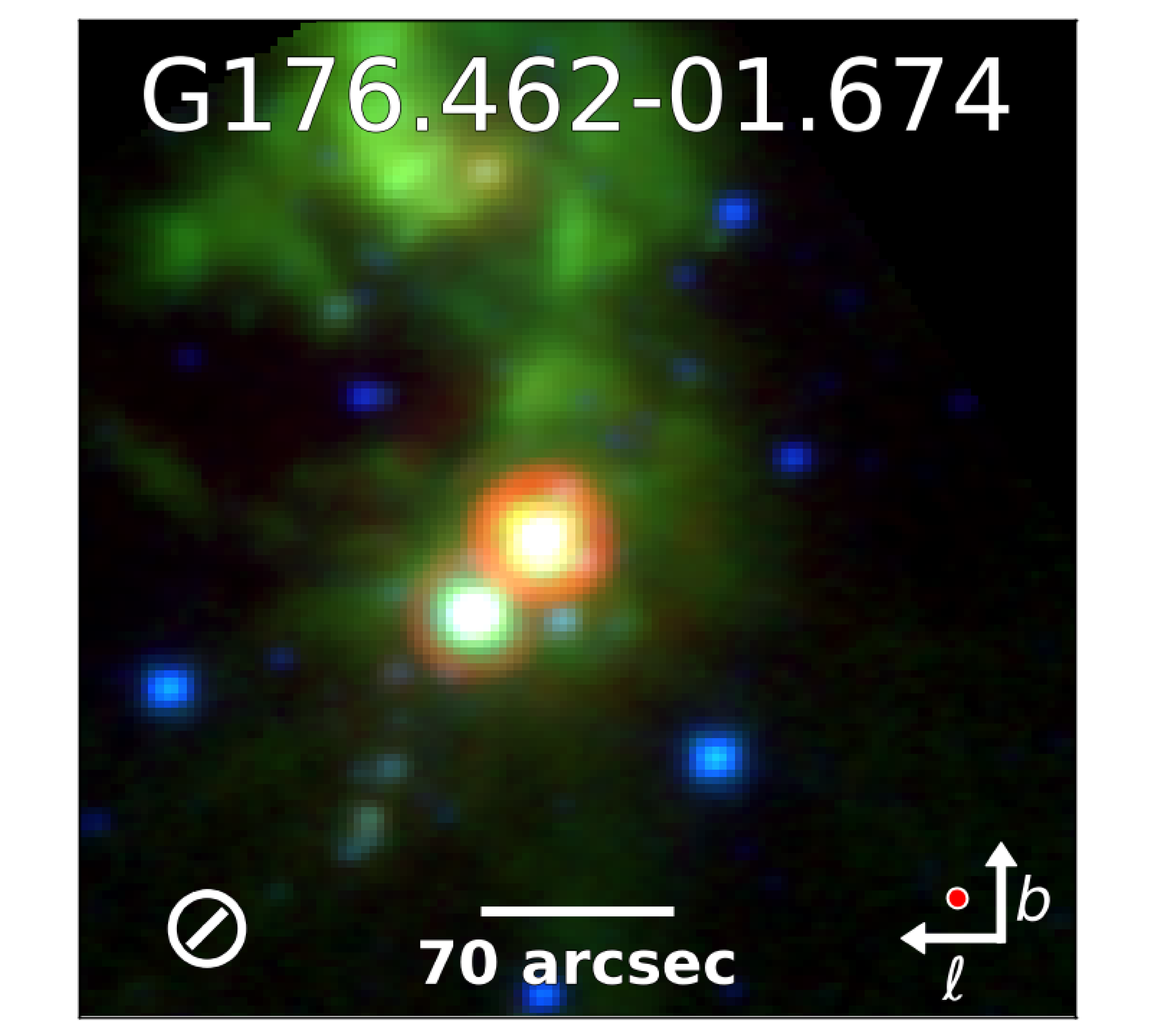}
\includegraphics[width=\figSize]{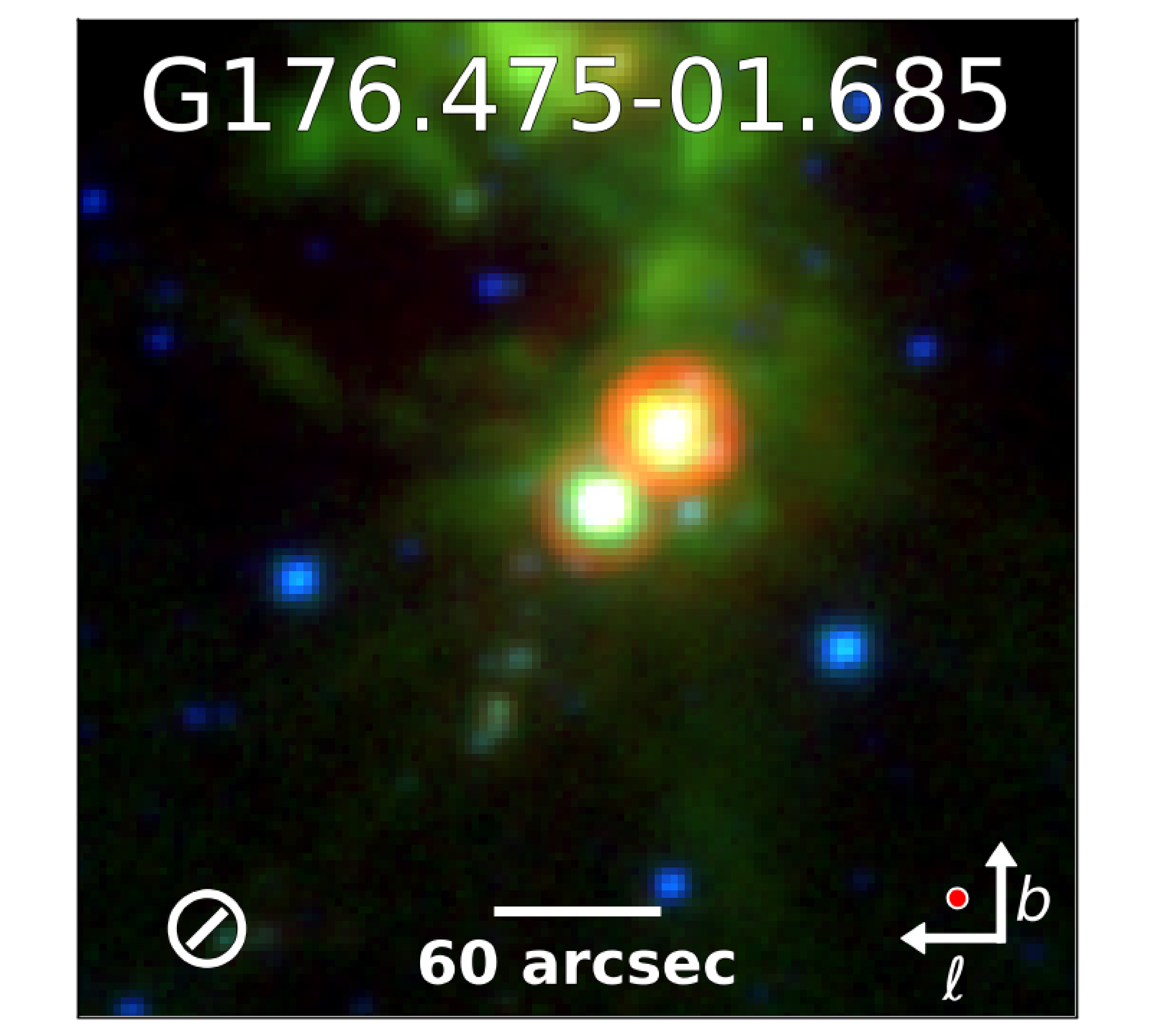}\\
\includegraphics[width=\figSize]{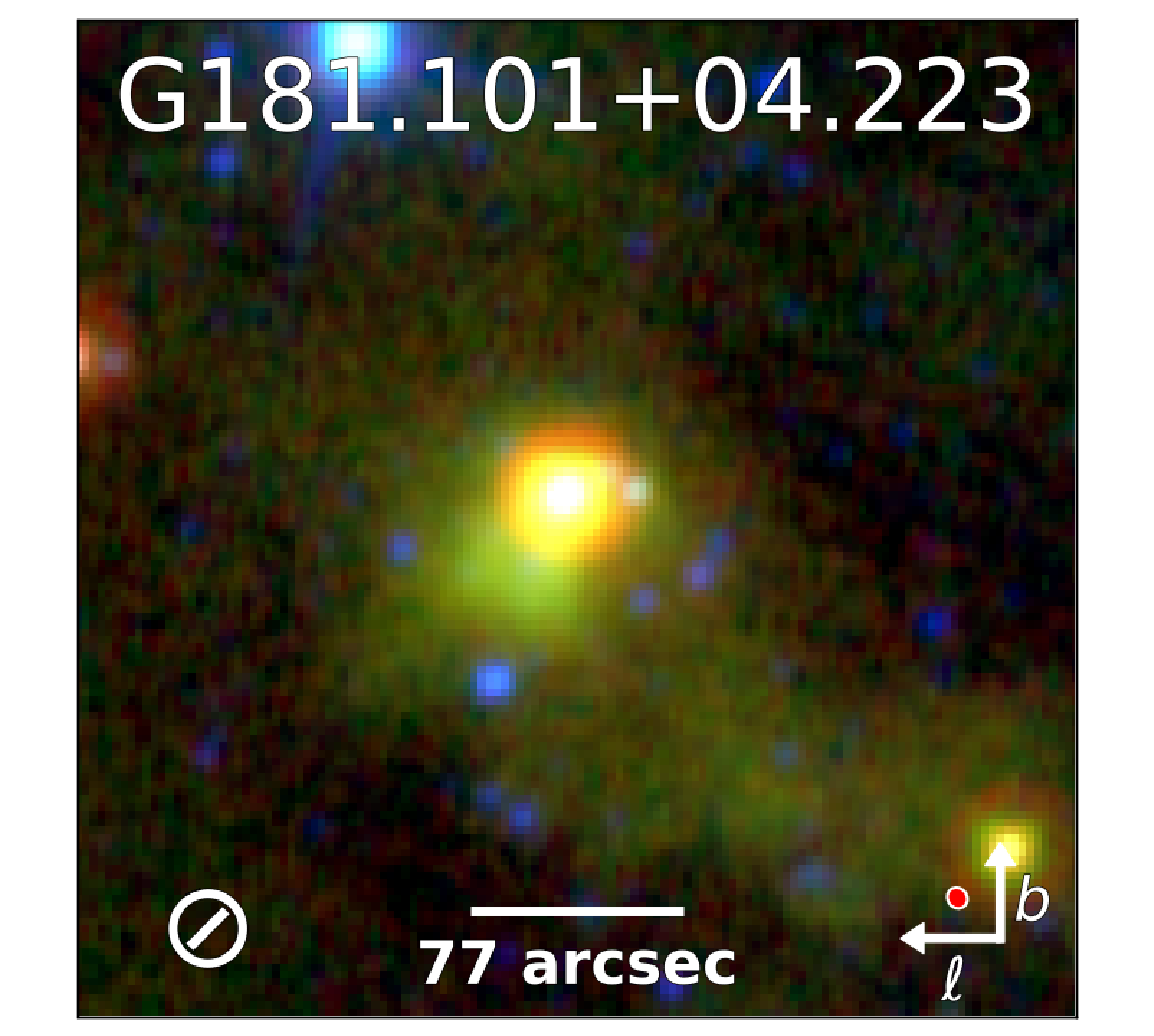}
\includegraphics[width=\figSize]{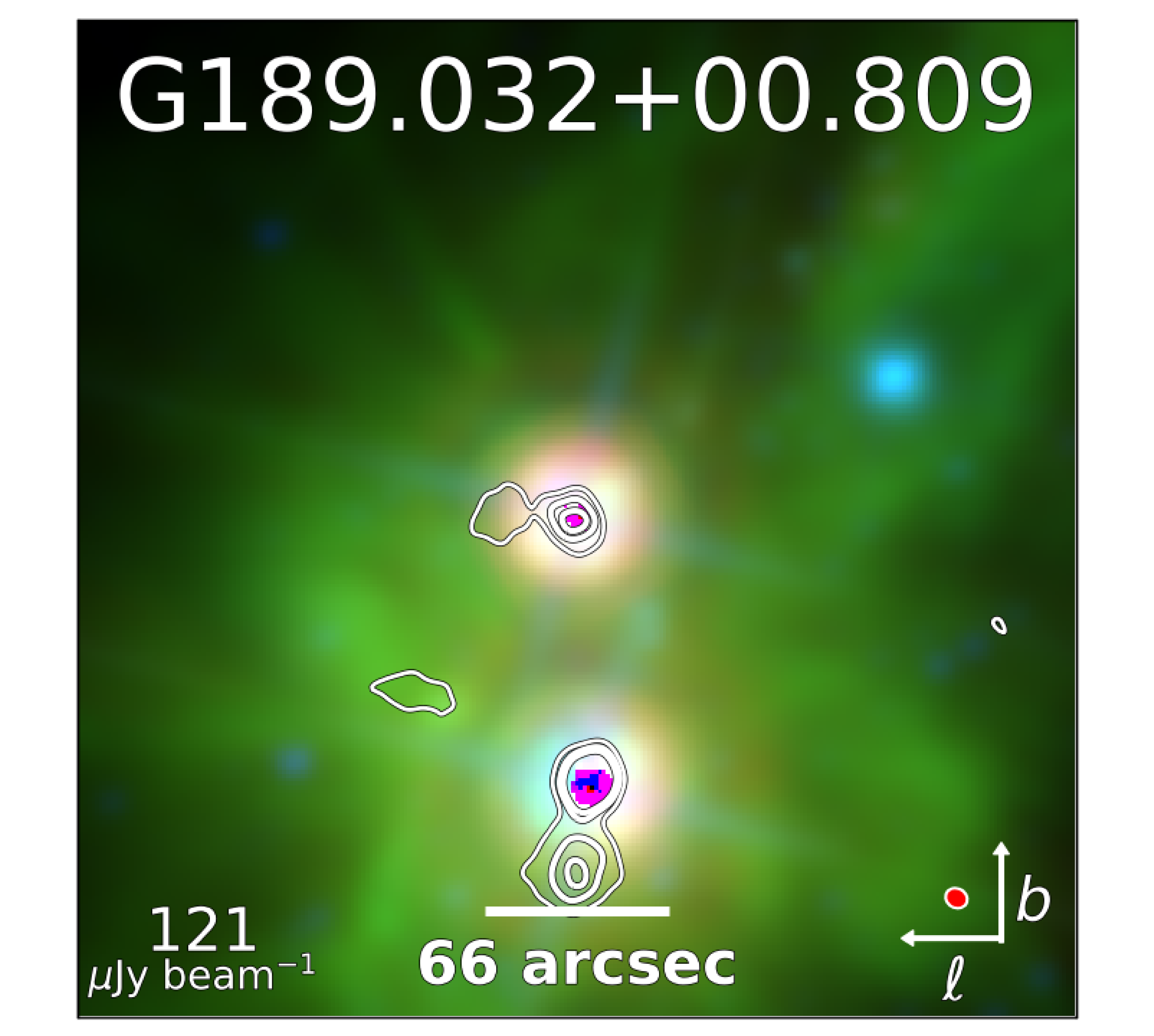}
\includegraphics[width=\figSize]{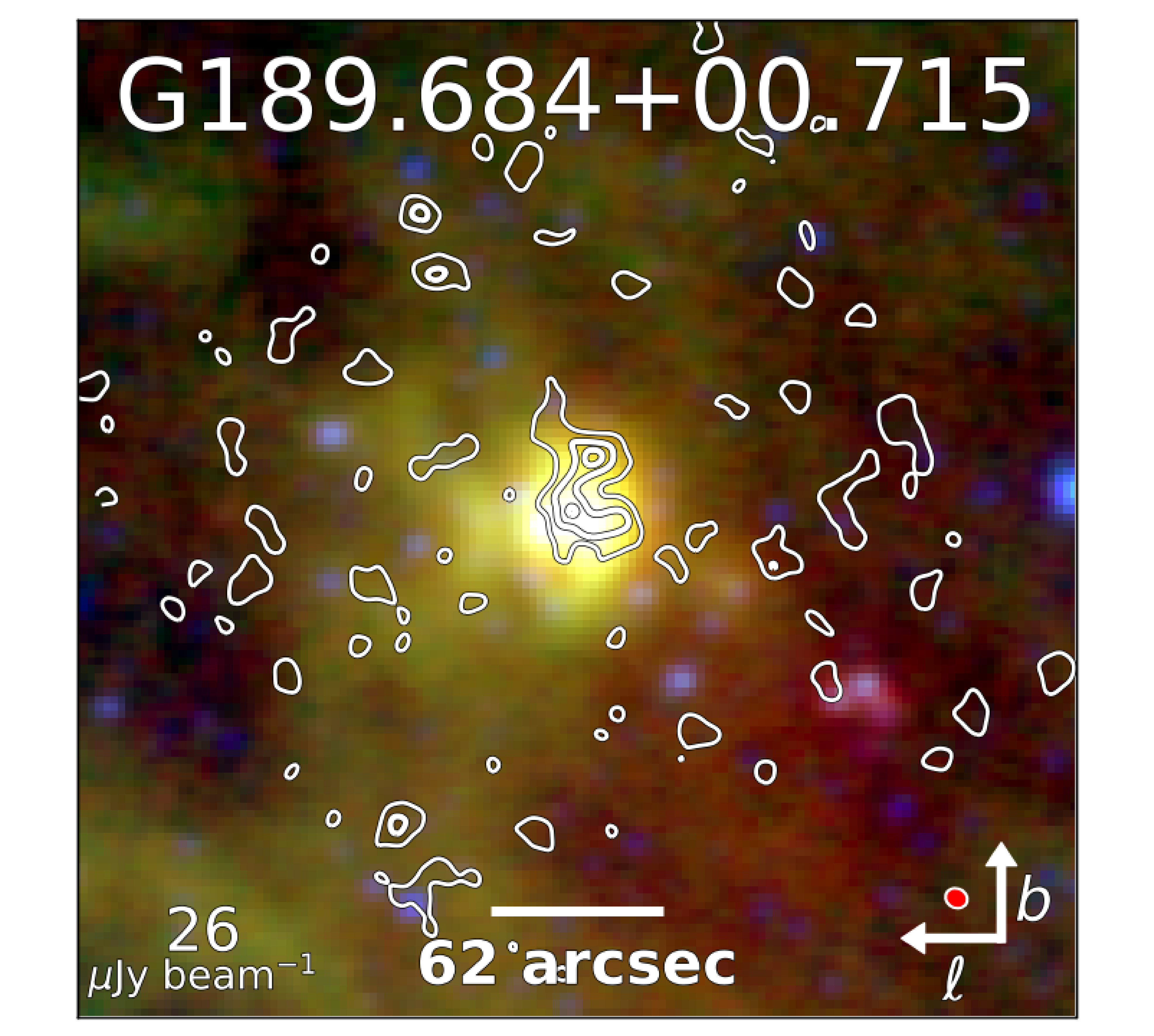}\\
\includegraphics[width=\figSize]{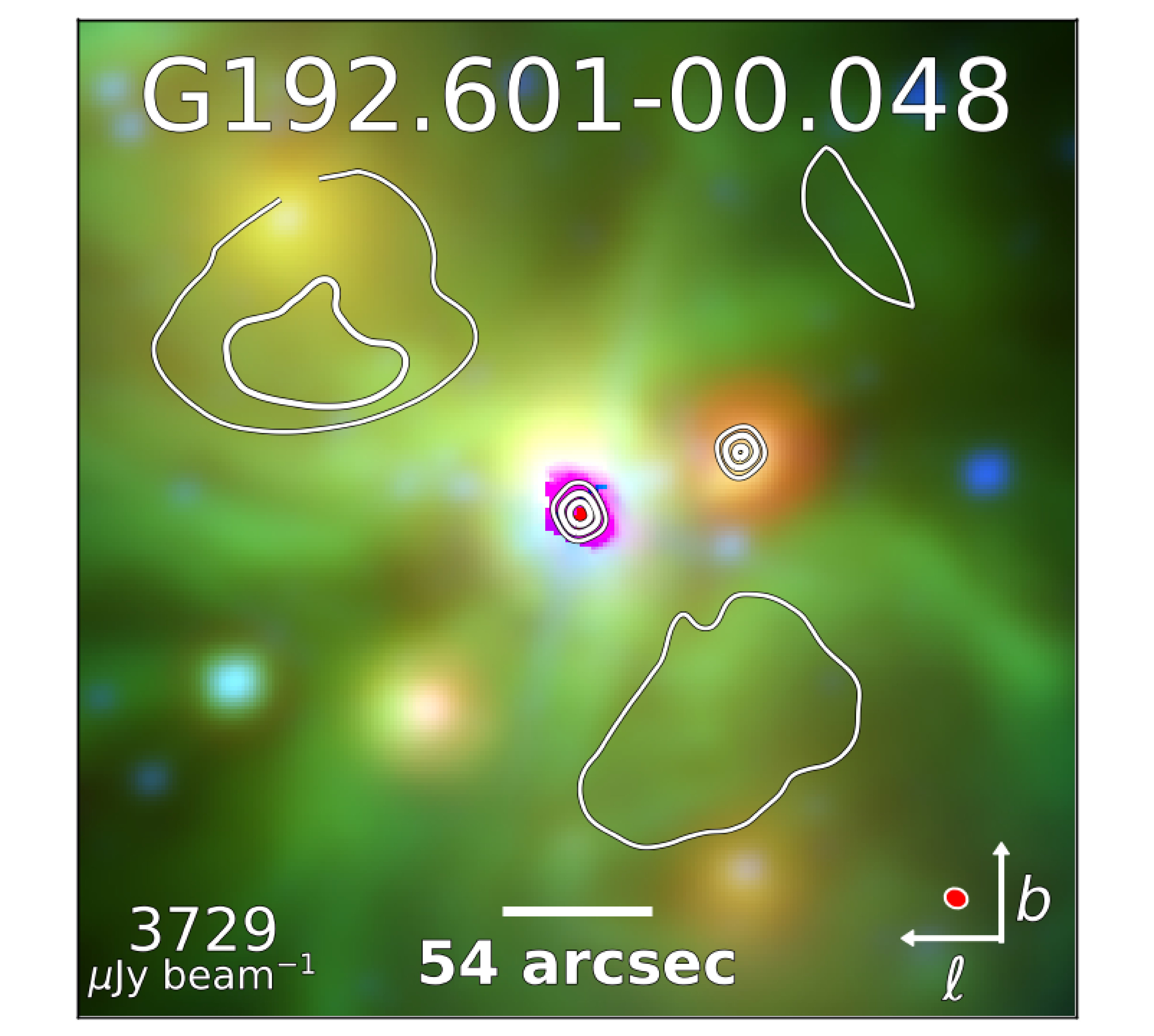}
\includegraphics[width=\figSize]{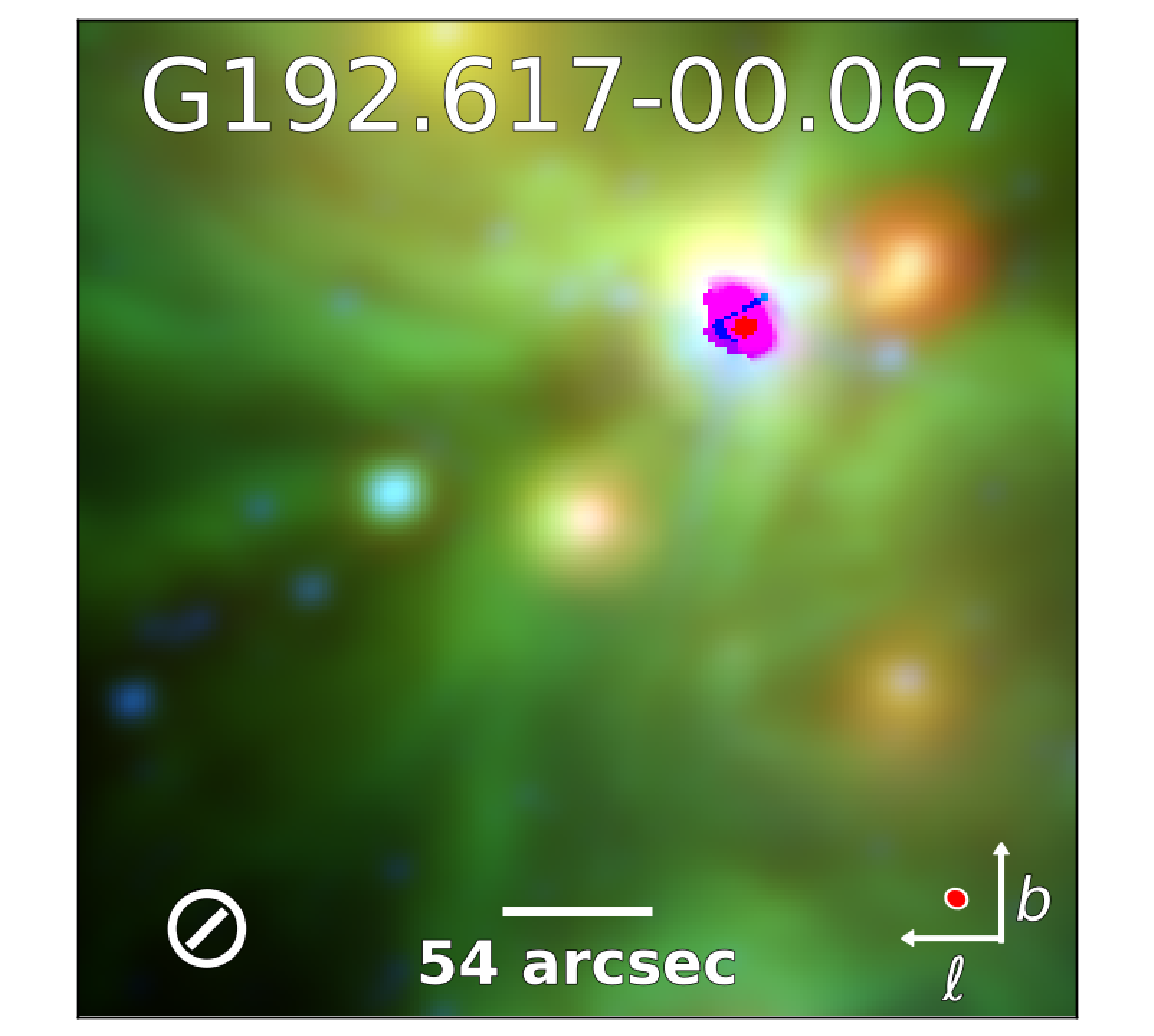}
\includegraphics[width=\figSize]{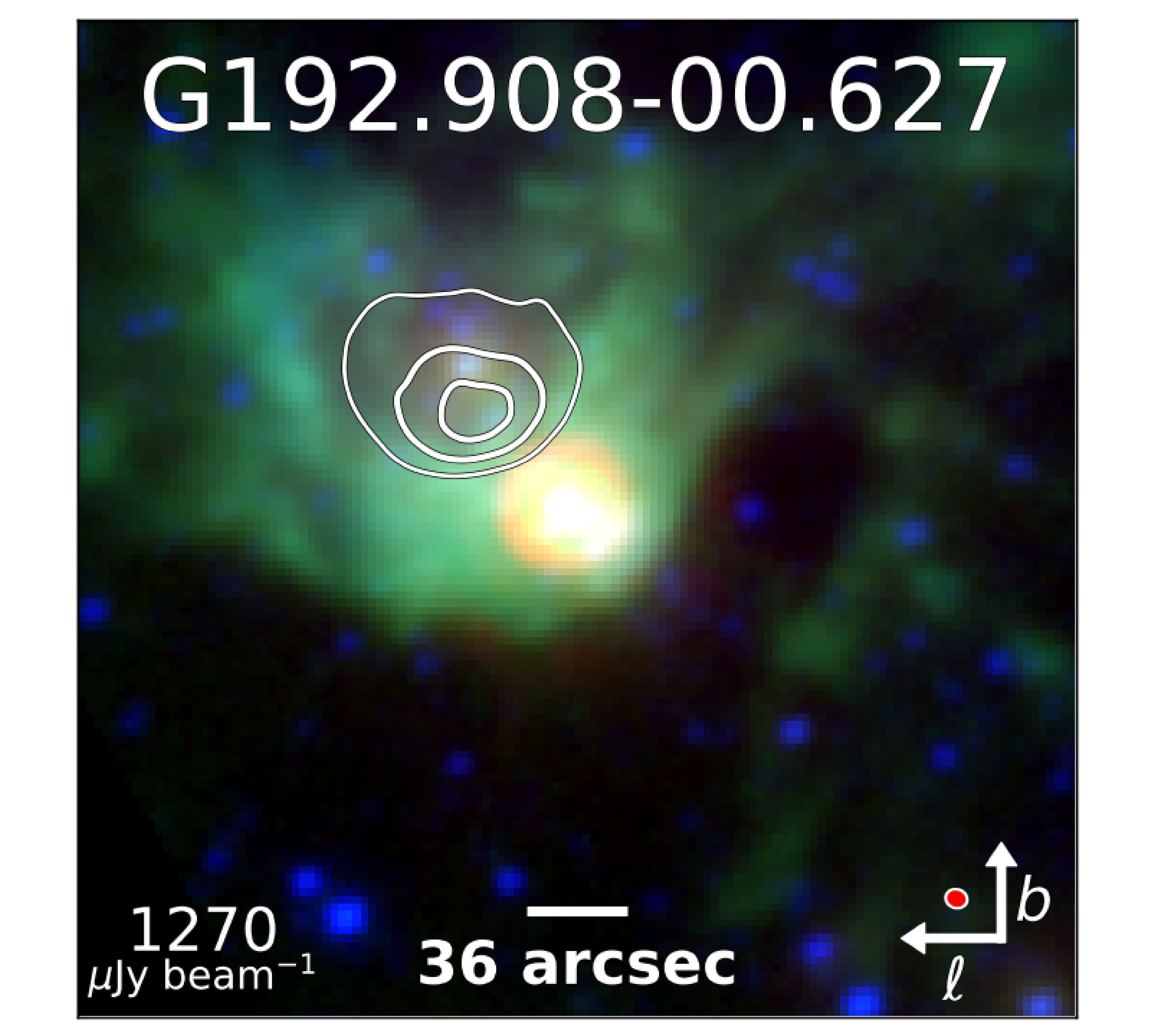}
\end{figure*}
\begin{figure*}[!htb]
\includegraphics[width=\figSize]{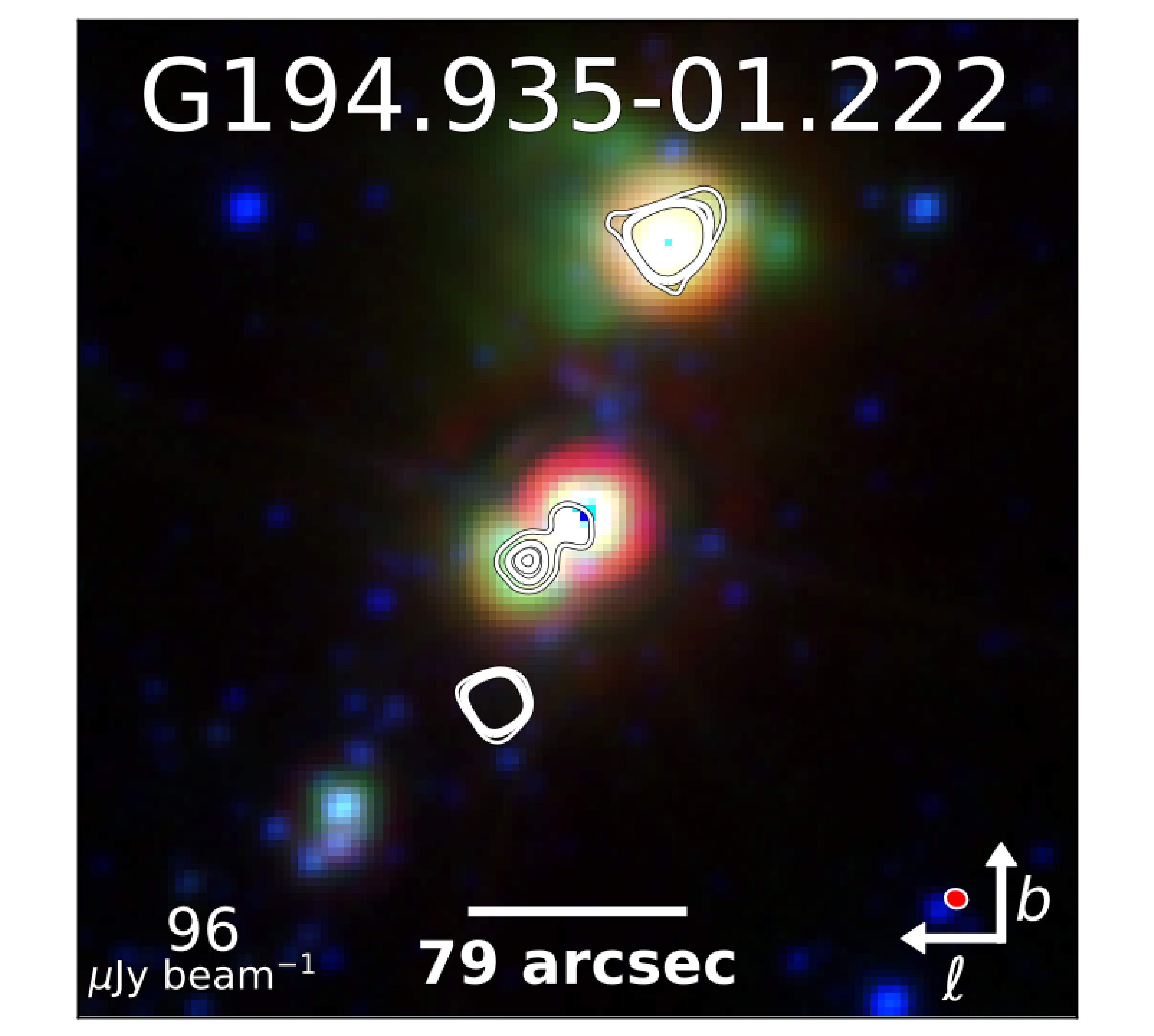}
\includegraphics[width=\figSize]{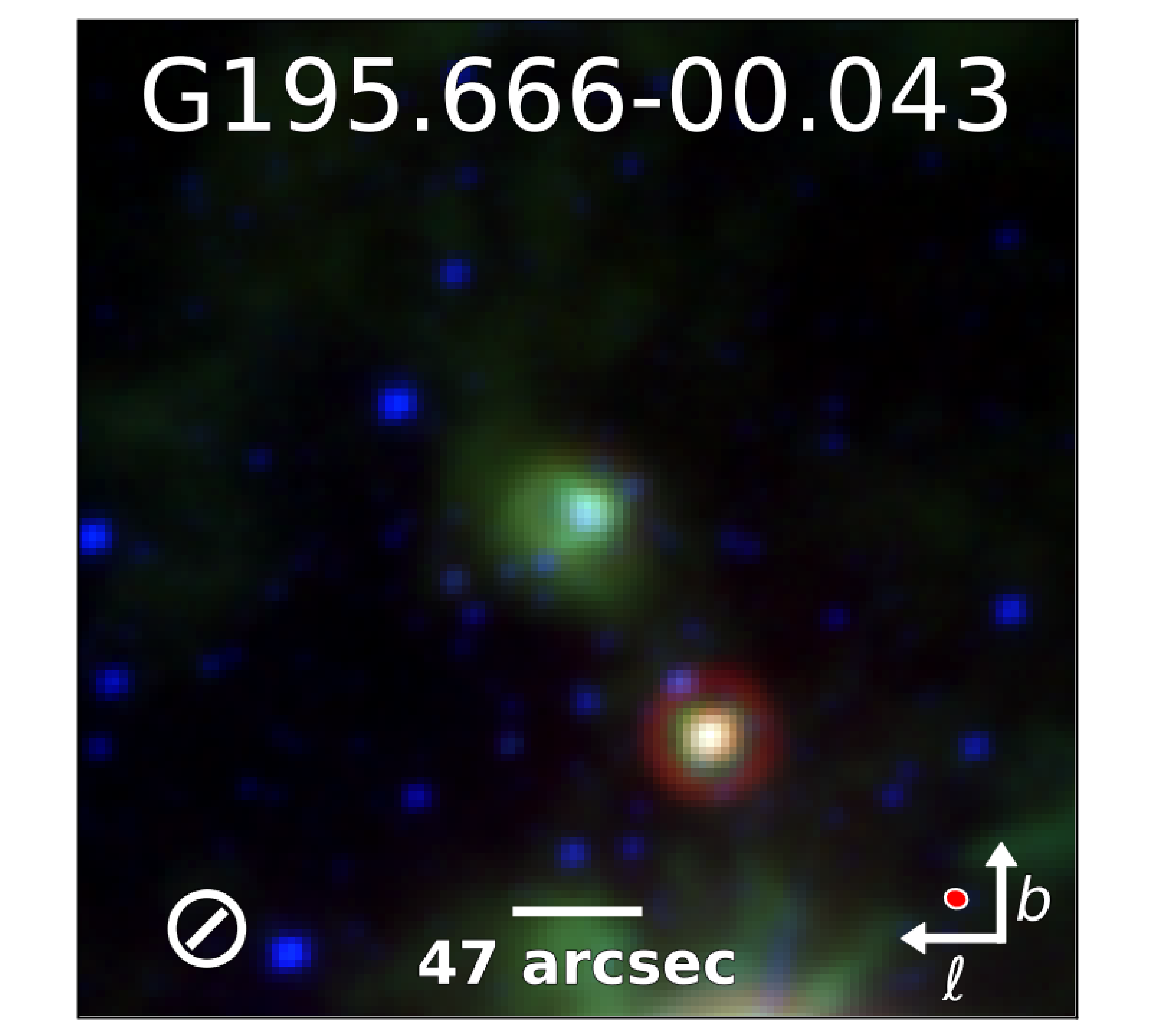}
\includegraphics[width=\figSize]{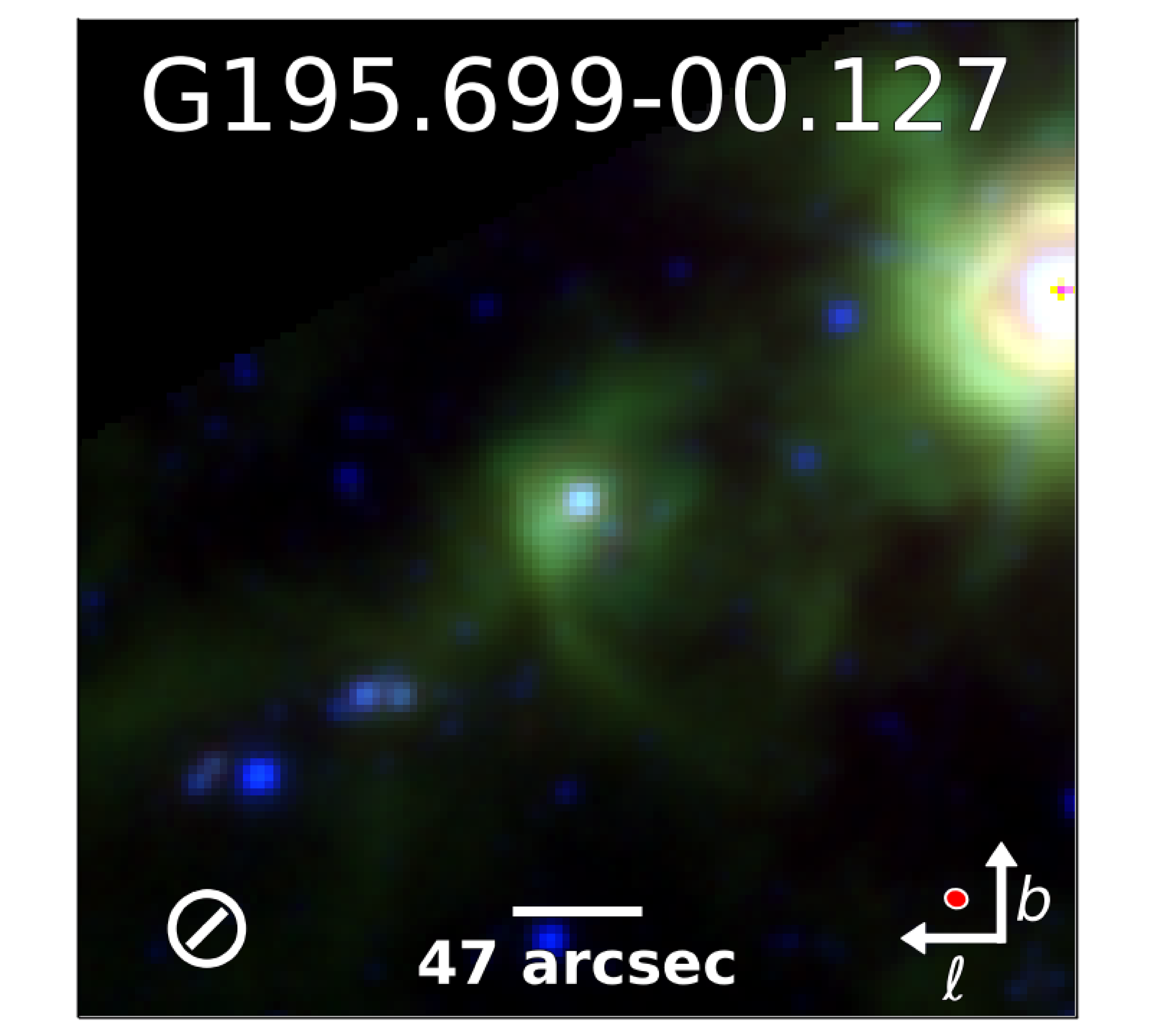}\\
\includegraphics[width=\figSize]{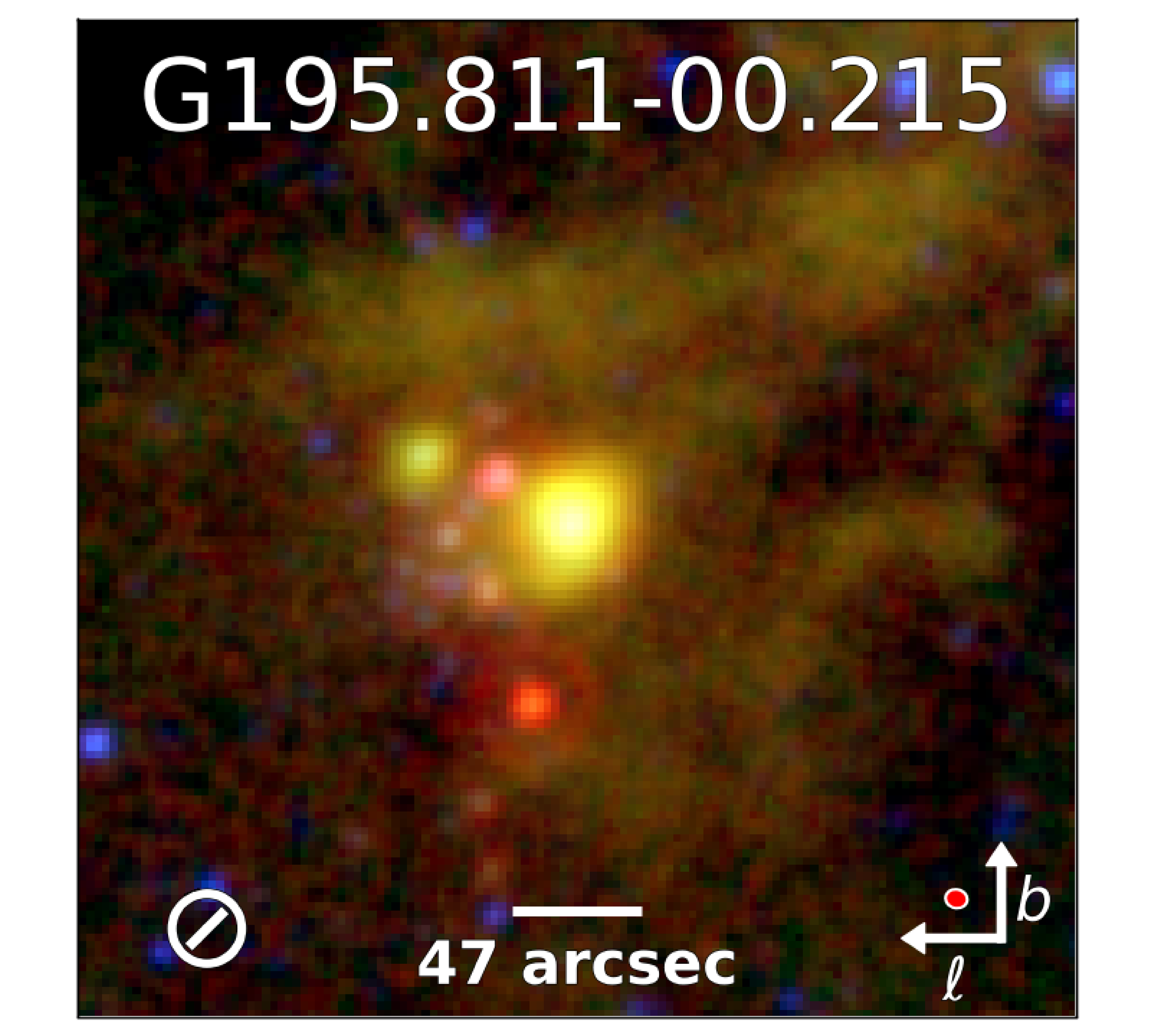}
\includegraphics[width=\figSize]{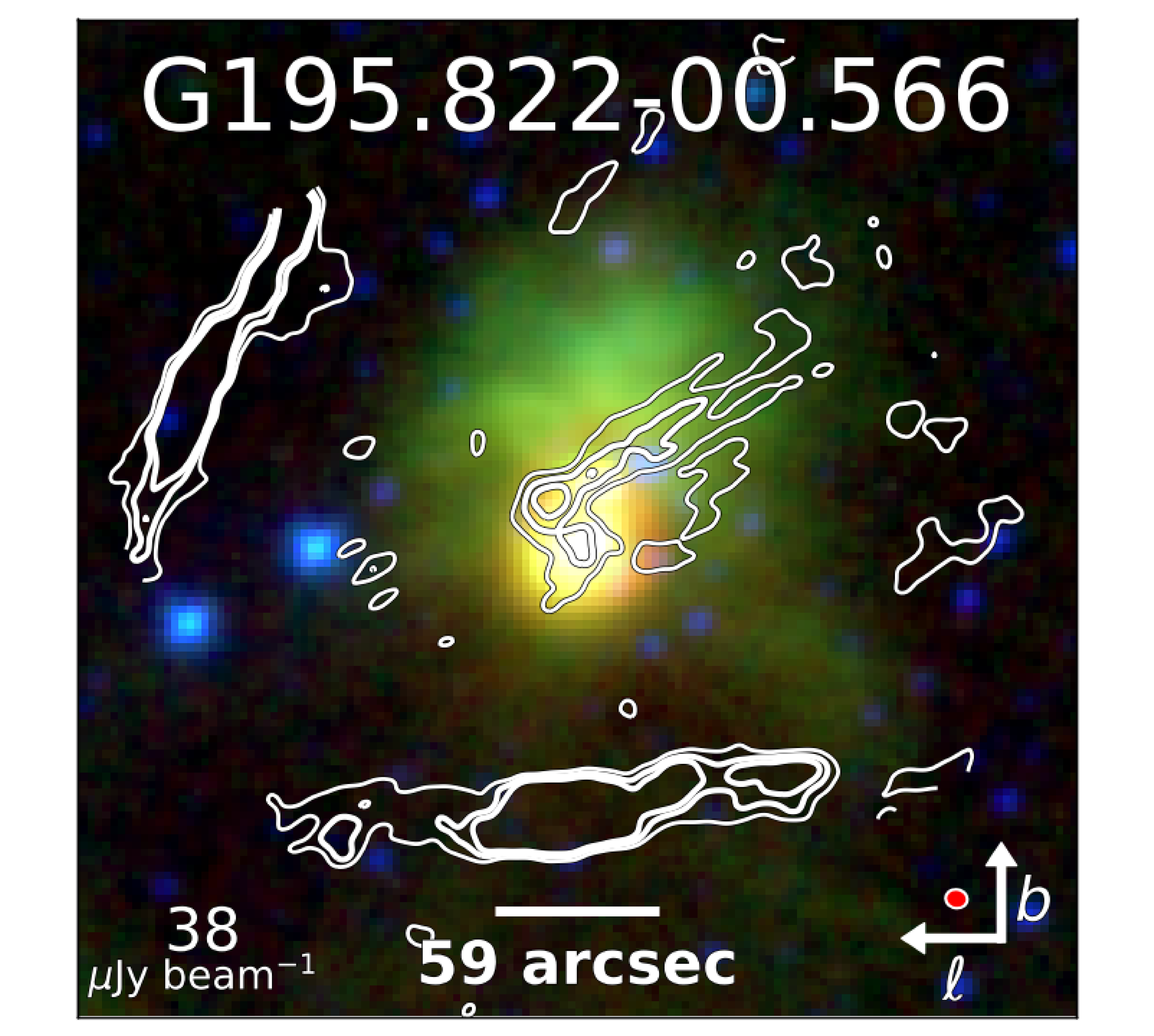}
\includegraphics[width=\figSize]{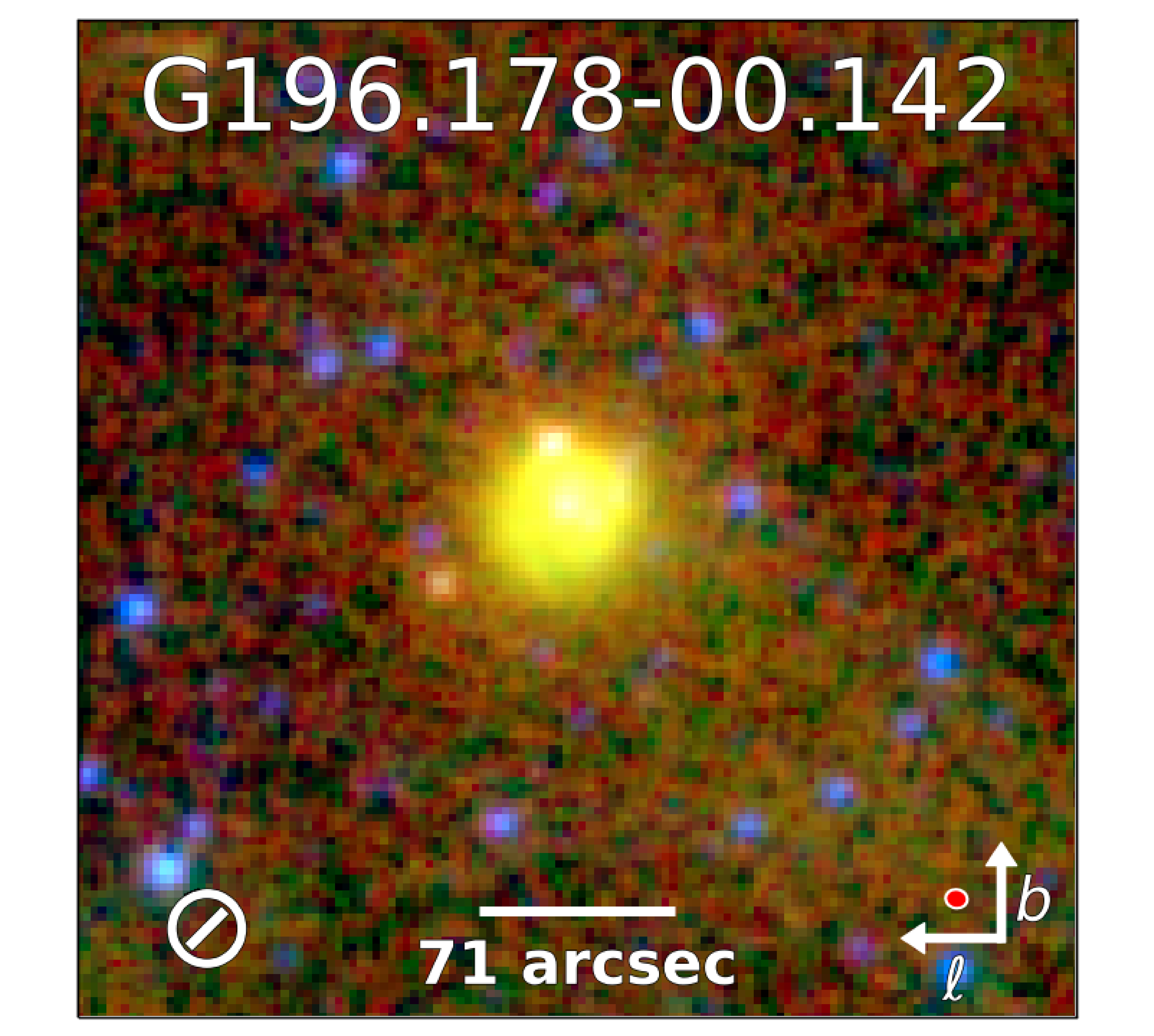}\\
\includegraphics[width=\figSize]{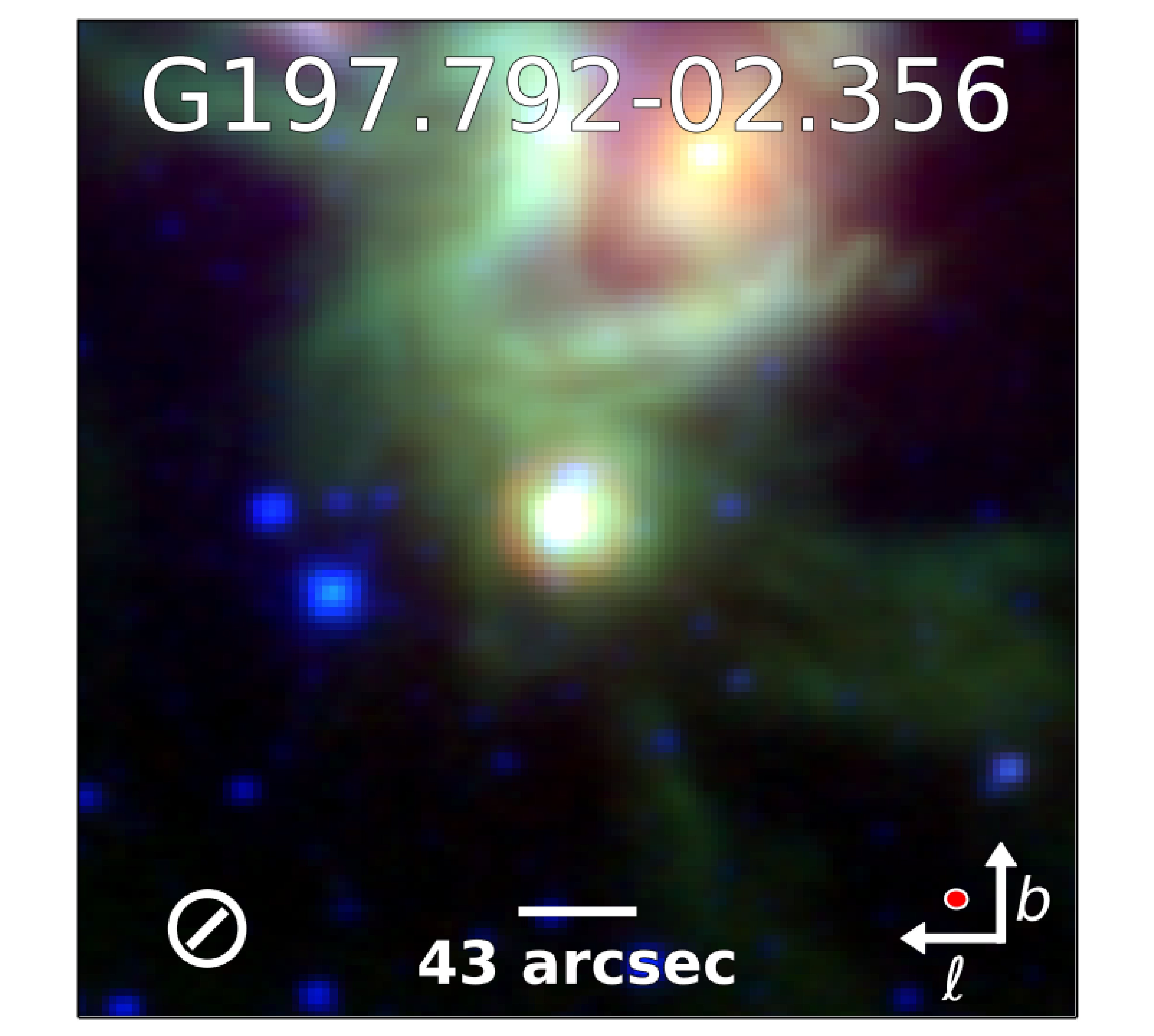}
\includegraphics[width=\figSize]{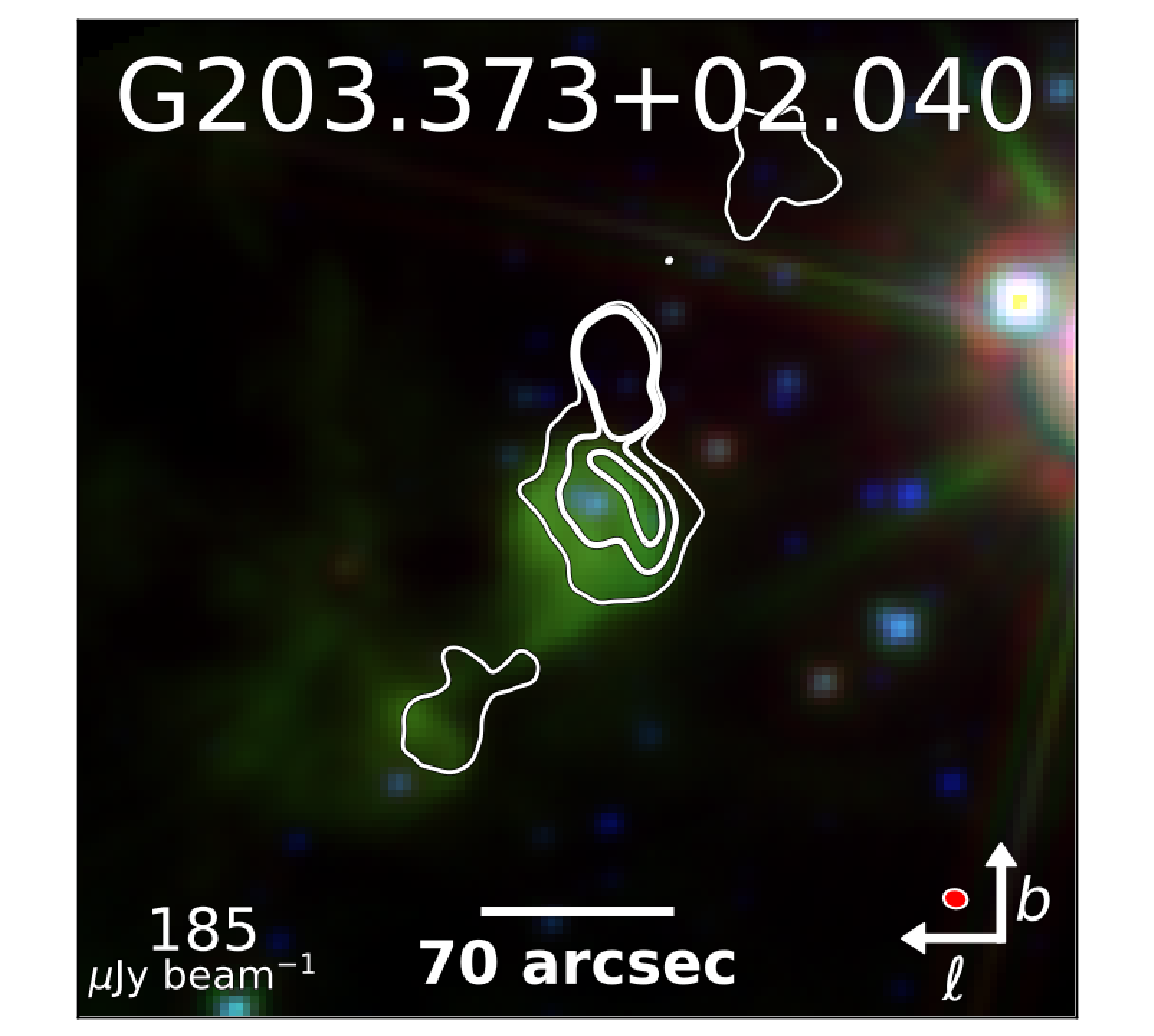}
\includegraphics[width=\figSize]{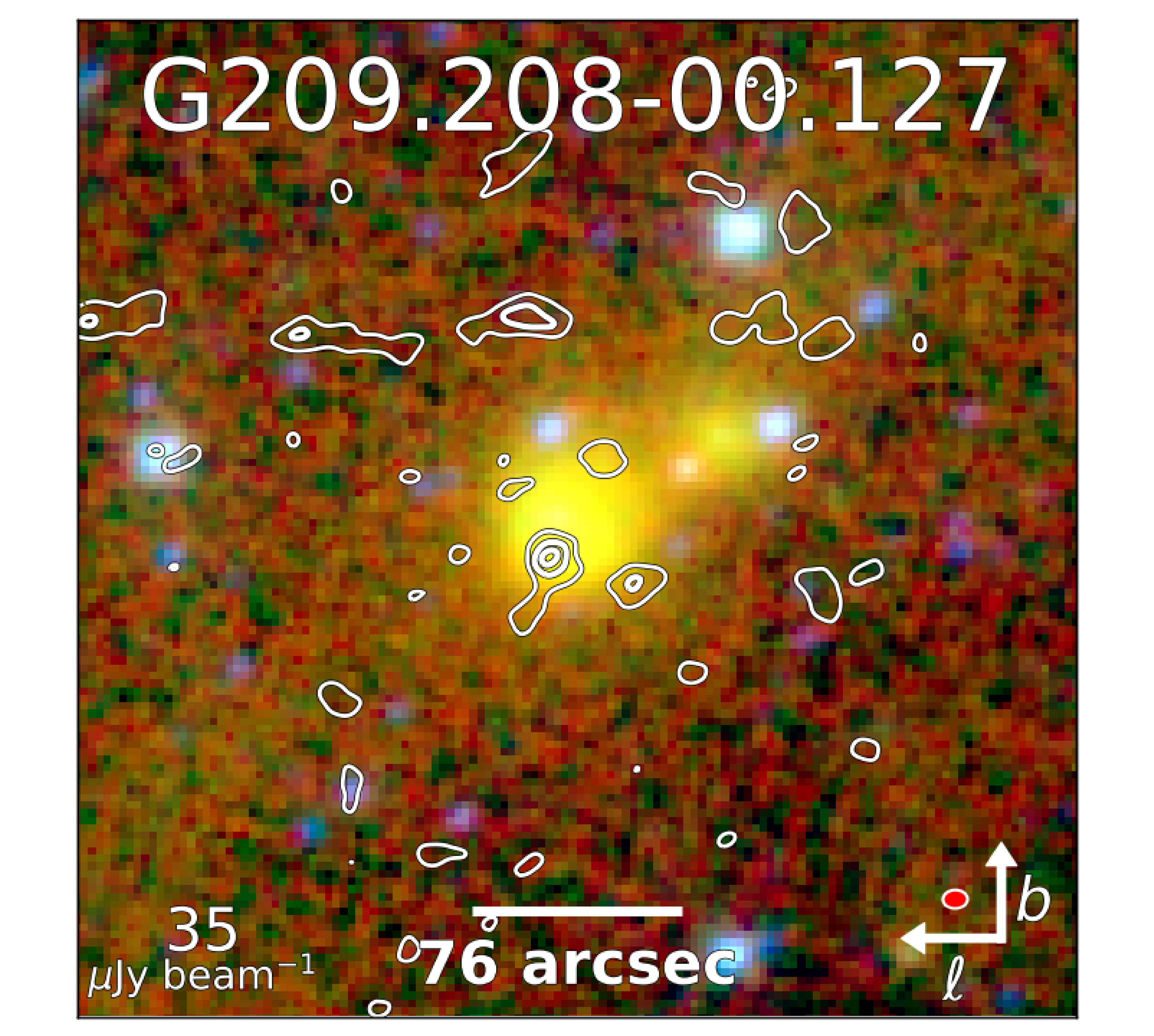}\\
\includegraphics[width=\figSize]{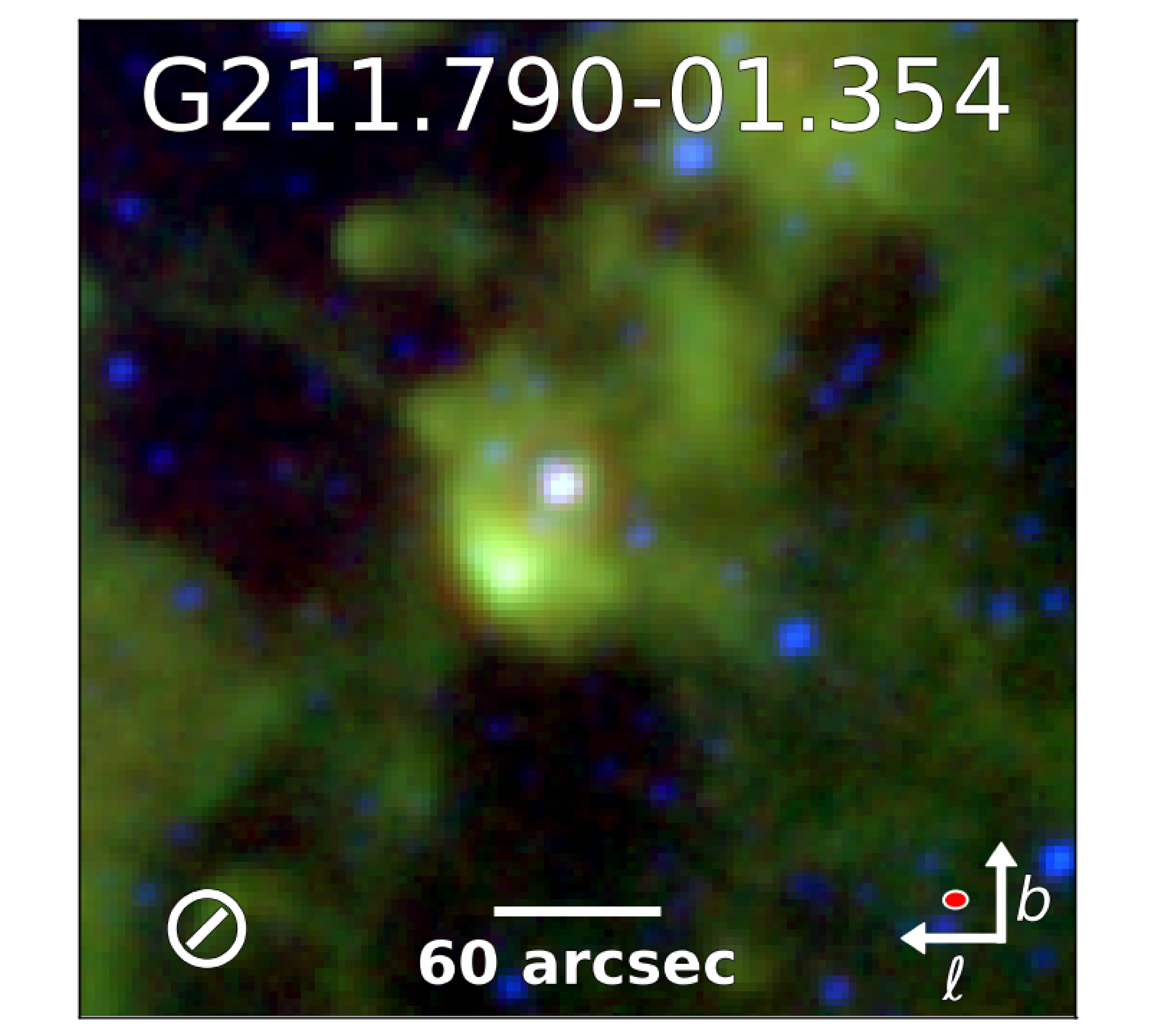}
\includegraphics[width=\figSize]{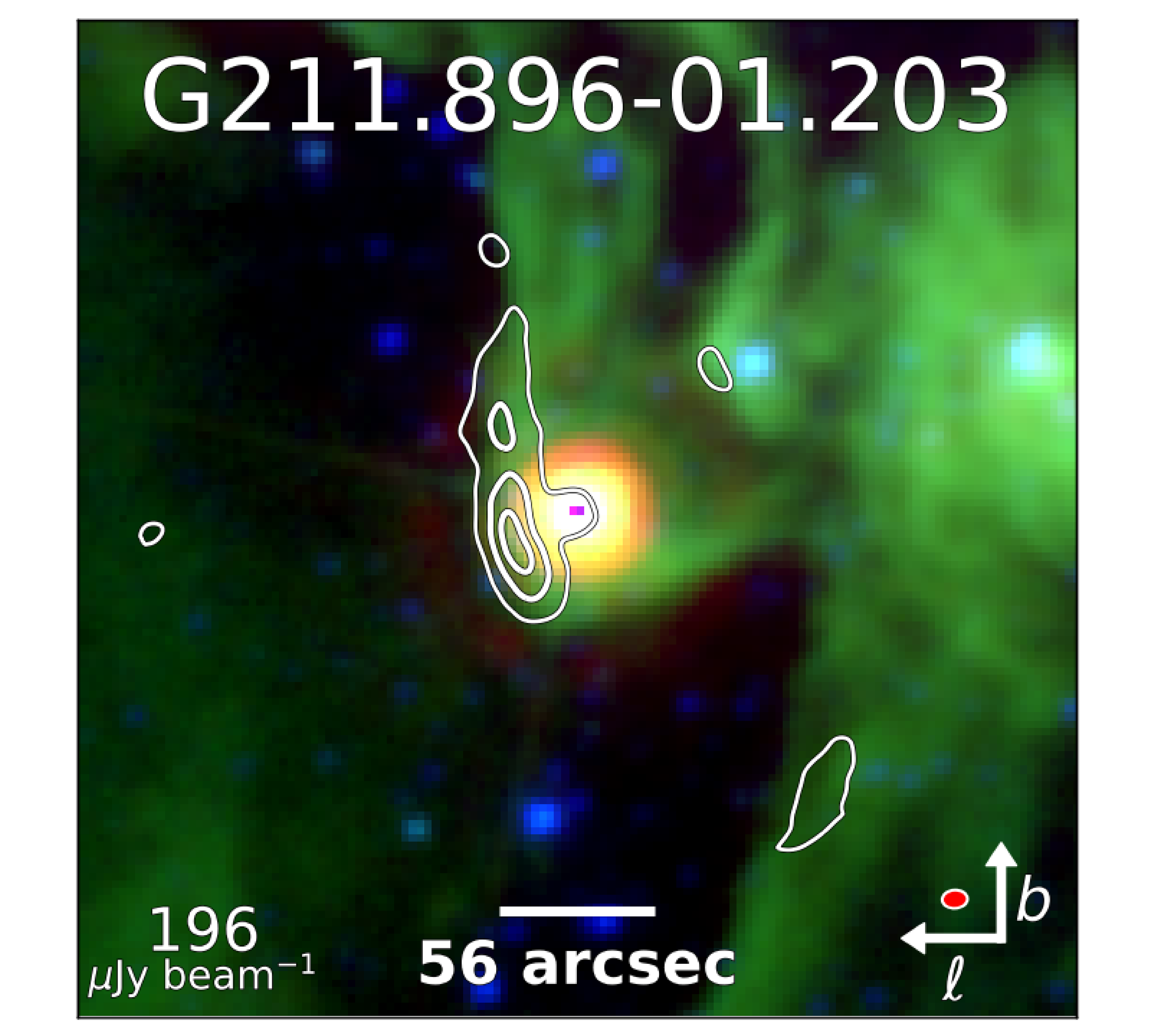}
\includegraphics[width=\figSize]{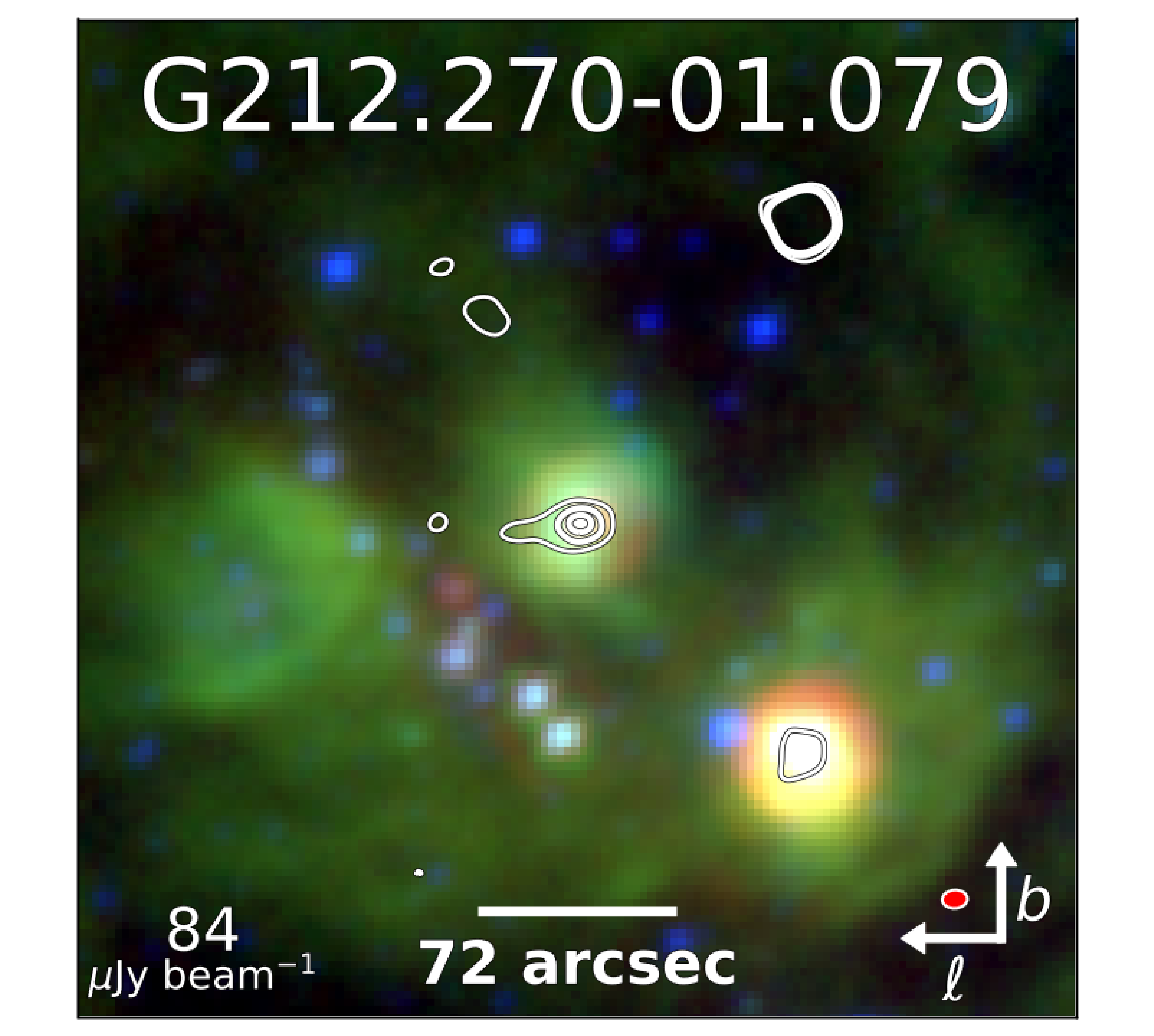}
\end{figure*}
\begin{figure*}[!htb]
\includegraphics[width=\figSize]{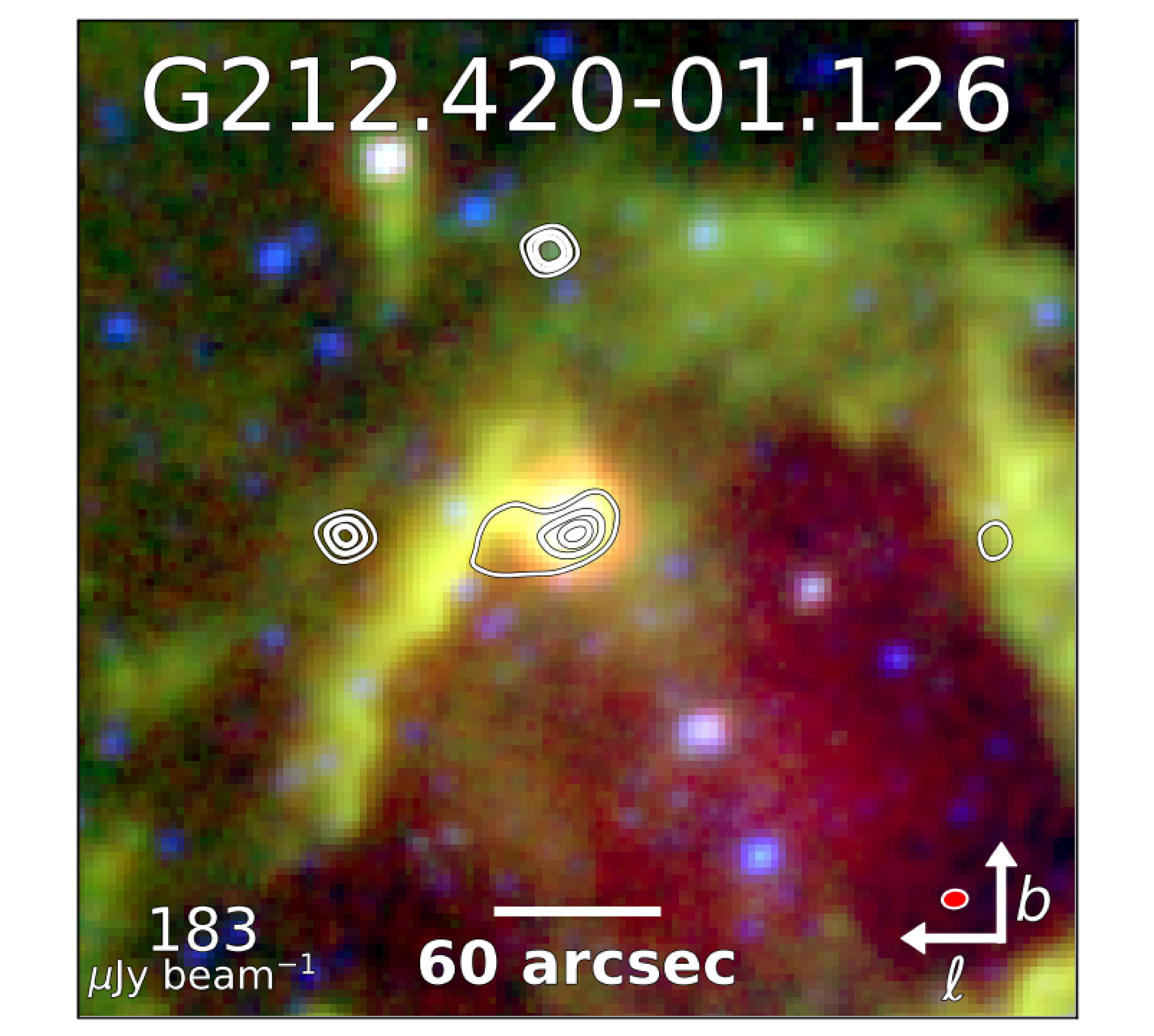}
\includegraphics[width=\figSize]{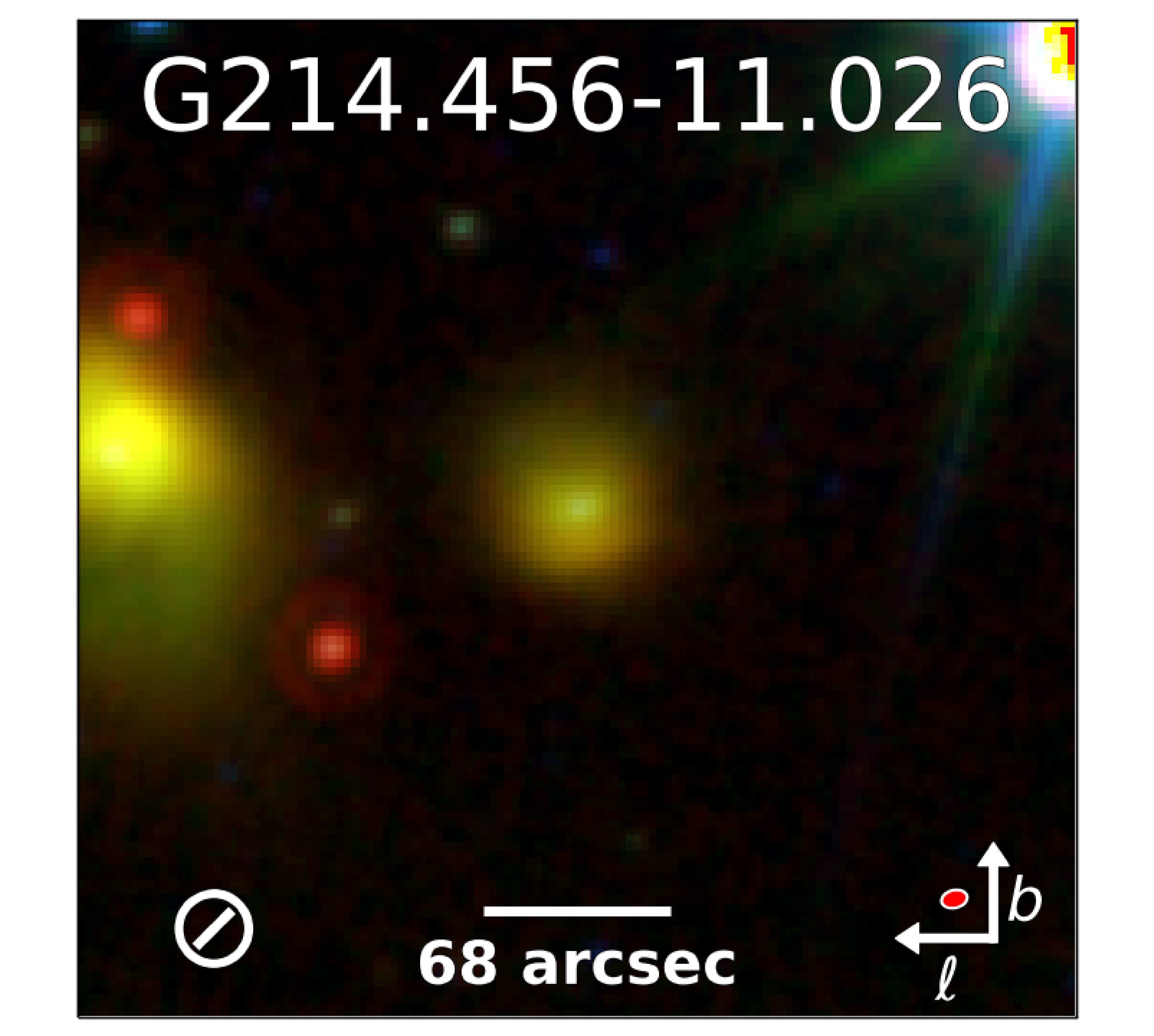}
\includegraphics[width=\figSize]{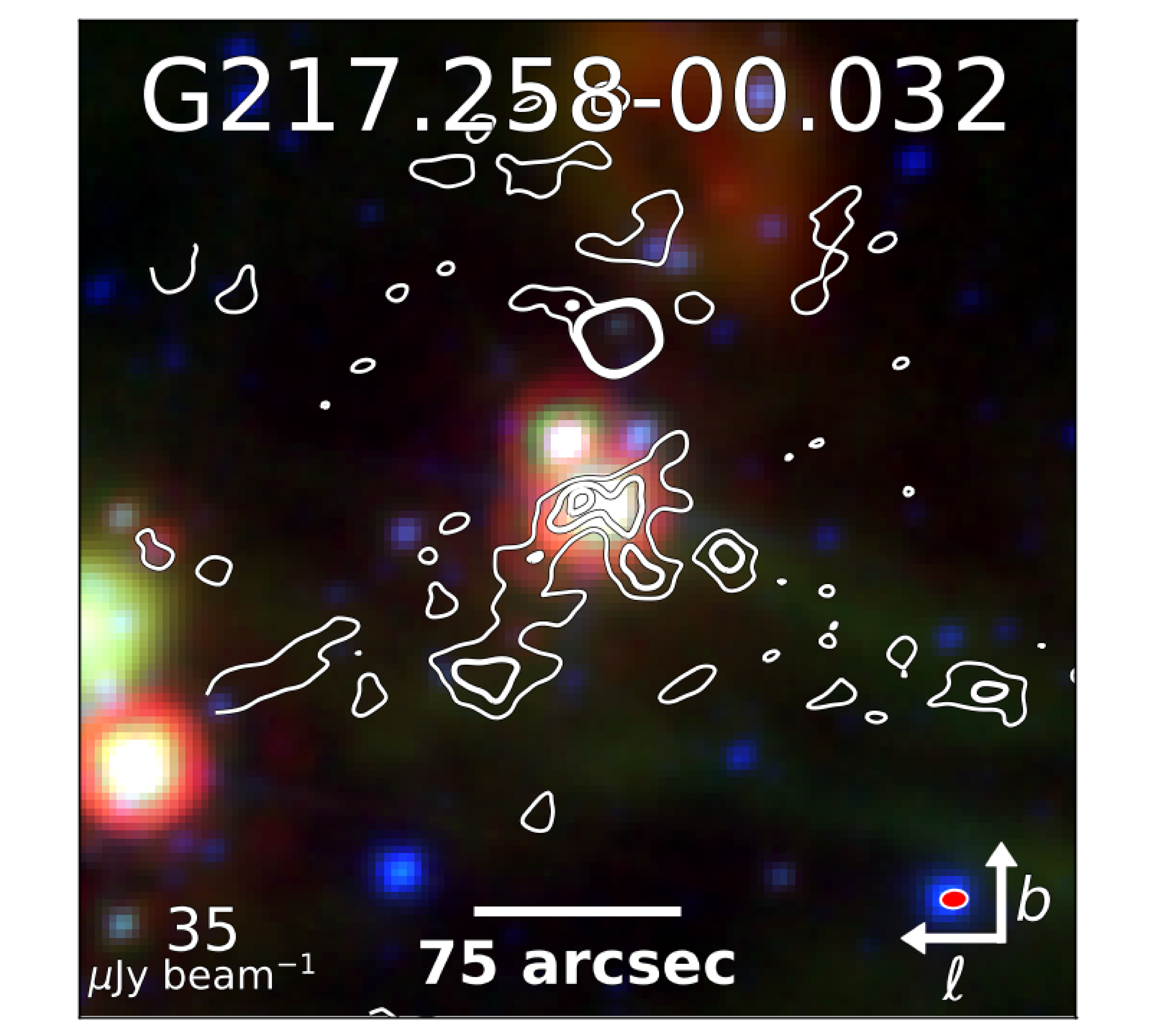}\\
\includegraphics[width=\figSize]{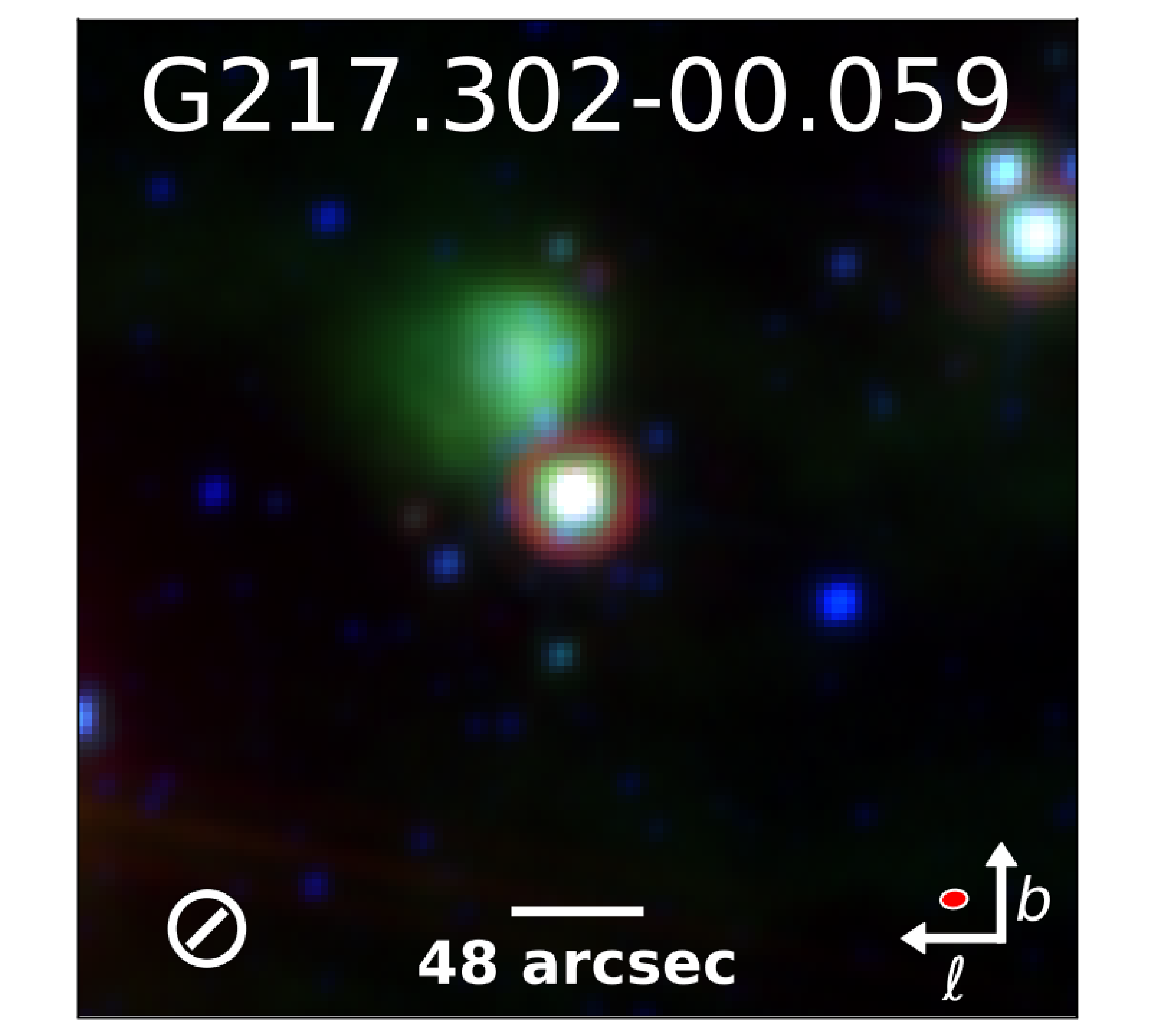}
\includegraphics[width=\figSize]{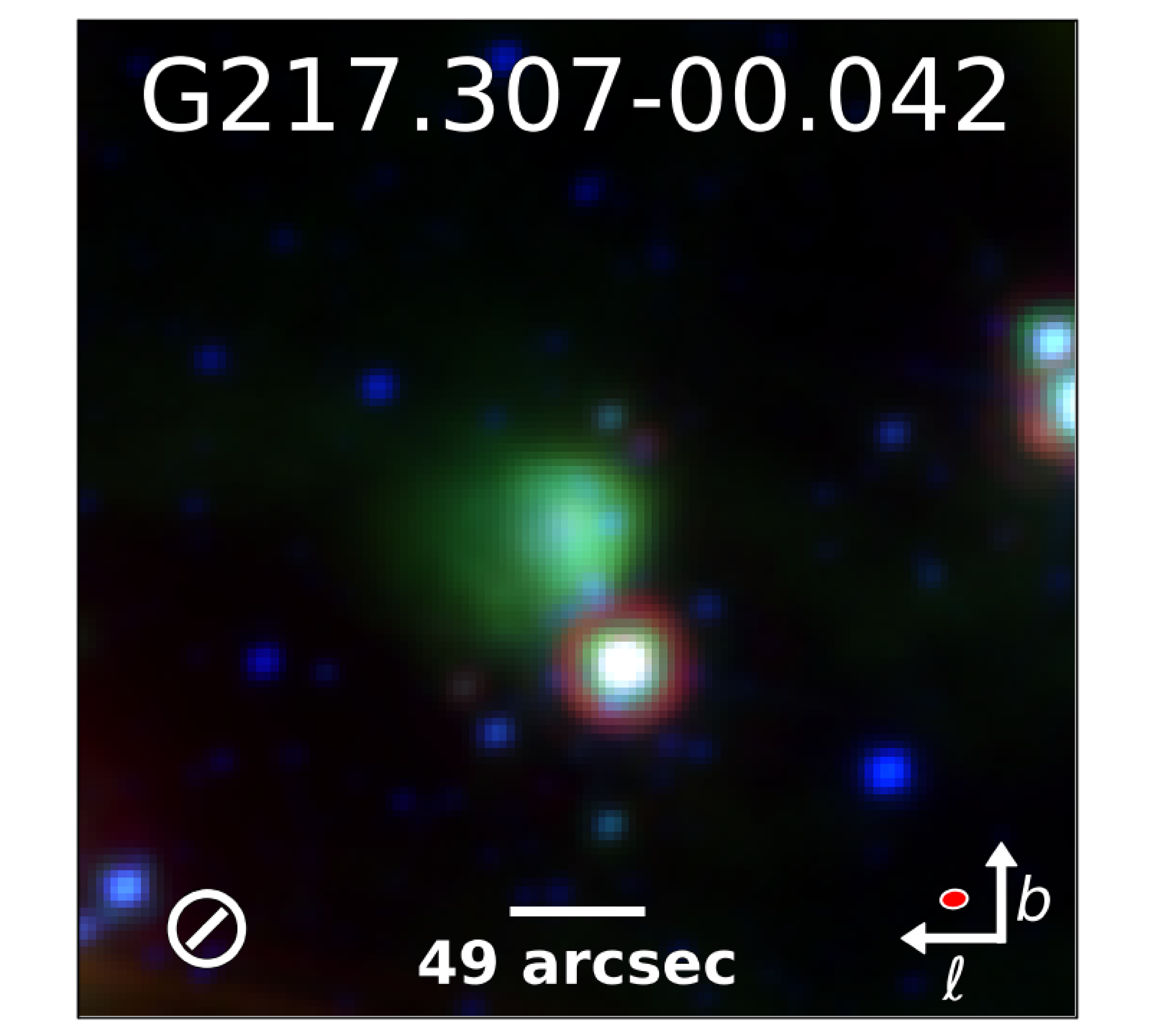}
\includegraphics[width=\figSize]{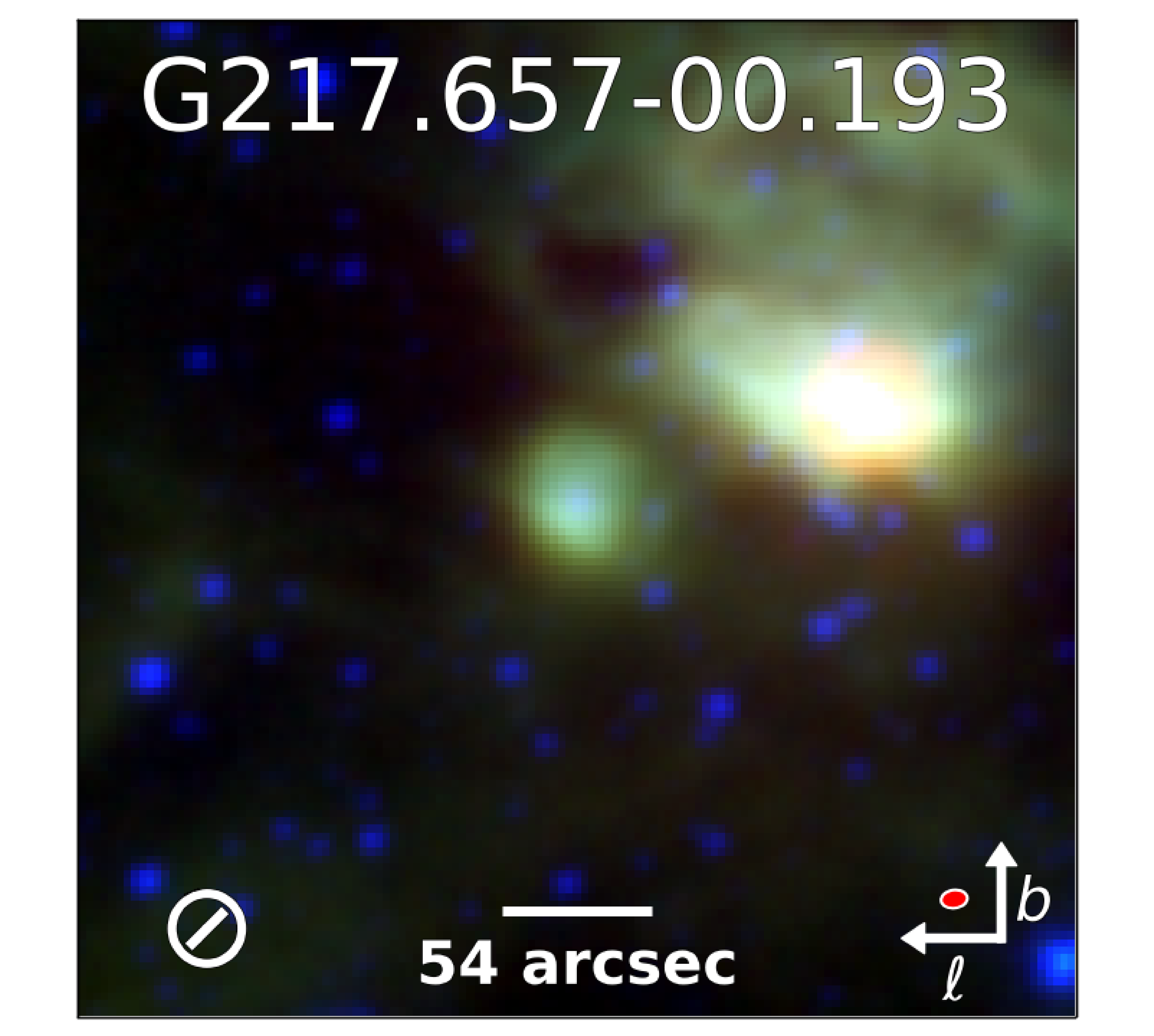}\\
\includegraphics[width=\figSize]{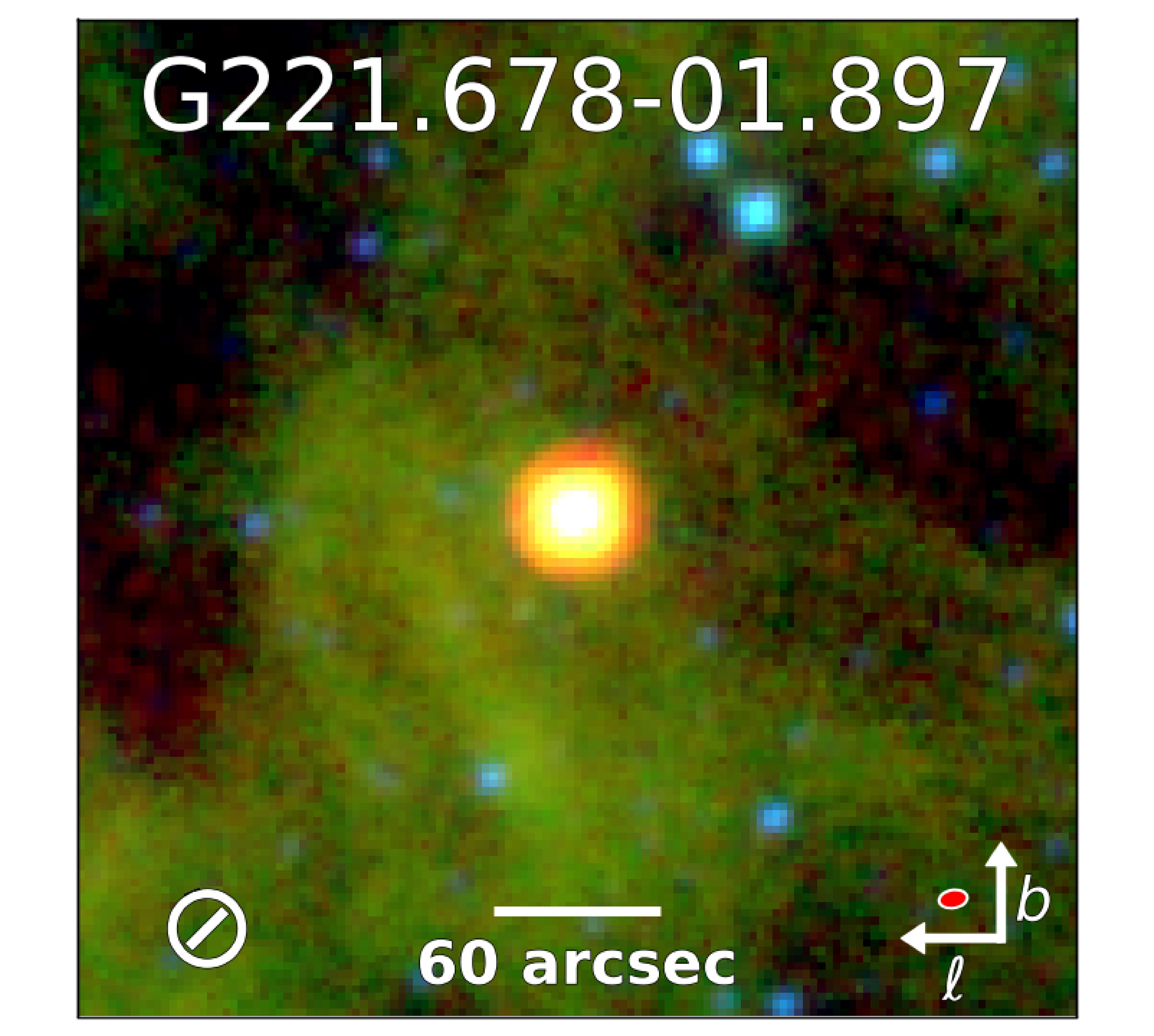}
\includegraphics[width=\figSize]{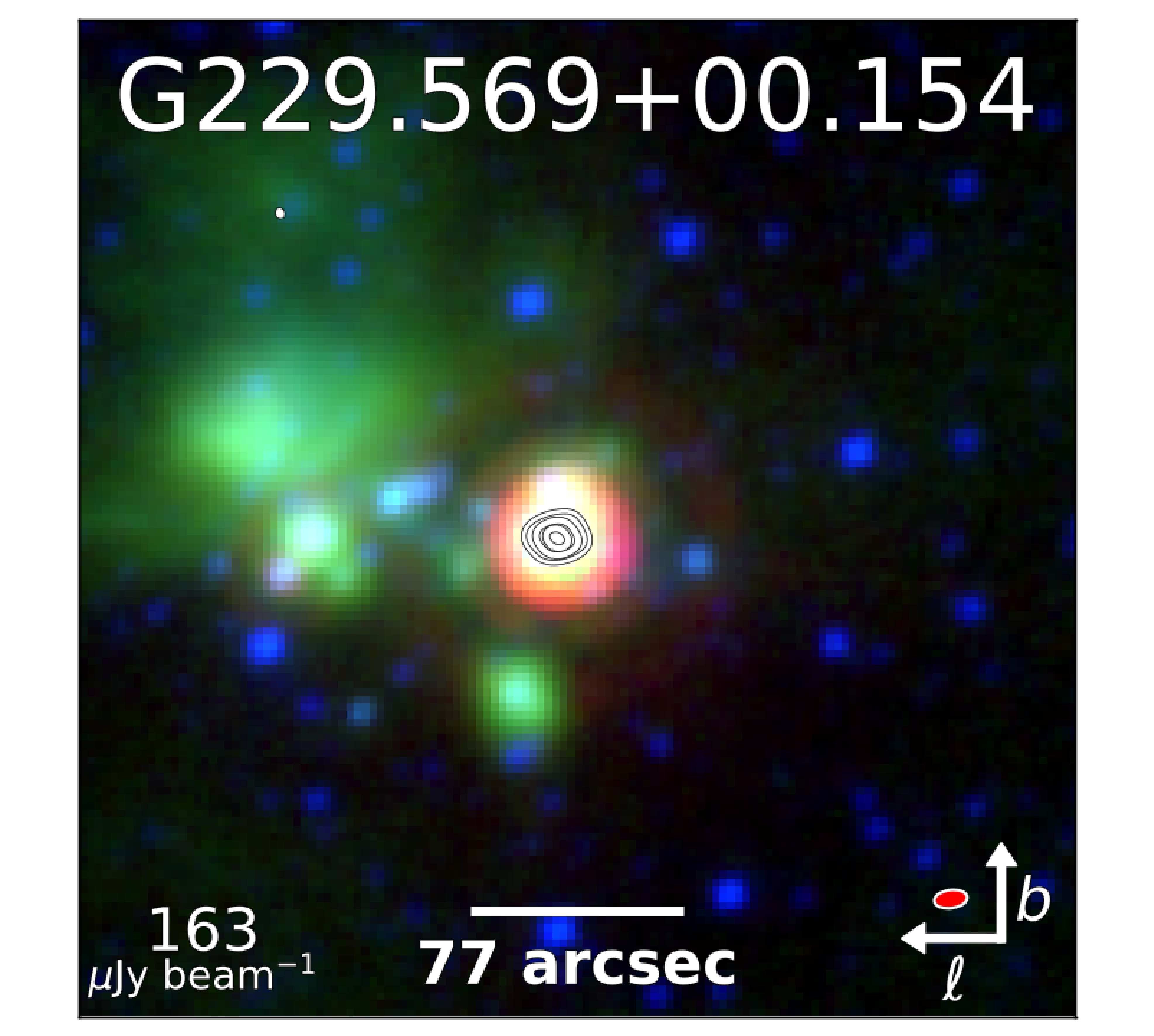}
\includegraphics[width=\figSize]{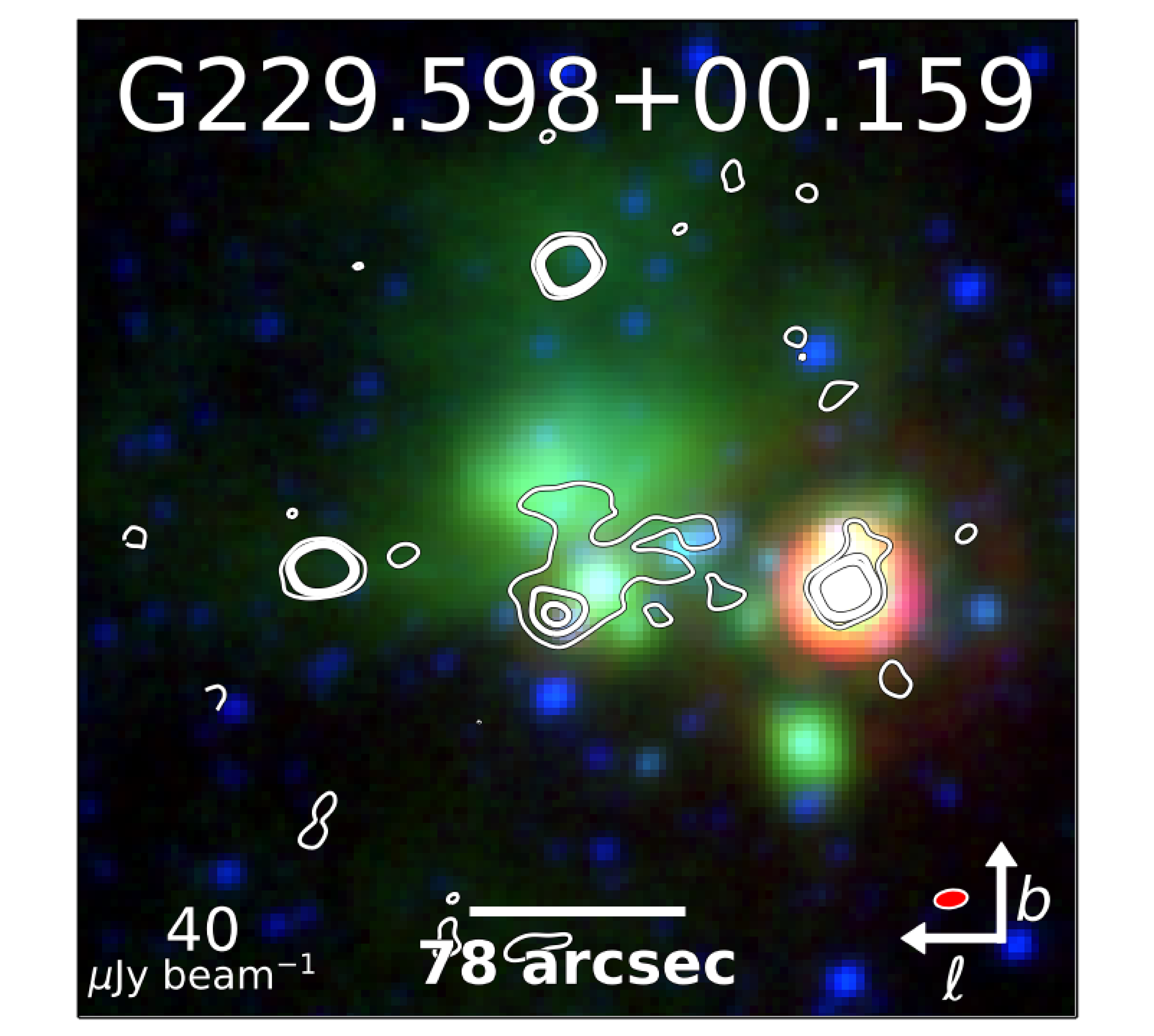}\\
\includegraphics[width=\figSize]{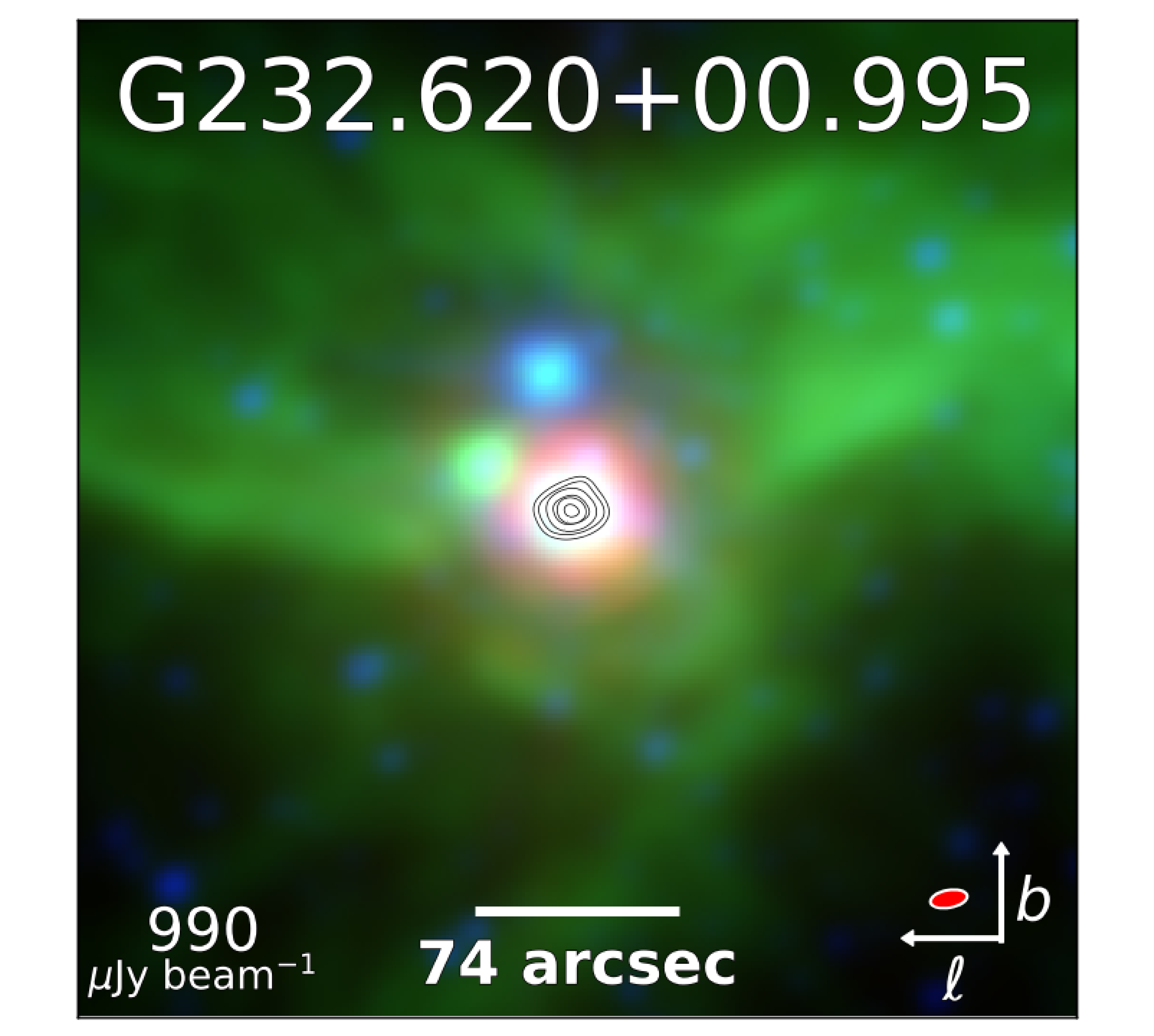}
\includegraphics[width=\figSize]{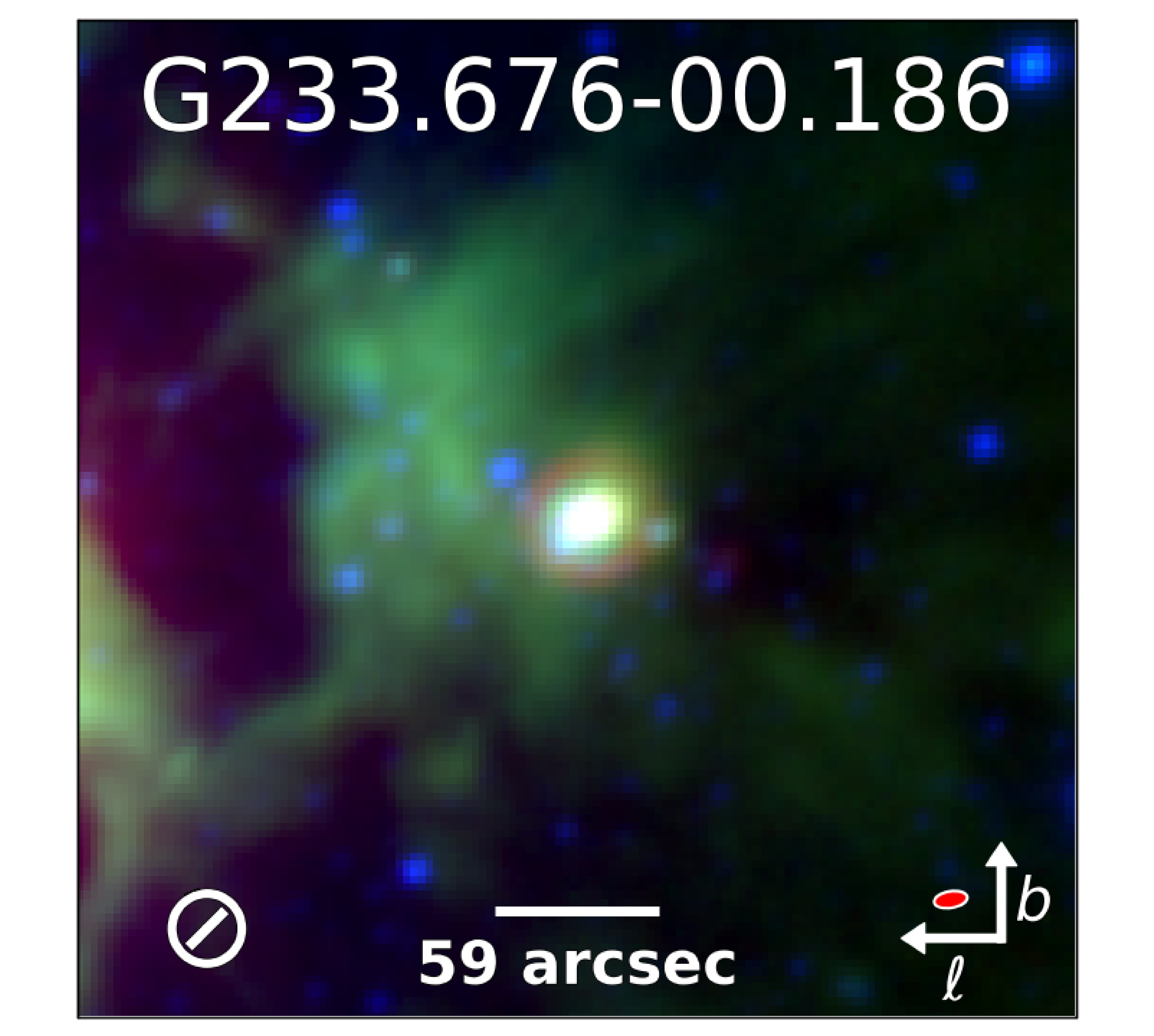}
\includegraphics[width=\figSize]{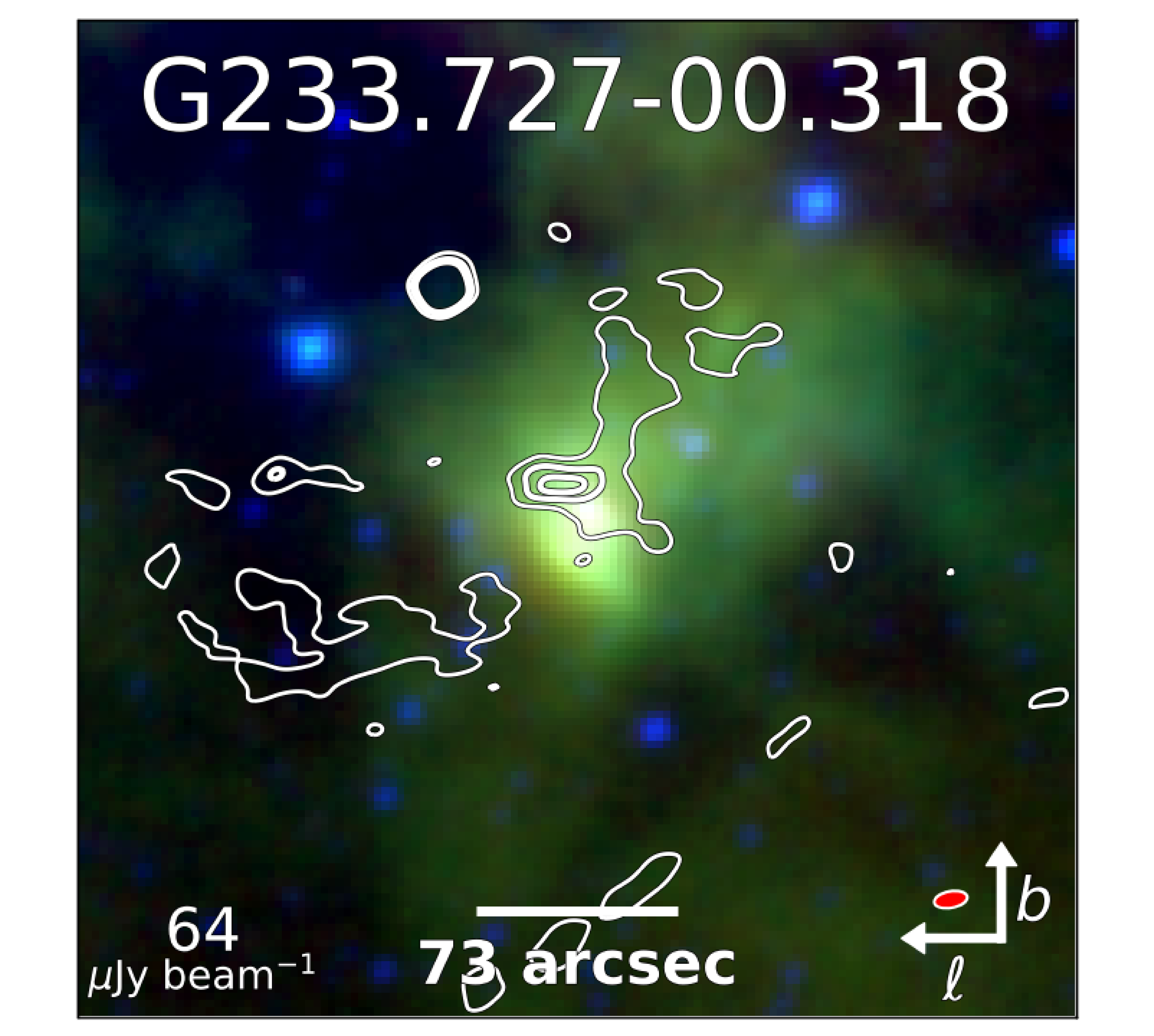}
\end{figure*}
\begin{figure*}[!htb]
\includegraphics[width=\figSize]{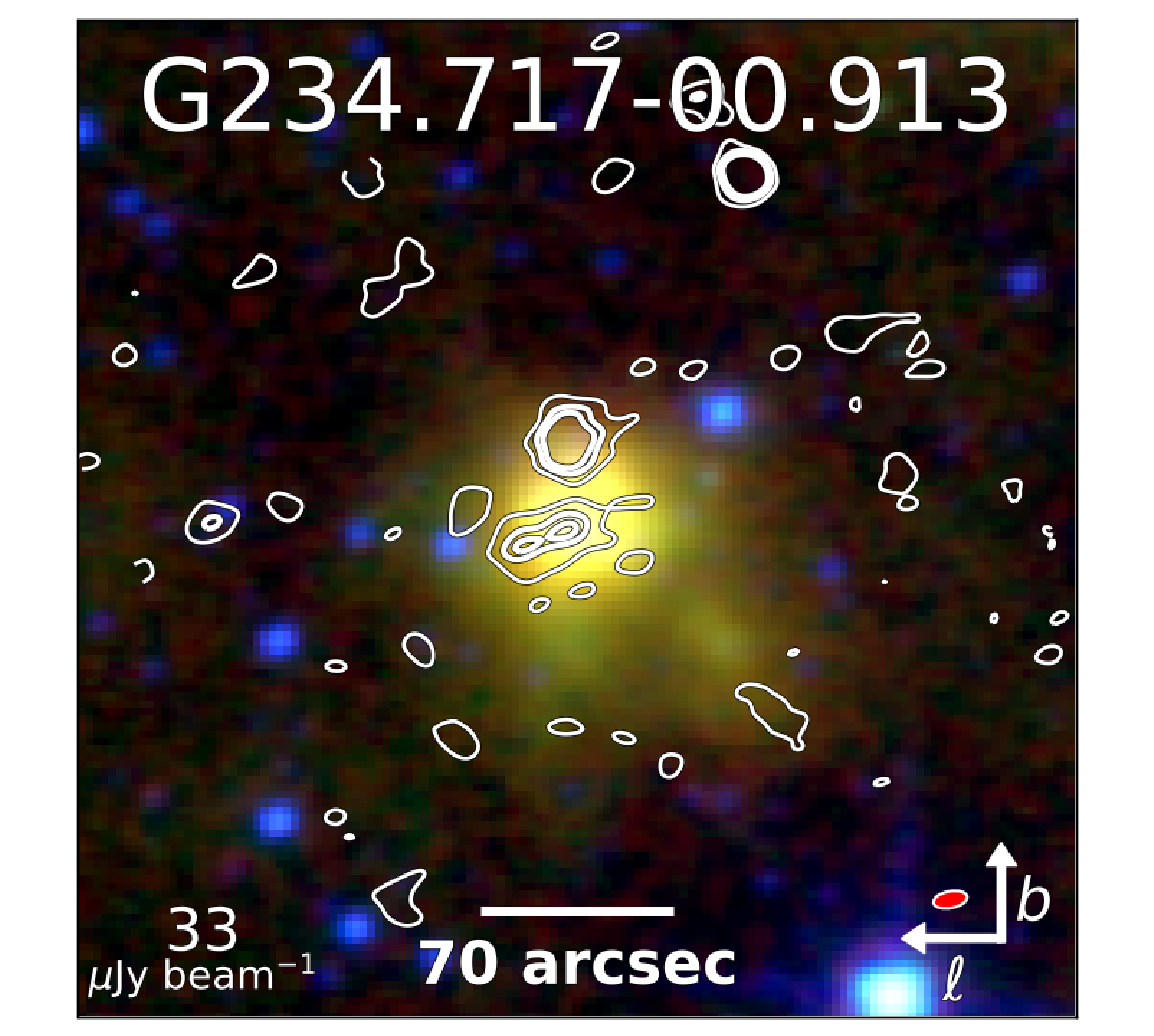}
\includegraphics[width=\figSize]{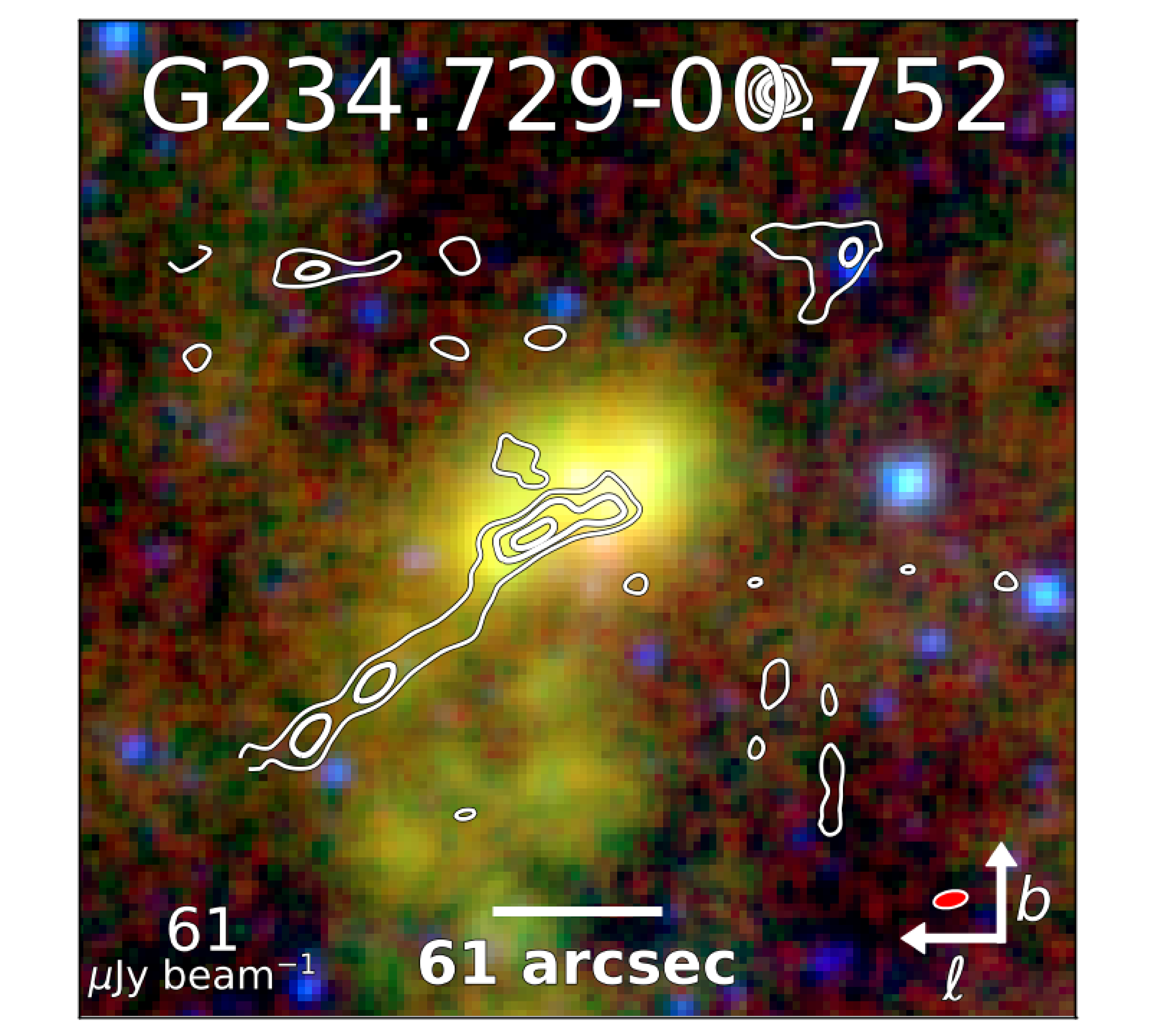}
\includegraphics[width=\figSize]{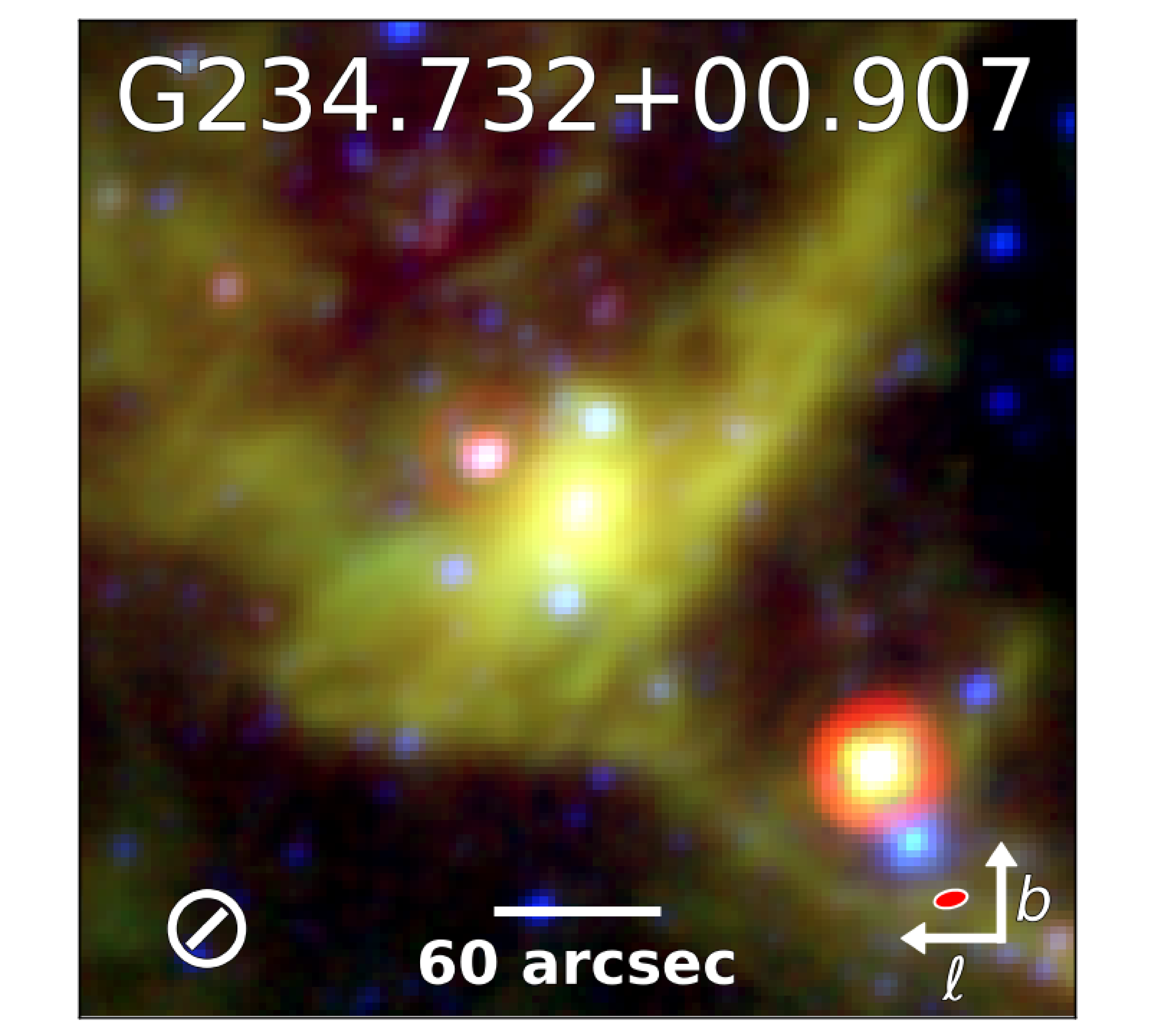}\\
\includegraphics[width=\figSize]{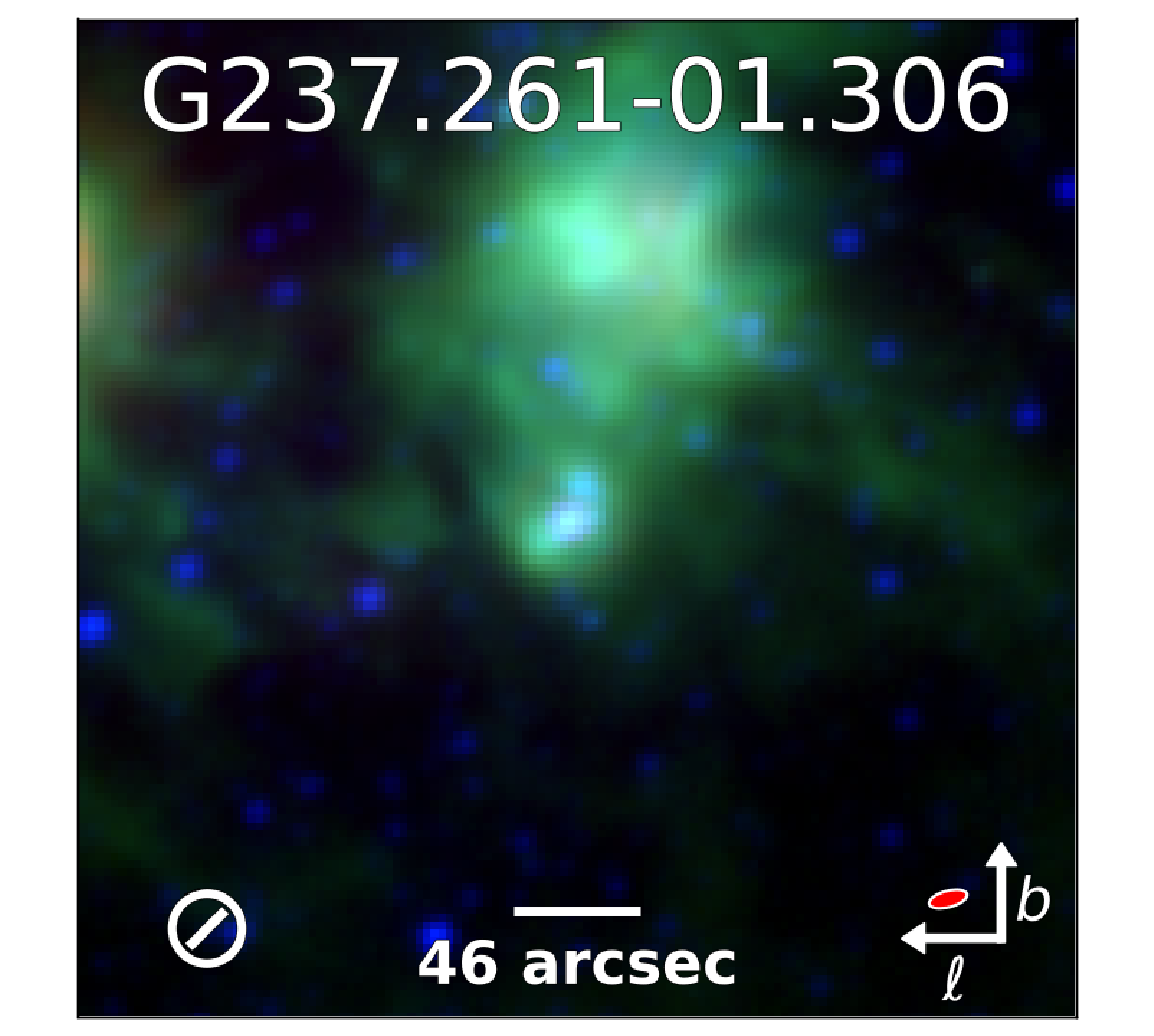}
\includegraphics[width=\figSize]{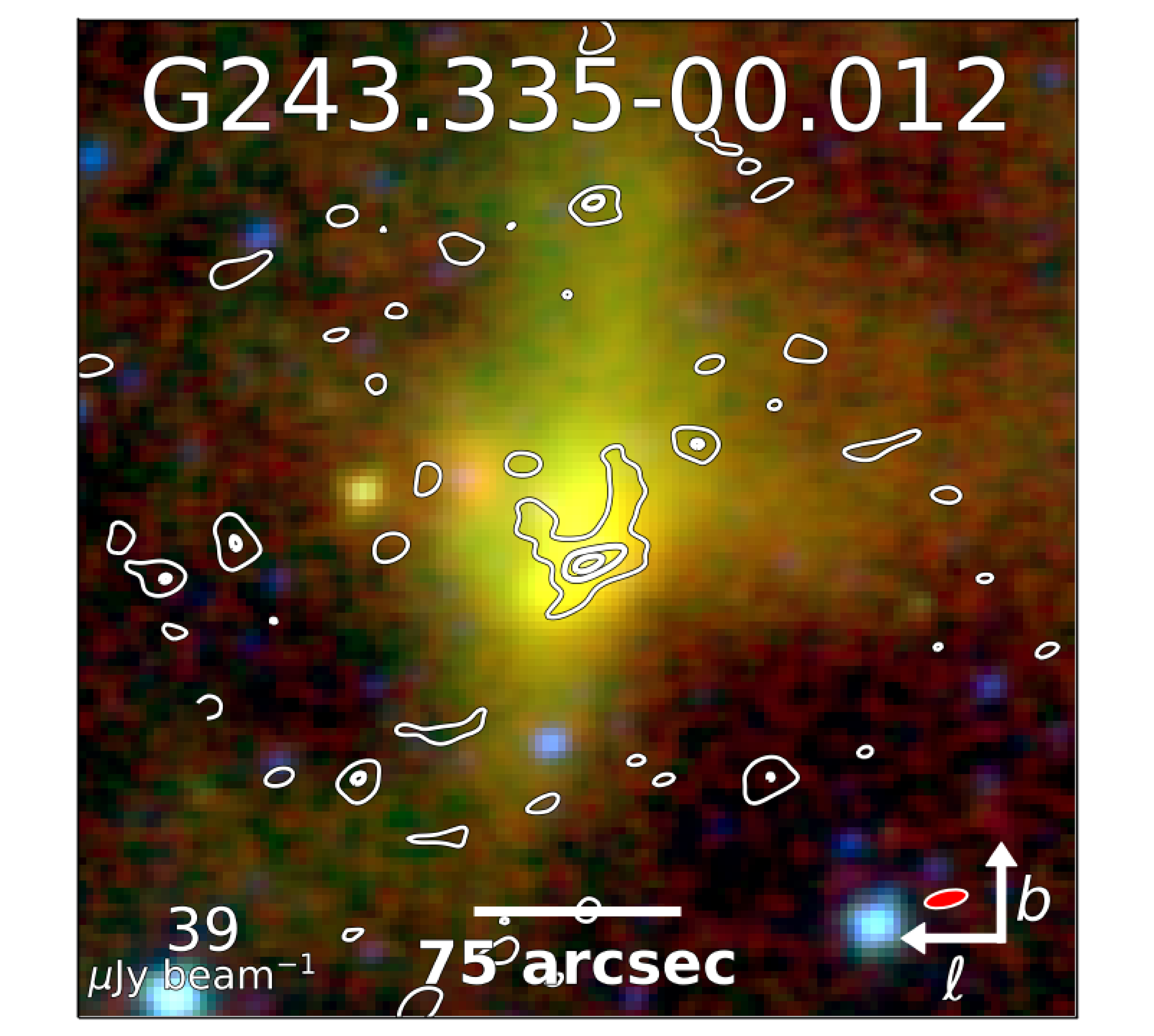}
\end{figure*}
\fi

\end{document}